\documentstyle[twoside,epsf]{article}

\catcode`\@=11
\long\def\@makefntext#1{
\protect\noindent \hbox to 3.2pt {\hskip-.9pt  
$^{{\eightrm\@thefnmark}}$\hfil}#1\hfill}		

\def\thefootnote{\fnsymbol{footnote}}
\def\@makefnmark{\hbox to 0pt{$^{\@thefnmark}$\hss}}	
	
\def\ps@myheadings{\let\@mkboth\@gobbletwo
\def\@oddhead{\hbox{}
\rightmark\hfil\eightrm\thepage}   
\def\@oddfoot{}\def\@evenhead{\eightrm\thepage\hfil
\leftmark\hbox{}}\def\@evenfoot{}
\def\sectionmark##1{}\def\subsectionmark##1{}}

\oddsidemargin=\evensidemargin
\renewcommand{\thefootnote}{\fnsymbol{footnote}}

\newcounter{sectionc}\newcounter{subsectionc}\newcounter{subsubsectionc}
\renewcommand{\section}[1] {\vspace{12pt}\addtocounter{sectionc}{1} 
\setcounter{subsectionc}{0}\setcounter{subsubsectionc}{0}\noindent 
	{\tenbf\thesectionc. #1}\par\vspace{5pt}}
\renewcommand{\subsection}[1] {\vspace{12pt}\addtocounter{subsectionc}{1} 
	\setcounter{subsubsectionc}{0}\noindent 
	{\bf\thesectionc.\thesubsectionc. {\kern1pt \bfit #1}}\par\vspace{5pt}}
\renewcommand{\subsubsection}[1] {\vspace{12pt}\addtocounter{subsubsectionc}{1}
	\noindent{\tenrm\thesectionc.\thesubsectionc.\thesubsubsectionc.
	{\kern1pt \tenit #1}}\par\vspace{5pt}}
\newcommand{\nonumsection}[1] {\vspace{12pt}\noindent{\tenbf #1}
	\par\vspace{5pt}}

\newcounter{appendixc}
\newcounter{subappendixc}[appendixc]
\newcounter{subsubappendixc}[subappendixc]
\renewcommand{\thesubappendixc}{\Alph{appendixc}.\arabic{subappendixc}}
\renewcommand{\thesubsubappendixc}
	{\Alph{appendixc}.\arabic{subappendixc}.\arabic{subsubappendixc}}

\renewcommand{\appendix}[1] {\vspace{12pt}
        \refstepcounter{appendixc}
        \setcounter{figure}{0}
        \setcounter{table}{0}
        \setcounter{lemma}{0}
        \setcounter{theorem}{0}
        \setcounter{corollary}{0}
        \setcounter{definition}{0}
        \setcounter{equation}{0}
        \renewcommand{\thefigure}{\Alph{appendixc}.\arabic{figure}}
        \renewcommand{\thetable}{\Alph{appendixc}.\arabic{table}}
        \renewcommand{\theappendixc}{\Alph{appendixc}}
        \renewcommand{\thelemma}{\Alph{appendixc}.\arabic{lemma}}
        \renewcommand{\thetheorem}{\Alph{appendixc}.\arabic{theorem}}
        \renewcommand{\thedefinition}{\Alph{appendixc}.\arabic{definition}}
        \renewcommand{\thecorollary}{\Alph{appendixc}.\arabic{corollary}}
        \renewcommand{\theequation}{\Alph{appendixc}.\arabic{equation}}
        \noindent{\tenbf Appendix \theappendixc #1}\par\vspace{5pt}}
\newcommand{\subappendix}[1] {\vspace{12pt}
        \refstepcounter{subappendixc}
        \noindent{\bf Appendix \thesubappendixc. {\kern1pt \bfit #1}}
	\par\vspace{5pt}}
\newcommand{\subsubappendix}[1] {\vspace{12pt}
        \refstepcounter{subsubappendixc}
        \noindent{\rm Appendix \thesubsubappendixc. {\kern1pt \tenit #1}}
	\par\vspace{5pt}}

\newcommand{\textlineskip}{\baselineskip=13pt}
\newcommand{\smalllineskip}{\baselineskip=10pt}

\def\eightcirc{
\begin{picture}(0,0)
\put(4.4,1.8){\circle{6.5}}
\end{picture}}
\def\eightcopyright{\eightcirc\kern2.7pt\hbox{\eightrm c}} 

\newcommand{\copyrightheading}[1]
	{\vspace*{-2.5cm}\smalllineskip{\flushleft
	{\footnotesize International Journal of Modern Physics A, #1}\\
	{\footnotesize $\eightcopyright$\, World Scientific Publishing
	 Company}\\
	 }}

\newcommand{\publisher}[2]{{\begin{center}\footnotesize\smalllineskip 
	Received #1\\
	Revised #2
	\end{center}
	}}

\def\abstracts#1#2#3{{
	\centering{\begin{minipage}{4.5in}\baselineskip=10pt\footnotesize
	\parindent=0pt #1\par 
	\parindent=15pt #2\par
	\parindent=15pt #3
	\end{minipage}}\par}} 

\newcommand{\bibit}{\nineit}
\newcommand{\bibbf}{\ninebf}
\renewenvironment{thebibliography}[1]
	{\frenchspacing
	 \ninerm\baselineskip=11pt
	 \begin{list}{\arabic{enumi}.}
	{\usecounter{enumi}\setlength{\parsep}{0pt}
	 \setlength{\leftmargin 12.7pt}{\rightmargin 0pt} 
	 \setlength{\itemsep}{0pt} \settowidth
	{\labelwidth}{#1.}\sloppy}}{\end{list}}

\newcounter{itemlistc}
\newcounter{romanlistc}
\newcounter{alphlistc}
\newcounter{arabiclistc}

\newcommand{\fcaption}[1]{
        \refstepcounter{figure}
        \setbox\@tempboxa = \hbox{\footnotesize Fig.~\thefigure. #1}
        \ifdim \wd\@tempboxa > 5in
           {\begin{center}
        \parbox{5in}{\footnotesize\smalllineskip Fig.~\thefigure. #1}
            \end{center}}
        \else
             {\begin{center}
             {\footnotesize Fig.~\thefigure. #1}
              \end{center}}
        \fi}

\newcommand{\tcaption}[1]{
        \refstepcounter{table}
        \setbox\@tempboxa = \hbox{\footnotesize Table~\thetable. #1}
        \ifdim \wd\@tempboxa > 5in
           {\begin{center}
        \parbox{5in}{\footnotesize\smalllineskip Table~\thetable. #1}
            \end{center}}
        \else
             {\begin{center}
             {\footnotesize Table~\thetable. #1}
              \end{center}}
        \fi}

\def\@citex[#1]#2{\if@filesw\immediate\write\@auxout
	{\string\citation{#2}}\fi
\def\@citea{}\@cite{\@for\@citeb:=#2\do
	{\@citea\def\@citea{,}\@ifundefined
	{b@\@citeb}{{\bf ?}\@warning
	{Citation `\@citeb' on page \thepage \space undefined}}
	{\csname b@\@citeb\endcsname}}}{#1}}

\newif\if@cghi
\def\cite{\@cghitrue\@ifnextchar [{\@tempswatrue
	\@citex}{\@tempswafalse\@citex[]}}
\def\citelow{\@cghifalse\@ifnextchar [{\@tempswatrue
	\@citex}{\@tempswafalse\@citex[]}}
\def\@cite#1#2{{$\null^{#1}$\if@tempswa\typeout
	{IJCGA warning: optional citation argument 
	ignored: `#2'} \fi}}

\def\pmb#1{\setbox0=\hbox{#1}
	\kern-.025em\copy0\kern-\wd0
	\kern.05em\copy0\kern-\wd0
	\kern-.025em\raise.0433em\box0}

\def\fnm#1{$^{\mbox{\scriptsize #1}}$}
\def\fnt#1#2{\footnotetext{\kern-.3em
	{$^{\mbox{\scriptsize #1}}$}{#2}}}

\def\fpage#1{\begingroup
\voffset=.3in
\thispagestyle{empty}\begin{table}[b]\centerline{\footnotesize #1}
	\end{table}\endgroup}

\def\runninghead#1#2{\pagestyle{myheadings}
\markboth{{\protect\footnotesize\it{\quad #1}}\hfill}
{\hfill{\protect\footnotesize\it{#2\quad}}}}
\font\tenrm=cmr10
\font\tenit=cmti10 
\font\tenbf=cmbx10
\font\bfit=cmbxti10 at 10pt
\font\ninerm=cmr9
\font\nineit=cmti9
\font\ninebf=cmbx9
\font\eightrm=cmr8

\def\mevc {\ifmmode {\rm MeV}/c \else MeV$/c$\fi}
\def\mevcc {\ifmmode {\rm MeV}/c^2 \else MeV$/c^2$\fi}
\def\gevc {\ifmmode {\rm GeV}/c \else GeV$/c$\fi}
\def\gevcc {\ifmmode {\rm GeV}/c^2 \else GeV$/c^2$\fi}
\def\ra   {\rightarrow}
\newcommand{\Pt} {\ifmmode p_{\rm t} \else $p_{\rm t}$\fi}
\newcommand{\Et} {\ifmmode E_{\rm t} \else $E_{\rm t}$\fi}
\newcommand{\Ds} {\ifmmode D_{\mbox{\sl s}}^{-}
                       \else $D_{\mbox{\sl s}}^{-}$\fi}
\newcommand{\Bs} {\ifmmode B_{\mbox{\sl s}}^{0}
                       \else $B_{\mbox{\sl s}}^{0}$\fi}
\newcommand{\Dsl} {\ifmmode D_{\mbox{\sl s}}^{-} \ell^+
                       \else $D_{\mbox{\sl s}}^{-} \ell^+$\fi}
\newcommand{\Bsh} {\ifmmode B_{\mbox{\sl s}}^H
                       \else $B_{\mbox{\sl s}}^H$\fi}
\newcommand{\Bsl} {\ifmmode B_{\mbox{\sl s}}^L
                       \else $B_{\mbox{\sl s}}^L$\fi}
\newcommand{\Bc} {\ifmmode B_{c}^{+}
                       \else $B_{c}^{+}$\fi}
\newcommand{\lxy} {\ifmmode L_{\rm xy} \else $L_{\rm xy}$\fi}
\newcommand{\dgam} {\ifmmode \Delta\Gamma \else $\Delta\Gamma$\fi}
\newcommand{\dgog} {\ifmmode \Delta\Gamma/\Gamma \else 
                            $\Delta\Gamma/\Gamma$\fi}
\newcommand{\phipi} {\ifmmode \phi \pi^-
                       \else $\phi \pi^-$\fi}
\newcommand{\kstark} {\ifmmode K^{*0} K^-
                       \else  $K^{*0} K^-$\fi}
\newcommand{\kstarpi} {\ifmmode K^{*0} \pi^-
                       \else  $K^{*0} \pi^-$\fi}
\newcommand{\ksk} {\ifmmode K^0_S K^-
                      \else $K^0_S K^-$\fi}
\newcommand{\phil} {\ifmmode \phi \mu^- \bar \nu
                      \else $\phi \mu^- \bar \nu$\fi}
\newcommand{\Dsmu} {\ifmmode D_{\mbox{\sl s}}^{-} \mu^+ 
                       \else$D_{\mbox{\sl s}}^{-} \mu^+$\fi}
\newcommand{\dm} {\ifmmode \Delta m \else $\Delta m$\fi}
\newcommand{\dmd} {\ifmmode \Delta m_d \else $\Delta m_d$\fi}
\newcommand{\dms} {\ifmmode \Delta m_{\mbox{\sl s}} \else 
			   $\Delta m_{\mbox{\sl s}}$\fi}
\newcommand{\eD} {\ifmmode \varepsilon{\cal D}^2 \else 
			  $\varepsilon{\cal D}^2$\fi}
\newcommand{\ptrel} {\ifmmode p_{\rm t}^{\rm rel} \else 
			     $p_{\rm t}^{\rm rel}$\fi}
\newcommand{\jpks} {\ifmmode J/\psi K^0_S \else 
			    $J/\psi K^0_S$\fi}
\newcommand{\stb} {\ifmmode \sin 2\beta \else 
			   $\sin 2\beta$\fi}
\newcommand{\sta} {\ifmmode \sin 2\alpha \else 
			   $\sin 2\alpha$\fi}

\renewcommand{\thefootnote}{\fnsymbol{footnote}}	

\begin{document}

\begin{titlepage}

\pagestyle{empty}
\begin{flushright}
CDF/PUB/BOTTOM/PUBLIC/4843 \\
FERMILAB-PUB-99/014-E \\
Version 2.0 \\
\today 
\end{flushright}
\vskip 2.0cm

\begin{boldmath}
\begin{center}
{\large\bf $B$ Lifetimes, Mixing and $CP$ Violation at CDF}
\end{center}
\end{boldmath}
\vskip 1.0cm

\begin{center}
Manfred Paulini\\
{\sl Lawrence Berkeley National Laboratory, Berkeley, CA 94720, USA}
\end{center}

\vspace{2.5cm}

\begin{abstract}
We review the status of bottom quark physics at the CDF
experiment. The measurements reported are based on about
110~pb$^{-1}$ of data collected  
at the Fermilab Tevatron $p\bar p$ Collider
operating at $\sqrt{s} = 1.8$~TeV. In particular, we review results on
$B$ hadron lifetimes, measurements of the time dependence of
$B^0\bar B^0$ oscillations, and a search for  
$CP$~violation in $B^0 \ra \jpks$ decays.
Prospects for future $B$~physics at CDF in the next
run of the Tevatron Collider starting in the
year 2000 are also given.
\end{abstract}

\vspace{3.5cm}

\centerline{(To appear in International Journal of Modern Physics A)}

\end{titlepage}

\thispagestyle{empty}
\vbox{}
\newpage

\runninghead{$B$ Lifetimes, Mixing and $CP$ Violation at CDF}
{$B$ Lifetimes, Mixing and $CP$ Violation at CDF}

\normalsize\textlineskip
\thispagestyle{empty}
\setcounter{page}{1}

\copyrightheading{}			

\vspace*{0.88truein}

\fpage{1}
\centerline{\bf \boldmath{$B$} LIFETIMES, MIXING AND 
	    \boldmath{$CP$} VIOLATION AT CDF}     
\vspace*{0.37truein}
\centerline{\footnotesize MANFRED PAULINI}
\vspace*{0.015truein}
\centerline{\footnotesize\it Lawrence Berkeley National Laboratory} 
\baselineskip=10pt
\centerline{\footnotesize\it Berkeley, California 94720, USA}
\publisher{(received date)}{(revised date)}

\vspace*{0.21truein}
\abstracts{
We review the status of bottom quark physics at the CDF
experiment. The measurements reported are based on about
110~pb$^{-1}$ of data collected  
at the Fermilab Tevatron $p\bar p$ Collider
operating at $\sqrt{s} = 1.8$~TeV. In particular, we review results on
$B$ hadron lifetimes, measurements of the time dependence of
$B^0\bar B^0$ oscillations, and a search for  
$CP$~violation in $B^0 \ra \jpks$ decays.
Prospects for future $B$~physics at CDF in the next
run of the Tevatron Collider starting in the
year 2000 are also given.
}{}{}




\setlength{\unitlength}{1.0mm}
\textheight=7.8truein
\setcounter{footnote}{0}
\renewcommand{\thefootnote}{\alph{footnote}}

\vspace*{1pt}\textlineskip	
\section{Introduction}
\vspace*{-0.5pt}
\runninghead{$B$ Lifetimes, Mixing and $CP$ Violation at CDF}
{Introduction}
\noindent
In 1977, the bottom quark was discovered as a resonance in the dimuon
invariant mass spectrum in 400 GeV proton-nucleus collisions at 
Fermilab\cite{yps}. Soon after the discovery of this new $b\bar b$ bound state 
with a mass of about 9.5~\gevcc, the $\Upsilon$~resonances 
were confirmed in $e^+e^-$ collisions at the
DORIS storage ring at DESY\cite{PlutoDasp}. Today, more than 20 years
later, all lowest mass bound states containing a $b$ quark have been
discovered and are 
experimentally well established. The pseudoscalar
$B$~meson states\fnm{a}\fnt{a}{Throughout the paper, unless otherwise
noted, references to a specific   
charge state are meant to imply the charge-conjugate state as well.} are   
$B^0 = |\,\bar b\, d\,\rangle$, 
$B^+ = |\,\bar b\, u\,\rangle$, 
$\Bs = |\,\bar b\, s\,\rangle$, and
$B^+_c = |\,\bar b\, c\,\rangle$, while the 
$\Lambda^0_b = |\,b\, d\, u\,\rangle$ is the $b$~baryon ground state
with lowest mass.  
The rest masses of the $B$ hadrons\cite{PDG} are between 5.3 \gevcc\ and 6.4
\gevcc, approximately six times the mass of a proton.
The lowest lying $B$~hadrons decay via the weak interaction. 

$B$~hadrons play a
special role among hadrons. The heaviest quark, the top
quark, decays weakly into a real $W$~boson and a $b$~quark before it
is able to form a meson with another antiquark through the strong
interaction. Therefore, hadrons containing a $b$ quark are the heaviest hadrons
experimentally accessible.
The principal interest in studying $B$ hadrons in the context of the Standard
Model\cite{GSW} arises from the fact that $B$ hadron decays provide
valuable information 
on the weak mixing matrix, the Cabibbo-Kobayashi-Maskawa (CKM)
matrix\cite{ckm1,ckm2}. 
In fact, $B$~decays measure five of the nine CKM matrix
elements: $V_{cb}$, $V_{ub}$, $V_{td}$, $V_{ts}$, and $V_{tb}$. The future
interest in $B$ physics certainly lies in the study of $CP$ violation in the
system of $B$~mesons.

Traditionally, $B$ physics has been the domain of $e^+ e^-$ machines
operating on the $\Upsilon(4S)$ resonance or the $Z^0$ pole.
But the UA\,1 collaboration has already shown that $B$ physics
is feasible in a hadron collider environment (for a review see
Ref.\cite{bfeasi}). Although $B$~physics did not play a significant
role in the considerations for the original 1981 technical design report of the
Collider Detector at Fermilab (CDF), several features in the CDF
design are advantageous for the studies of $B$~decays. These features include
a large solenoidal magnetic tracking volume, a well segmented
calorimeter for the detection of electrons, and muon chambers that
allow low momentum muon detection. However, the device that made 
$B$~physics possible at CDF in a competitive way, allowing for a broad
$B$~physics program, is a silicon micro-vertex
detector, installed in 1992. 

There are several motivations for pursuing $B$~physics 
at the Fermilab Tevatron $p\bar p$ Collider
operating at $\sqrt{s} = 1.8$~TeV.
The primary reason is demonstrated by comparing the 
$B$~production cross section in $e^+e^-$ collisions, which is about 1~nb   
at the $\Upsilon(4S)$ and about 6~nb at the $Z^0$ pole, to 
the large $b$~quark production cross section at a hadron collider. 
At the Tevatron, $\sigma_{b}$ is 
$\sim\!50~\mu$b within the central detector region of rapidity less than one. 
This is a huge cross section which
results in about $5\cdot 10^9\ b\bar b$ pairs being produced in 
100~pb$^{-1}$ of data. However, the total
inelastic cross section at the Tevatron is still about three orders of
magnitude larger than the $b$~cross section. This puts 
certain requirements on the trigger system used to find $B$~decay products,
as will be further discussed later. 

In addition to high rates of $B$~hadrons, the Tevatron offers other
features 
worth noting. First, in contrast to an $e^+ e^-$
machine operating at the $\Upsilon(4S)$, all $B$~hadron species are
produced at a hadron collider. Second, the
transverse momentum spectrum for $B$ hadrons scales with the $B$~mass
and is significantly harder for heavy hadrons than that for light
hadrons. As a result,  
the $B$ hadrons are Lorentz-boosted at all rapidities, including the
central detector region where the production rate is the highest with
an average transverse $B$ momentum around 4-5~\gevc. Third, the hard
$B$~hadron momentum spectrum can be exploited to improve the
signal to background ratio in finding $B$~decay products. For all
momenta, $b$~production accounts 
for about 0.2\% of the total $p\bar p$ inelastic cross section, while
at high momenta the ratio of $b$~jet to inclusive jet production is
close to 2\%. Finally, the initial $p\bar p$~state is a $CP$~symmetric
state where we  
expect equal rates of $B$ and $\bar B$ hadrons to be produced, at
least in the central detector region.

In this article, we review
recent $B$ physics results at CDF 
concentrating on $B$~lifetime measurements, proper time dependent
measurements of $B^0\bar B^0$ oscillations, and the search for $CP$
violation in $B^0 \ra \jpks$. 
The measurements reported here are based on about
110~pb$^{-1}$ of data collected at the Fermilab Tevatron Collider.
Prospects for future
$B$~physics at CDF in the next
run of the Tevatron starting in the
year 2000 are also discussed.
The outline of this paper is as follows: In Section~2, we give a brief
overview of
heavy quark production in $p\bar p$ collisions. In Section~3, 
we describe the experimental environment including 
the Tevatron Collider, as well as the CDF detector. We focus on the 
collection of $B$~physics datasets at CDF in Sec.~4, 
emphasizing the CDF trigger scheme. In Section~5, we highlight several
features of $B$~physics in a hadron collider environment and
describe some of the ways $B$~decays
are studied at CDF. Section~6 is devoted to the measurements of 
$B$~hadron lifetimes, where the lifetimes of
all weakly decaying $B$~mesons as well as the $\Lambda^0_b$~baryon 
are measured at CDF. We then review several   
proper time dependent measurements of $B^0\bar B^0$ oscillations in
Section~7 and discuss various
$B$~tagging methods to identify the $B$~flavour in hadronic
collisions. The search 
for $CP$~violation using the current data set of $B^0 \ra \jpks$
decays is summarized in Section~8. An 
outlook for future $B$~physics at CDF in the next run
of the Tevatron Collider, starting in the 
year 2000, is given in Sec.~9. Finally, we offer our conclusions in Section~10.

\section{Heavy Quark Production in Hadronic Collisions}
\runninghead{$B$ Lifetimes, Mixing and $CP$ Violation at CDF}
{Heavy Quark Production in Hadronic Collisions}
\noindent
In this section, we give a short introduction to heavy quark production in
$p\bar p$ collisions. 
With the expression ``heavy quark'' we will mainly
refer to $b$~quarks.
The discussion will be rather qualitative and
cannot serve as a complete review of heavy quark production in
hadronic collisions. It is meant to give an idea of the main issues,
introducing some of the nomenclature often used in the literature. 
The interested reader is referred to reviews in 
references\cite{nason,frixione,mlm_lec}.
Here, we first discuss the parton model and parton
distribution functions. Then, we summarize heavy quark production in
lowest order QCD, as well as next-to-leading order QCD calculations and
discuss hadronization of heavy quarks. 
Finally, we briefly compare theoretical predictions of $b$~quark
production to measurements at~CDF.  

\subsection{Parton model and parton distribution functions}
\noindent
In a static picture, a proton is a bound state of three quarks
$|\,u\,u\,d\,\rangle$ with a radius of about 1~fm. However, in a hadronic
collision such as that found at the Tevatron, a proton is better
characterized as a beam 
of free partons: Three constituent quarks (valence quarks), virtual
gluons, and quark-antiquark pairs (sea quarks)\cite{nason}. The
different partons  
don't necessarily divide up the beam energy equally. The distribution of
partons within the proton is described by the so-called
parton distribution function (PDF) $F_i^a(x,Q^2)$ which is the
number density of
parton $i$ (quark or gluon) carrying the
momentum fraction $x$ of the hadron $a$ (proton or antiproton) when
probed at momentum transfer $Q^2$. 

Information on parton distributions comes from measurements
of deep inelastic lepton or nucleon scattering, such as $ep$, $eN$,
$\mu N$, or $\nu N$.  
Once parton distributions have been measured at some value of $Q^2$
(normally at low $Q^2$),
and the running coupling constant $\alpha_s(Q^2)$ of the strong
interaction has been determined, QCD permits to compute the parton
distributions at higher values of $Q^2$ based on the formalism provided by
the Altarelli-Parisi equations\cite{altpar} and the evolution
calculations pioneered by Dokshitzer, Gribov, and Lipatov\cite{dgl}.
Modern representations and predictions of parton distribution functions, used
for comparison with hadron collider data, come, for example, from the
Martin-Roberts-Stirling (MRS) group\cite{mrsd0,mrs} or from a collaboration of
theorists and experimentalists called the
CTEQ collaboration\cite{cteq}. 

\begin{figure}[tbp]
\centerline{
\put(42,14){\large\bf (a)}
\put(71,14){\large\bf (b)}
\epsfysize=5.3cm
\epsffile[65 160 545 633]{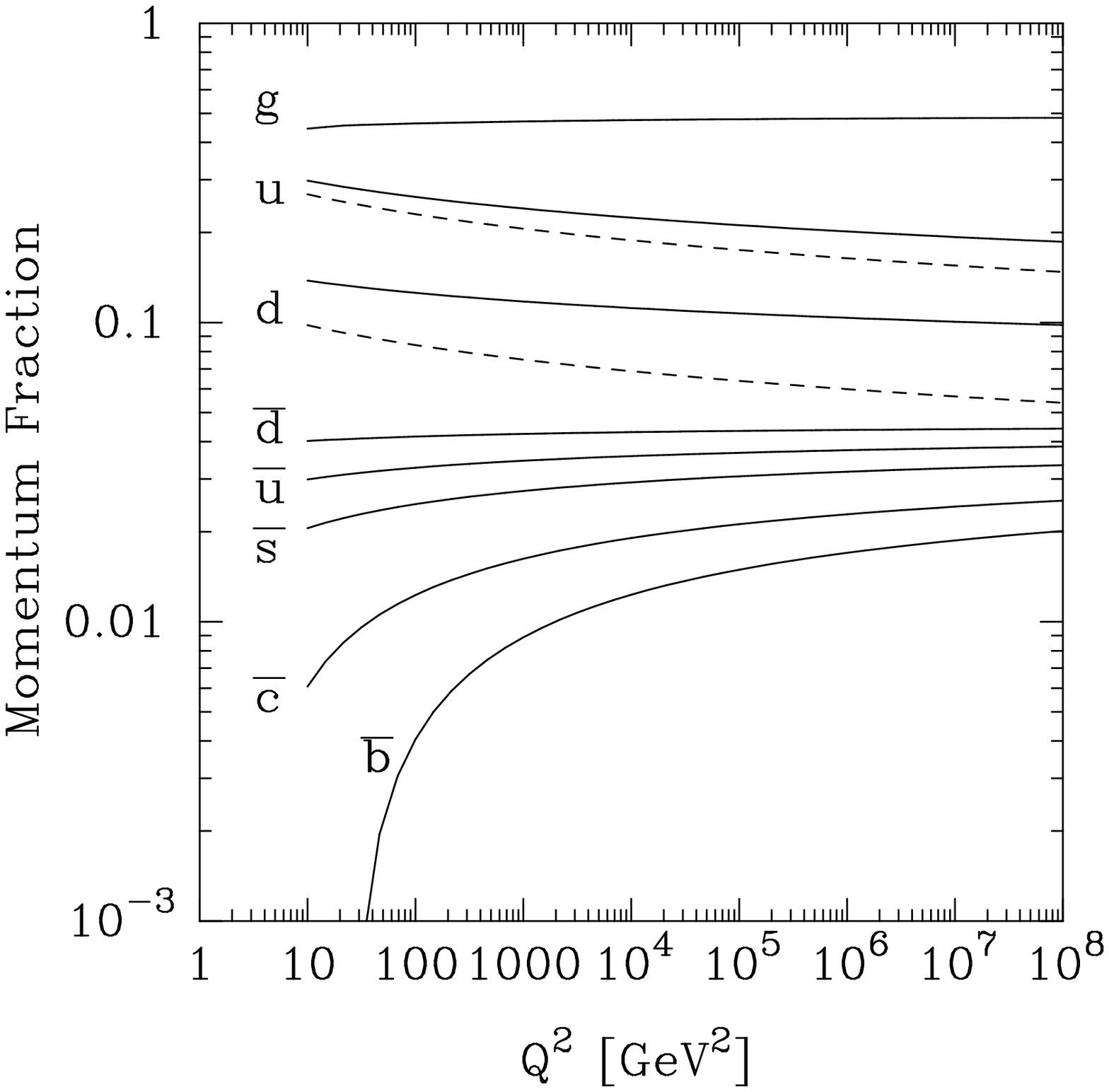}
\epsfysize=5.1cm
\epsffile[65 225 525 555]{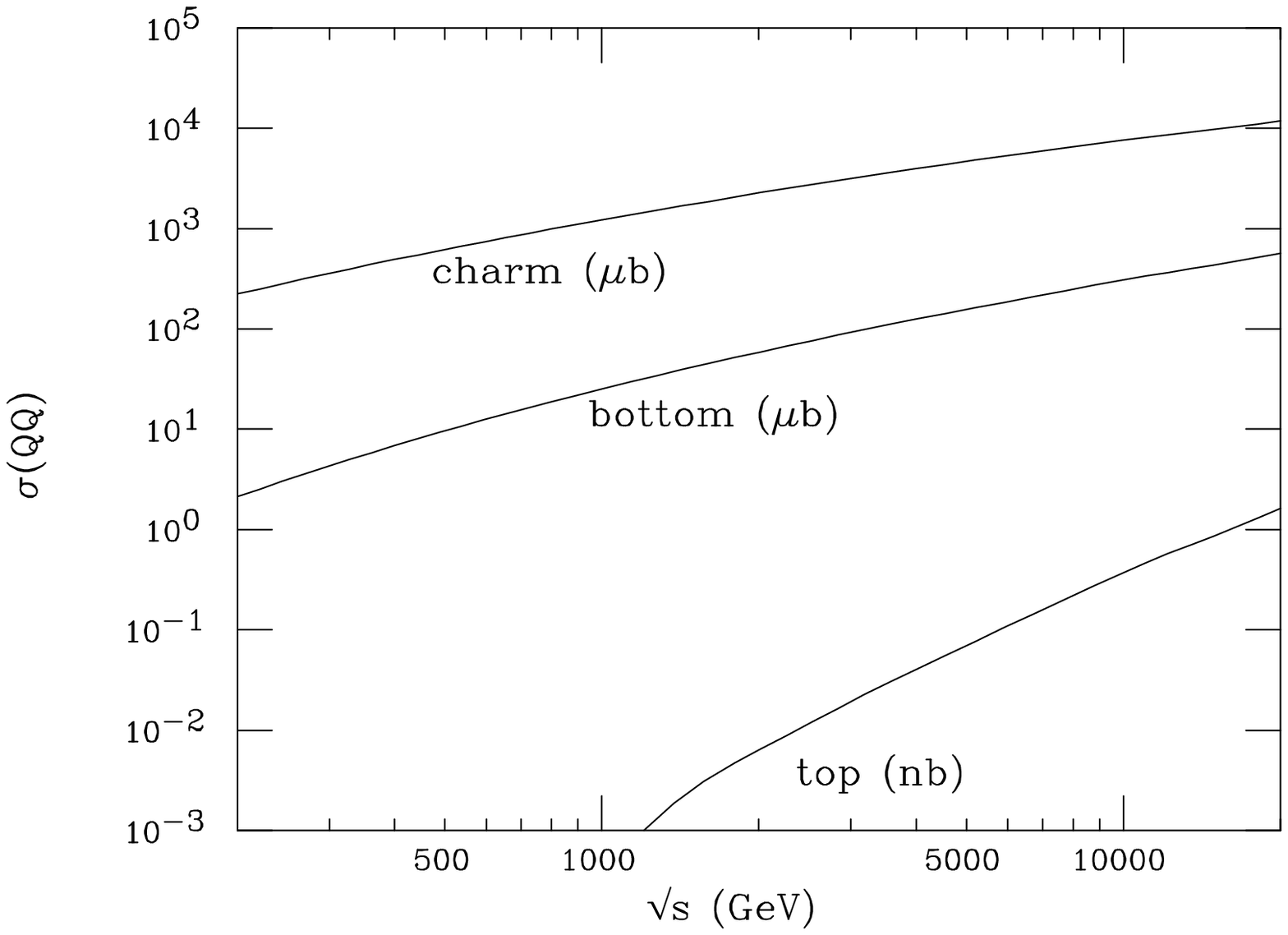}
}
\vspace*{0.2cm}
\fcaption{(a) $Q^2$ evolution of the momentum
fractions carried by the various parton species in the proton for the
CTEQ4 parton distributions\cite{cteq4}.
(b) Total production cross sections for charm, bottom, and top quark
pairs in $p\bar p$ collisions\cite{mlm_lec} 
as a function of the hadronic centre-of-mass energy~$\sqrt{s}$.}  
\label{pdfs}
\end{figure}

As an example, 
Figure~\ref{pdfs}a) illustrates the flavour content of the proton as
measured by the momentum fraction $\int_0^1 {\rm d}x\,xF_i(x,Q^2)$
carried by each parton species $i$, as obtained for the CTEQ4 parton
distributions\cite{cteq4}. Gluons carry about half of the proton's
momentum almost independently of $Q^2$, while the momentum is shared
more and more equally among the quark and antiquark flavours as 
$Q^2$ increases.

\subsection{QCD predictions for hard scattering in 
\boldmath{$p\bar p$} collisions}
\noindent
In a $p\bar p$ interaction at the Tevatron, heavy quarks $\cal Q$ are produced 
in the hard collision of two partons, one from each hadron. In terms of the
rapidity $y$, defined~as 
\begin{equation}
y = \frac{1}{2} \ln \left( \frac{E+p_z}{E-p_z} \right)
\end{equation}
and transverse momentum \Pt, the relativistically invariant phase
space volume element of the final state heavy quark with four-momentum
$(E,p_x,p_y,p_z)$ is 
\begin{equation}
\frac{{\rm d}^3 p}{E} = {\rm d}y\ {\rm d}^2 \Pt.
\end{equation}
The differential partonic cross section $\hat \sigma_{ij}$ per
invariant phase  
space volume for the production of a heavy quark in a given
parton-parton subprocess can be written as 
\begin{equation}
\frac{{\rm d}^3 \hat \sigma_{ij}}{{\rm d}^3 p/E} = 
\frac{E\,{\rm d}^3 \hat \sigma_{ij}}{{\rm d}^3 p}(x_i p_a, x_j p_b, p;
m_{\cal Q}, \Lambda, \mu_R, \mu_F), 
\end{equation}
where $x_i p_a$ and $x_j p_b$ are the momenta of the incoming partons
$i$ and $j$, $p$ is the momentum of the outgoing heavy quark, while $p_a$ and
$p_b$ are the momenta of the colliding hadrons. The remaining
variables are parameters of the theory:
$m_{\cal Q}$ is the mass of the heavy quark, 
$\Lambda$ determines the coupling strength of the
strong interaction expressed through $\alpha_s$,
$\mu_R$ is the renormalization scale, which is related to the energy scale 
used in the evaluation of $\alpha_s$, and $\mu_F$ is the factorization
scale used in the evolution of the parton densities.
In QCD, the partonic
cross section can 
be expressed as a perturbative expansion in $\alpha_s$, provided
$m_{\cal Q}$ is sufficiently large. 

The cross section for heavy quark production in the collision of two
hadrons $(a + b \ra {\cal Q} + X)$ is then obtained by convoluting the
partonic cross section above with parton distribution functions
$F_i(x,Q^2)$, where it is common to equate the momentum transfer 
$Q^2 = m_{\cal Q}^2 + \Pt^2$ with $\mu^2 = \mu_R^2 = \mu_F^2$. 
Note, the uncertainty on the choice of the $\mu$-scale,
which is not extractable from the data,
is one of the large sources of uncertainties in QCD predictions of
heavy quark production. The perturbative QCD formula for the inclusive
production of a heavy quark in a hadron-hadron collision is then given
as
\begin{equation}
\frac{E\,{\rm d}^3 \sigma}{{\rm d}^3 p} = 
\sum_{ij} \int {\rm d}x_i\, {\rm d}x_j \left( 
\frac{E\,{\rm d}^3 \hat \sigma_{ij}}{{\rm d}^3 p}(x_i p_a, x_j p_b, p;
m_{\cal Q}, \Lambda, \mu)\right)
F_i^a(x_i,\mu^2) F_j^b(x_j,\mu^2). 
\label{eq:hqxsec}
\end{equation}
The corrections to Eq.~(\ref{eq:hqxsec}) are suppressed by powers of
the heavy quark mass. Finally, the total cross section for the production of a
heavy quark is obtained by integrating
Eq.~(\ref{eq:hqxsec}) over momentum $p$.
As a further illustration, 
Figure~\ref{pdfs}b) compares the total production cross sections for
charm, bottom, and top quark pairs in $p\bar p$~collisions\cite{mlm_lec} as a
function of the hadronic centre-of-mass energy $\sqrt{s}$.
Note the different units used for top quarks compared to charm and bottom.

\subsubsection{$b\bar b$ production in leading order QCD} 
\begin{figure}[tbp]
\centerline{
\hspace*{0.5cm}
\put(-10,16){\large\bf (a)}
\put(47,16){\large\bf (b)}
\epsfxsize=4.0cm
\epsffile{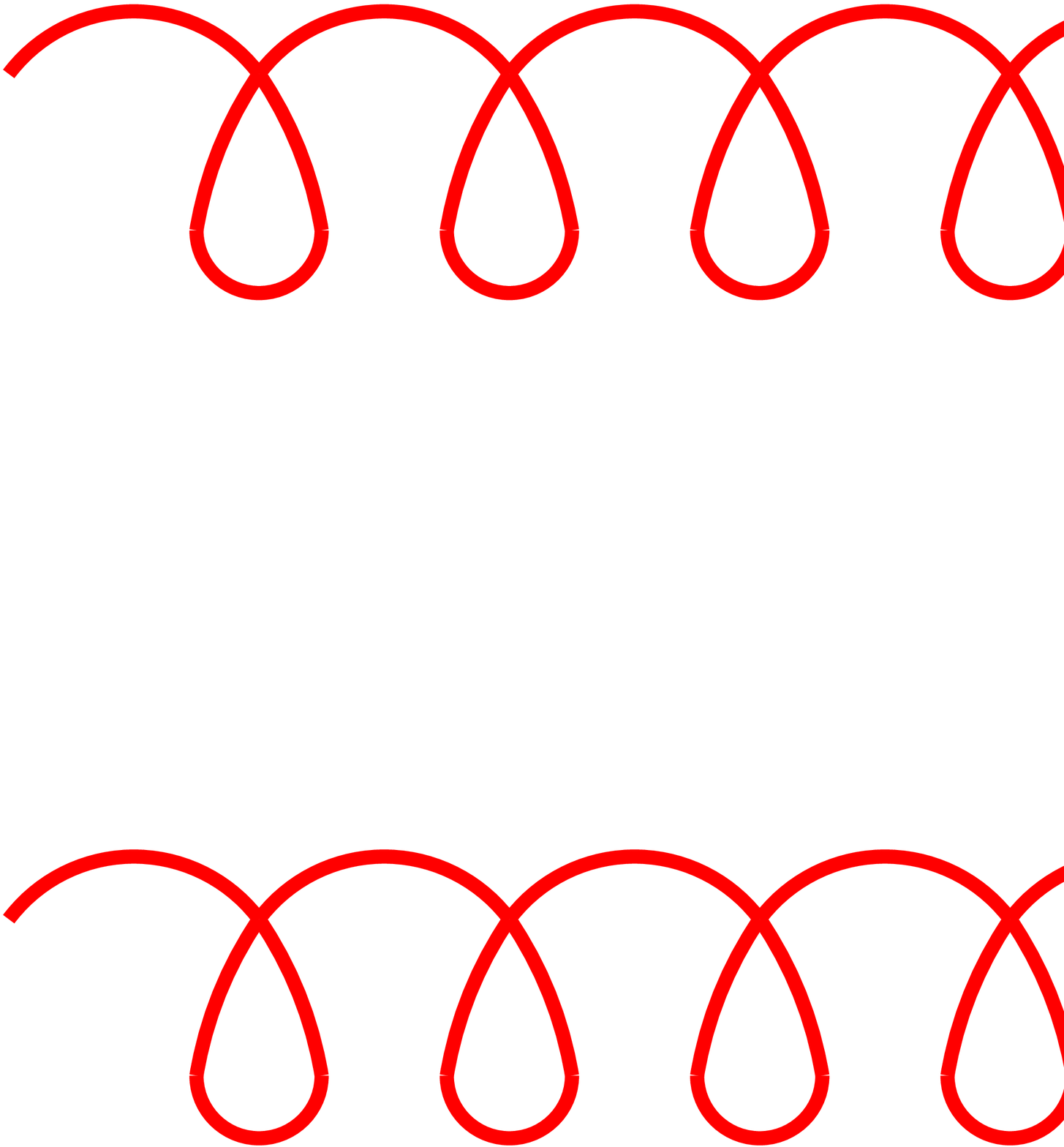}
\hspace*{1.6cm}
\epsfxsize=4.0cm
\epsffile{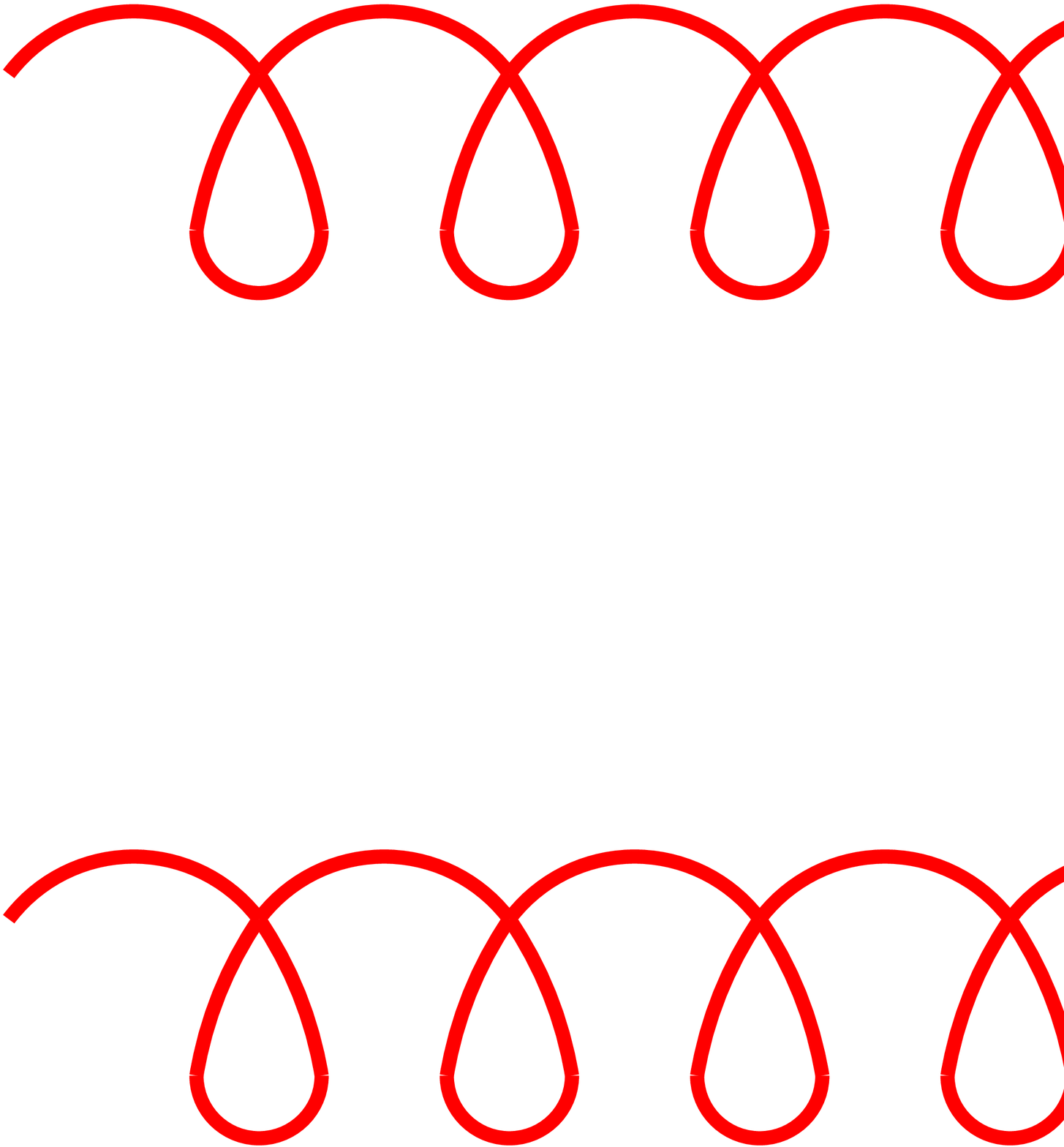}
}
\vspace*{0.5cm}
\centerline{
\hspace*{0.5cm}
\put(-10,20){\large\bf (c)}
\put(47,20){\large\bf (d)}
\epsfxsize=4.0cm
\epsffile{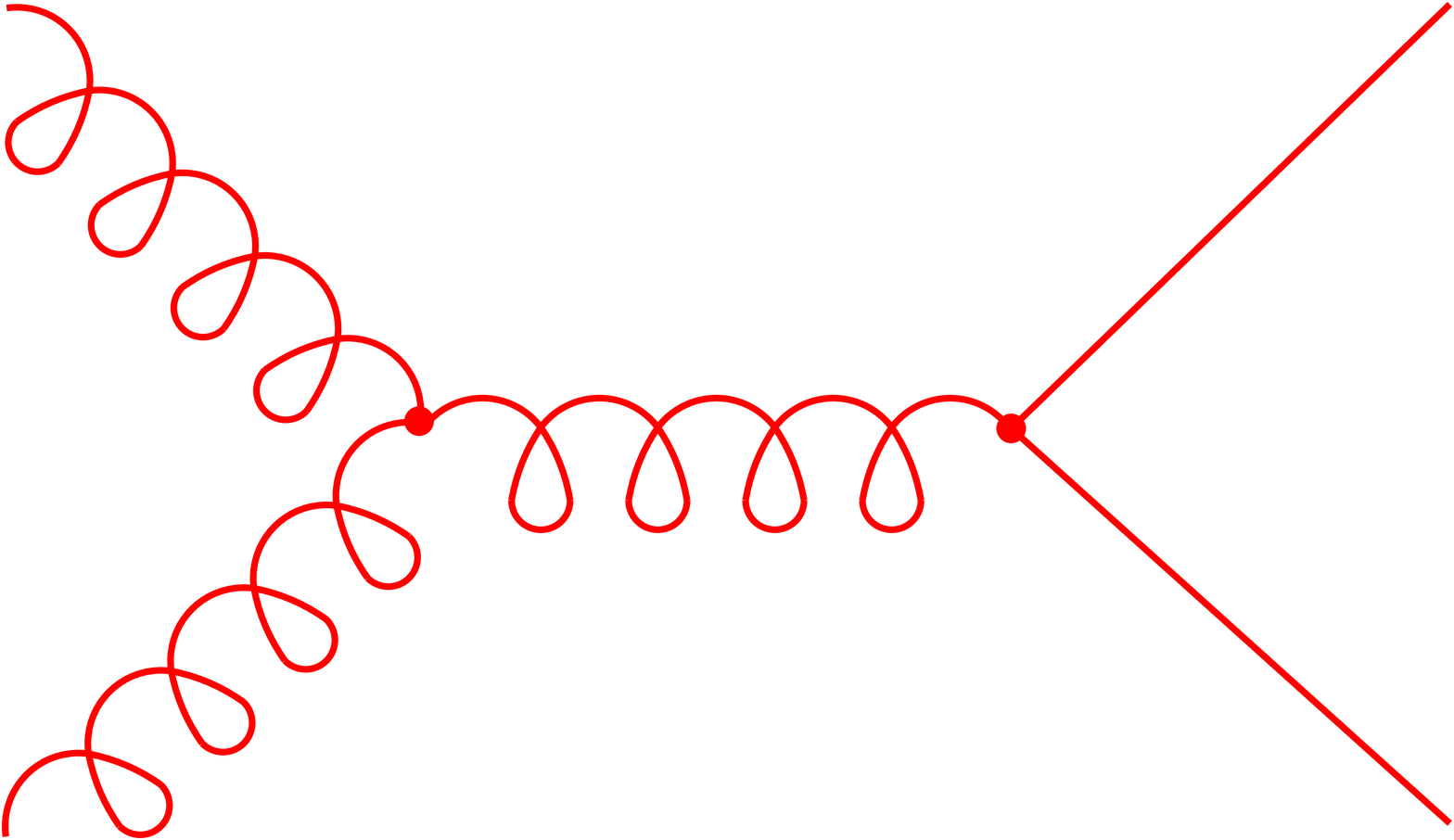}
\hspace*{1.6cm}
\epsfxsize=4.0cm
\epsffile{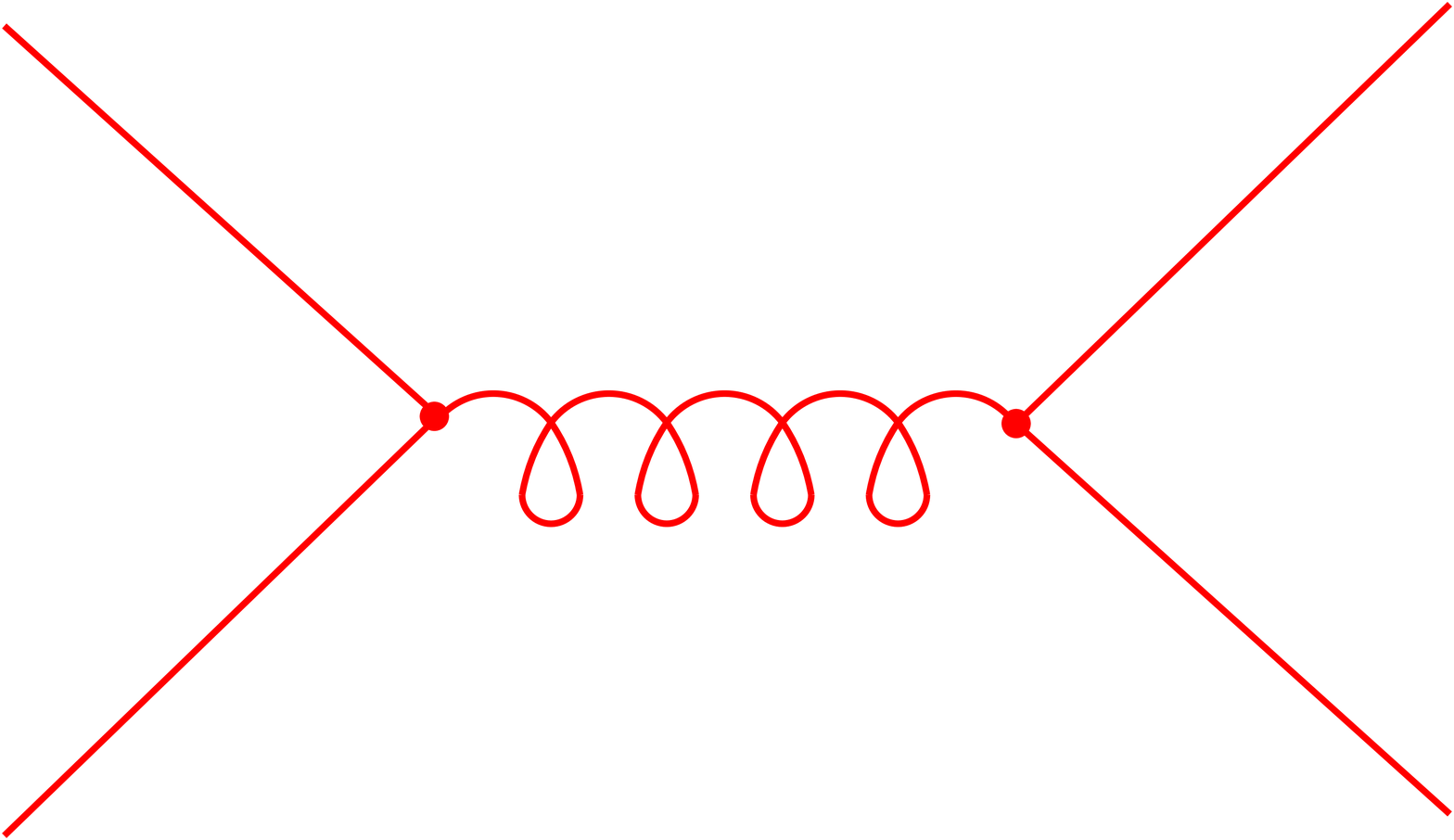}
}
\vspace*{0.2cm}
\fcaption{
Lowest order ${\cal O}(\alpha_s^2)$ diagrams for heavy quark
production through (a)-(c) gluon-gluon fusion and (d) quark-antiquark
annihilation.}  
\label{bprodlo}
\end{figure}

\noindent
The leading order (LO) $\alpha_s^2$ diagrams for heavy flavour production
are the $2 \ra 2$ processes of gluon-gluon fusion $g + g \ra {\cal Q} + 
\bar {\cal Q}$ as shown in Fig.~\ref{bprodlo}a-c) and quark-antiquark
annihilation $q + \bar q \ra {\cal Q} + \bar {\cal Q}$ displayed 
in Fig.~\ref{bprodlo}d), respectively. 
In the latter process, the ${\cal Q}\bar{\cal Q}$ pair is always in a
colour octet state while in $g\,g \ra {\cal Q} + \bar {\cal Q}$ both
colour singlet and octet are allowed.
The gluon-gluon fusion process is the dominant production
mechanism for $b$~quarks at the Tevatron, while top quarks are mainly
produced from the quark-antiquark annihilation process. The lowest
order ${\cal O}(\alpha_s^2)$ matrix elements 
in the $\alpha_s$ expansion and the cross sections for these processes 
have been available in the literature for some time\cite{gor}.
 
We briefly summarize some of the phenomenological consequences of the 
${\cal O}(\alpha_s^2)$ processes.
First, we note that the partonic cross section
$\hat \sigma$ is proportional to $\alpha_s^2/m_{\cal Q}^2$.
The average transverse momentum of the
heavy quark grows approximately with its mass $\langle\,\Pt({\cal Q})\,\rangle
\sim m_{\cal Q}$, meaning that the average $b$ quark transverse momentum
is about 4-5~\gevc. In addition, the \Pt\ distribution falls rapidly
to zero as \Pt\ becomes larger than the heavy quark mass. 
Furthermore, 
the heavy quark and antiquark are produced back-to-back in the
parton-parton centre-of-mass rest frame and are correspondingly
back-to-back in the plane transverse to the colliding hadron
beams. 
In addition, the rapidity distribution of the 
${\cal Q} \bar {\cal Q}$ pair has a typical bell shape becoming wider
and flatter as the partonic energy grows. This means, 
heavy quark production is larger in the
central region falling at higher rapidities. 
The rapidity difference between the $\cal Q$ and $\bar {\cal Q}$ tends
to be of order unity. But for a fixed value of \Pt, the production
rate is highly suppressed when the rapidity
difference becomes large, which means that $\cal Q$ and $\bar{\cal Q}$  
tend to be produced with similar rapidity.

\subsubsection{$b\bar b$ production in next-to-leading order QCD} 
\noindent
The next-to-leading order (NLO) terms in the $\alpha_s$ expansion were
originally considered `corrections' to the leading order terms. But, it
was soon recognized that the higher order corrections could be
large\cite{higho}. It was noticed that the process $g + g \ra g + g$
with $g \ra {\cal Q} +\bar {\cal Q}$, which is formally of order
$\alpha_s^3$, can 
be as important as the lowest order processes because the cross
section for the production of gluons $g + g \ra g + g$ is about a
hundred times larger than the LO cross section for the process 
$g + g \ra {\cal Q} + \bar {\cal Q}$. The complete calculations of
next-to-leading 
order corrections to heavy quark production in hadronic collisions
have been performed by several authors\cite{nde,nloqcd}.
It has been shown that
the ${\cal O}(\alpha_s^3)$ terms are actually larger than the lowest order
processes for
$b$ and $c$ quark production, if the hadron-hadron centre-of-mass
energy is much larger 
than $m_{\cal Q}$ which is the case for the Tevatron.

\begin{figure}[tb]
\centerline{
\put(0,15){\large\bf (a)}
\put(0,-10){\large\bf (b)}
\put(0,-36){\large\bf (c)}
\put(0,-62){\large\bf (d)}
\hspace*{0.9cm}
\epsfysize=2.6cm
\epsffile{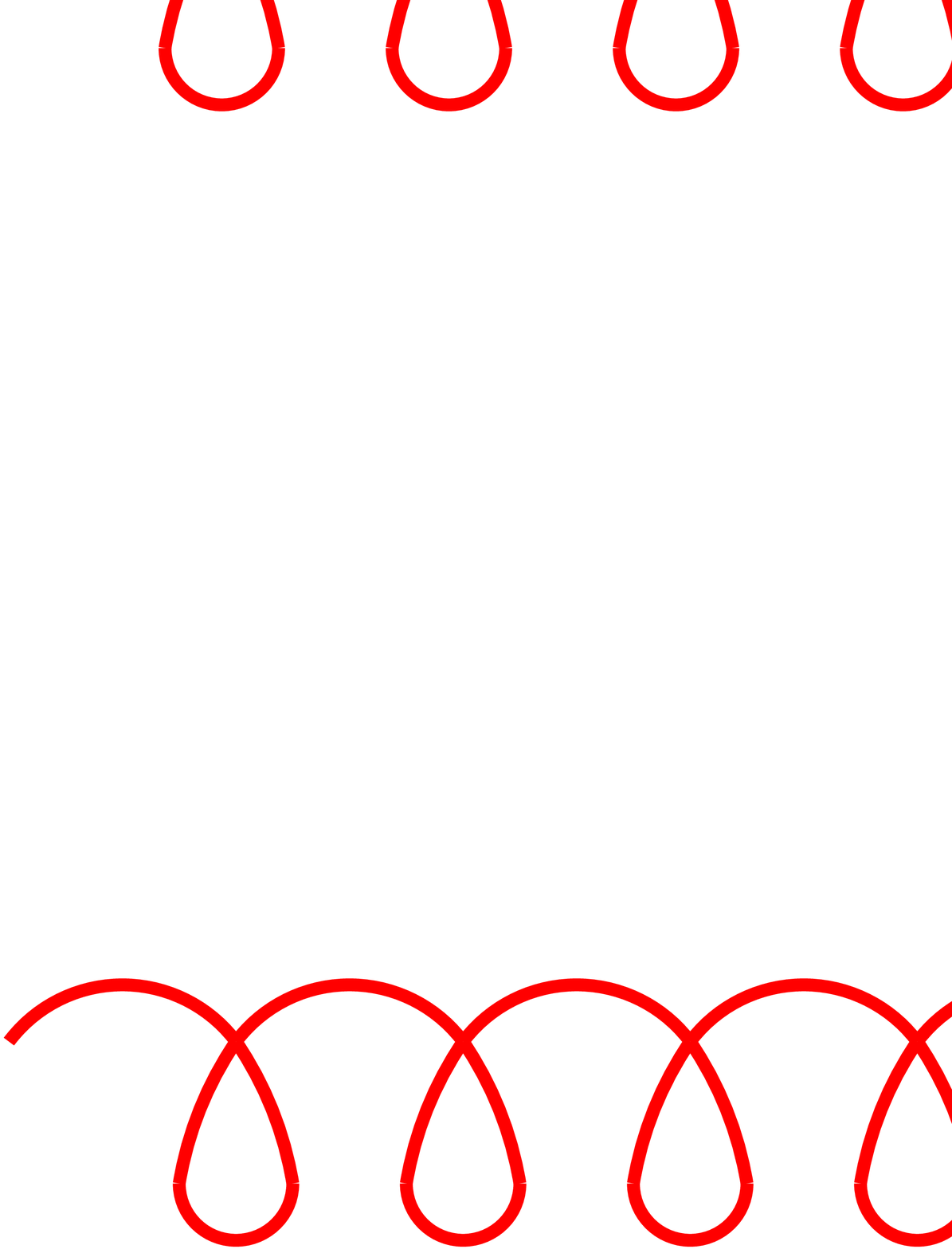}
\hspace*{0.4cm}
\epsfysize=2.6cm
\epsffile{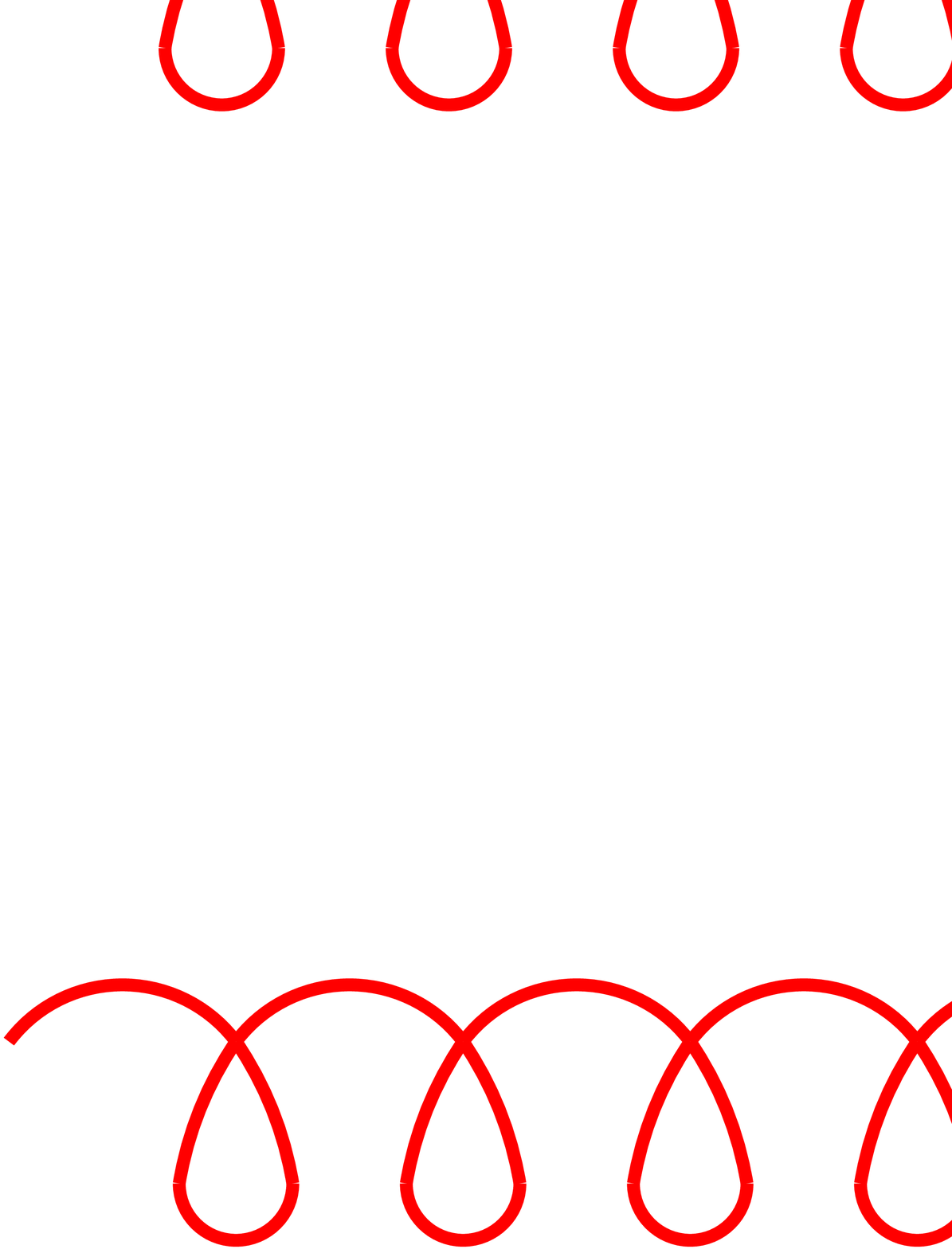}
\hspace*{0.4cm}
\epsfysize=2.6cm
\epsffile{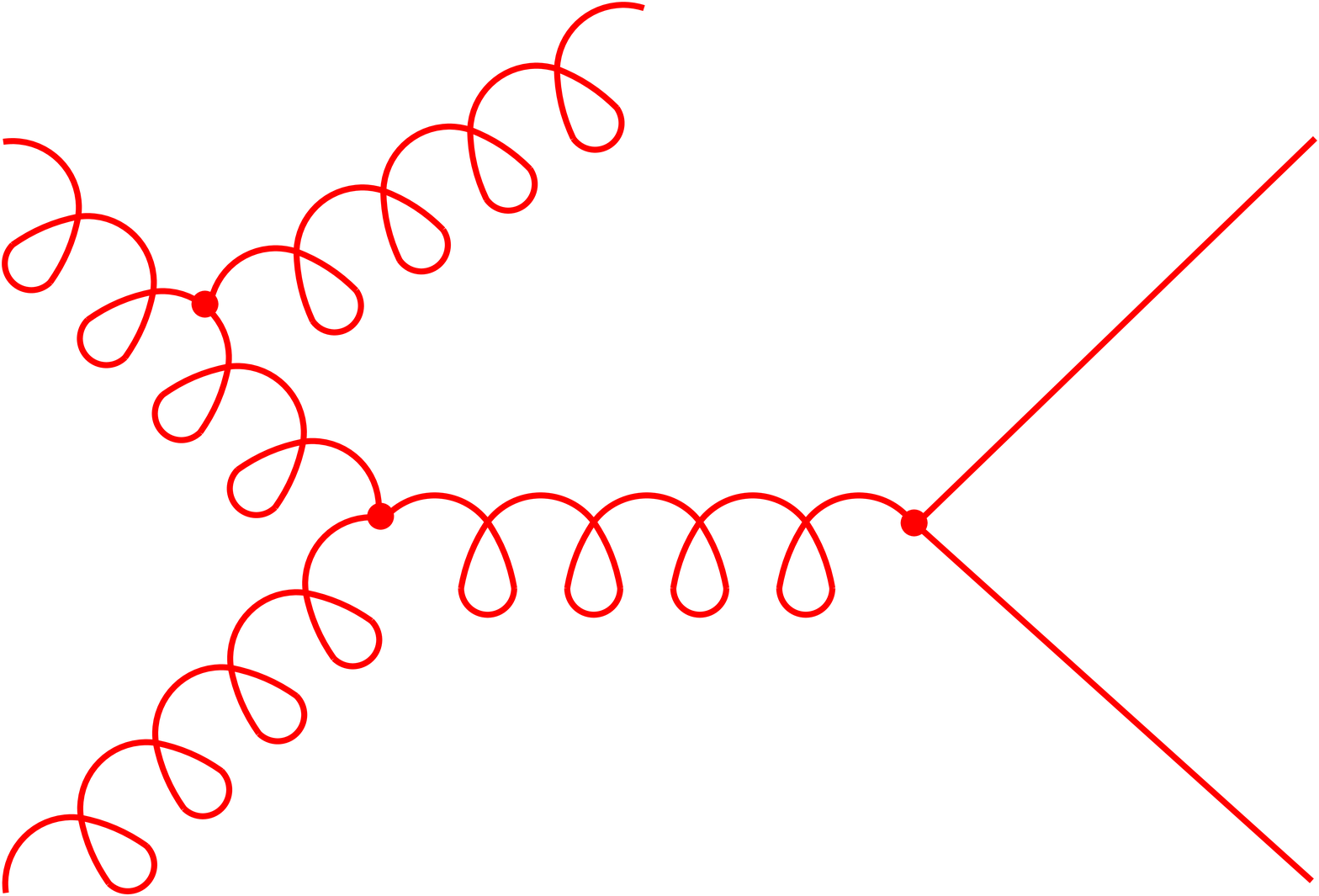}
}
\vspace*{-5.9cm}
\centerline{
\hspace*{0.6cm}
\epsfysize=2.1cm
\epsffile{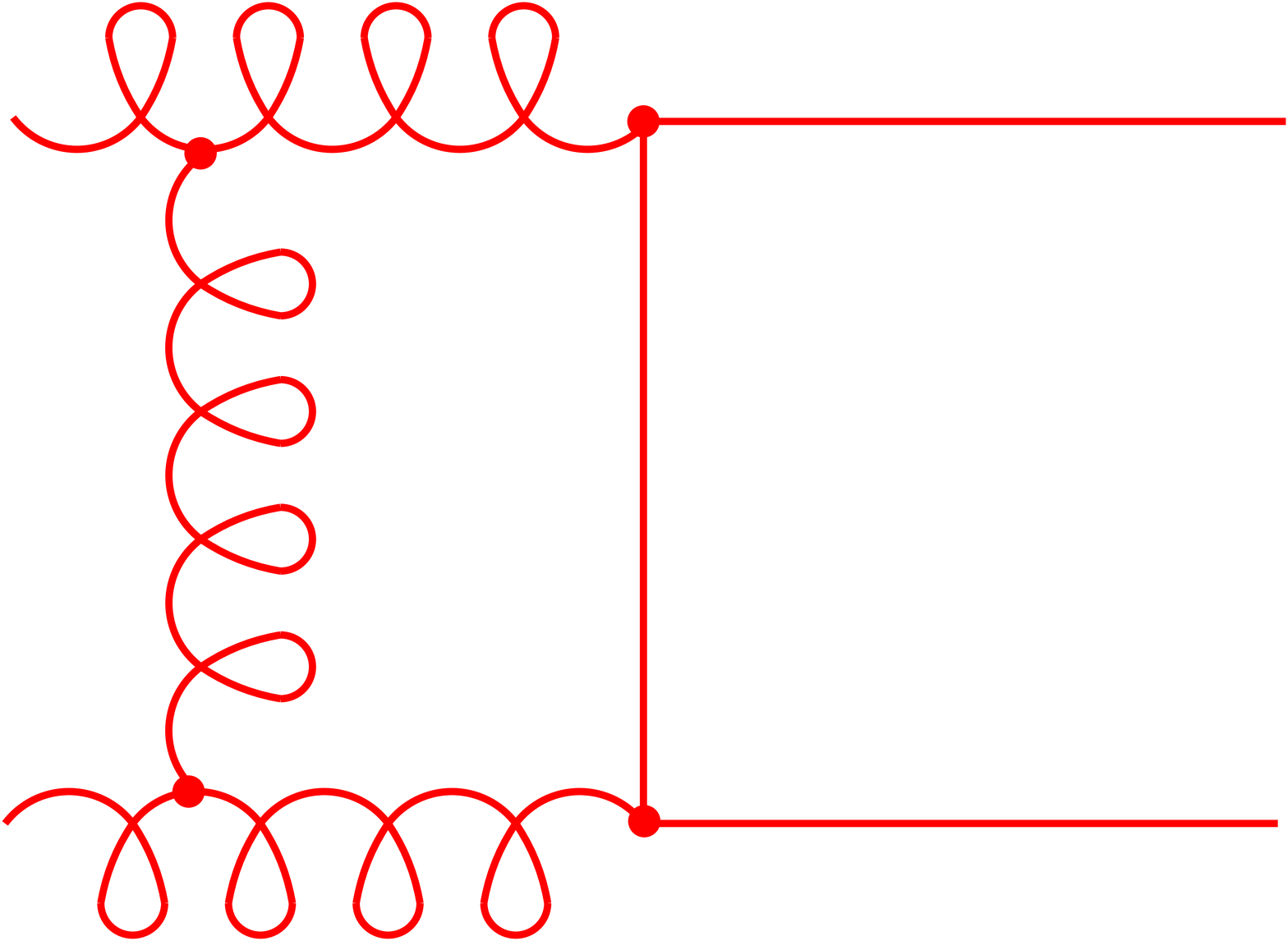}
\hspace*{0.3cm}
\epsfysize=2.1cm
\epsffile{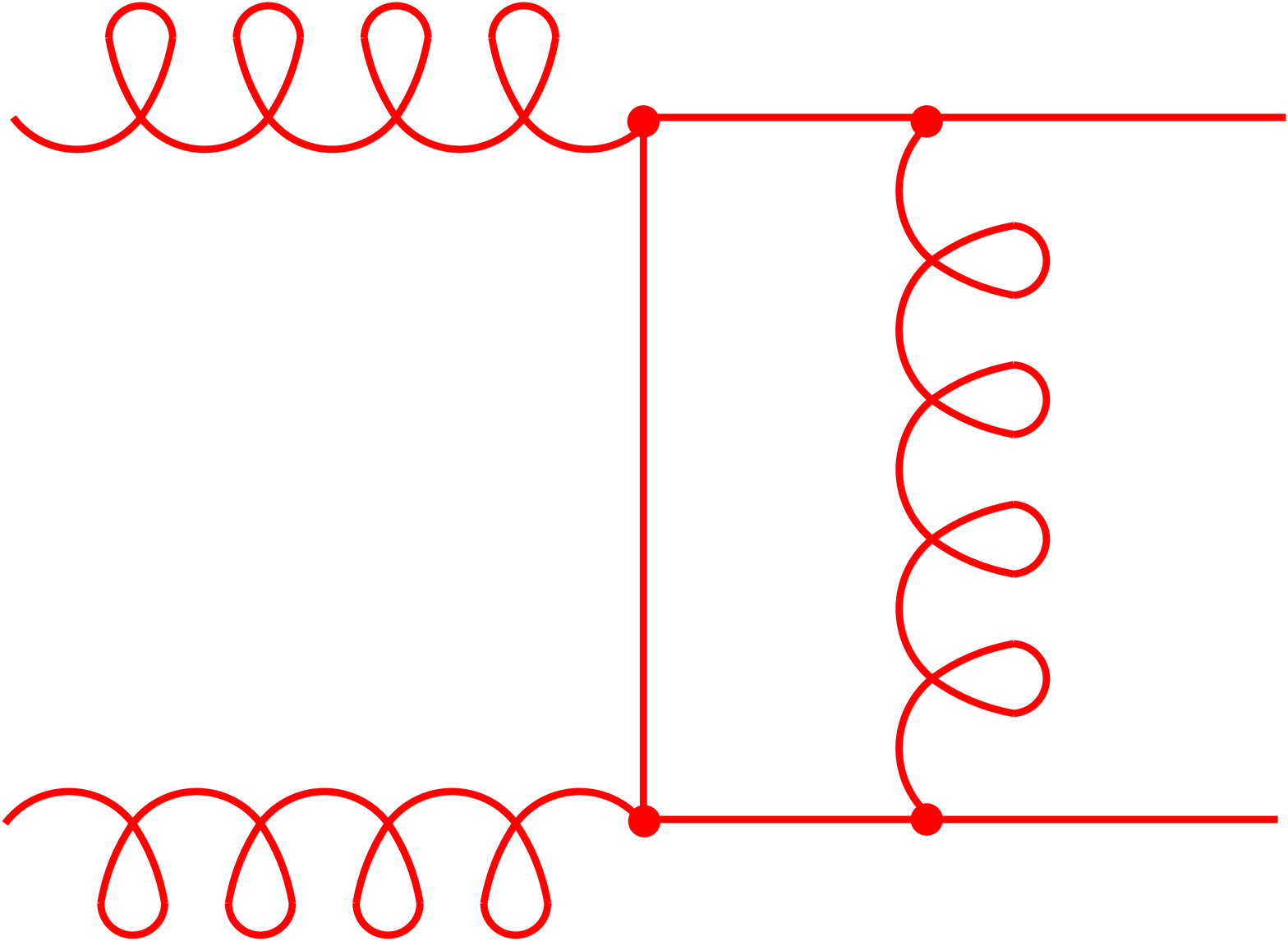}
\hspace*{0.4cm}
\epsfysize=2.2cm
\epsffile{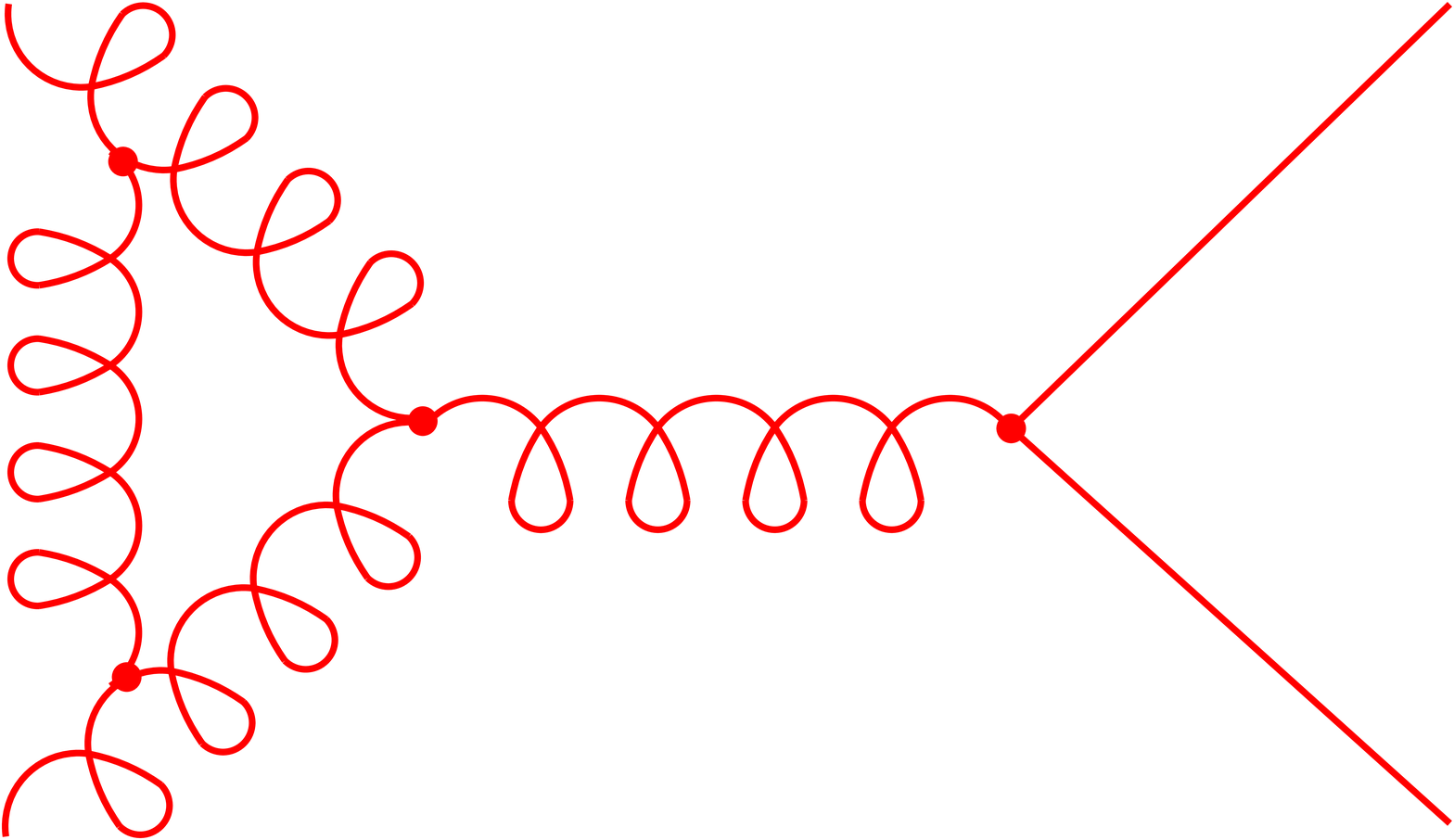}
}
\vspace*{0.5cm}
\centerline{
\epsfysize=2.2cm
\epsffile{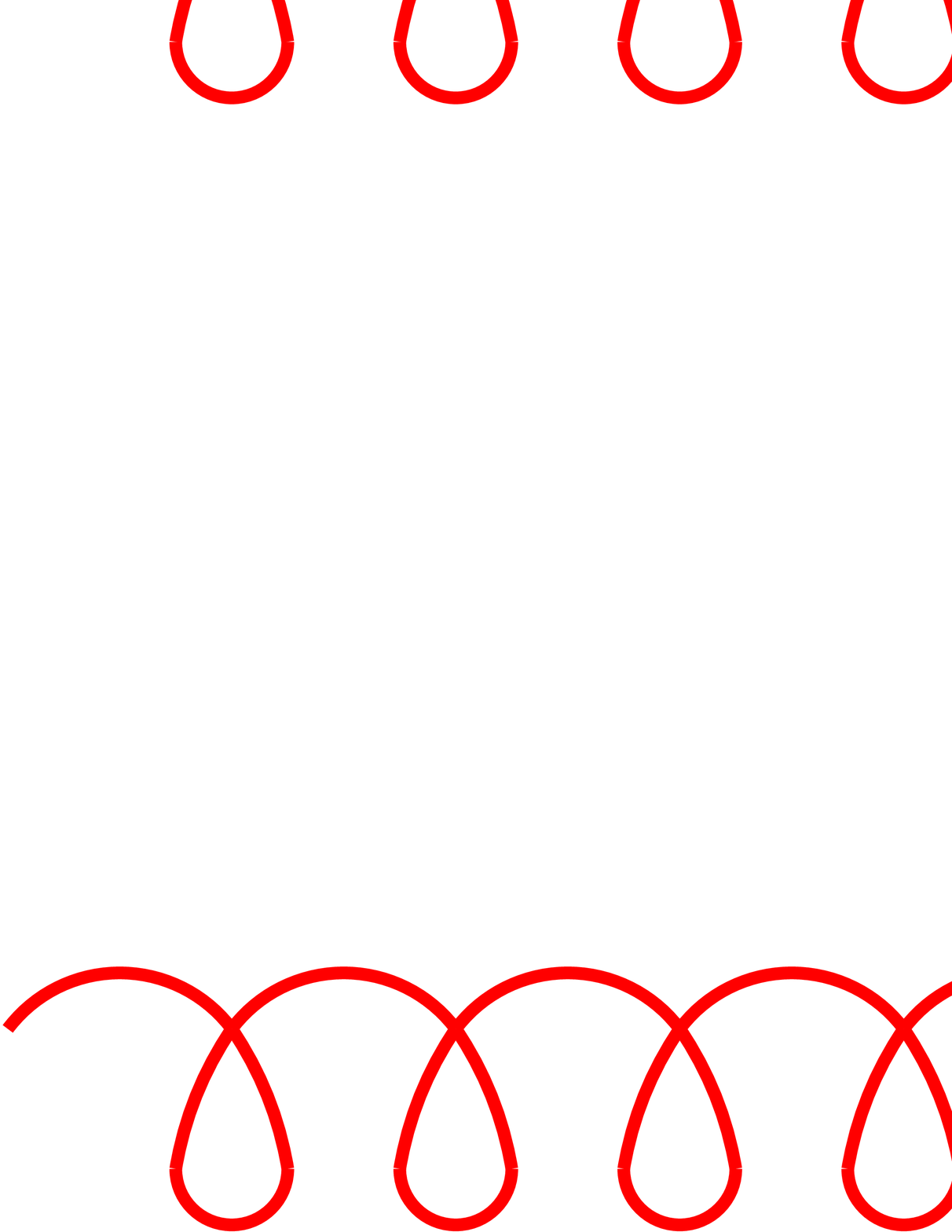}
\hspace*{0.5cm}
\epsfysize=2.2cm
\epsffile{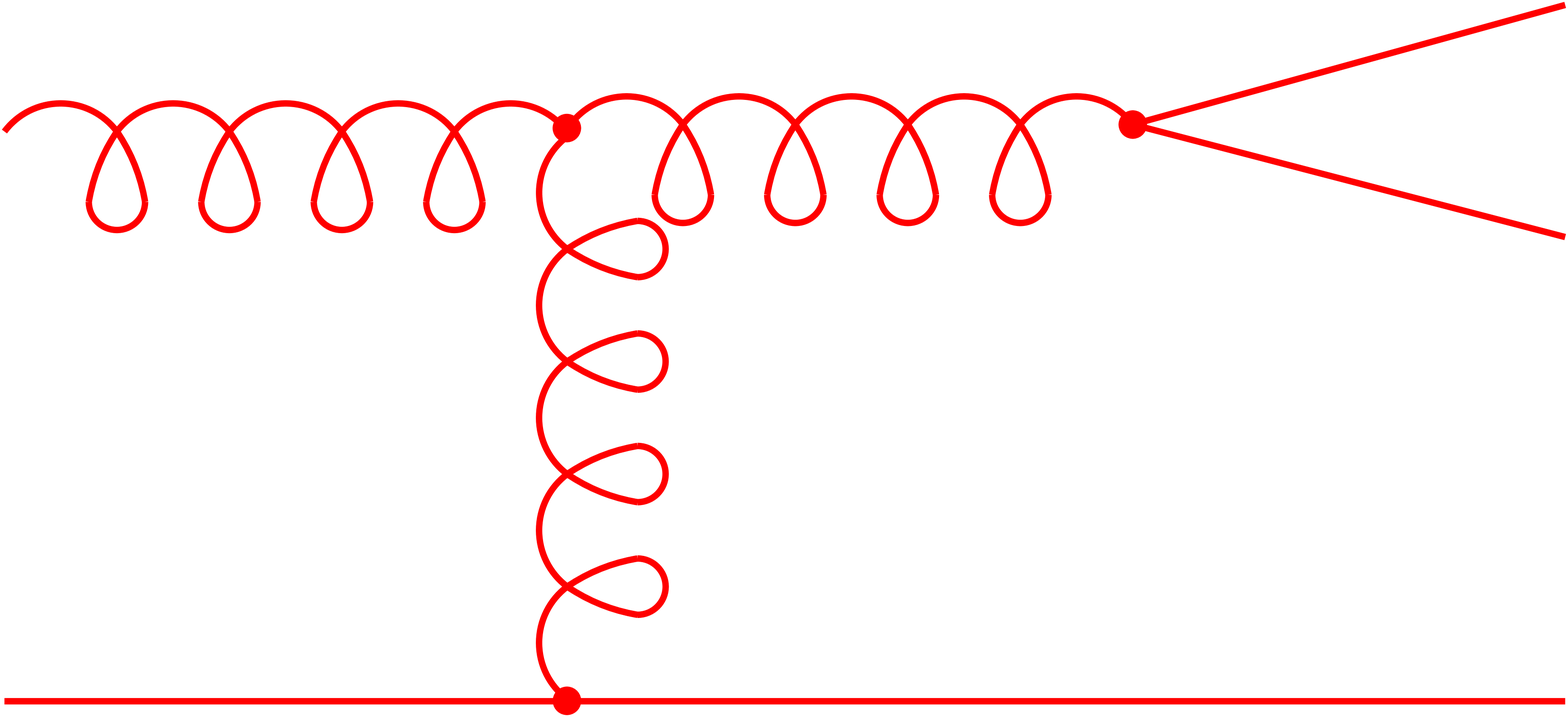}
}
\vspace*{0.5cm}
\centerline{
\epsfysize=2.2cm
\epsffile{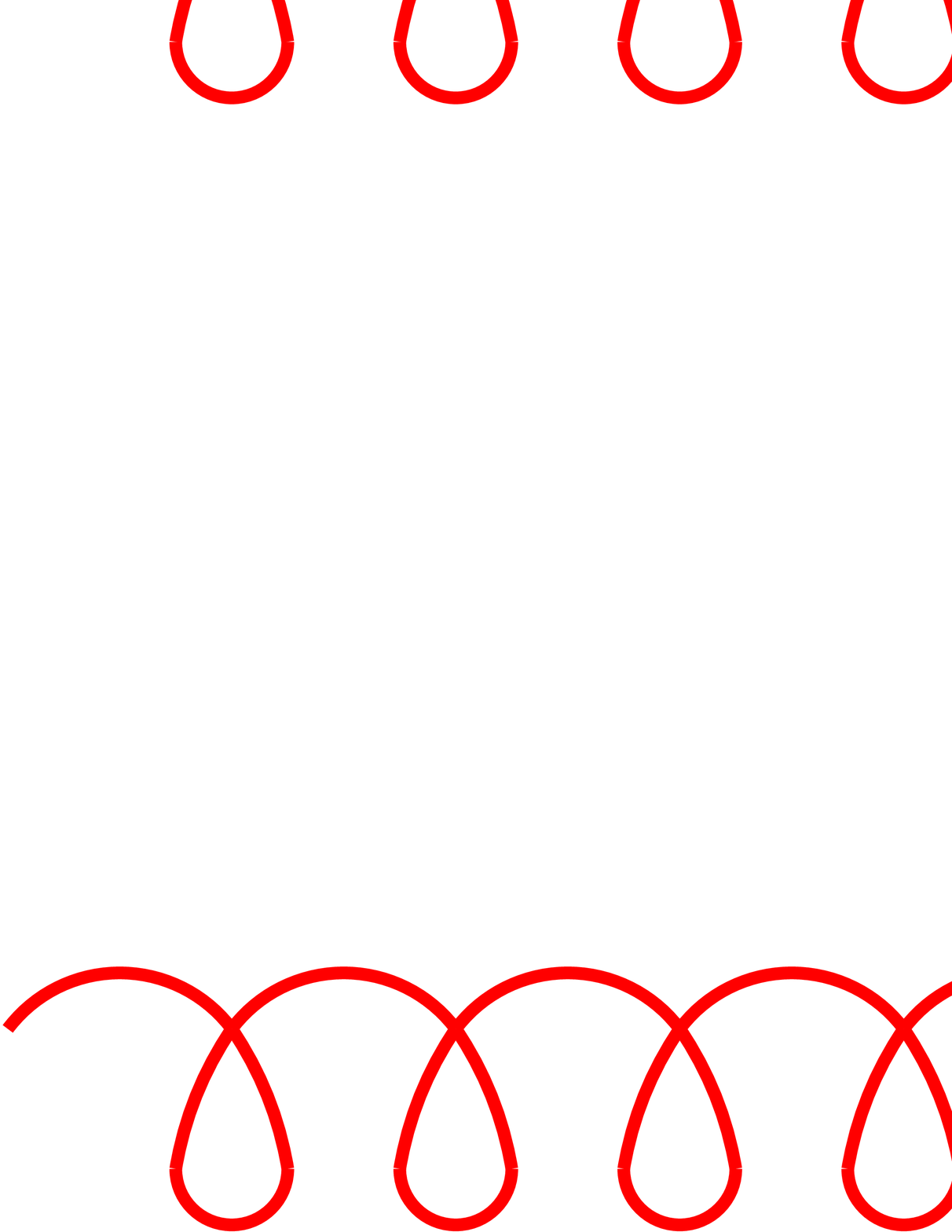}
\hspace*{0.5cm}
\epsfysize=2.2cm
\epsffile{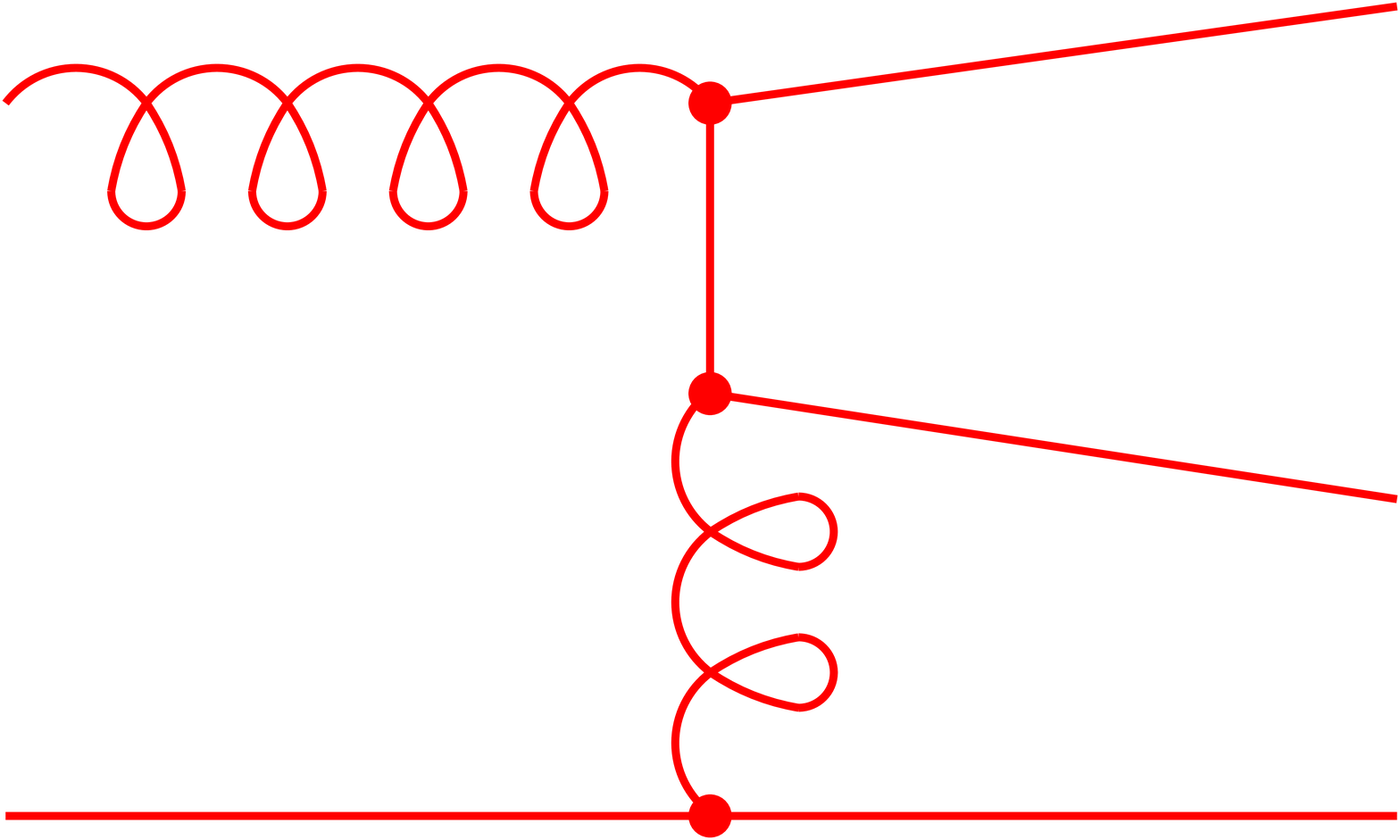}
}
\vspace*{0.3cm}
\fcaption{
Examples of diagrams for heavy quark production at next-to-leading order: 
(a) Real emission diagrams, (b) virtual emission diagrams,
(c) gluon splitting, and (d) flavour excitation.}
\label{bprodnlo}
\end{figure}

Examples of order $\alpha_s^3$ diagrams are displayed in
Figure~\ref{bprodnlo}. These processes include 
real emission matrix elements (Fig.~\ref{bprodnlo}a) and the
interference of virtual matrix elements with the leading order
diagrams shown in Fig.~\ref{bprodnlo}b).
Other NLO contributions come from 
gluon splitting diagrams (Fig.~\ref{bprodnlo}c) or the flavour
excitation process as displayed in
Fig.~\ref{bprodnlo}d). In the gluon splitting case, 
the probability to find a heavy quark from a gluon with large \Pt\
has a logarithmic increase. Phenomenologically, 
the ${\cal Q}\bar {\cal Q}$ pair  
is produced close in phase space and will often appear as a single jet.
In the flavour excitation process, the heavy quark is considered to be
already present with a certain heavy quark density in the incoming
hadron. It is excited by the exchange 
of a gluon with the other hadron and appears on mass-shell in the
final state. In the case of a flavour excitation process, only one of
the quarks from the ${\cal Q}\bar {\cal Q}$ pair is usually at high \Pt. 

In summary, next-to-leading order 
processes are an important contribution to $b$~quark production in 
$p\bar p$~collisions at $\sqrt{s} = 1.8$~TeV. These processes are the source of
$b\bar b$ pairs that can be close together in phase space and do not
necessarily appear
in the central detector region as compared to the LO processes which
result in back-to-back quark pairs. 
However, there is evidence (see Sec.~2.4.) that NLO processes are not
sufficient to 
obtain accurate estimates since large $\mu$-scale dependences are
still present. These are an indication that higher order corrections
could be large. 
 
\subsection{Hadronization of \boldmath{$b$} quarks}
\noindent
Once $b$~quarks are produced through the initial hard scattering, 
the process of forming 
$B$~hadrons follows and is called hadronization or fragmentation. It is a low
$Q^2$~process and is beyond the reach of perturbative QCD
calculations. The hadronization process is therefore described by
semi-empirical models inspired by theory. A commonly used approach is
the string fragmentation model\cite{sjostr}. In a naive picture, we
can imagine a ``cloud'' of gluons acting as a string 
between the $b$ and $\bar b$
quark pair. As the quark and antiquark separate, the string stretches
until it breaks and a new $q\bar q$ pair is created
out of the vacuum to form the new ends of the string. 
These new strings also stretch and break, producing more
quark-antiquark pairs until there is no longer sufficient energy
available to generate new $q\bar q$~pairs. The particles produced
in this hadronization process, along with the $B$~hadron, are usually
referred to as fragmentation particles.  
The fraction of the initial $b$~quark momentum transfered to the
$B$~hadron is commonly described by a so-called fragmentation function,
suggested, for example, by Peterson et al.\cite{Peterson} for the case
of a heavy 
quark $\cal Q$ forming a hadron together with a light quark $\bar q$:
\begin {equation}
\frac{{\rm d}N}{{\rm d}z} \propto \frac{1}{z} \cdot \left(1-\frac{1}{z}-
\frac{\epsilon_b}{1-z}\right)^{-2}.
\end{equation}
Here, $\epsilon_b$ is the so-called Peterson fragmentation parameter,
related to the ratio of the effective light and heavy quark
masses $\epsilon_b \sim (m_{\bar q}/m_{\cal Q})^2$. The variable $z$ is
originally defined as $z = (E + p_{\parallel})_{{\cal Q}\bar q} /
(E+p_{\cal Q})$
where $p_{\parallel}$ is the projection of the momentum of the hadron
onto the direction of the heavy quark before hadronization. This
variable is often approximated with experimentally better accessible
parameters such as $x_p = p/p_{\rm max}$ or $x_E = E_{\rm hadron} / 
E_{\rm beam}$ used in $e^+e^-$ annihilation.

\subsection{Comparison of heavy quark production with CDF data}
\noindent
\begin{figure}[tbp]
\centerline{
\epsfysize=6.8cm
\epsffile[47 187 564 578]{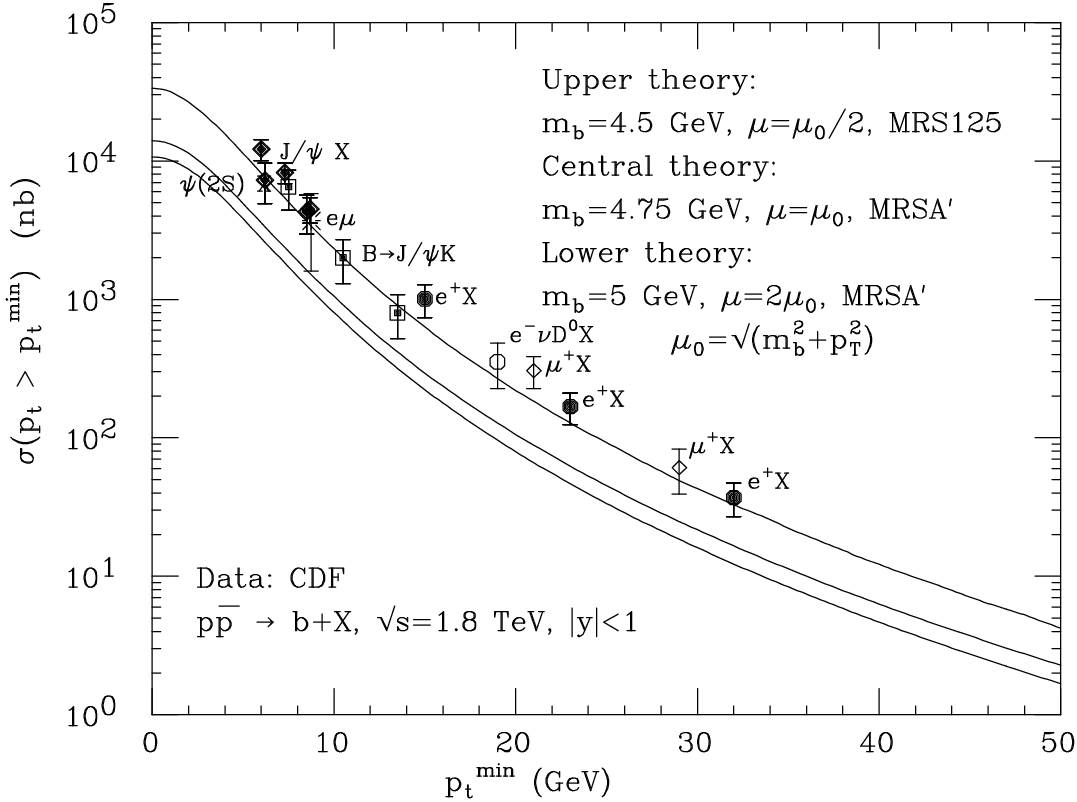}
}
\vspace*{0.2cm}
\fcaption{Integrated cross section for $b$ quark production versus
\Pt\ compared to the results of next-to-leading order QCD 
predictions\cite{frixione}.}
\label{cdfbxsec}
\end{figure}
To conclude our review of heavy quark production, we compare in 
Figure~\ref{cdfbxsec}, taken from Ref.\cite{frixione}, various CDF
measurements of $b$~quark production with next-to-leading order
predictions.
The distribution
most commonly studied by hadron collider experiments is the $b$~quark
differential \Pt~spectrum, integrated
above a given \Pt~threshold $\Pt^{\rm min}$ and
within a fixed rapidity range which is usually
set to $y_{\rm max}=1$ at CDF:
\begin{equation}
\sigma(\Pt > \Pt^{\rm min}) = \int_{|y|<y_{\rm max}} {\rm d}y 
\int_{\Pt > \Pt^{\rm min}} {\rm d}\Pt 
\frac{{\rm d}^2\sigma}{{\rm d}y\,{\rm d}\Pt}.
\end{equation}
Different
techniques, involving primarily CDF's silicon vertex detector to
improve background rejection,
are used to obtain the data points from, e.g.
high-\Pt\ $J/\psi$ mesons reconstructed through $\mu^+\mu^-$
or high-\Pt\ leptons accompanied by a nearby charm meson ($D^0$).
The measurements are
compared to next-to-leading order QCD calculations which agree very well
in shape with the data. However, the data are higher by a factor of almost
three than the default prediction based on 
$\mu = \mu_0 = \sqrt{\Pt^2 + m_b^2}$. Using variations on, for
example, $m_b$, $\mu$, and the parton distribution function, theory is able to
accommodate the data as indicated by the upper curve in Fig.~\ref{cdfbxsec}.
Since the production of 
$b$~quarks and $B$~hadrons is not subject of this
review, we refer to the literature\cite{frixione,kniehl} for a
further discussion of this issue.

\section{Experimental Environment}
\runninghead{$B$ Lifetimes, Mixing and $CP$ Violation at CDF}
{Experimental Environment}
\noindent
In this section, we summarize the experimental environment beginning
with the Tevatron Collider. We then describe the
CDF~detector, emphasizing the device that made a broad $B$~physics
program possible at CDF, the silicon micro-vertex~detector. 

\subsection{Tevatron collider}
\noindent
The Fermilab Tevatron, with a circumference of 6.28 km, is a proton-antiproton
collider operating at a centre-of-mass energy of $\sqrt{s} = 1.8$~TeV. 
The Tevatron delivered its first physics collisions in
1987. In this article, however, we will concentrate on the data taking
period referred to as Run\,I which started in May~1992 and ended
in Feb.~1996. During that time, the
Tevatron operated with six bunches of protons and six bunches of
antiprotons crossing every 3.5 $\mu$s at CDF's interaction
region. 

\begin{figure}[tbp]
\centerline{
\epsfxsize=9.5cm
\epsffile[25 240 600 560]{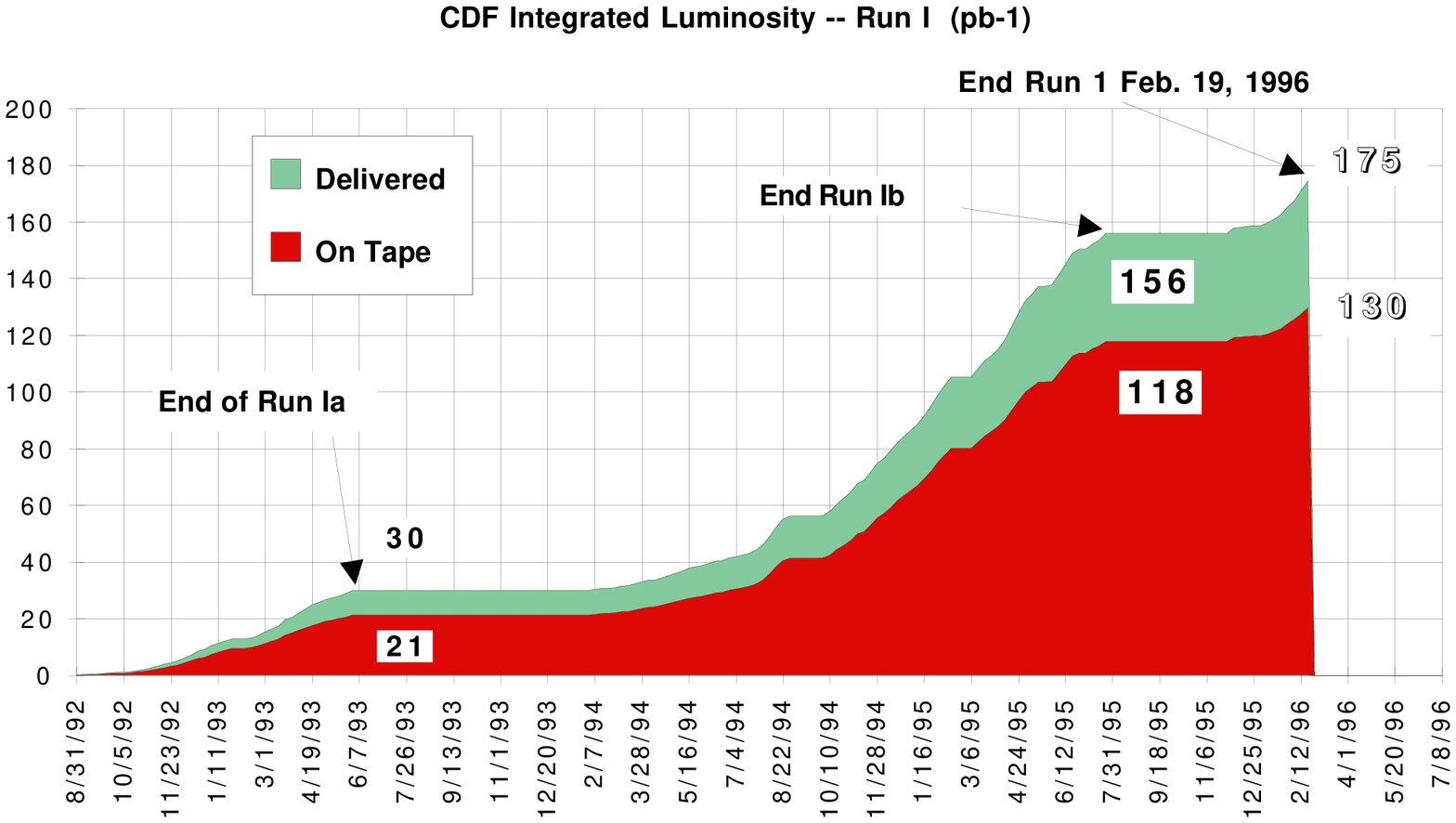}
}
\vspace*{0.2cm}
\fcaption{
Time profile of the luminosity delivered by the Tevatron Collider and
accumulated by the CDF experiment throughout Run\,I.}
\label{cdflumi}
\end{figure}

Figure~\ref{cdflumi} shows a time profile of the delivered and
accumulated luminosity at CDF throughout Run\,I. 
During this period, a total integrated luminosity of about
175~pb$^{-1}$ was delivered to the CDF experiment.
During Run\,I, the highest instantaneous luminosities
reached, were around $2.5\cdot10^{31}$~cm$^{-2}$s$^{-1}$. 
The Run\,I running period was divided up into a Run\,Ia,
from May~1992 through June~1993, a Run\,Ib, from Dec.~1993 to
July~1995, and a Run\,Ic, from Dec.~1995 to Feb.~1996. 
The collected integrated luminosities of data used for physics
analyses were
about 19.3~pb$^{-1}$  for Run\,Ia,
approximately 90~pb$^{-1}$ for Run\,Ib, and about
10~pb$^{-1}$ for Run\,Ic. 
However, the Run\,Ic data taking period was dedicated to the
accumulation of very specialized trigger datasets and is usually not
included in Run\,I physics analyses, resulting in about 110~pb$^{-1}$ of
data used for physics results from Run\,I.
All measurements presented in this paper refer to this Run\,I luminosity,
unless otherwise~noted.

\subsection{CDF detector}
\noindent
The CDF detector is a multi-purpose apparatus
designed to study 1.8~TeV $p\bar p$~collisions produced by the
Fermilab Tevatron Collider. It has both azimuthal and forward-backward
symmetry. A superconducting solenoid, 4.8~m in length and 1.5~m in radius, 
generates a 1.4~T magnetic field containing tracking chambers used
to detect charged particles and measure their momenta. Surrounding
the solenoid, sampling calorimeters measure electromagnetic
and hadronic energies of jets, electrons, and photons. Outside the calorimeters
are drift chambers for muon detection.

\begin{figure}[tbp]
\centerline{
\epsfxsize=11.5cm
\epsffile[100 145 640 454]{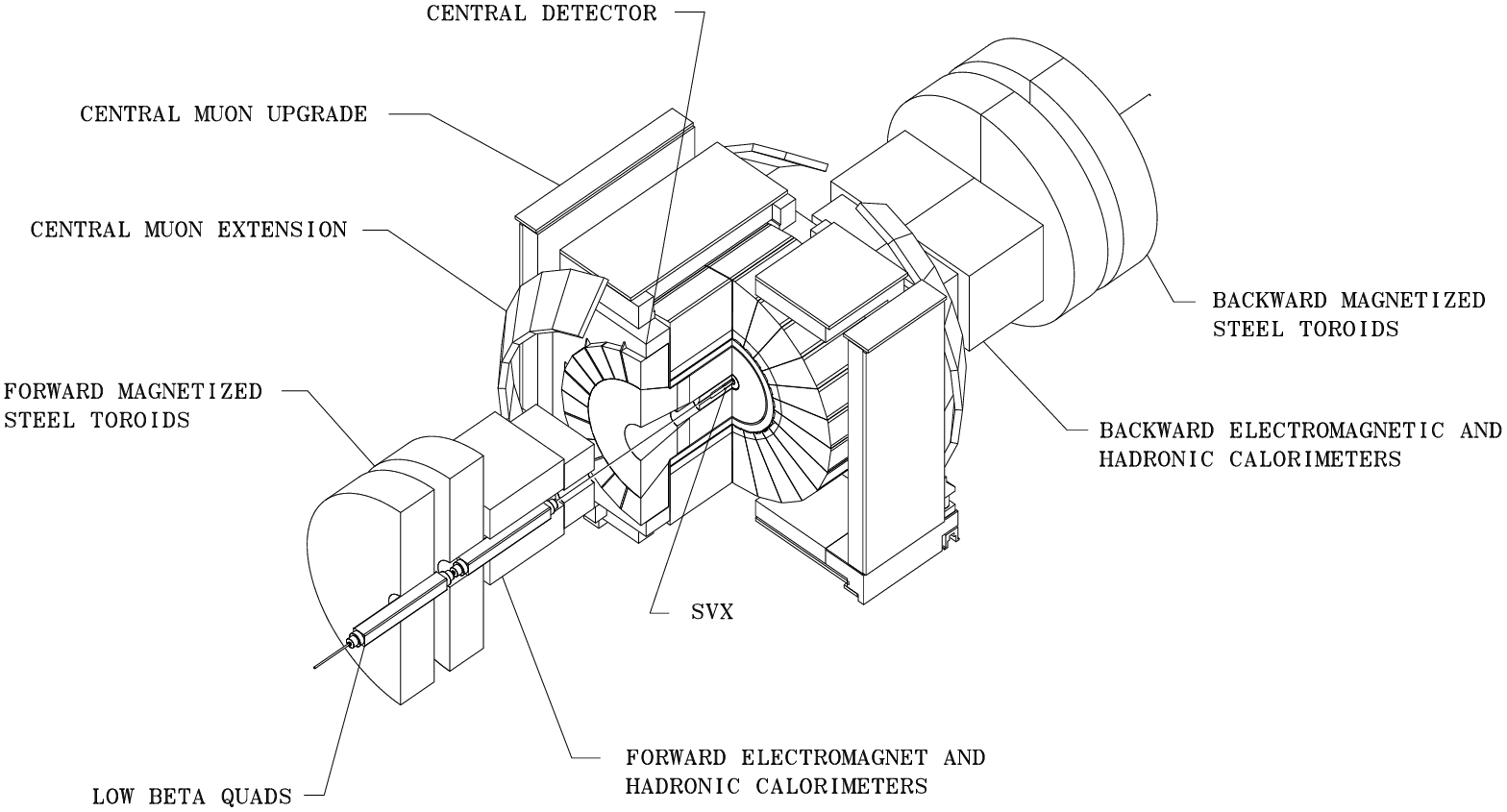}
}
\vspace*{0.2cm}
\fcaption{
Schematic view of the CDF detector.}
\label{cdf_det}
\end{figure}

\begin{figure}[tbp]
\centerline{
\epsfxsize=9.5cm
\epsffile[90 495 517 767]{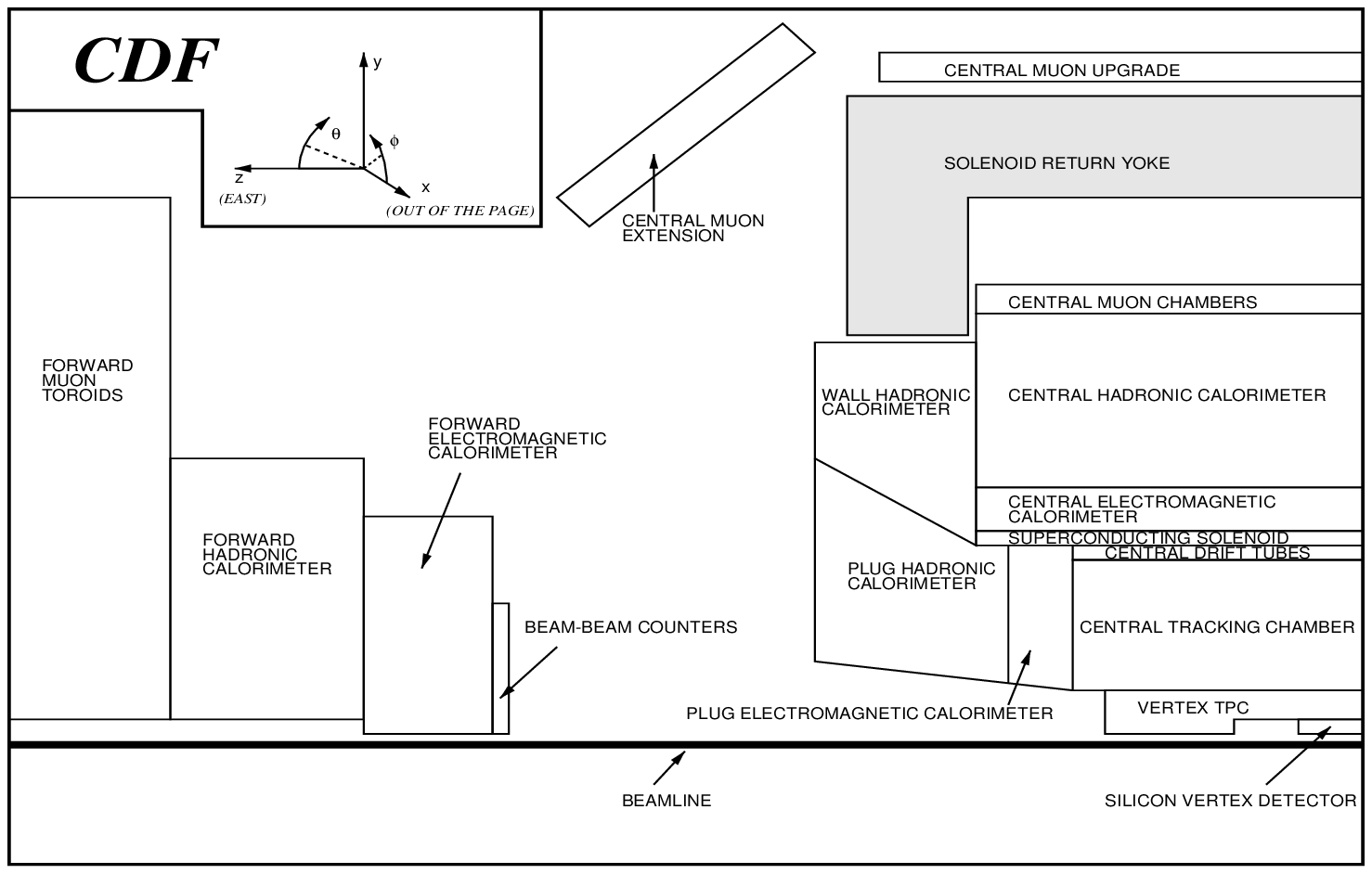}
}
\vspace*{0.2cm}
\fcaption{
Side-view quarter cross section of the CDF detector. For a full view, the
image is to be reflected to the left and below. The $p\bar p$ interaction
point is at the lower-right corner of the figure.}
\label{cdf_quarter}
\end{figure}

A schematic view of the CDF experiment can be found in
Figure~\ref{cdf_det} while a side-view quarter cross section of the CDF
detector is displayed in Fig.~\ref{cdf_quarter}. 
The CDF experiment uses a coordinate system with the $z$-axis along the
proton beam direction, the $y$-axis pointing vertically upwards, and
the $x$-axis pointing horizontally out of the Tevatron ring. 
Throughout this article, $\varphi$ is the azimuthal angle, 
$\theta$ is the polar angle measured from the proton direction, 
and $r$ is the radius perpendicular to the beam axis. The
pseudorapidity $\eta$ is defined as $\eta=-\ln[\,\tan(\theta/2)\,]$.
The transverse momentum \Pt\ is the component of the track
momentum~$p$ transverse to the $z$-axis ($\Pt = p \cdot \sin\theta$)
while $\Et = E\cdot\sin\theta$ with $E$ being
the energy of the calorimeter cluster.
A more complete description of 
the CDF detector can be found elsewhere\cite{cdf_det}. 
We summarize here only those detector features most relevant to $B$~physics.

Three devices inside the 1.4 T solenoidal magnetic field 
are used for the tracking of charged particles: 
The silicon vertex detector (SVX), a set of vertex time projection
chambers (VTX), and the central tracking chamber (CTC).
The SVX is described in more detail in Section~3.2.1.
The VTX, which is located outside the SVX up to a radius of
22~cm, reconstructs track segments in the $r\,z$-plane up to 
$|\eta| < 3.25$. The VTX is
used to determine the $z$-position of the primary interaction vertex
with a resolution of about 0.2~cm on average. 

Surrounding the SVX
and VTX is the CTC, located between radii of 30~cm and 132~cm. The
CTC is a 3.2~m long cylindrical drift chamber that contains 
84~layers of sense wires grouped into nine 
alternating super-layers of axial and stereo wires with a stereo angle
of $3^{\circ}$. 
The CTC provides three-dimensional tracking and
covers the pseudorapidity interval $|\eta|$ less~than~about~1.1.
The outer 54 layers of the CTC are instrumented to record the specific
ionization d$E$/d$x$ of charged particles. 

Outside the solenoid are electromagnetic (CEM) and hadronic (CHA)
calorimeters ($|\eta|<1.1$) which employ a projective tower geometry
back to the nominal interaction point with a 
segmentation of $\Delta\eta \times \Delta\varphi \sim 0.1 \times 15^{\circ}$.
The sampling medium is composed of scintillators layered with lead
absorbers in the electromagnetic calorimeter and steel in the CHA. 
The energy resolution for the CDF
central calorimeter is 
$\sigma(E_{\rm t}) / E_{\rm t} = [(13.5\% / \sqrt{E_{\rm t}})^2 
+ (2\%)^2]^{1/2}$ 
for electromagetic showers and
$\sigma(E_{\rm t}) / E_{\rm t} = [(50\% / \sqrt{E_{\rm t}})^2
+ (3\%)^2]^{1/2}$ 
for hadrons with $E_{\rm t}$ measured in GeV.
A layer of proportional chambers (CES), with wire and strip readout, 
is located six radiation lengths deep in the CEM calorimeters
approximately near shower maximum for electromagnetic showers.
The CES provides a measurement of electromagnetic shower profiles 
in both the $\varphi$- and $z$-directions.
Proportional chambers located between the solenoid and the CEM
comprise the central preradiator detector (CPR) which samples the
early development of electromagnetic showers in the material of the
solenoid coil, providing information in $r$-$\varphi$ only. Finally, 
plug and forward calorimeters instrument the region of $1.1 < |\eta| <
4.2$. These consist of gas proportional chambers as active media, while lead
and iron are used as absorber~materials. 

The central calorimeters act as hadron absorber for the muon
detection system. Four of its layers of planar drift 
chambers (CMU) are located beyond the central calorimeters.
The CMU system covers $|\eta| \leq 0.6$ and can be reached by muons
with \Pt\ in excess of 1.4~\gevc. 
To reduce the probability of misidentifying penetrating hadrons as muon
candidates in the central detector region,
four additional layers of drift chambers (CMP) were added in 1992 and
are located 
behind 0.6~m of steel outside the CMU system. Approximately 84\% of the
solid angle for $|\eta| \leq 0.6$ is covered by the CMU detector, 63\%
by the CMP, and 53\% by both. To reach these two detectors,
particles produced at the primary interaction vertex, with a polar
angle of $90^{\circ}$, must traverse material totaling 5.4 and 8.4 pion
interaction lengths, respectively. 
An additional set of muon chambers (CMX) is located in the pseudorapidity
interval $0.6 < |\eta| < 1.0$ to extend the polar acceptance of the
muon system. Approximately 71\% of the solid angle for 
$0.6 < |\eta| < 1.0$ is covered by the  
free-standing conical arches of the CMX. Finally, the forward muon
system (FMU) is a magnetic spectrometer consisting of three planes of
drift chambers surrounding two 1~m thick iron toroids located $\pm
10$~m from the interaction point. 

\subsubsection{CDF silicon micro-vertex detectors}
\begin{figure}[tbp]
\centerline{
\epsfxsize=6.6cm
\epsffile[90 210 610 585]{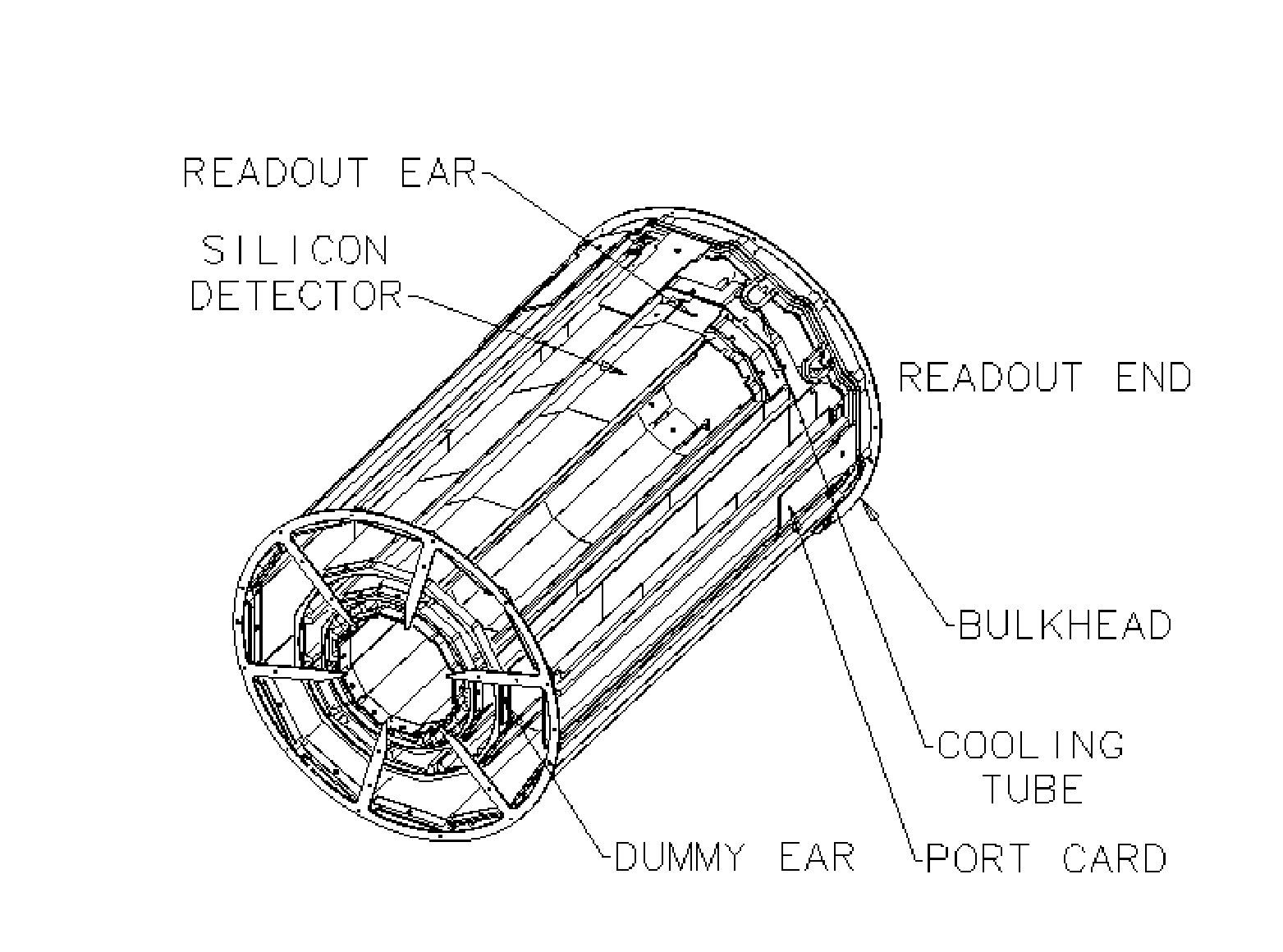}
\epsfxsize=6.0cm
\epsffile[80 60 738 568]{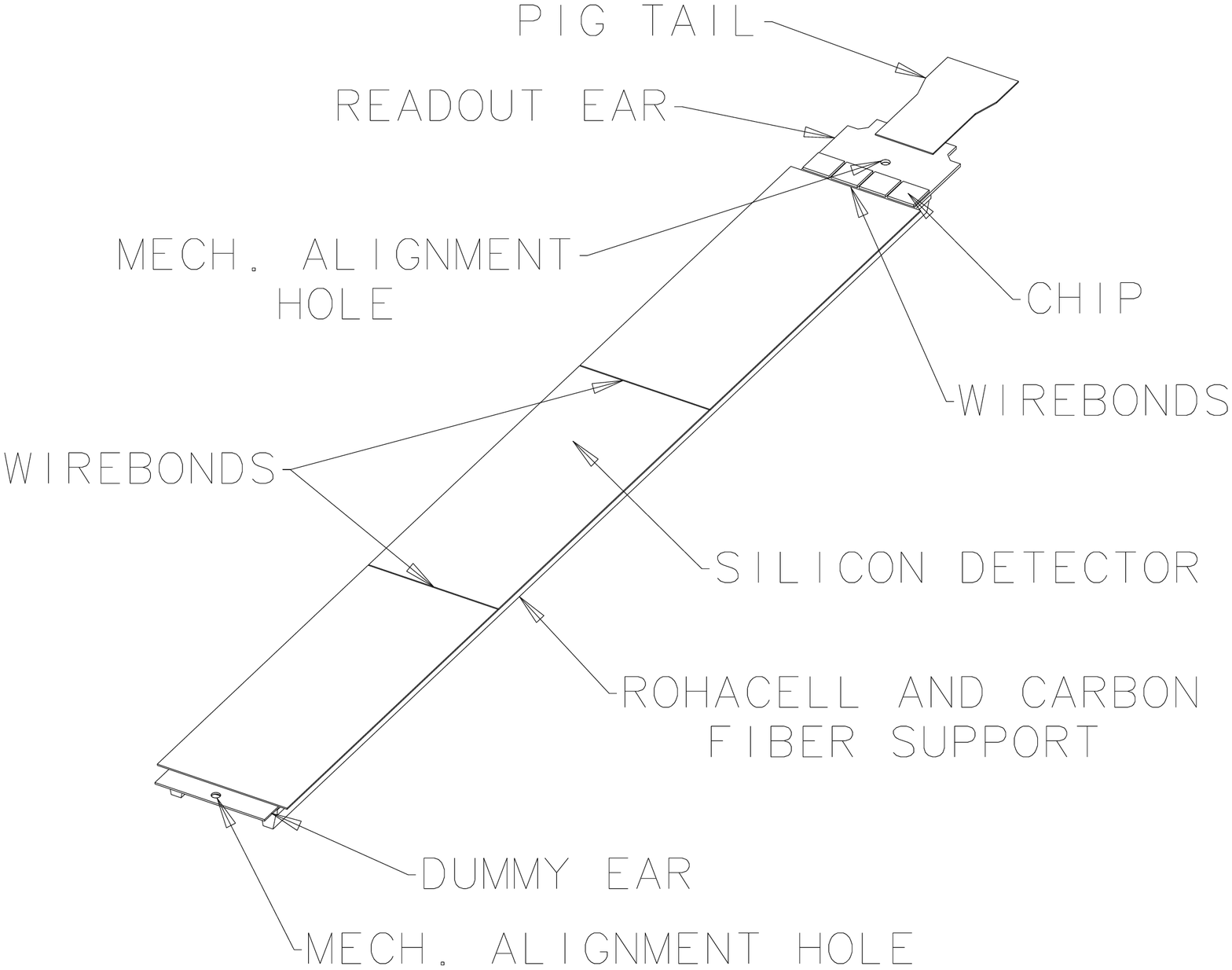}
\put(-70,45){\large\bf (a)}
\put(-7,45){\large\bf (b)}
}
\vspace*{0.2cm}
\fcaption{
(a) Isometric view of one of the SVX cylindrical modules.
(b) Sketch of a ladder detector.}
\label{svxdet}
\end{figure}

\noindent
Surrounding the 1.9~cm radius beryllium beampipe is the
CDF silicon micro-vertex detector (SVX)\cite{cdf_svx}
originally installed at CDF in 1992.
The SVX is 51~cm long and consists of two identical cylindrical modules 
which meet at $z = 0$, with a gap of 2.15~cm between them. 
A sketch of one of these modules is shown in Fig.~\ref{svxdet}a).
Both SVX modules 
consist of four layers of silicon micro-strip detectors
segmented into twelve $30^{\circ}$ wedges. The basic detector element
is called a ladder and is shown in Fig.~\ref{svxdet}b). There are 96
such ladders in the complete detector. The four layers of the SVX are
located at radii of 2.9~cm, 4.2~cm, 5.7~cm, and 7.9~cm from 
the beamline. Axial micro-strips, with a 60~$\mu$m pitch on the three
innermost layers and a 55~$\mu$m pitch on the outermost layer, 
provide precision spatial measurements in the $r\,\varphi$-plane
transverse to the beam.
The geometric acceptance of the SVX is about $60\%$ of the $p\bar p$
interactions, as it covers only $\pm~25$~cm from the nominal
interaction point, whereas the luminous region of the Tevatron beam 
has an RMS of $\sim\!30$~cm~along~$z$.

Because of radiation damage to the SVX readout chip, the performance
of the SVX deteriorated over the course of the Run\,Ia data taking
period. At the beginning of Run\,Ib, the SVX was replaced with a new
silicon vertex detector (SVX$^{\prime}$)\cite{cdf_svxp} which is
equipped with a radiation hard readout chip able to tolerate more
than 1~MRad of radiation. The 
typical charge gain of the readout chip is around 21~mV/\,f\,C, with
typical values for the noise of around 1300 electrons (10.8 ADC counts),
compared to about
2200 electrons for the original SVX detector. One improvement in noise
level resulted from the fact that SVX$^{\prime}$ is AC coupled, in
comparison to SVX which was DC coupled. This allowed the
SVX$^{\prime}$ detector to be operated in double sample and hold mode,
with one charge integration, resulting in a reduction in noise by a
factor of $\sqrt{2}$ compared to SVX which was operated in quadruple
sample \& hold mode, with two successive charge integrations 
for the determination of the outgoing~signal.

\begin{figure}[tbp]
\centerline{
\put(12,48){\large\bf (a)}
\put(75,48){\large\bf (b)}
\epsfxsize=6.3cm
\epsffile[20 145 540 690]{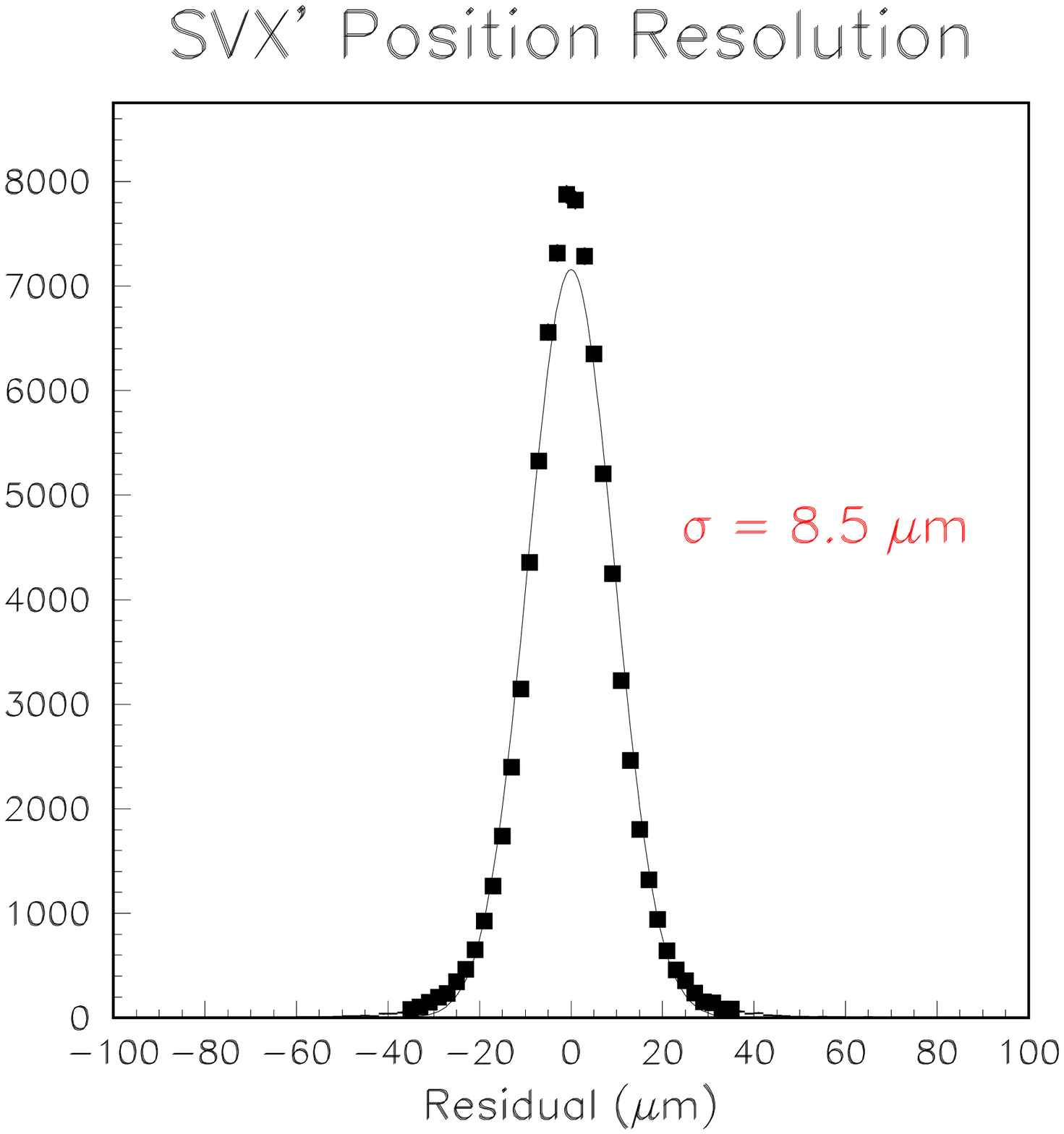}
\epsfxsize=6.3cm
\epsffile[20 155 525 645]{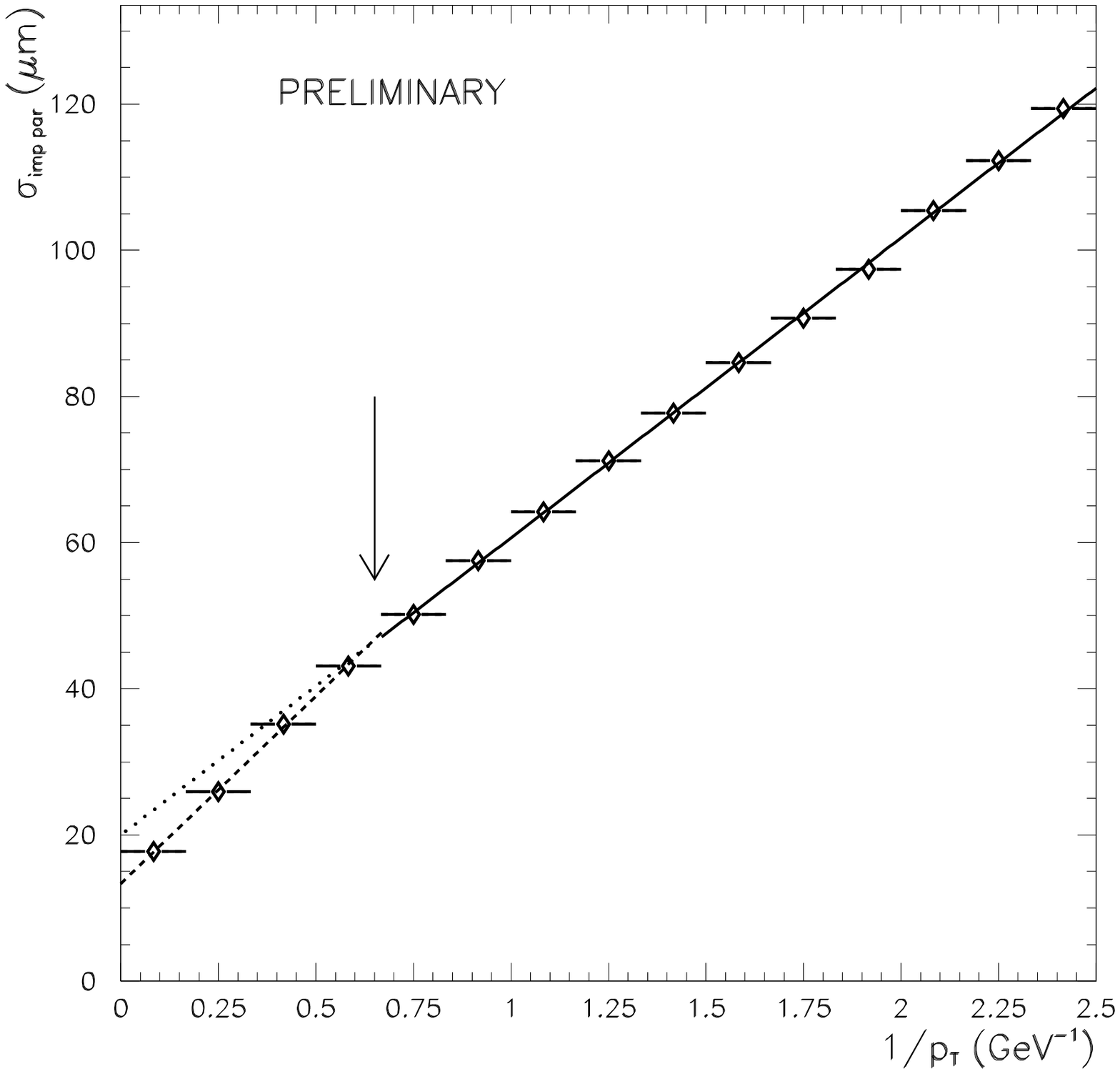}
}
\vspace*{0.2cm}
\fcaption{
(a) SVX$^{\prime}$ residual track position resolution.
(b) Track impact parameter resolution versus~$1/\Pt$.}
\label{svximp}
\end{figure}

The SVX$^{\prime}$ position resolution is found to be better than
10~$\mu$m after alignment of the detector, using track data as can
be seen in Fig.~\ref{svximp}a). Here, the residual distribution of the
distance of track intersection from reconstructed cluster centroids on
a SVX$^{\prime}$ layer is plotted. The track impact parameter resolution is
measured to be 
about $(13 + 40/\Pt)~\mu$m with \Pt\ given in \gevc. In
Figure~\ref{svximp}b), the SVX track impact parameter resolution is
shown versus $1/\Pt$. Note the contribution 
from multiple scattering at low \Pt,
while the impact parameter resolution at high \Pt\ is dominated by the
intrinsic detector resolution.
The \Pt\ resolution of the CTC combined with the SVX is 
$\sigma(\Pt)/\Pt = [(0.0066)^2 + (0.0009\,\Pt)^2]^{1/2}$, with \Pt\
measured in \gevc, while, for CTC tracks alone, the resolution is  
$\sigma(\Pt)/\Pt=[(0.0066)^2+(0.002\,\Pt)^2]^{1/2}$.

\section{Collection of \boldmath{$B$} Physics Data}
\runninghead{$B$ Lifetimes, Mixing and $CP$ Violation at CDF}
{Collection of $B$ Physics Data}
\noindent
In this section, we describe the collection of datasets used for $B$
physics analyses.
We begin with the description of the $B$ physics trigger data
sets, followed by a summary of the selection requirements used
to
identify leptons, hadrons, and jets. At the end of this section, we 
describe the Monte Carlo simulation of CDF's $B$~physics data.

\subsection{\boldmath{$B$} physics triggers} 
\noindent
The total inelastic $p\bar p$ cross section at the Tevatron is 
about three orders of magnitude larger than the $b$~production cross
section, putting certain requirements on the trigger system in terms
of finding $B$~decay products. In addition, the cross
section for $b$~quark 
production is a steeply falling cross section, as seen in
Figure~\ref{cdfbxsec}. It drops by almost two orders of magnitude
between a $b$~quark \Pt\ of about 8~\gevc\ and 25~\gevc. 
In order to find $B$~decay products in hadronic collisions, it is desirable
to go as low as possible in the decay products transverse
momentum, exploiting as much as possible of the steeply falling
$b$~cross section. Of course, the limiting factor is the bandwidth of
CDF's data acquisition system. Throughout Run\,I, the CDF
collaboration was able 
to maintain low \Pt-thresholds, without increasing the 
deadtime of the experiment during data taking.

In Run\,I, all $B$ physics triggers at CDF are based on leptons.
Inclusive single lepton triggers ($e$ and $\mu$) and dilepton triggers
(dimuon and $e\mu$) both exist. CDF uses a three-level
trigger system to identify events to be written to tape. The first two
levels are hardware based triggers. The Level~1 trigger uses
information from detector subsystems, such as the muon chambers or the
calorimeter, reducing the beam crossing rate from about 300 kHz to
about 2~kHz. At Level~2, information from different detector subsystems
is combined, such as a track stub in the CTC with hits in the muon
chambers. This reduces the trigger rate to about 30~Hz.
Level~3 is a software trigger based on the offline reconstruction code,
optimized for computational speed, resulting in an event rate to tape of
up to 10~Hz.

\subsubsection{Inclusive single lepton trigger}
\noindent
The inclusive single lepton trigger identifies events with at least
one electron or one muon, providing datasets enriched in events from
semileptonic $B$~decays.
Inclusive electrons are selected at Level~1 by the presence of a single
calorimeter tower above a threshold of 6-8~GeV depending on run
conditions, while inclusive muons require 
the presence of a track in the CMU as well as the CMP. 
At Level~2, both of these triggers demand a charged track
with $\Pt>7.5$~\gevc\ reconstructed in the $r\,\varphi$-plane of the CTC by
the central fast tracker (CFT)\cite{cft}, 
a hardware track processor, which uses fast timing information from the
CTC as input. The momentum resolution of the CFT is
$\sigma(\Pt)/\Pt^2 = 3.5\%$ with a high efficiency.
In the case of the electron trigger, this track has to be matched to a
cluster in the electromagnetic calorimeter with transverse energy
$E_{\rm t}>8.0$~GeV.
In the case of the muon trigger, this track must be
matched to a reconstructed track-segment in both the CMU and CMP.
The trigger efficiency for a single lepton turns on at $\Pt \sim 6$~\gevc, 
rises to about 50\% at a transverse momentum of 
$\sim\!8$~\gevc, and typically plateaus at over 90\% at $\Pt \sim 10$~\gevc\
depending on the exact trigger conditions.

At Level~3, a computer farm is used 
to fully reconstruct the data, including three-dimensional track
reconstruction in the CTC. However, the fast algorithm used for
tracking is only efficient 
for particles with $\Pt>1.4$~\gevc.
In the third level of the trigger, more stringent electron and muon
selection criteria, similar to those described in 
Section~4.2, are applied. During Run\,I, about $7.5 \cdot 10^6$ electron
trigger events and about $2.5 \cdot 10^6$ inclusive muon trigger events
were recorded by CDF. Table~\ref{nbtrig} details the numbers of events
recorded by the different trigger streams for the Run\,Ia and Run\,Ib
data taking periods. The numbers of events do not necessarily scale
with the integrated luminosities of Run\,Ia and Run\,Ib. This is due to
different trigger thresholds and prescale factors, which accept only
every second or third event triggered at high instantaneous luminosities,
to prevent an increased~deadtime.

\begin{table}[tbp]
\tcaption{Numbers of events recorded by the different $B$~physics
trigger streams for the Run\,Ia and Run\,Ib
data taking periods.}  
\centerline{\footnotesize\smalllineskip
\begin{tabular}{l|cc}
\hline
 Trigger Stream & Run\,Ia & Run\,Ib  \\
\hline
 & & \\
 \vspace*{-0.6cm} \\
 Single electrons & $1.9 \cdot 10^6$ & $5.7 \cdot 10^6$ \\
 Single muons & $4.1 \cdot 10^5$ & $2.1 \cdot 10^6$ \\
 Dimuons & & \\
 \ \ \ \ \ $m_{\mu\mu} < 2.8$~\gevcc\ & 
		$5.2 \cdot 10^5$ & $8.0 \cdot 10^5$ \\ 
 \ \ \ \ \ $2.8~\gevcc\ < m_{\mu\mu} < 3.4$~\gevcc\ & 
		$2.2 \cdot 10^5$ & $1.0 \cdot 10^6$ \\ 
 \ \ \ \ \ $3.4~\gevcc\ < m_{\mu\mu} < 4.1$~\gevcc\ & 
		$2.5 \cdot 10^5$ & $7.4 \cdot 10^5$ \\ 
 \ \ \ \ \ $8.5~\gevcc\ < m_{\mu\mu} < 11.3$~\gevcc\ & 
		$8.1 \cdot 10^4$ & $2.8 \cdot 10^5$ \\ 
 \ \ \ \ \ $m_{\mu\mu} > 4.0$~\gevcc\ & 
		$5.1 \cdot 10^5$ & $1.9 \cdot 10^6$ \\ 
 Electron-muons & $2.4 \cdot 10^5$ & $5.0 \cdot 10^5$ \\ 
\hline
\end{tabular}}
\label{nbtrig}
\end{table}

\begin{figure}[tbp]
\centerline{
\put(31,51){\large\bf (a)}
\put(73,51){\large\bf (b)}
\put(115,51){\large\bf (c)}
\epsfxsize=4.2cm
\epsffile[15 10 380 510]{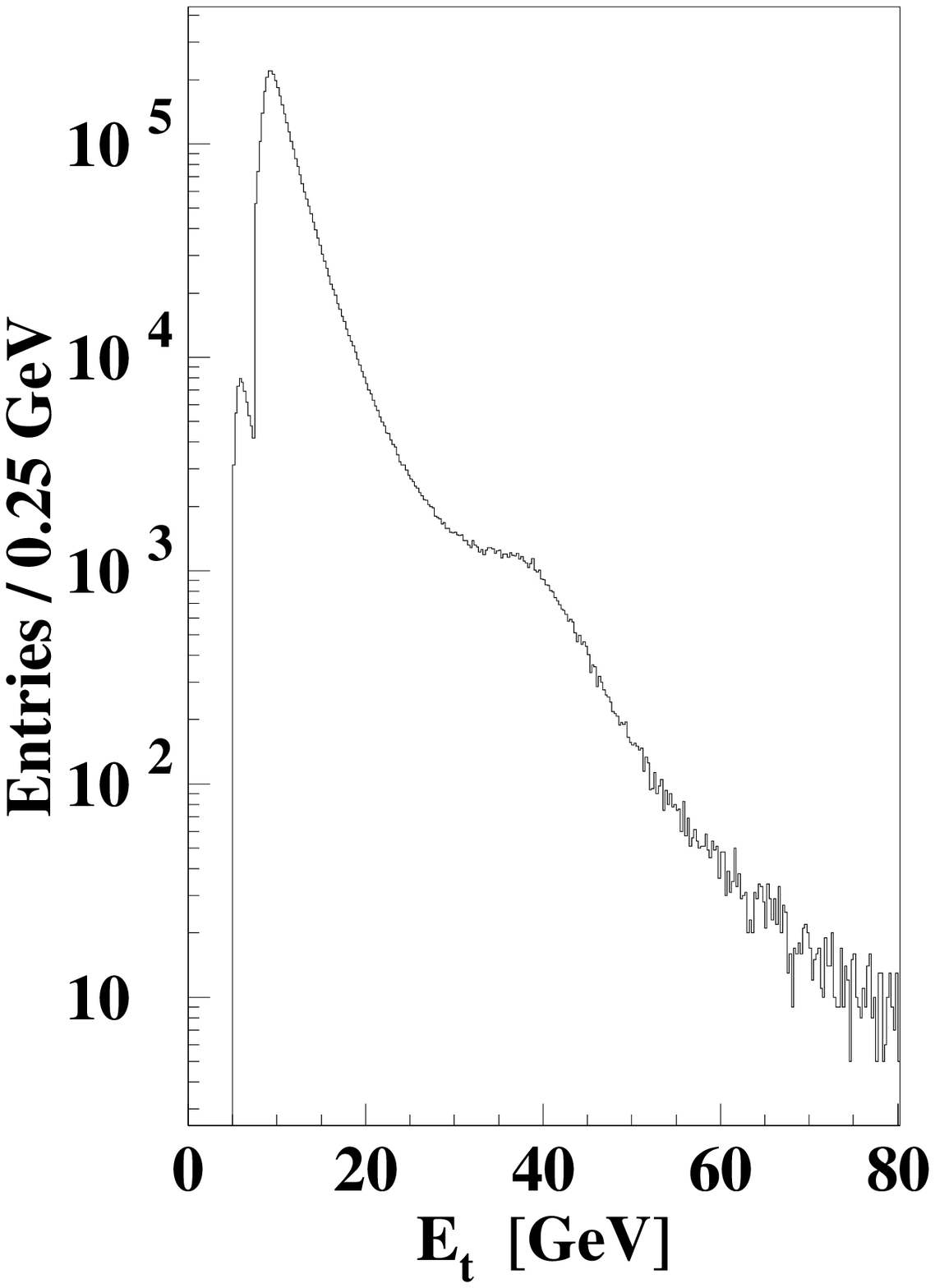}
\epsfxsize=4.2cm
\epsffile[15 10 380 510]{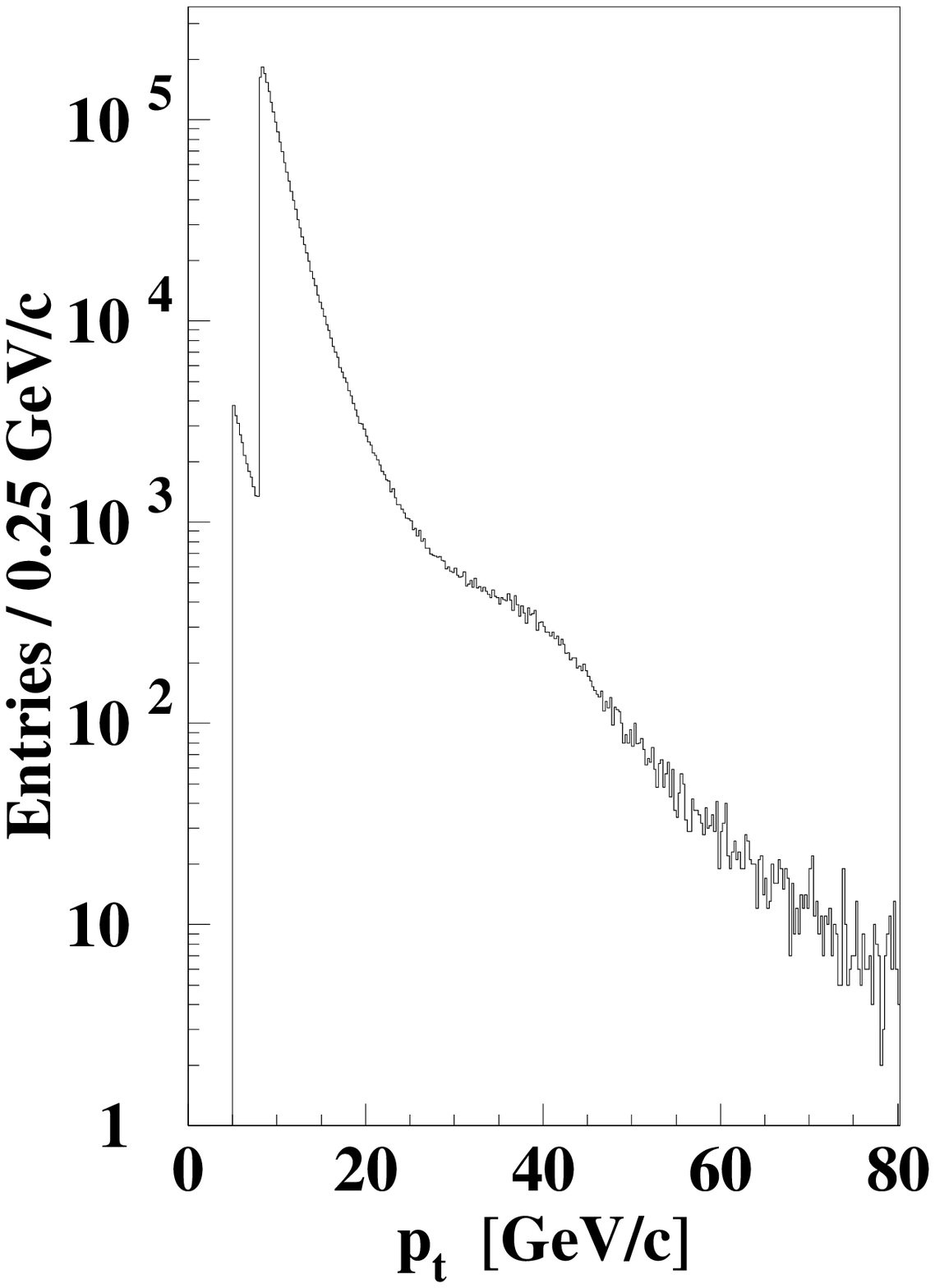}
\epsfxsize=4.2cm
\epsffile[15 10 380 510]{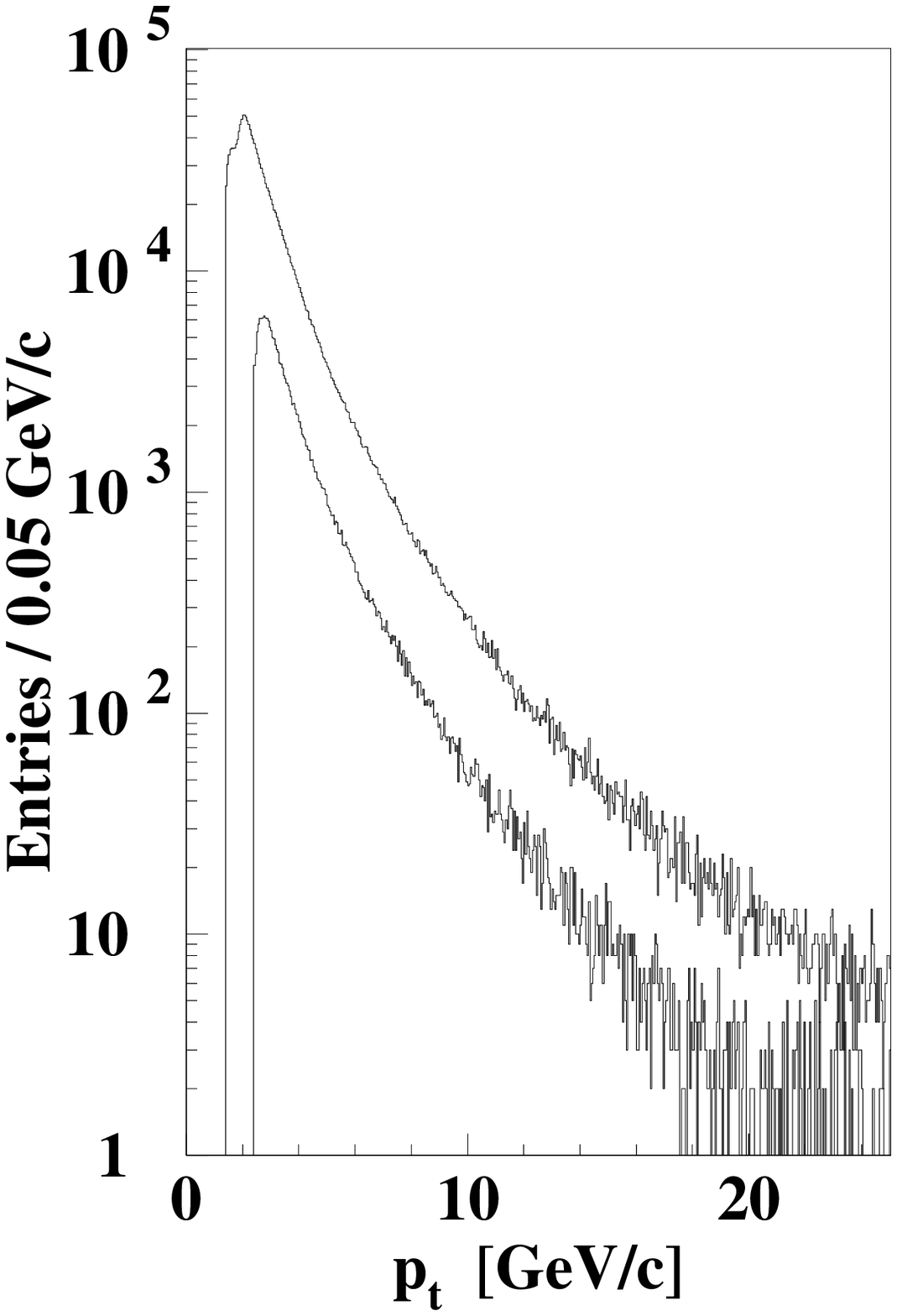}
}
\vspace*{0.2cm}
\fcaption{Distribution of 
(a) transverse energy spectrum of inclusive single electrons and \Pt\
spectrum of (b) inclusive single muons and (c) muons from dimuon
trigger data. The upper curve in (c) represents Run\,Ib events while
the lower distribution is obtained from Run\,Ia data.} 
\label{trigpt}
\end{figure}

As an example, Figure~\ref{trigpt}a) shows the transverse energy spectrum of
electrons from the single electron triggers recorded in Run\,Ib. The
trigger turn-on is visible at $\Et \sim 8$~GeV followed by an
exponentially falling energy spectrum. The shoulder around 40~GeV originates
from a different physics process. These are electrons from $W$ boson
decays $W \ra e \nu$, demonstrating how large the $b$~cross section is
compared to other physics processes like $W$~production. The
distribution in Fig.~\ref{trigpt}a) is obtained by identifying
electrons with $\Et > 5$~GeV in single electron trigger events. The
enhancement of electrons at 5~GeV below the trigger turn-on results
from additional volunteer electrons in these events, which are not the
trigger electrons. For comparison, Figure~\ref{trigpt}b) shows the
\Pt-spectrum of inclusive single muons. 

\subsubsection{Dimuon trigger}
\begin{figure}[tbp]
\vspace*{0.1cm}
\centerline{
\put(10,7){\large\bf (a)}
\put(73,7){\large\bf (b)}
\hspace*{0.4cm}
\epsfxsize=5.6cm
\epsffile[57 58 367 368]{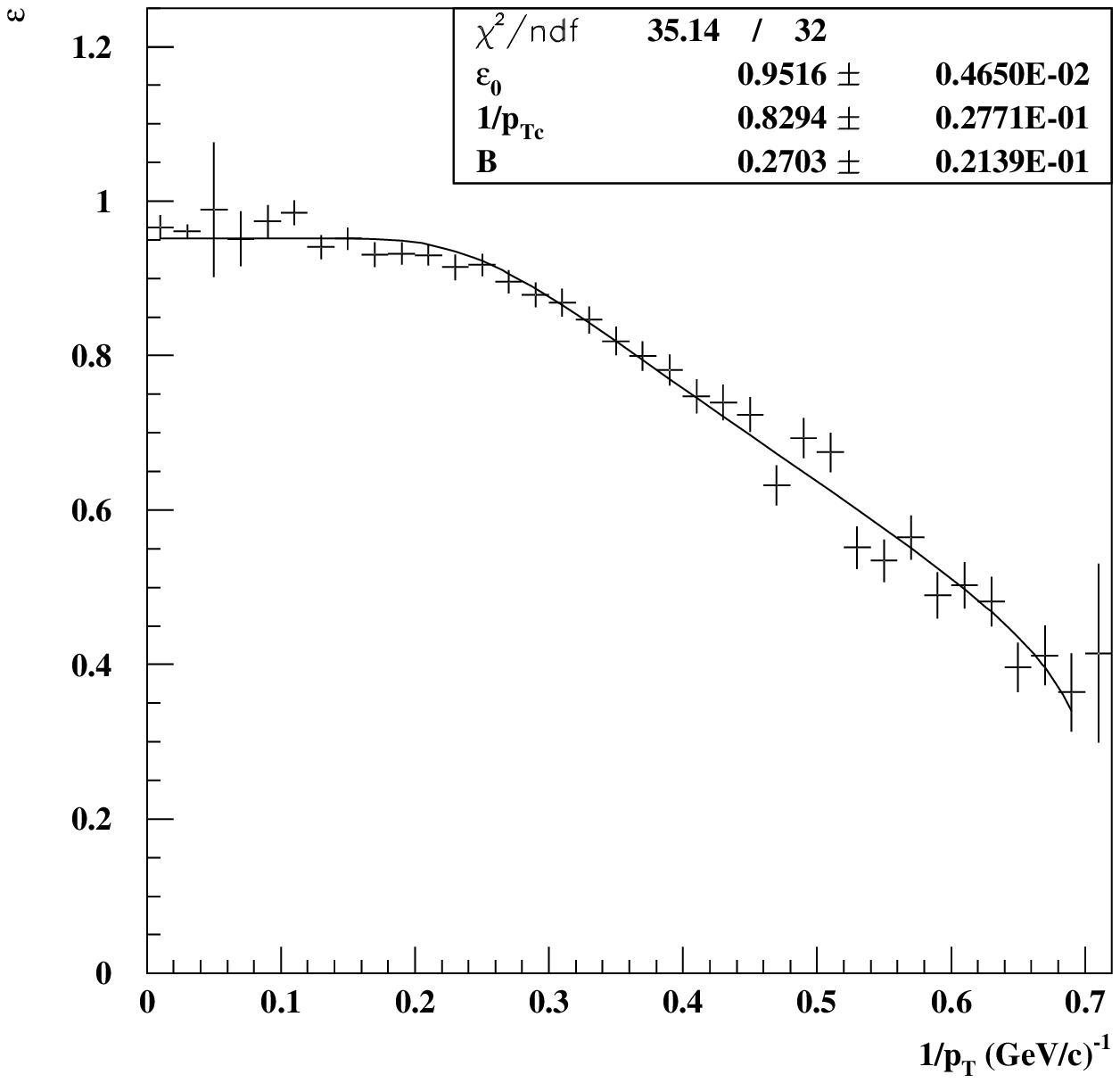}
\hspace*{0.6cm}
\epsfxsize=5.6cm
\epsffile[57 58 410 411]{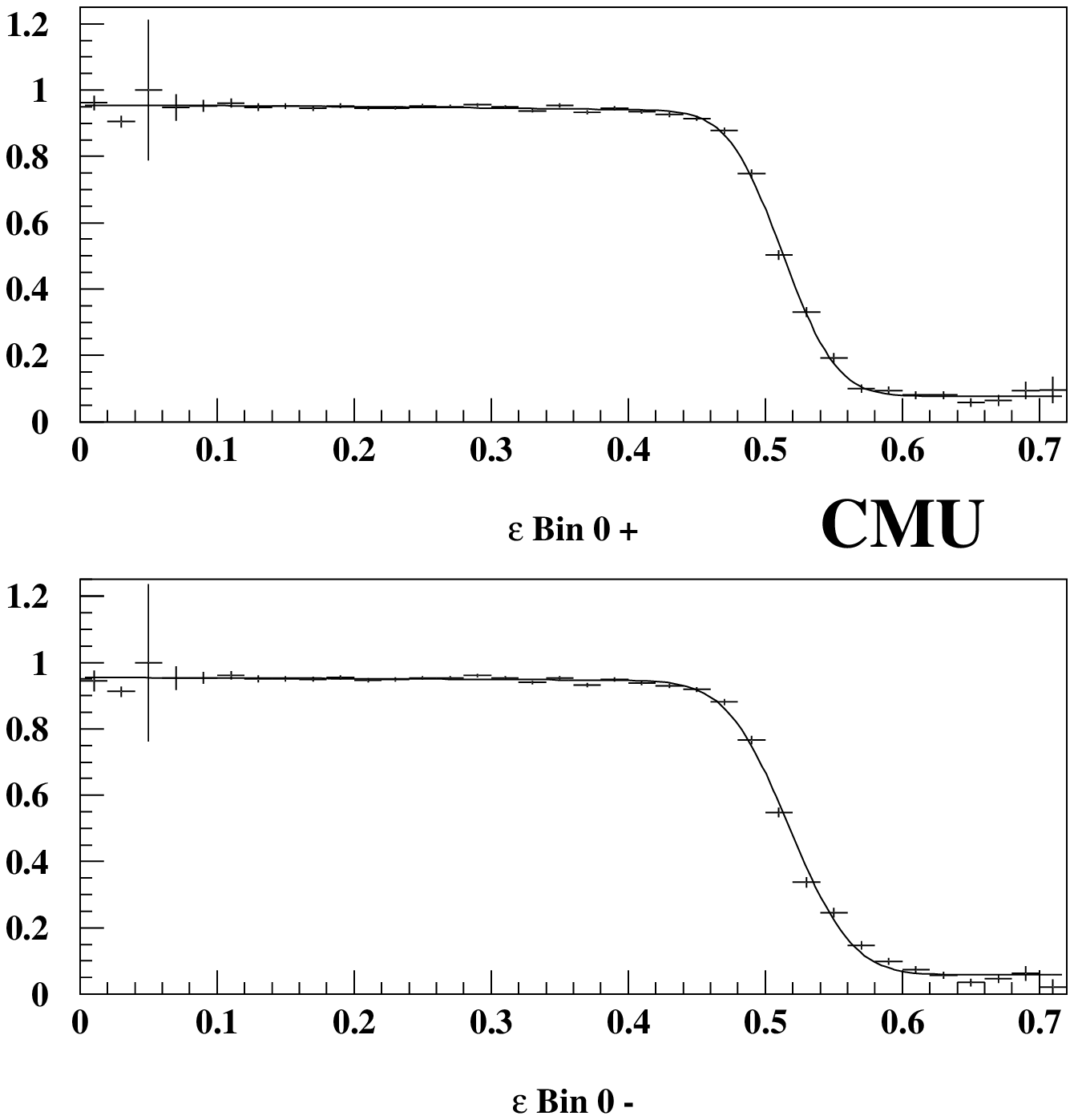}
}
\vspace*{0.7cm}
\fcaption{(a) Level~1 and (b) Level~2 trigger turn-on efficiency for CMU 
muons plotted versus $1/\Pt$ using $J/\psi \ra \mu\mu$ and $Z^0 \ra
\mu\mu$ data. The Level~2 efficiency is displayed for positive (top)
and negative (bottom) muons separately.}
\label{trigturnon}
\end{figure}

\noindent
The collection of dimuon trigger data requires two muon
candidates be observed in the muon system at Level~1. 
The trigger efficiency for a muon at Level~1 rises from about 50\% at
$\Pt = 1.6$~\gevc\ to 90\% at $\Pt \sim 3.1$~\gevc\ with a plateau of
$\sim\!95\%$. As an illustration of the trigger turn-on,
Figure~\ref{trigturnon}a) shows the Level~1 efficiency for CMU 
muons plotted versus $1/\Pt$ using $J/\psi \ra \mu\mu$ and $Z^0 \ra
\mu\mu$ data. The range-out at low momenta is predicted
to occur for $\Pt < 1.4$~\gevc. 
The second level trigger
requires that at least one of the muon tracks is matched in $\varphi$
to a track found by the CFT. The efficiency for
finding a track with the CFT rises from 50\% at \Pt\ of about
1.9~(2.6)~\gevc\ to 90\% at $\Pt \sim 2.2$~(3.1)~\gevc\ and
reaches a plateau of $\sim\!94\%$ ($93\%$). The numbers given in
brackets refer to the Run\,Ia settings where the dimuon trigger was
operated with a slightly higher momentum threshold.
Figure~\ref{trigturnon}b) shows the Level~2 efficiency for CMU 
muons plotted versus $1/\Pt$ again using $J/\psi \ra \mu\mu$ and $Z^0 \ra
\mu\mu$ data. The Level~2 efficiency is displayed for positive and negative
muons separately, indicating no charge dependence of the L2~trigger turn on.

At Level~3, the events are again fully reconstructed, requiring two
CTC tracks to be matched 
with two tracks in the muon chambers. For offline reconstruction the
dimuon data are split into different samples, according to the
dimuon invariant mass, as detailed in Table~\ref{nbtrig} which again shows
the numbers of events separately for the Run\,Ia and Run\,Ib
data taking periods. The data with 
$2.8 < m_{\mu\mu} < 3.4$~\gevcc\ are the source of CDF's 
$J/\psi \ra \mu^+\mu^-$ events, while the data with
$3.4 < m_{\mu\mu} < 4.1$~\gevcc\ contain $\psi(2S) \ra
\mu^+\mu^-$ events. Dimuon decays of the $\Upsilon$ resonances are
contained in the sample with
$8.5 < m_{\mu\mu} < 11.3$~\gevcc.
Events with $m_{\mu\mu} <2.8$~\gevcc\ are a source
of double-semileptonic decays $b \ra c\, \mu X$ followed by $c \ra \mu
X$, while the high-mass dimuon stream with $m_{\mu\mu} > 4.0$~\gevcc\
provides events where the two muons originate from the decay of a
$b$ and $\bar b$~quark.  

As a further illustration, Figure~\ref{trigpt}c) shows the
\Pt-spectrum of muons from the dimuon trigger data. 
The upper curve in Fig.~\ref{trigpt}c) represents Run\,Ib events, while
the lower distribution is obtained from Run\,Ia data. 
The higher muon momentum threshold in Run\,Ia is clearly visible and
the trigger turn-on is softer compared to the single muon events 
(see Fig.~\ref{trigpt}b).

\subsubsection{Electron-muon trigger}
\noindent
CDF also triggered on events containing an electron and a muon. The
trigger is a combination of the single electron and muon
triggers. This sample has a principal electron $E_{\rm t}$ threshold
of about 5~GeV and a principal muon \Pt~threshold of about 3.0~\gevc\
for Run\,Ia data and $\sim\!2.5~\gevc$ for Run\,Ib. Because of the higher
lepton threshold, the $e\mu$-data comprise a significantly smaller
dataset compared to the dimuon data as shown in Table~\ref{nbtrig}. 

\subsection{Lepton identification}
\subsubsection{Electron identification}
\noindent
The identification of electron candidates reconstructed after data
collection uses information from both the
calorimeters and the tracking chambers. Here, we describe typical
selection criteria used for electron identification at CDF. The
longitudinal shower profile must be consistent with 
an electron
shower with a leakage energy from the CEM into the CHA of less than 4\%,
if one track is pointing to the calorimeter tower, or less than 10\% if
more than one track is pointing to the calorimeter tower. The
lateral shower profile of the CEM cluster has to be 
consistent with that from test beam electrons. 
Additionally, a $\chi^2$ comparison of the CES shower profile with that
of test beam electrons has to result in $\chi^2 < 10$.
For the association of a single high \Pt\
track with the calorimeter shower, based on the position matching at
the CES plane, it is required that $r|\Delta \varphi| < 1.5$ cm and
$|\Delta z\sin \theta| < 3$~cm. In addition, we demand the $E_{\rm T}$
of the electron candidate, reconstructed offline, to be greater than
6~GeV (5~GeV) for single electron ($e\mu$) data. Usually  
electron candidates from photon conversion,
are removed from the analysis by looking for
oppositely charged 
tracks which have a small opening angle with the electron
candidate. 

\subsubsection{Muon identification}
\noindent
The typical reconstruction of muon candidates at CDF is very similar
for the single 
muon and dimuon datasets. Usually, a $\chi^2$ characterizing the
separation between the track segment in
the muon chamber and the extrapolated CTC track is computed, where
the uncertainty in this $\chi^2$ variable is dominated by
the multiple scattering in the detector material. Typical selection
requirements are $\chi^2 < 9$
in the $r\,\varphi$-view (CMU, CMP, and CMX) and $\chi^2 < 12$ in the
$r$-$z$ view (CMU and CMX). A minimum energy deposition of 0.5~GeV in
the hadron calorimeter is required for a muon candidate, as expected
for a minimum ionizing track.
The transverse muon momentum
reconstructed offline is usually required to be greater than 7.5~\gevc\ for
inclusive muons. For the
dimuon sample, this cut is typically placed at $\Pt > 2$~\gevc\ for
each muon candidate. 

\subsection{Charged particle selection} 
\noindent
Charged particles are identified as tracks in CDF's central tracking
chamber. To ensure a good track reconstruction, quality cuts
requiring a minimum number of hits in the CTC are usually imposed.
Typical requirements are at least five hits in
two or more CTC axial superlayers and at least two hits in more than
one CTC stereo superlayer. In addition, tracks are often required to be
reconstructed in the SVX with hits in at least three out of the four silicon
layers. In many analyses, a $\chi^2$ is calculated and defined as
the increase in the track fit $\chi^2$, when the SVX hits are included
in the CTC track fit. This $\chi^2$ divided by the number of
SVX hits usually has to be less than six to reject badly
measured tracks. The track quality requirements are generally applied to
hadrons as well as leptons. 
For optimal vertex 
resolution, hadron and lepton candidate tracks are often required
to be reconstructed in the SVX detector. 

\subsubsection{Hadron selection} 
\noindent
Usually, all reconstructed tracks are assigned the desired particle
hypotheses, due to the lack of a sophisticated particle identification at
CDF. Applying cuts on the tracks transverse momenta are often the only
means to reduce combinatorial background. The minimum \Pt\ cut on
tracks is usually set at 0.4~\gevc. However, the specific
ionization information d$E$/d$x$ from the CTC 
provides a $\pi/K$ separation of about 
one~$\sigma$ for tracks with \Pt\ greater than about 1.2~\gevc.
Therefore, d$E$/d$x$ information from the CTC is sometimes used to
help identify hadrons. Because of the large
Landau tail of the ionization distribution, the 80\% truncated mean
of the measured charges from the CTC sense wires is
taken as the best estimator of the track d$E$/d$x$. The probabilities,
$P(i)$, for
a track to be consistent with the $i = e$, $\mu$, $\pi$, $K$, or $p$
hypotheses are then calculated using the measured d$E$/d$x$ value and
the predictions for the assumed particle hypotheses. A likelihood ratio, 
$\ell h_{{\rm d}E/{\rm d}x}^K$, 
for a track being, for example, a
kaon is defined to be the ratio of $P(K)$ divided by the
sum of the probabilities of all particle hypotheses. 
Usually, this likelihood ratio is required to be greater than 0.01.
  
The kinematic selection criteria used in a particular
analysis are often optimized by maximizing the quantity 
$N_S / \sqrt{N_S + N_B}$,  
where $N_S$ is the predicted number of signal events based on Monte
Carlo calculations (see Sec.~4.5) and $N_B$ is the observed
number of background events estimated e.g. from the signals sideband regions.

\subsection{Jet reconstruction}
\noindent
Some analyses require the reconstruction of jets which are usually
formed from charged particles in $B$ physics analyses, rather than from
the more commonly used calorimeter clusters. A jet in a $B$~event is
often of low momentum and its energy is therefore not very precisely
measured in the calorimeter. In addition, there is a
difference between jets associated with electrons compared to jets
containing a muon, since electrons deposit much more energy in the
calorimeter than muons. If calorimeter clusters were used to find jets,
electrons would be associated with jets much more often than muons for any
minimum jet energy requirement. Therefore, jets are formed in
$B$~physics analyses from tracks using a cone clustering algorithm.
Usually, tracks with $\Pt > 1.0$~\gevc\ are used as jet seeds. If two
seeds are within a cone with  
radius $\Delta R = \sqrt{(\Delta\eta)^{2} + (\Delta\varphi)^{2}} < 0.7$, 
the momenta of the seeds are added together to form a new seed. After
all possible seed merging, lower momentum tracks (usually $0.4 <
\Pt < 1.0~\gevc$) that are within $\Delta R < 0.7$ of a seed are added
to form the final jet. A jet can, in principle, consist of a single
track with $\Pt > 1$~\gevc.

\subsection{Simulation of heavy flavour production and decay}
\noindent
Monte Carlo (MC) generation programs are used to simulate heavy flavour
production in $p\bar p$ interactions. These programs are
supplemented by phenomenological models of hadronization and decays of
unstable particles. They are usually used in conjunction with programs that
simulate the detector response to the final state
particles. MC programs are necessary
tools to calculate the geometrical and kinematic acceptance for
observing heavy flavour events and are also used to estimate 
backgrounds for particular analyses.
While considerable effort has gone into developing and tuning simulations
for $e^{+}e^{-}$ collisions, the state of the art is somewhat
less well developed for the more complex high energy hadron-hadron
collisions.

At CDF, two types of Monte Carlo simulations are used.
Calculations depending only on the production and decay
of $B$~hadrons employ
a Monte Carlo generator that simulates only a single 
$B$~hadron and its decay products.
Situations which depend upon the fragmentation particles
resulting from the hadronization of the $b$~quark,
as well as the ``underlying event'' particles from the $p\bar p$
scattering, use a full event generator like the PYTHIA
program package\cite{pythia}. In the following, we describe both Monte
Carlo generators as well as the CDF detector simulation.

\subsubsection{Simulation of a single $B$ hadron}
\noindent
The simulation of a single $B$ hadron begins with a model of $b$~quark
production based on a next-to-leading order QCD
calculation\cite{nloqcd}. This calculation employs the MRSD0 parton
distribution function\cite{mrsd0} with $m_b = 4.75$~GeV, 
$\Lambda = 215$~MeV, and a renormalization scale of 
$\mu = \mu_0 = \sqrt{\Pt^2 + m_b^2}$
to model the kinematics of the
initial state partons. Usually, $b$~quarks are generated in the rapidity
interval $|y_b| < 1.0$ with a minimum \Pt\ for the $b$~quark 
chosen in a way to avoid any biases in the $B$~meson kinematic
distributions after the application of the kinematic cuts used in an 
analysis. The $b$~quarks are then fragmented into $B$~hadrons,
with no additional hadronization products, according to
a model using the Peterson fragmentation function\cite{Peterson} with
a Peterson parameter of $\epsilon_b = 0.006$ (see Sec.~2.3).
The bottom and charm hadrons are decayed into the various final
states using branching ratios and decay kinematics governed by the
world average masses and 
lifetimes of the involved particles\cite{PDG}. The decay of bottom
hadrons is accomplished using the QQ~program\cite{cleoqq} 
developed by the CLEO Collaboration, extended to include \Bs~mesons
and $b$~baryons.  

\subsubsection{Monte Carlo simulation of the whole event using PYTHIA}
\noindent
The PYTHIA Monte Carlo generator\cite{pythia} 
is used in instances 
where more than just a single decaying $B$ meson is required.
PYTHIA simulates a complete $p\bar{p}$ interaction: The $b\bar{b}$ pair, 
the hadronization products, and the remaining beam fragments 
from the $p\bar p$ scattering (``underlying event'').  
PYTHIA exploits an improved string fragmentation model
tuned to experimental data, mostly from high energy $e^+e^-$ collisions.
PYTHIA generation at CDF uses most of the typical default parameters
including the CTEQ2L\cite{cteq2l} parton distribution functions.
The $b$ quarks are again fragmented using the Peterson fragmentation
model\cite{Peterson}. 
However, the actual $B$ decay performed 
by PYTHIA is suppressed and instead, the QQ program\cite{cleoqq} is invoked.

The PYTHIA generator is controlled by a series 
of parameters whose default values are adjusted to achieve
good agreement with primarily high energy $e^+e^-$ data.
Discrepancies between the ``default'' PYTHIA 
and CDF $p\bar{p}$ data 
are apparent, especially when considering particle production
that does not originate from the $b$ hadronization but from 
the ``underlying event''.
The fidelity of the  ``default'' PYTHIA 
generator is studied 
by comparing generated semileptonic $B \ra D \ell X$ Monte Carlo data after
detector simulation to real CDF data. 
This comparison studied track multiplicities 
in $\Delta R$ and $\Delta \varphi$ intervals 
around the $B \ra D \ell$ direction and
found the data to have a higher
multiplicity of underlying event tracks 
than PYTHIA predicts.
A good description of the charged particle
multiplicities and \Pt~distributions is obtained
by adjusting several PYTHIA parameters.
The properties of multiple interactions and
beam remnants are controlled primarily through the 
multiple interaction cross section [PARP(31)], the model for
their generation [MSTP(82)], the ratio of $gg$ and
$q\bar q$ multiple interactions [PARP(85,86)], and the
width of the Gaussian $p_T$ spread  of
particles produced in the breakup of color strings [PARJ(21)].
Once these parameters are adjusted to obtain agreement
with the data away from the $b$~jets, 
it is assumed that the underlying event is well modeled.
In addition, the Peterson parameter PARJ(55) is modified
so that the generated multiplicity of tracks inside 
a cone of $|\Delta R| < 1$ around the $b$~quark matches the observed one.
Table~\ref{tunepythia} lists the default and tuned values 
of the relevant PYTHIA parameters.  More details of the tuning
procedure may be found in Ref.\cite{sst_prd}.

\begin{table}[tbp]
\tcaption{
The PYTHIA Monte Carlo parameters modified from their
defaults to agree with $B \ra D \ell X$ data at CDF.}
\centerline{\footnotesize\smalllineskip
\begin{tabular}{r|cc|l}
\hline
 Parameter & Default & Tuned & Description\\ 
\hline
 MSTP(82)  &1        &3      & model of multiple interactions\\
 PARP(85)  &0.33     &1.0    & fraction of color-connected
 	$gg$ multiple interactions \\
 PARP(86)  &0.66     &1.0    & total fraction of $gg$
 	multiple interactions\\
 MSTP(33)  &No       &Yes    & multiply cross sections by PARP(31)\\
 PARP(31)  &1.00     &1.66   & increase cross sections by 66\%\\
 PARJ(21)  &0.36     &0.6    & $\sigma^{\rm frag}(\Pt)$\\
 MSTJ(11)  &4        &3      & use Peterson fragmentation for $b, c$\\
 PARJ(55)  &0.006    &0.0063 & Peterson fragmentation parameter $\epsilon_b$\\
\hline
\end{tabular}}
\label{tunepythia}
\end{table}

\subsubsection{Detector simulation}
\noindent
The output of the MC generation is often simply passed 
through the standard fast simulation of the CDF detector 
which is based on parametrizations and simple models
of the detector response determined from data or test beam measurements. 
The detector response is often parametrized as a function of the
particle kinematics. 
The inclusive lepton or dilepton trigger usually introduces a strong 
kinematic bias on an analysis. This bias must be well modeled 
in the MC simulation to obtain the proper relative reconstruction efficiencies.
Often, an empirical approach is taken rather than simulating
the trigger directly.
The trigger is either modeled by a simple parametrization of
the CFT trigger, depending on the lepton \Pt, or by
an error function parametrization of the ratio of 
the observed lepton \Pt\ distribution in the data compared
to that generated by the simulation. 
For example, the efficiency parametrization of the single electron
trigger follows the functional form
\begin{equation}
A \cdot {\rm freq}\left(  \frac{ \Pt - B_1  }{ C_1 } \right)
         \cdot {\rm freq}\left(  \frac{ \Pt - B_2  }{ C_2 } \right),
\end{equation}
where $A = 0.927$, $B_1 = 6.18$ GeV/$c$, $C_1 = 4.20$,
$B_2 = 7.48$ GeV/$c$, $C_2 = 0.504$, and freq is the normal frequency function.
After the simulation of the CDF detector, the same selection criteria
applied to the data are usually imposed on the Monte Carlo events.

\section{Features of \boldmath{$B$} Physics at a Hadron Collider: A Brief Tour}
\runninghead{$B$ Lifetimes, Mixing and $CP$ Violation at CDF}
{Features of $B$ Physics at a Hadron Collider: A Brief Tour}
\noindent
In this section, we highlight some of the features of $B$~physics at a
hadron collider. We also make an attempt to describe, in an
illustrative way, how CDF studies $B$~decays. We will emphasize some of
the tools used to find  
$B$~decay products in hadronic collisions, omitting
technical details which may be found later.

\begin{table}[tbp]
\tcaption{
Comparison of important features of different experiments studying
$B$~physics.}  
\centerline{\footnotesize\smalllineskip
\begin{tabular}{c|ccc} 
\hline
 & & & \\
 \vspace*{-0.6cm} \\
 & $e^+e^- \ra \Upsilon(4S) \ra B^{ }\bar B$ & $e^+e^- \ra Z^0 \ra b\bar b$ 
 & $p\bar p \ra b\bar b X$ \\ 
\hline 
 & & & \\
 \vspace*{-0.6cm} \\
Accelerator & CESR, DORIS & LEP, SLC & Tevatron \\
Detector & ARGUS, CLEO & ALEPH, DELPHI & CDF, D\O \\
 & & L3, OPAL, SLD &  \\
$\sigma(b\bar b)$ & $\sim\!1$ nb & $\sim\!6$ nb & $\sim\!50~\mu$b  \\
$\sigma(b\bar b):\sigma(had)$  & 0.26 & 0.22 & $\sim\!0.001$  \\ 
$B^0,\ B^+$ & yes & yes & yes \\
$\Bs,\ B_c^+,\ \Lambda_b^0$ & no & yes & yes \\
Boost $<\beta\gamma>$  & 0.06 & 6 & $\sim$\,2\,-\,4 \\
$b\bar b$ production  & both $B$ at rest & $b\bar b$ back-to-back & 	
	$b\bar b$ not back-to-back \\ 
Multiple events   & no & no & yes  \\ 
Trigger  & inclusive & inclusive & leptons only  \\ 
\hline
\end{tabular}}
\label{bproducers}
\end{table}

To set the stage, we first compare current
producers of $B$~hadrons. Table~\ref{bproducers} summarizes some of
the important features of $B$~physics experiments and the accelerators
at which they operate. There are three main approaches to produce
$B$~hadrons. First, 
$e^+e^- \ra \Upsilon(4S) \ra B\bar B$ at the CLEO experiment, located at the
CESR storage ring at Cornell and the DORIS storage ring at DESY, where
the ARGUS experiment operated until 1993. Second,
$e^+e^- \ra Z^0 \ra b\bar b$ at the four experiments ALEPH,
DELPHI, L3, and OPAL, located at the LEP storage ring at CERN, as well as the
SLD detector at the SLC   
Collider at SLAC. Finally, $p\bar p \ra b\bar b X$ at the Tevatron
where the CDF and D\O\ detectors are operated. 
As discussed earlier, the main motivation for studying $B$ physics at
a hadron collider is the large $b$~quark cross section 
(see Tab.~\ref{bproducers}). 

\begin{figure}[tbp]
\centerline{
\epsfxsize=6.3cm
\epsffile{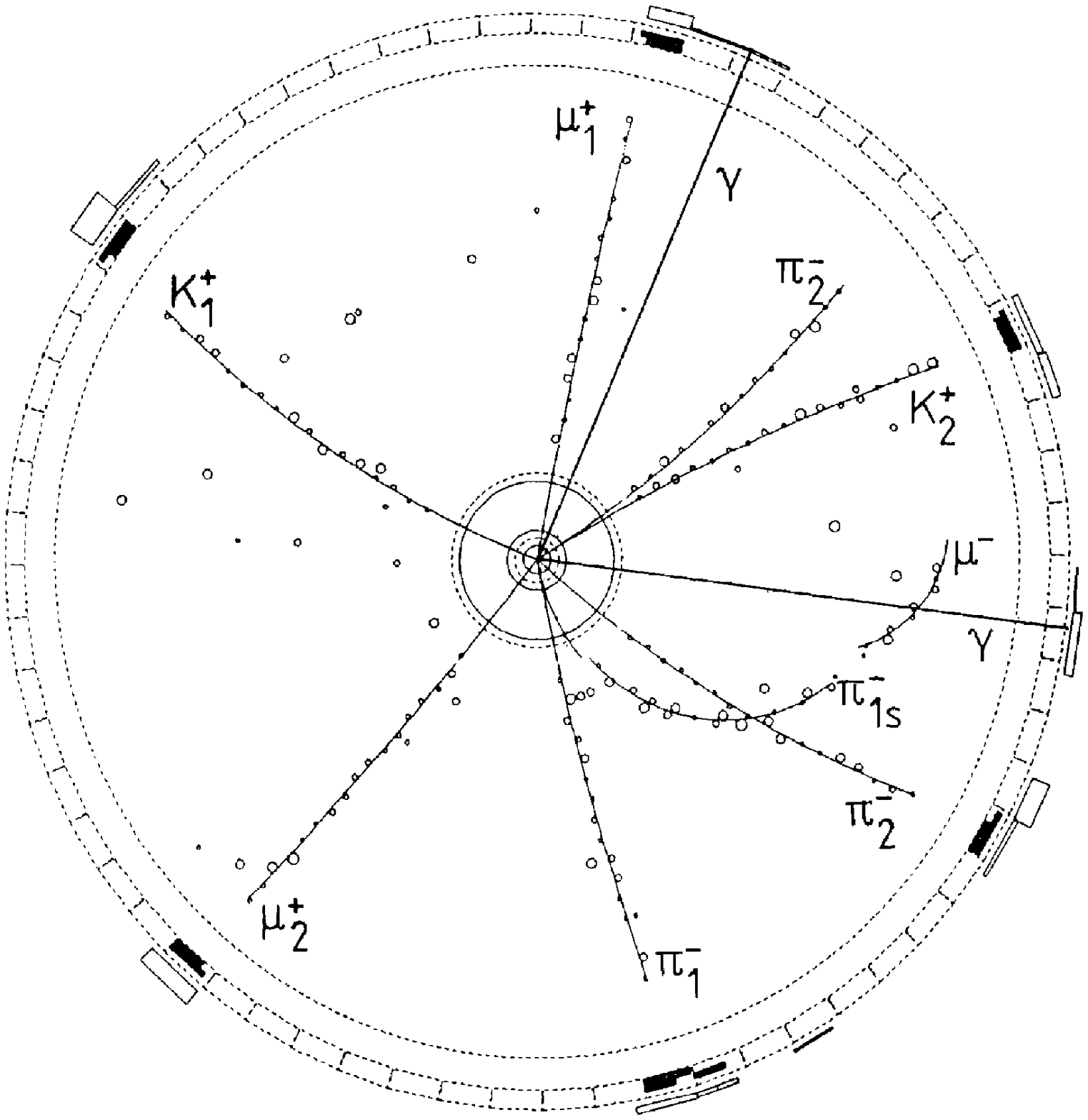}
\epsfclipon
\epsfxsize=6.3cm
\epsffile[70 140 550 630]{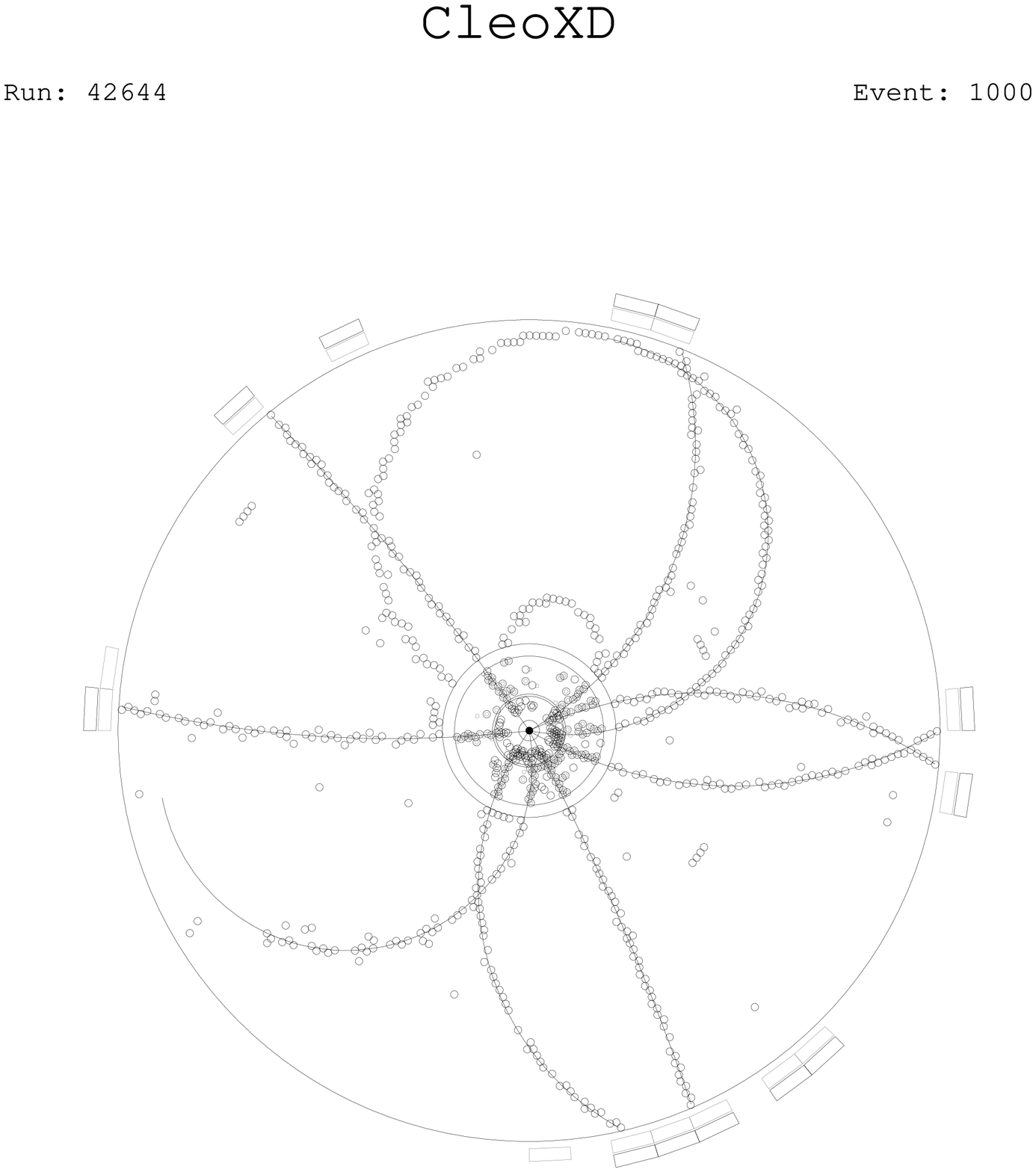}
\epsfclipoff
}
\vspace*{0.2cm}
\fcaption{
A typical $B$ event at the $\Upsilon(4S)$ in the $r\,\varphi$-view from 
the ARGUS experiment (left) and the CLEO detector (right).}
\label{evt_ups}
\end{figure}

\begin{figure}[tbp]
\centerline{
\epsfysize=5.5cm
\epsffile[20 20 570 570]{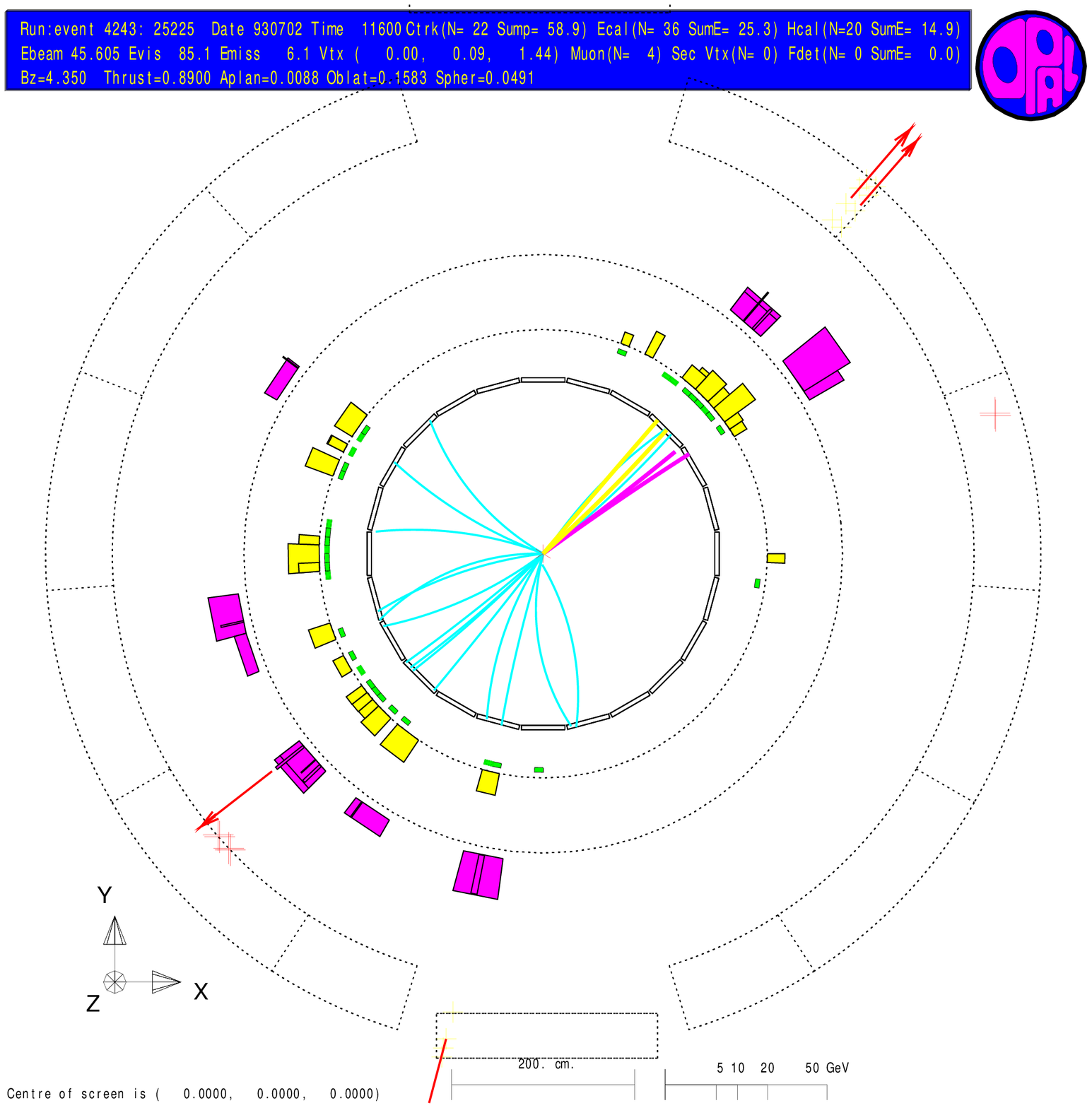}
\epsfysize=5.5cm
\epsffile[1 35 710 570]{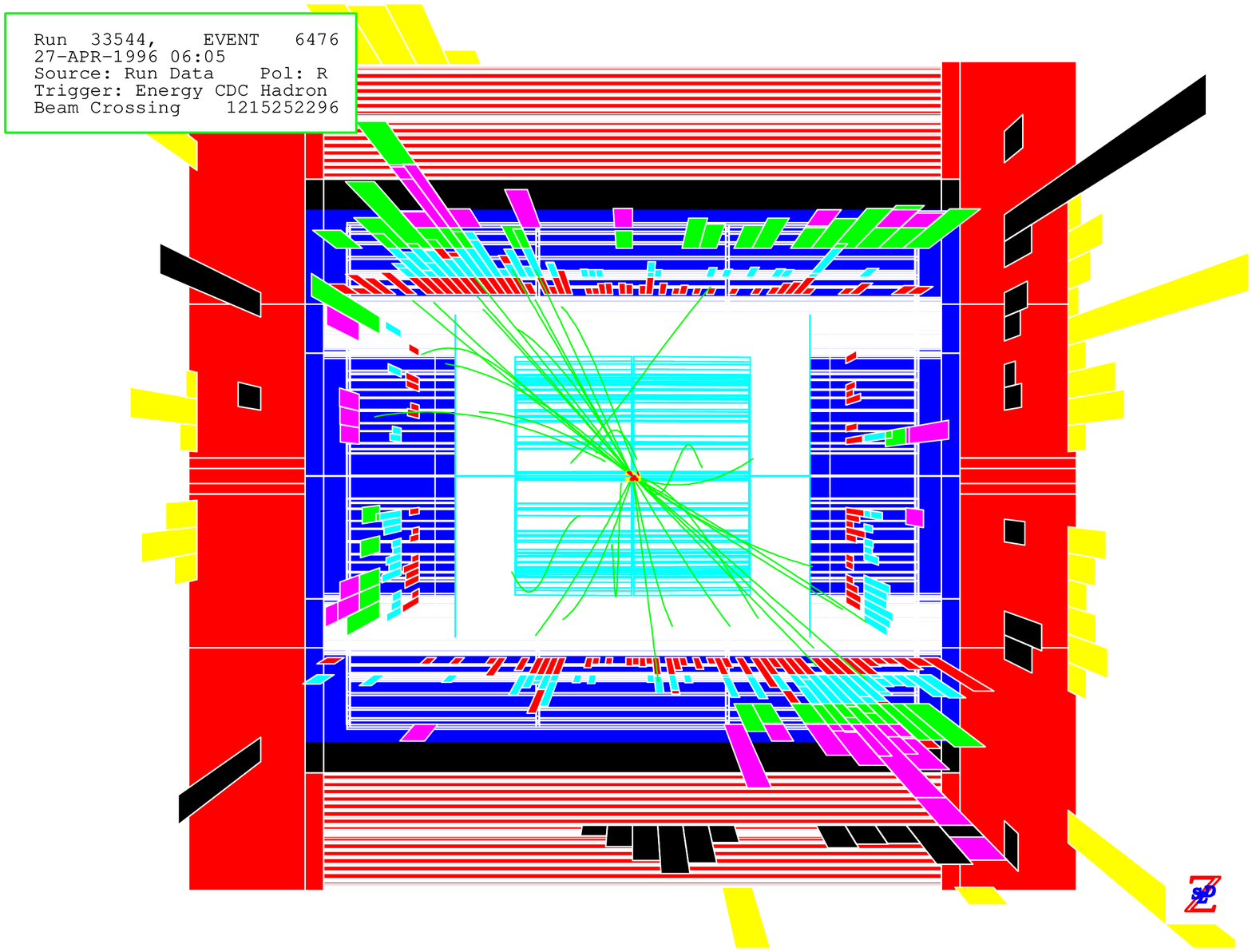}
}
\vspace*{0.2cm}
\fcaption{
A typical $B$ event at the $Z^0$ pole recorded with 
the OPAL detector (left) in the $r\,\varphi$-view and the SLD experiment
(right) in the $rz$-view.}
\label{evt_z}
\end{figure}

Figure~\ref{evt_ups} shows examples of typical $B$~events at the
$\Upsilon(4S)$ recorded with the ARGUS detector (left) and the CLEO
experiment (right). At the $\Upsilon(4S)$ resonance, only $B^0\bar B^0$ or
$B^+B^-$ pairs are produced nearly at rest, resulting in a spherical event
shape with an average charged particle multiplicity of about ten tracks. 
At the LEP or SLC accelerators, $b\bar b$ quark pairs are produced from
the decay of the $Z^0$~boson where both quarks share half of the energy
of the $Z^0$ resonance of 91.2~GeV. This results in two $b$~jets
being back-to-back as seen in Figure~\ref{evt_z}, where examples
from the OPAL (left) and SLD (right) experiments are displayed. The
average boost of $B$~hadrons at the $Z^0$~resonance is $\beta\gamma \sim 6$. 
  
\begin{figure}[tbp]
\centerline{
\epsfxsize=6.6cm
\epsffile[25 160 570 600]{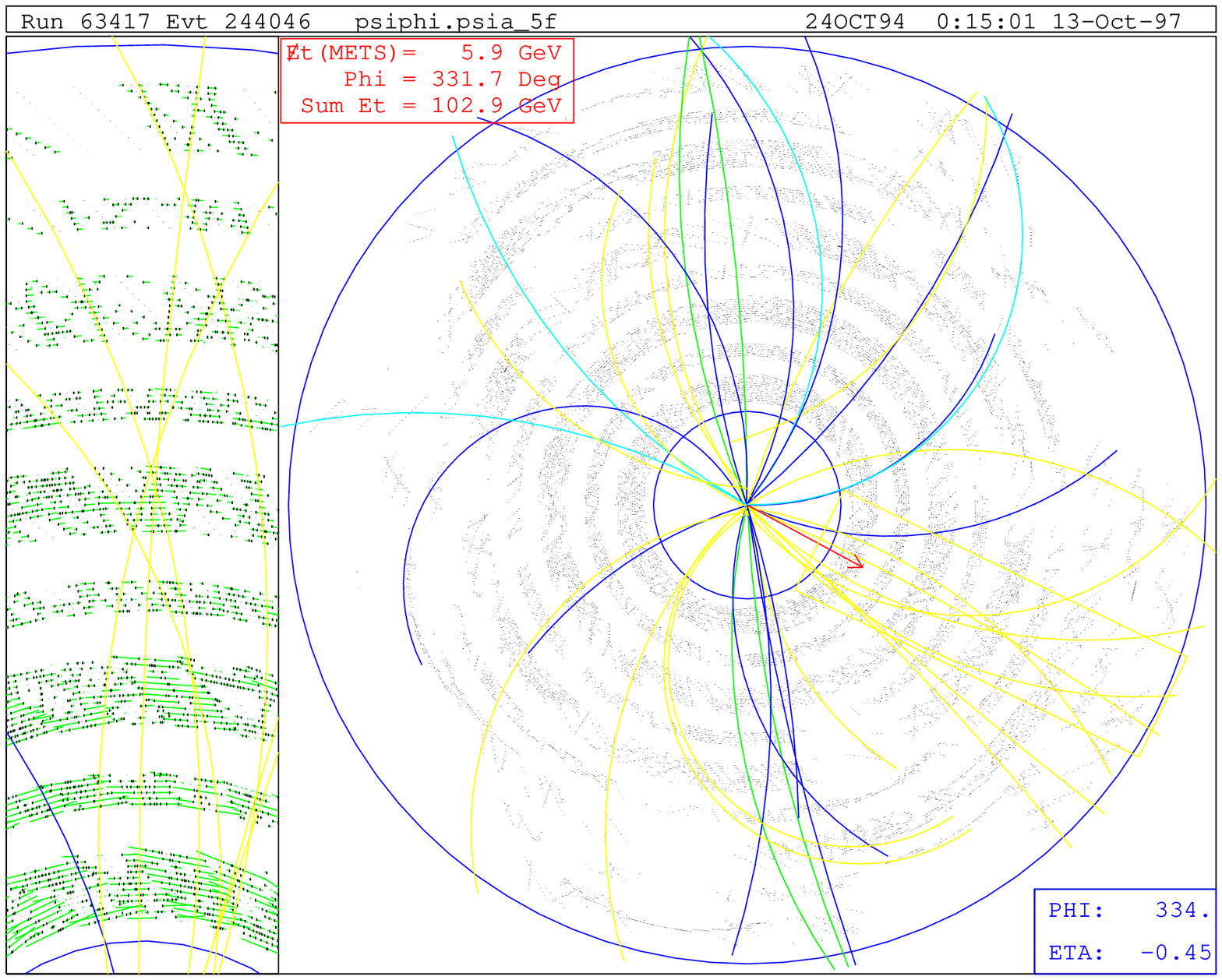}
\hspace*{0.1cm}
\epsfxsize=5.9cm
\epsfysize=5.2cm
\epsffile[5 175 570 590]{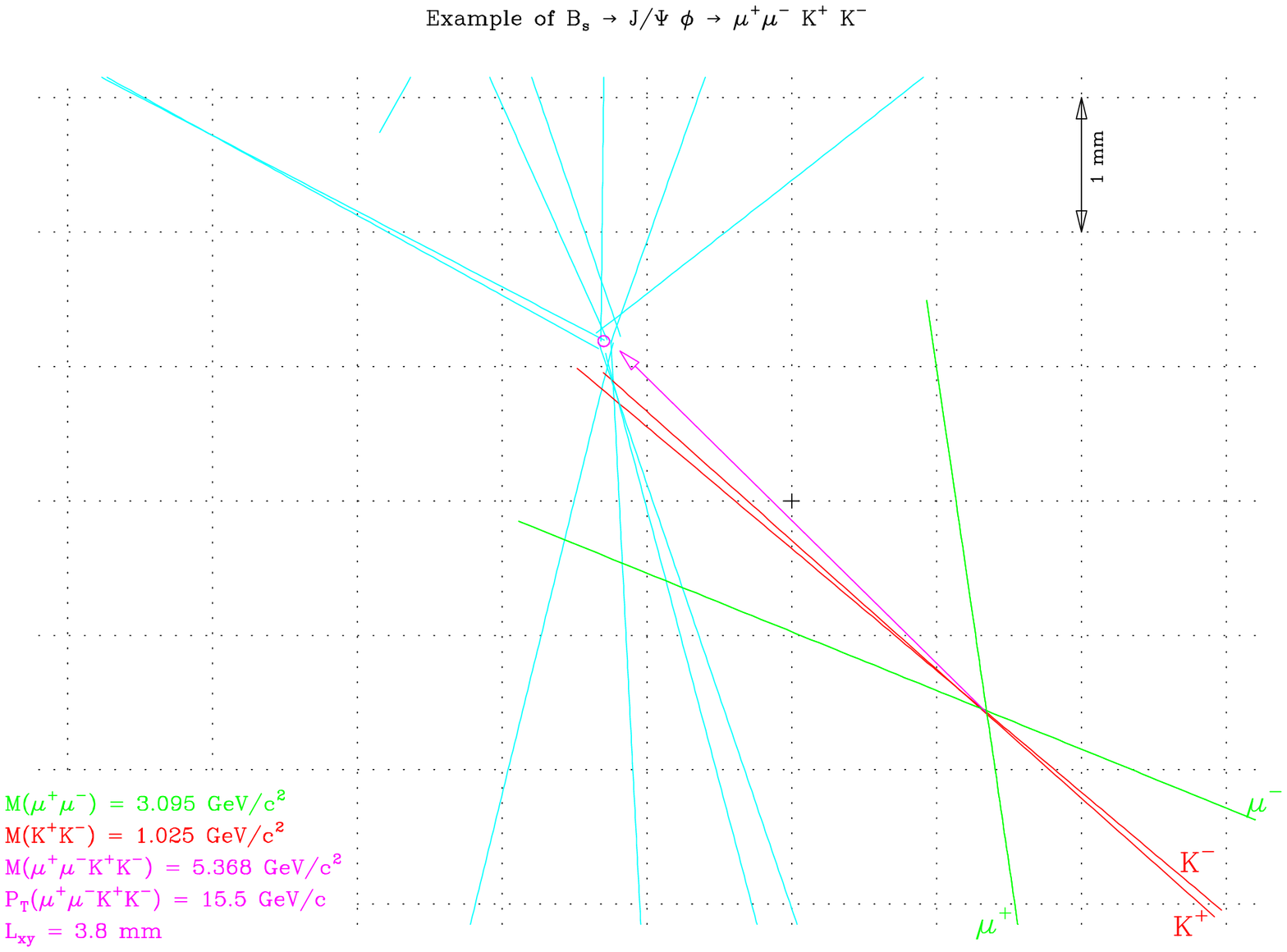}
}
\vspace*{0.2cm}
\fcaption{
A typical $B$ event at the Tevatron recorded with 
the CDF detector (left). On the right hand side, the decay vertex of a 
$\Bs \ra J/\psi \phi$ decay with $J/\psi \ra \mu^+\mu^-$ and $\phi \ra
K^+K^-$ is clearly separated from the primary interaction vertex.} 
\label{evt_cdf}
\end{figure}

Figure~\ref{evt_cdf} represents a typical $B$~event from the Tevatron,
shown on the left hand side in the $r\,\varphi$-view of the CDF central
drift chamber. No well defined jet structure is visible; the average
multiplicity is about 40 charged tracks including tracks from the
``underlying event'' particles.
It might appear challenging to find the $B$~decay products in this
quite messy environment of a hadronic collision. One way to extract
$B$~decays in a $p\bar p$ collision is illustrated on the right hand
side of Fig.~\ref{evt_cdf}. The relatively long lifetime of a
$B$~hadron and a particle boost of about $\beta\gamma \sim$ 2\,-\,4,
depending on the trigger which recorded the event, results in a 
$B$~decay vertex which is clearly separated from the primary $p\bar p$
interaction vertex. In the example shown on the right hand
side of Fig.~\ref{evt_cdf}, the decay vertex of a 
$\Bs \ra J/\psi \phi$ decay with $J/\psi \ra \mu^+\mu^-$ and $\phi \ra
K^+K^-$ is clearly visible and about 3.8~mm separated from the primary
interaction vertex.  

\begin{figure}[tbp]
\centerline{
\put(50,52){\large\bf (a)}
\put(74,52){\large\bf (b)}
\epsfxsize=6.3cm
\epsffile[20 1 555 515]{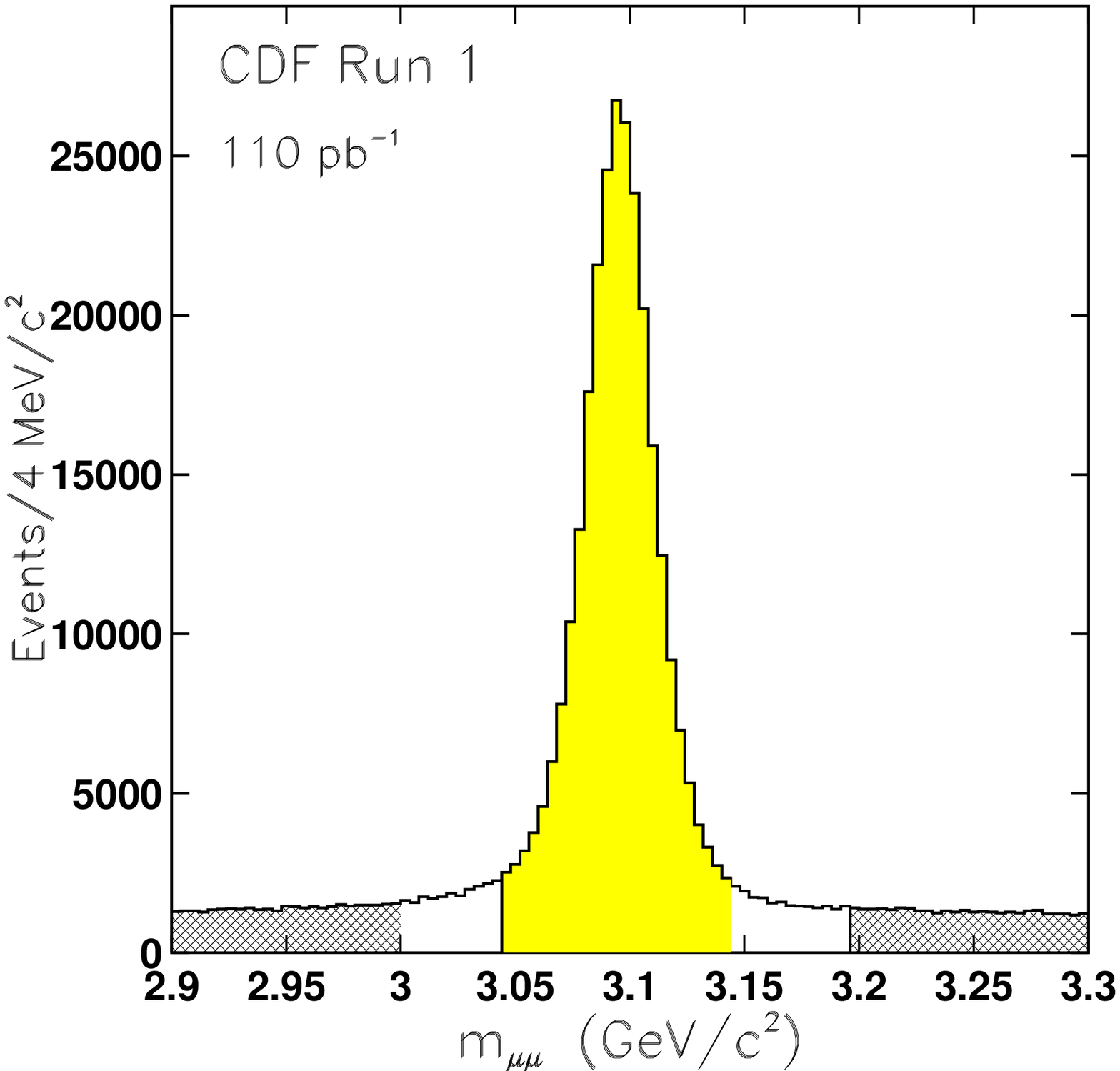}
\epsfxsize=6.3cm
\epsffile[15 150 535 650]{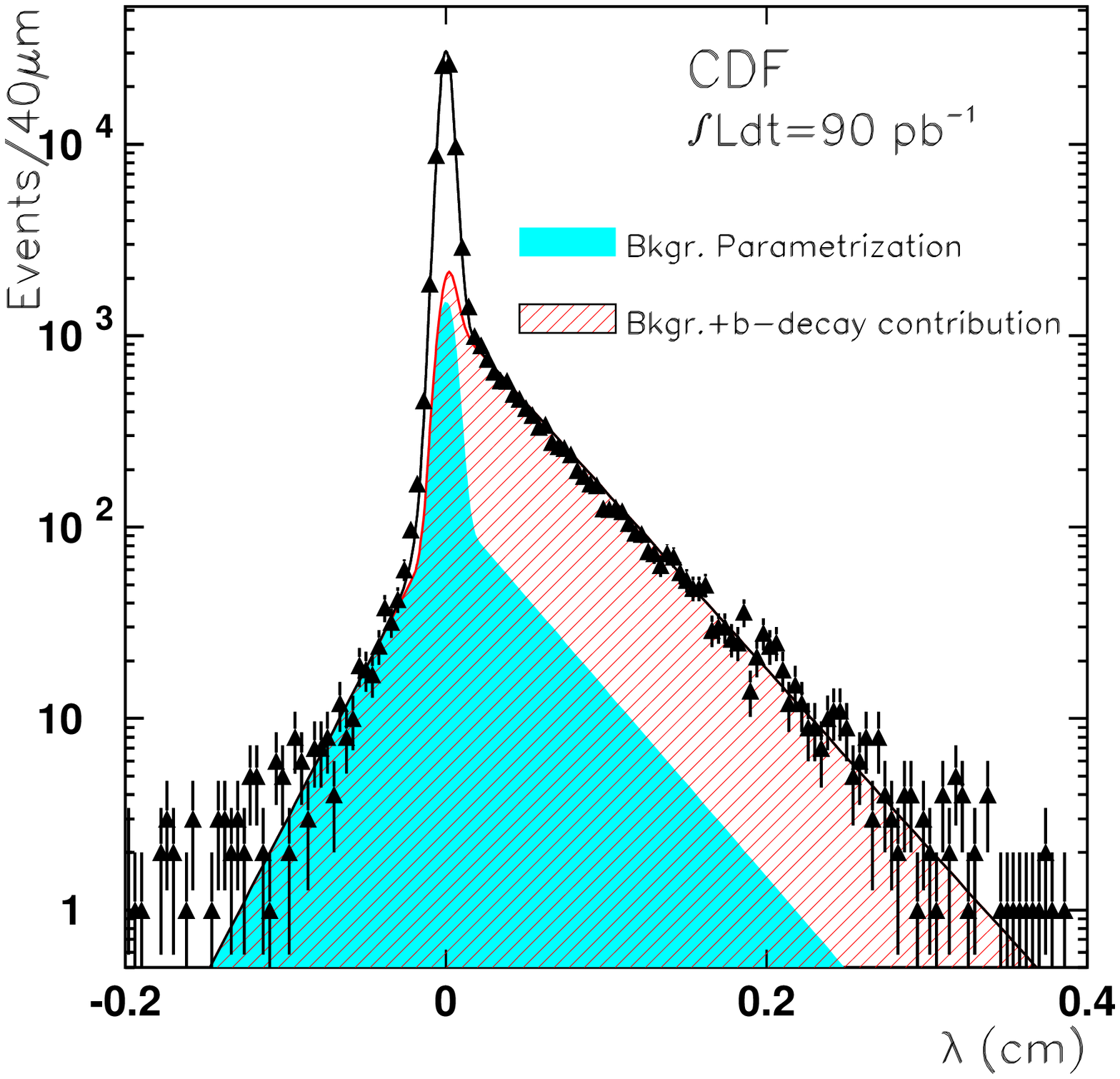}
}
\vspace*{0.2cm}
\fcaption{
(a) Invariant mass distribution of oppositely charged muon pairs from
the CDF dimuon trigger.
(b) Decay length distribution of the signal events indicated as shaded
area in (a) with the result of a fit superimposed (see text).}
\label{jpsi_mass_tau}
\end{figure}

In the following, we give a few more
examples on how $B$~decays
are studied at CDF and which essential 
tools are used for $B$~physics in hadron collisions. In Section~4.1,
we already pointed out the 
importance of the trigger to select events containing $B$~hadrons.
Another important feature for $B$~physics at a hadron collider is
the good tracking capability such as that at CDF. The central tracking chamber
together with the SVX provide excellent track momentum resolution
which translates into an excellent invariant mass resolution, as illustrated
in Figure~\ref{jpsi_mass_tau}. Here, the dimuon invariant mass is
displayed for muons from the dimuon trigger stream. A prominent $J/\psi$ peak
is visible with about $(243,000\pm540)$ $J/\psi$ signal candidates from
the Run\,I data on low background. For this distribution,
where both muons are reconstructed in the SVX, the mass resolution of
the $J/\psi$ peak is about 16~\mevcc. 

In addition to excellent tracking, superb vertexing is the other
essential feature of successful $B$~physics studies at a hadron collider.
This is demonstrated in Figure~\ref{jpsi_mass_tau}b) where the two muons of
the $J/\psi$ signal candidates (light shaded area in
Fig.~\ref{jpsi_mass_tau}a) are vertexed using tracking information
from the SVX. The two-dimensional distance between the
primary $p\bar p$ interaction vertex and the reconstructed dimuon
vertex is plotted. This distribution shows several features: A
prominent peak at zero decay length results from prompt $J/\psi$
candidates which  
are produced at the primary interaction vertex and constitute about 80\% of all
$J/\psi$ candidates. The width of this peak reveals information about
the vertexing resolution which is on average $40$-$50~\mu$m, for this
sample. At positive decay lengths, $J/\psi$ mesons from
$B$~hadron decays are described by an exponential slope. At a distance
of about 100~$\mu$m from the 
primary interaction vertex, mainly $J/\psi$'s from $B$~decays remain. 
There is also a small exponential slope at negative decay lengths
where the particle seems to decay, before the point where
it is produced. These events result from the combinatorial background
underneath the $J/\psi$ signal. This is indicated by events from the
$J/\psi$ sidebands (dark shaded regions in Fig.~\ref{jpsi_mass_tau}a) 
which describe well the distribution at negative decay lengths as 
seen by the dark shaded area in Figure~\ref{jpsi_mass_tau}b).

\begin{figure}[tbp]
\centerline{
\put(53,56){\large\bf (a)}
\put(54,25){\large\bf (b)}
\epsfxsize=6.5cm
\epsffile[10 10 545 520]{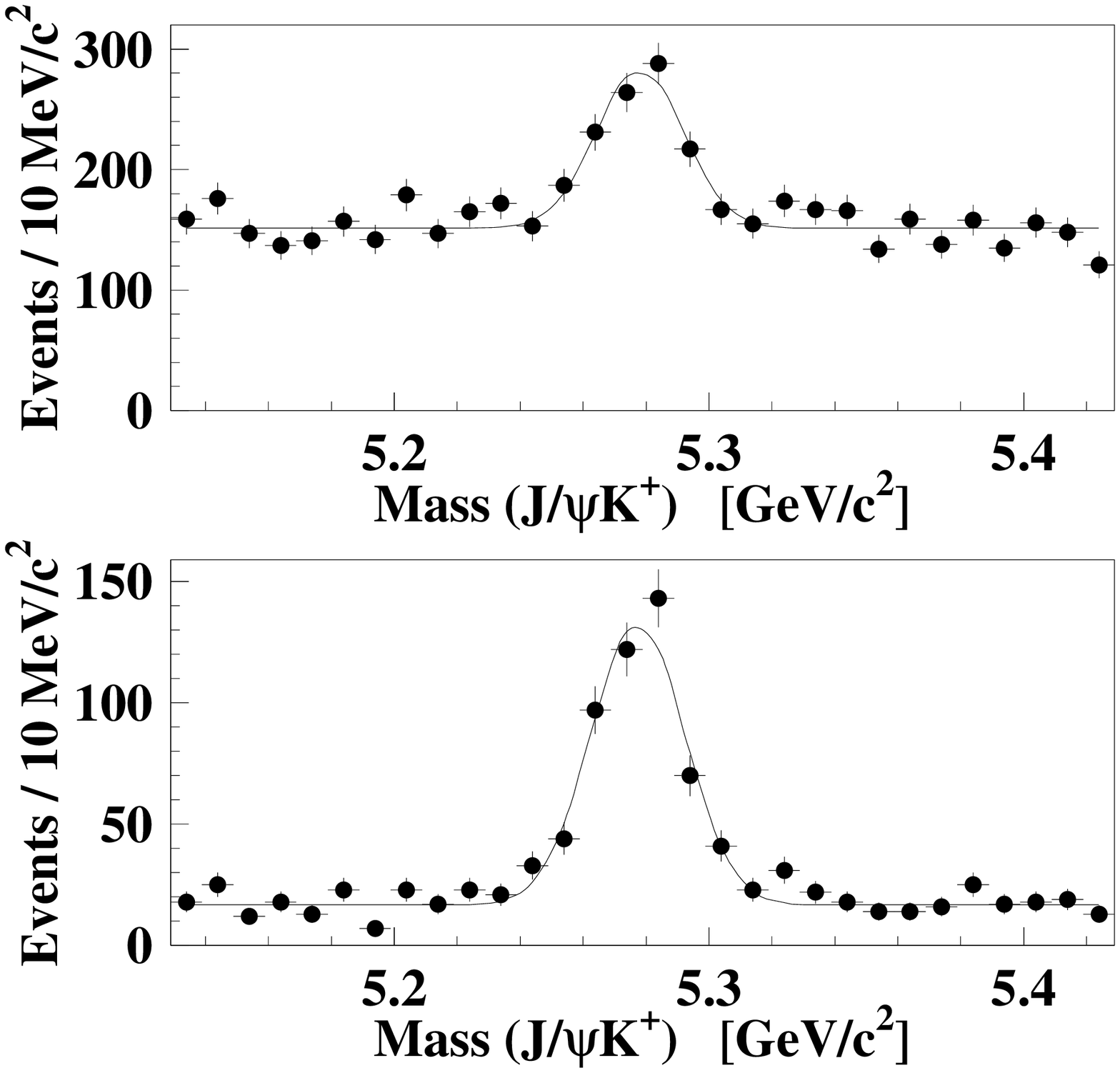}
}
\vspace*{0.2cm}
\fcaption{
Invariant mass distribution of reconstructed $B^+ \ra J/\psi K^+$
candidates with $J/\psi \ra \mu^+\mu^-$ (a) without a requirement on the
$B$~decay length and (b) with a cut on
the $B$~decay length of greater than 100~$\mu$m.} 
\label{b_mass_cut}
\end{figure}

In order to fully reconstruct $B$~mesons, for example through the
decay $B^+ \ra J/\psi K^+$, the $J/\psi$
candidates from the signal region of Fig.~\ref{jpsi_mass_tau}a) are
paired with another track in the event, assumed to be a kaon.
A $J/\psi K^+$ invariant mass distribution, such as the one in
Figure~\ref{b_mass_cut}a), is observed with a $B^+$ signal on a large
background. This background can be drastically reduced if a displaced
$B$~vertex is required. This is demonstrated in Fig.~\ref{b_mass_cut}b)
with a cut on the $B$ decay length of greater than
100~$\mu$m. A clear $B$~signal can now be found on a small background.  

\begin{figure}[tbp]
\centerline{
\put(3,54){\large\bf (a)}
\put(76,54){\large\bf (b)}
\epsfxsize=6.0cm
\epsffile{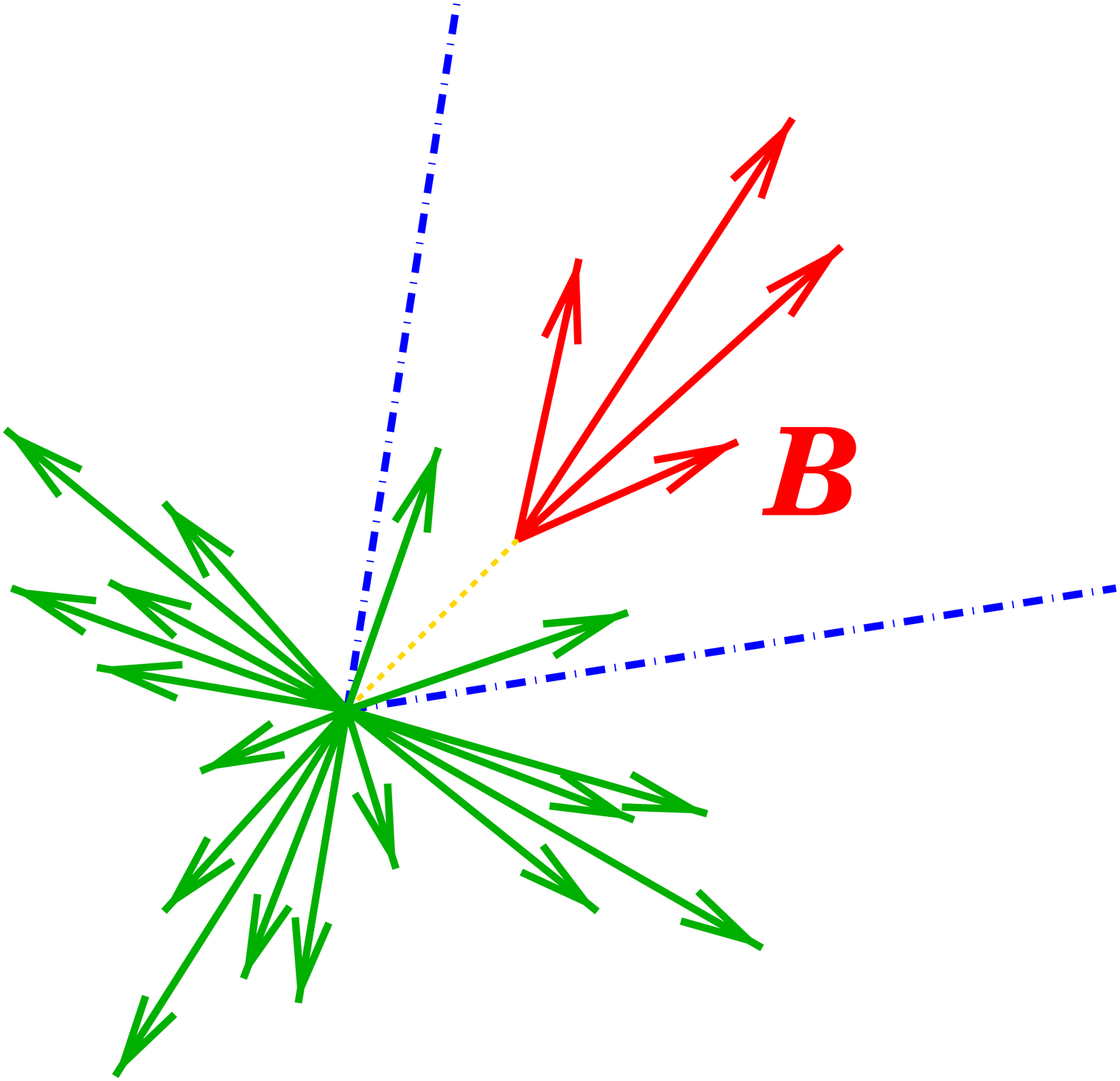}
\hspace*{0.1cm}
\epsfxsize=6.5cm
\epsffile[15 10 545 515]{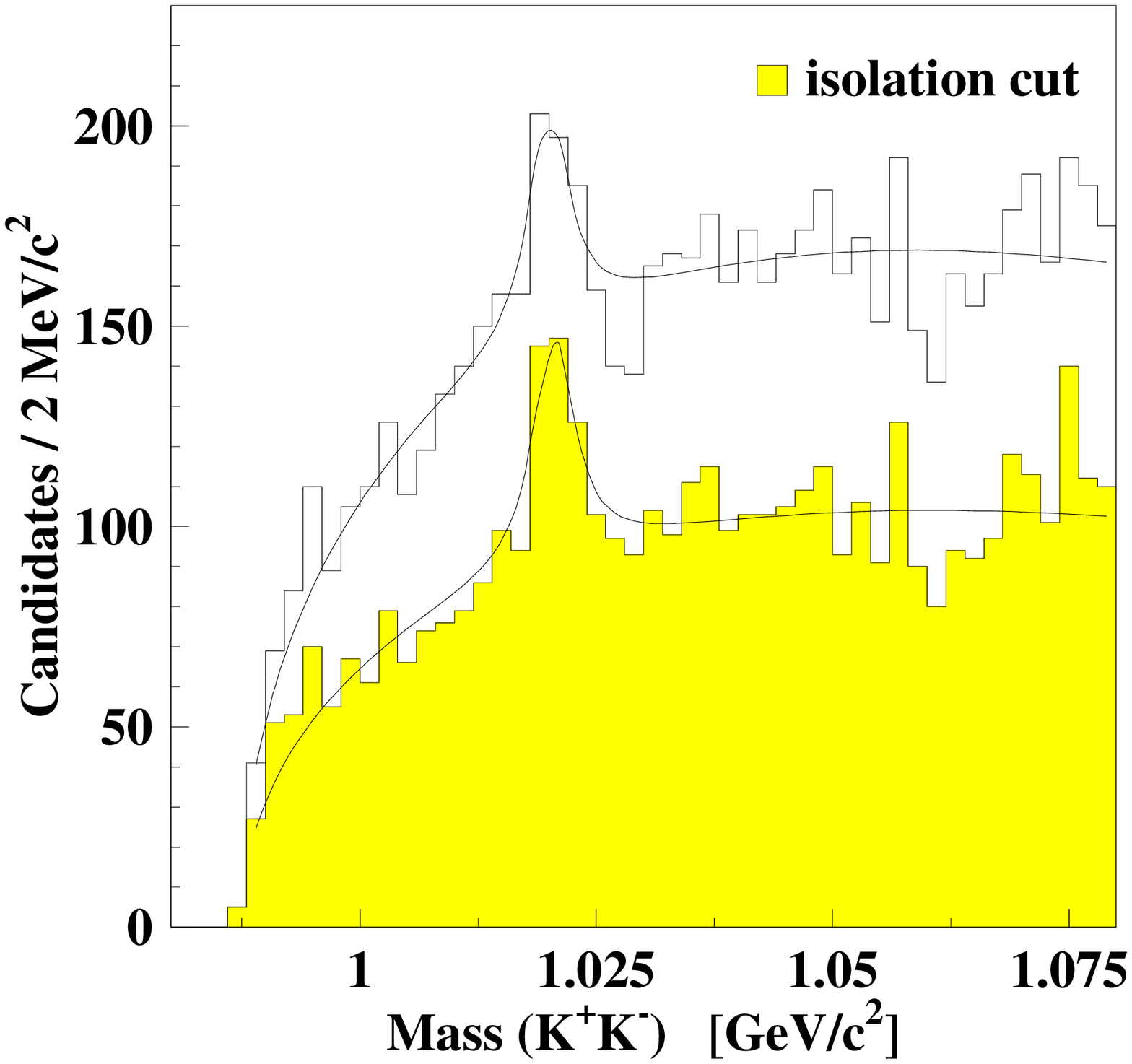}
}
\vspace*{0.2cm}
\fcaption{
(a) Illustration of the isolation of $B$ mesons in hadronic collisions. 
(b) Invariant $K^+K^-$ mass distribution from 
$\Bs \ra \Ds \mu^+ \nu$ decays with 
$\Ds \ra \phil$. The upper distribution shows the $K^+K^-$ invariant mass
before an isolation requirement is applied
resulting in the shaded $K^+K^-$ mass distribution.}
\label{iso_cut}
\end{figure}

Another tool in separating $B$~decays in hadronic
collisions, can be obtained from the fact that $B$~decays are often
isolated. This means there is usually not much track activity in the
vicinity of a $B$ hadron decay as illustrated in Fig.~\ref{iso_cut}a). 
This fact can be exploited by a track based isolation quantity $I$. It
is typically defined as the \Pt\ of the reconstructed $B$~candidate
divided by the scalar sum of the transverse momenta of all tracks
within a cone in $\eta\,\varphi$-space around the $B$~candidate
direction, including the tracks from the $B$~decay. A typical cone
radius is $\Delta R < 1$ and a typical cut value requires more
than 50\% of all momentum within the cone to be carried by the $B$~candidate. 
The effect of the isolation cut is demonstrated in
Fig.~\ref{iso_cut}b) on a $\phi 
\ra K^+ K^-$ signal from a $\Bs \ra \Ds \mu^+ \nu$ decay with 
$\Ds \ra \phil$. The upper distribution shows the $K^+K^-$ invariant mass
before a cut on $I > 0.5$ is applied to obtain the
lower distribution. The combinatorial background is drastically
reduced with a high efficiency on the $\phi$ signal.    
Finally, the significantly harder momentum spectrum of particles from
$b$ hadrons, compared to light hadrons, can be used to place high
\Pt-cuts of several \gevc\ on $B$~candidates. This also results in a
strong reduction of combinatorial background from light quark jet
production. 
 
To summarize, the essential features to extract $B$ hadron decays in
hadronic collisions at CDF are excellent tracking in the CTC and SVX,
the superb vertexing capabilities of CDF's silicon vertex detector
exploiting the long lifetimes of $B$~hadrons by requiring decay
vertices to be displaced from the primary interaction vertex,
the harder momentum spectrum of particles from
$b$ hadrons, and the fact that $B$ mesons are often isolated.

\section{Measurements of \boldmath{$B$} Hadron Lifetimes}
\runninghead{$B$ Lifetimes, Mixing and $CP$ Violation at CDF}
{Measurements of $B$ Hadron Lifetimes}
\noindent
CDF has measured the lifetimes of all weakly decaying $B$~mesons
($B^0$, $B^+$, \Bs, and $B^+_c$) as well as the lifetime of the
$\Lambda^0_b$ baryon. 
In this section, we discuss the $B$~lifetime measurements at CDF. 
After a short introduction in the theory of $B$~lifetimes, we describe
experimental techniques of 
$B$~lifetime measurements including the determination of the primary
$p\bar p$ interaction vertex, the measurement of the $B$~hadron decay
length, and a description of the fitting procedure. We
then review the lifetime measurements of the individual 
$B$~species at CDF emphasizing crucial aspects of a particular
measurement rather than describing an analysis~in~detail. 

\subsection{Introduction to \boldmath{$B$} hadron lifetimes}
\begin{figure}[tbp]
\centerline{
\epsfxsize=6.3cm
\epsffile{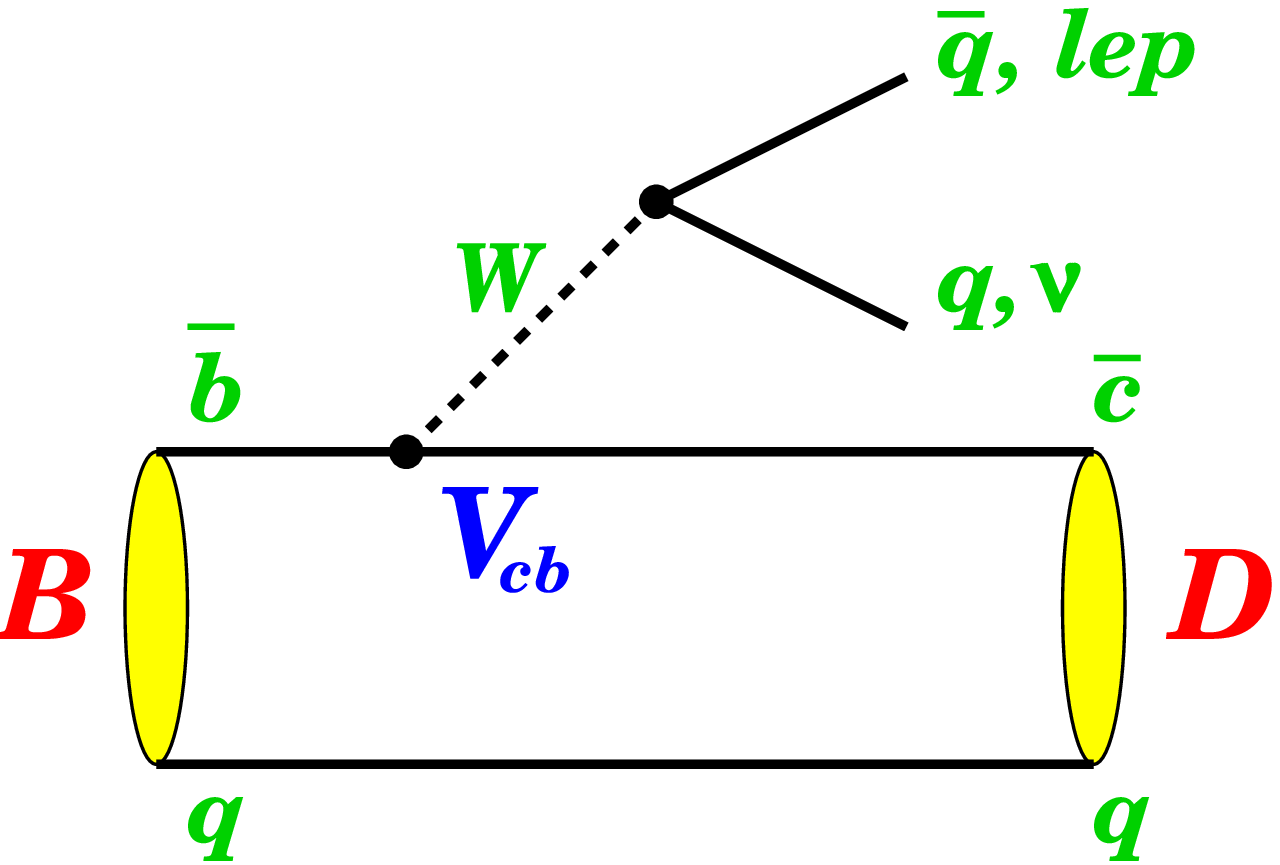}
}
\vspace*{0.2cm}
\fcaption{
Sketch of the spectator model of a $B$~hadron decay.}
\label{blife_spectator}
\end{figure}

\noindent
The lifetimes of $B$~hadrons are fundamental properties of these
particles and are often needed to determine other Standard Model\cite{GSW}
quantities like the CKM matrix element $|V_{cb}|$ or to measure the
time dependence of $B\bar B$ oscillations. 
In the simple spectator model of a $B$~hadron decay, 
where the $b$~quark decays into a $c$~quark emitting a $W$~boson which
couples to a $\ell\nu$ or $q\bar q$ pair, as illustrated in
Figure~\ref{blife_spectator}, the $\bar q$~quark within the $B$~hadron
only acts as a spectator and the $b$~quark decays as a free
particle. In this case, the $B$~hadron lifetime would be
given in analogy to the muon lifetime as
\begin{equation}
\Gamma = \frac{1}{\tau} = \frac{G_F^2\, m_b^5}{192\,\pi^3} \cdot
|V_{cb}|^2 \cdot {\cal F},
\end{equation}
where $\cal F$ is a phase space factor. Here, we have neglected 
$b \ra u$~transitions resulting in a
term with $|V_{ub}|^2$ which is small. In the spectator
model, all hadrons containing a $b$~quark would have the same
lifetime. However, this picture does not hold for 
the prediction of charm hadron lifetimes which are measured\cite{PDG}
to be
\begin{equation}
\tau(D^-) \sim 2.5\,\tau(D^0) \sim 2.5\,\tau(\Ds) \sim 5.0\,\tau(\Lambda_c^-).
\end{equation}

\begin{figure}[tbp]
\centerline{
\hspace*{0.5cm}
\put(-10,16){\large\bf (a)}
\put(54,16){\large\bf (b)}
\epsfxsize=5.0cm
\epsffile{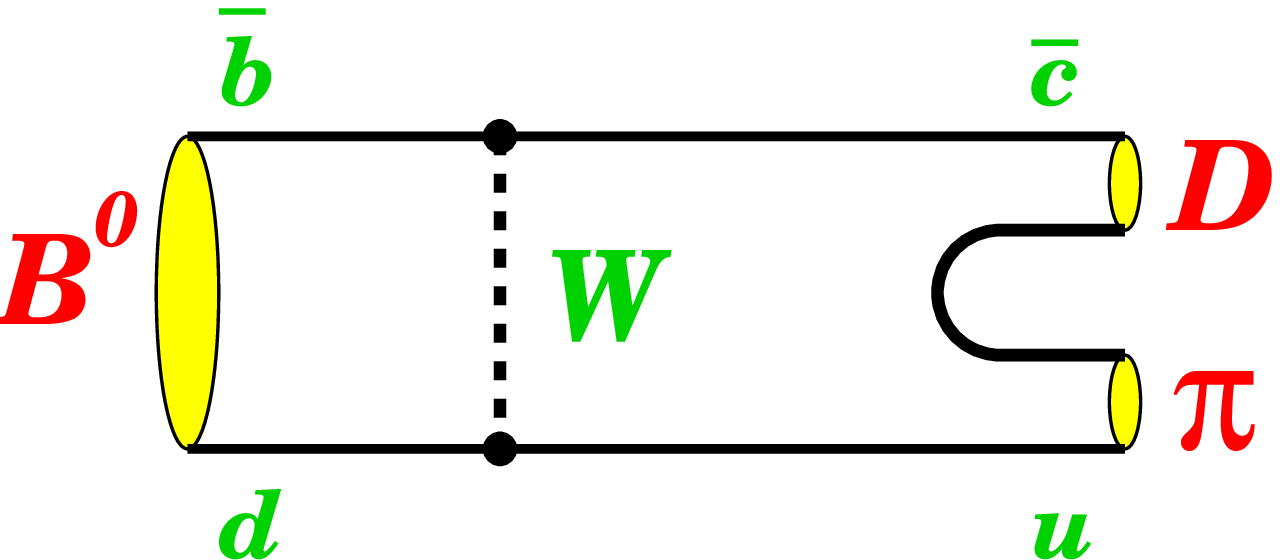}
\hspace*{1.2cm}
\epsfxsize=5.0cm
\epsffile[1 -35 380 95]{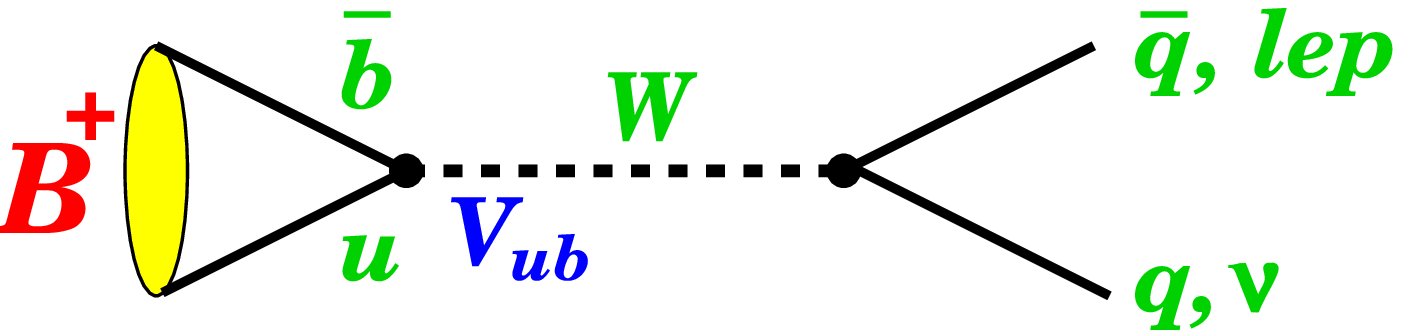}
}
\vspace*{0.5cm}
\centerline{
\hspace*{0.5cm}
\put(-10,23){\large\bf (c)}
\put(54,23){\large\bf (d)}
\epsfxsize=5.0cm
\epsffile{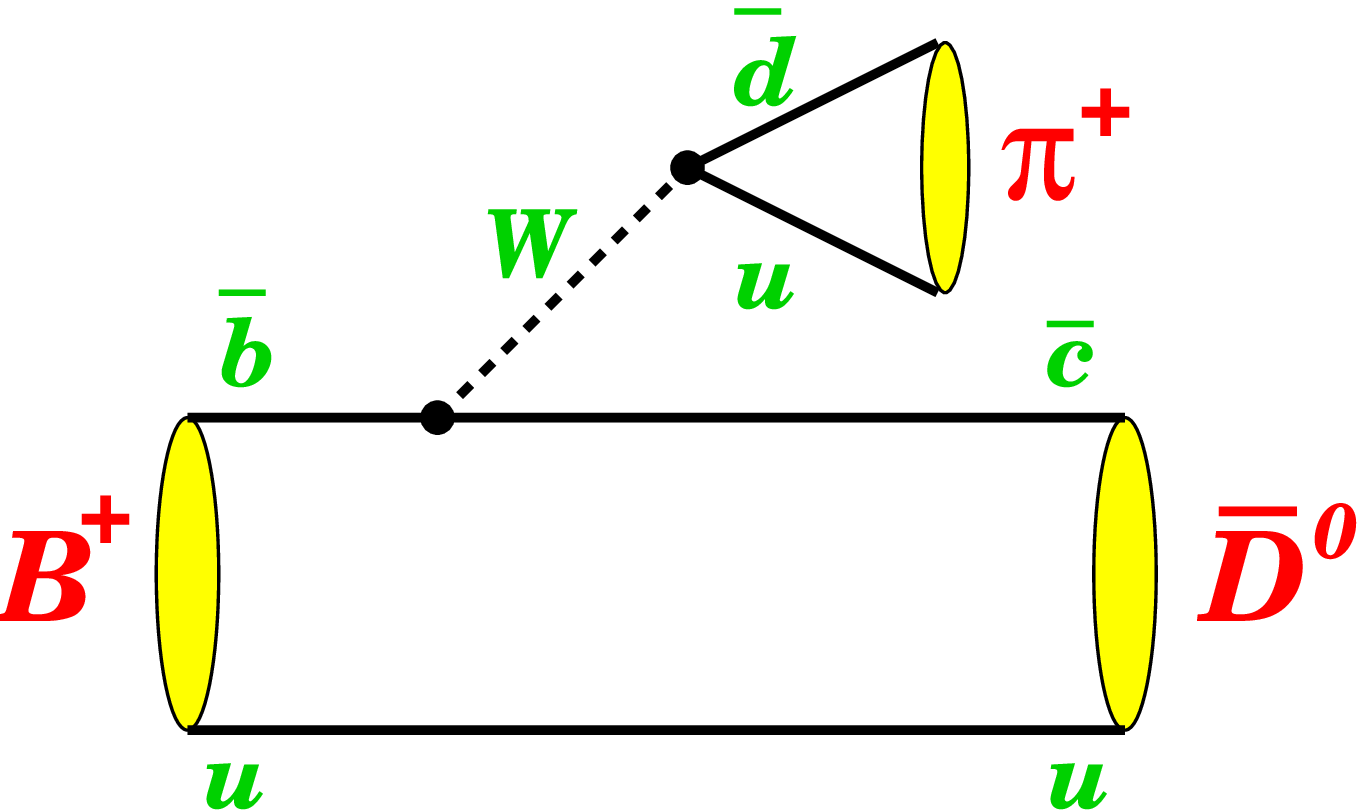}
\hspace*{1.2cm}
\epsfxsize=5.0cm
\epsffile{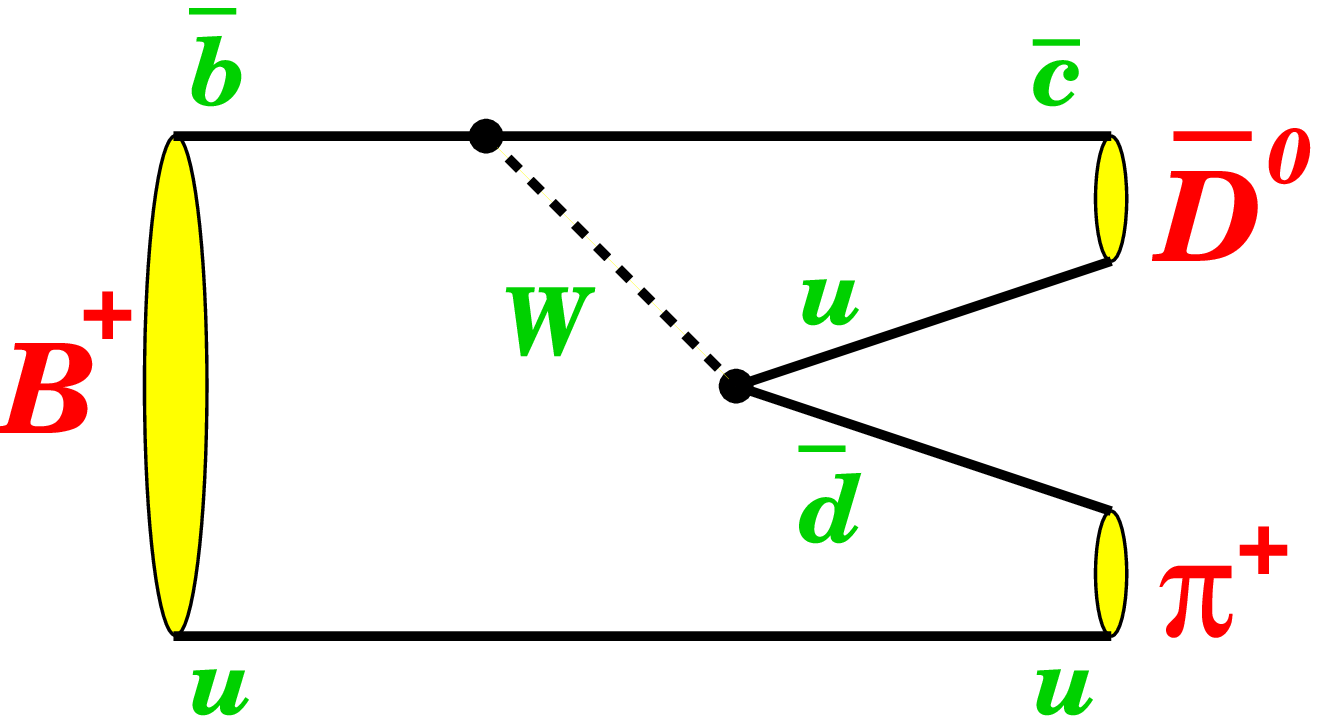}
}
\vspace*{0.2cm}
\fcaption{
Illustration of (a) $W$~exchange process, (b) $B^+$~annihilation
diagram as
well as final state inter\-ference through (c) external $W$~emission and
(d) internal $W$~emission.}
\label{bnonspec}
\end{figure}

Possible causes for these lifetime differences originate from
non-spectator effects, playing an important role in the decay of
charm hadrons. These non-spec\-tator decays include contributions
like the $W$~exchange process for neutral $B$~mesons, shown in
Figure~\ref{bnonspec}a), or the $B^+$~annihilation diagram displayed in 
Fig.~\ref{bnonspec}b). Other non-spectator mechanisms are caused 
by so-called final state Pauli interference effects. 
An example is shown in Fig.~\ref{bnonspec} for the decay $B^+ \ra \bar
D^0 \pi^+$ which can occur through a so-called external $W$~emission
(Fig.~\ref{bnonspec}c) or a so-called internal $W$~emission
(Fig.~\ref{bnonspec}d). There is a destructive interference between
both diagrams which causes the $B^+$ decay width to be smaller than
the $B^0$~width. This is one of the reasons the $B^+$~lifetime is
predicted to be larger than the $B^0$~lifetime.
Measurements of the
lifetimes of individual $B$~hadron species can therefore probe $B$~decay
mechanisms beyond the simple spectator model decay picture. However, 
among bottom hadrons, the lifetime differences are expected to be 
smaller than in the charm system due to the heavier bottom quark mass.

In the past few years, 
the heavy quark expansion technique has been applied extensively
to the calculations of inclusive decay rates of heavy hadrons, 
for both spectator and non-spectator decays. It provides quantitative
predictions for lifetime differences among the heavy hadrons.
It is generally believed that 
a lifetime difference of order 5-10\% should exist 
between the $B^+$ and $B^0$ meson.
Reference\cite{bigi} predicts that the $B^+$ meson
lifetime should be longer than the $B^0$ meson lifetime
by about 5\%.
However, Reference\cite{neubert} questions some assumptions made in
Ref.\cite{bigi} and states that the sign of the 
deviation from unity cannot be predicted reliably.
There is agreement that the
models expect a much smaller difference of 
less than about 1\% between the $B^0$ and \Bs\
lifetimes. 

Since this subject seems to be controversial, it might be best solved by
precisely measuring all the $B$~hadron lifetimes.
Several direct measurements of $B^+$ and $B^0$ meson lifetimes
have been performed by $e^+e^-$ experiments
(see Ref.\cite{PDG} for an overview) and by CDF. 
The precision of current measurements now
approaches the level where the predicted small differences
can be seen and improvements in these measurements
will provide a strong test of $B$~hadron decay mechanisms.

\subsection{Experimental techniques}
\noindent
In general, two experimental methods have been employed to measure $B$~hadron
lifetimes. The first is based on the signed track impact parameter, 
the distance of closest approach of the track trajectory extrapolated
to the $B$~hadron production point. The average impact parameter is
proportional to the lifetime of the $B$~hadron. The advantage of using
the impact parameter is that it is fairly insensitive to the boost of
the $B$~hadron: A $B$~hadron with a large Lorentz boost will travel
further, but the decay products will come out at a smaller angle leaving
the impact parameter unchanged. To extract the $B$~hadron lifetime, a
Monte Carlo model is used to reproduce the observed impact parameter
distribution as a function of the $B$~lifetime. Impact parameter
measurements typically use leptons from semileptonic $B$~decays. In
fact, the first measurement of a $b$~lifetime\cite{blife_MarkII}
used the signed impact parameter of leptons.

The second method for measuring $B$~hadron lifetimes is based
on the decay length, which is the distance from the $B$ hadron
production point to the $B$~hadron decay point. The decay length $L$
is related to the proper decay time $t$ in the $B$~restframe by the Lorentz
boost $\beta\gamma$ as
$L = \beta\gamma \,\, c\,t$. Unlike the impact parameter method, it is
necessary to know the boost value. In the case of a fully
reconstructed $B$ hadron decay, its boost value is determined as
$\beta\gamma = p_B / m_B$, where $p_B$ is the $B$~hadron momentum and
$m_B$ is the $B$~hadron mass. If the $B$~hadron is only partially
reconstructed, the boost value must be inferred using a Monte Carlo model.

All $B$~hadron lifetime results at CDF are based on the decay length
measurement using fully reconstructed $B$~hadrons as well as partially
reconstructed $B$~decays, which usually comprise higher statistics samples.
In the following, we give an overview of how $B$~lifetimes are
measured at CDF, using the more complicated example of a partially
reconstructed $B$~decay. We first describe the Tevatron beam profile and the
determination of the 
$B$~hadron production point, the primary event vertex. Then, the
$B$~hadron decay length reconstruction using CDF's silicon vertex
detector is subject of Section~6.2.3, while Sec.~6.2.4. gives a
description of the lifetime fit procedure used 
to extract the $B$~hadron lifetime from the measured decay length.

\subsubsection{Tevatron $p\bar p$ beam profile}
\begin{figure}[tbp]
\centerline{
\put(52,55){\large\bf (a)}
\put(117,57){\large\bf (b)}
\epsfxsize=6.3cm
\epsffile[25 155 525 645]{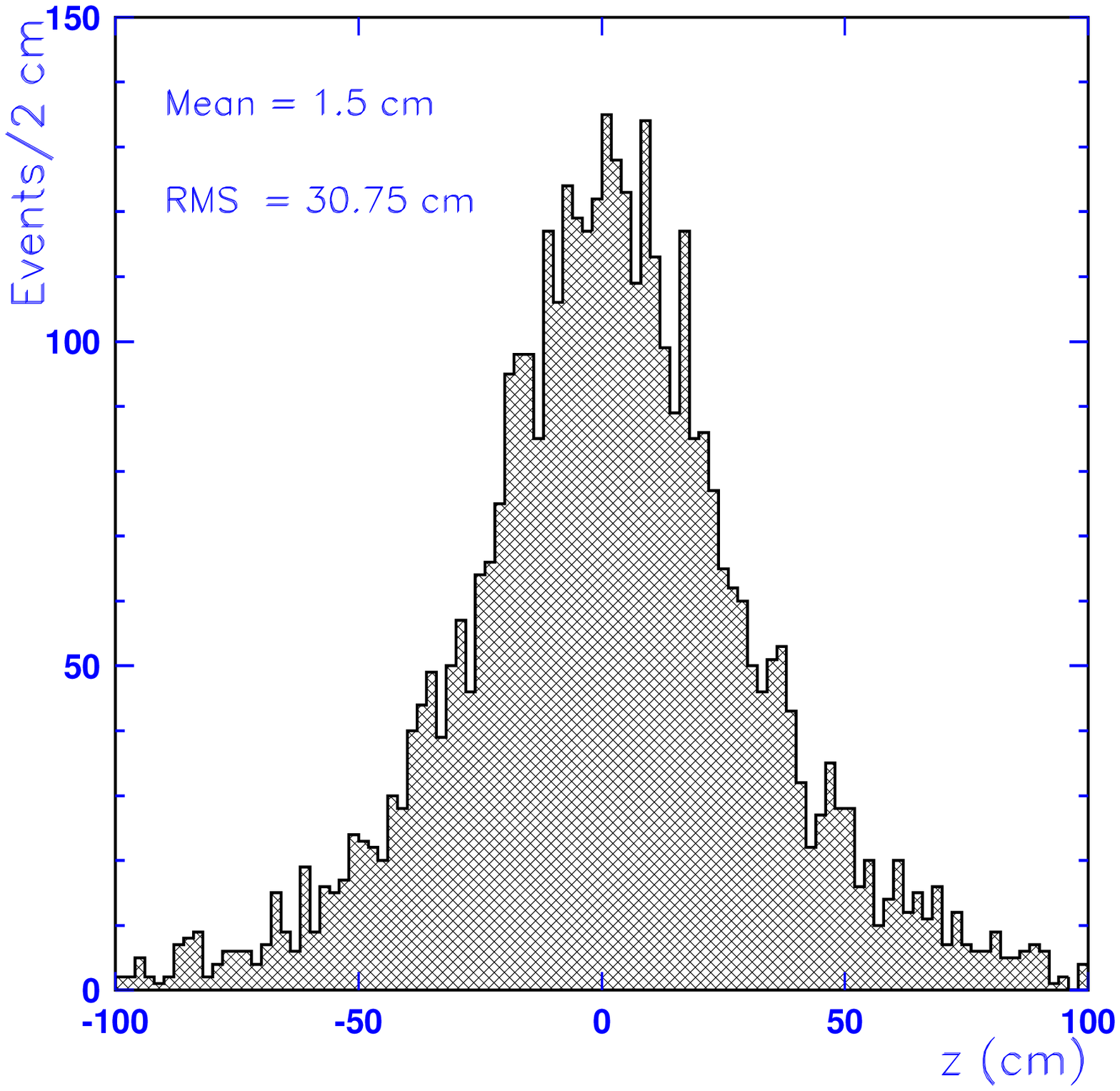}
\epsfxsize=6.3cm
\epsffile[25 155 525 645]{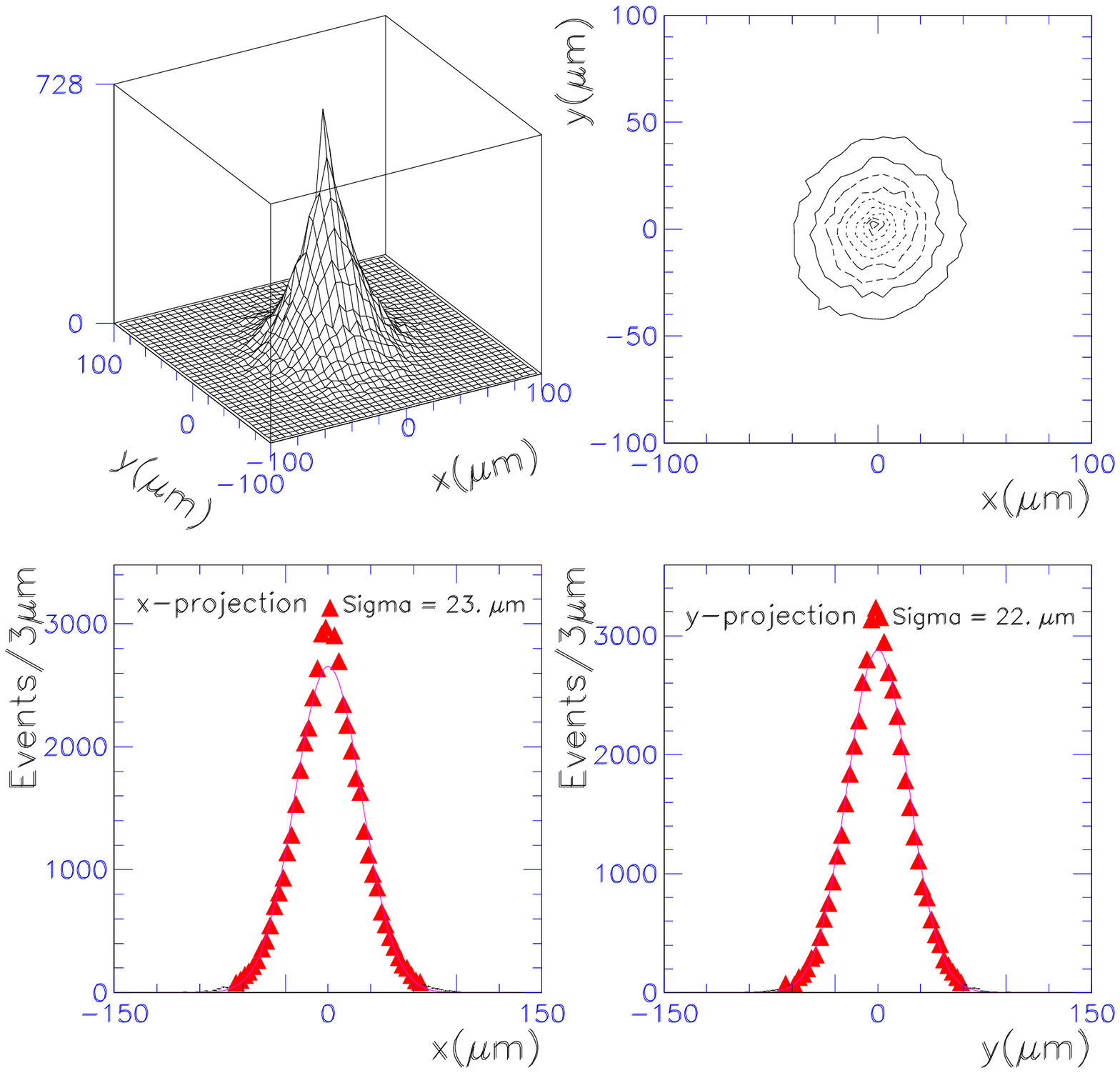}
}
\vspace*{0.2cm}
\fcaption{
(a) Distribution of primary $p\bar p$ interaction vertices along the
proton direction ($z$) for a typical data acquisition run during the
1994-1995 running period.   
(b) 
The two-dimensional distribution of the beam spot for a typical data
acquisition run is shown in the
upper two diagrams. The $x$- and $y$-projections, respectively, are
shown in the lower two plots.}
\label{beamline}
\end{figure}

\noindent
With a proton and antiproton bunch length of about 0.4~m, the 
primary $p\bar p$ interaction vertices are distributed along the beam direction
according to approximately a Gaussian function, with a 
width of $\sim\!30$~cm as shown in Fig.~\ref{beamline}a).
Near the interaction region, the $p$ and $\bar p$~beams follow straight
lines but could have an offset and slope with respect to the $z$-axis
of the tracking detectors. The profile of the beam for a typical data
acquisition run is shown in Fig.~\ref{beamline}b). The
deviation of reconstructed primary vertices from the calculated
average beam position is plotted in the transverse plane.
To ensure that the spread
of the beam, instead of the resolution of the vertex fit, is the
dominant contribution to the width of the observed distribution,
only vertices with high track multiplicities are used. The upper two plots
show the two-dimensional distribution of the beam spot for a
typical Run\,I data acquisition run. The lower two plots show the $x$-
and $y$-projections, respectively, with a fit to a Gaussian
distribution superimposed. This shows that the
transverse profile of the Tevatron beam was roughly Gaussian and
circular with a width $\sigma$ of $\sim\!25~\mu$m in both the $x$- and
$y$-directions. 

\subsubsection{Primary event vertex}
\noindent
The $B$~hadron lifetime measurements reported in this article are based
on measuring the 
distance between the primary $p\bar p$ event vertex and the secondary
$B$~decay vertex in the transverse plane exploiting CDF's silicon
vertex detector. First, the $z$-position of the primary interaction
vertex is identified, 
using the tracks reconstructed in the VTX detector and then 
the transverse position of the primary event vertex is determined, using  
the run average beam position.
The tracks reconstructed in the VTX detector, when
projected back to the beam axis, determine the longitudinal
location of the primary interaction with an accuracy of about
0.2~cm along the beam direction.
On average during Run\,I, the number of reconstructed interaction
vertices in a given event follows a Poisson distribution with a mean
of about 2.5.  
For $B$~hadron lifetime measurements, the $z$-location of 
the primary event vertex is usually determined by choosing the 
$p\bar p$ interaction vertex, recorded by the VTX which is closest to
the intercept of the 
$B$~candidate with the beamline. Usually,
the $z$-coordinates of all tracks from the $B$~decay are required to be
within five~cm of the $z$-position of this primary vertex, in order to exclude
tracks from other $p\bar p$~interactions in the event.

The transverse position of the primary event vertex is determined using  
the average beam position through the detector, together with the
knowledge of the longitudinal primary vertex position from the VTX.
The average beam position is calculated offline for each
data acquisition run. It is found
that the average beam trajectory is stable over the period of which a given
$p\bar p$~beam is stored in the Tevatron Collider. 
For $B$~lifetime measurements,
usually only events from data runs with a sufficiently
large number of collected events are considered in order to allow 
a good determination of the run averaged beamline.

In $B$ lifetime analyses, the primary
vertex is usually not measured on an 
event-by-event basis because the presence of a second $b$ quark decay
in the event, coupled with the low multiplicity in e.g.~semileptonic
$B$~decays can lead to a systematic bias in the lifetime determination. 
The algorithm that determines the primary event vertex on an
event-by-event basis, 
calculates first the $z$-position of the primary $p\bar p$~interaction
vertex from the VTX in the same way described above.
The transverse position of the primary vertex is then determined 
for each event by a weighted fit of all SVX tracks, with a
$z$-coordinate within 5~cm of  
the $z$-vertex position obtained from the VTX. At first, all SVX tracks
are fit to originate from a common vertex. 
Tracks which have large impact parameters, with respect to this vertex,
are then removed and the fit is repeated in an iterative process until
no more tracks are removed. At least five tracks must remain for a
successful primary vertex reconstruction to occur. The uncertainty in the fitted primary
vertex coordinates transverse to the beam direction ranges from about
10-35~$\mu$m depending on the number of tracks used in the fit and the event
topology. 

\subsubsection{Decay length measurement}
\begin{figure}[tbp]
\centerline{
\epsfxsize=10.0cm
\epsffile{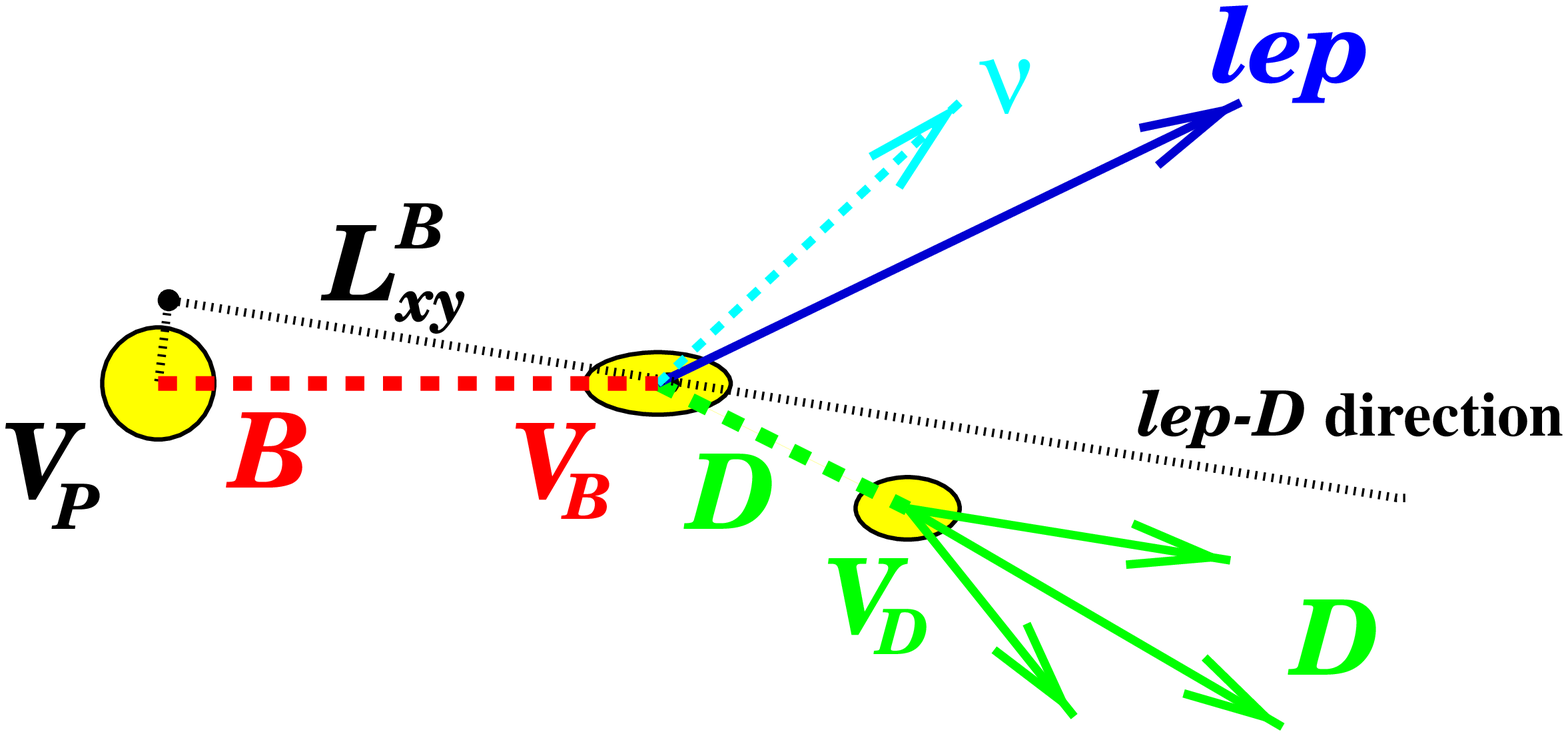}
}
\vspace*{0.2cm}
\fcaption{
Schematic representation of a semileptonic $B \ra D \ell \nu$ decay.}
\label{bdecay_sketch}
\end{figure}

\noindent
In the following, we describe the reconstruction of the $B$~hadron
decay length on the more complicated example of a partially
reconstructed $B$~decay. 
We shall point out the difference to the simpler case of a fully
reconstructed $B$~hadron. As our example, we
choose a semileptonic decay $B \ra D \ell \nu$ which would represent a
lifetime measurement using, for
example, $B^0 \ra D^- \ell^+ \nu$ or $\Bs \ra \Dsl \nu$ decays. A schematic
representation of such a semileptonic $B \ra D \ell \nu$ decay is
displayed in~Figure~\ref{bdecay_sketch}.

Usually, the $D$~candidate is searched for in a cone in
$\eta\,\varphi$-space around the lepton candidate and fully reconstructed.
The tracks forming the $D$ candidate are then 
refit with a common vertex constraint referred to as the tertiary vertex
and marked as $V_D$ in Fig.~\ref{bdecay_sketch}. 
The secondary vertex, where the $B$ decays to a lepton and a $D$
(referred to as $V_B$), is obtained by simultaneously 
intersecting the trajectory of the lepton track with the flight path 
of the $D$ candidate. If the $D$~meson is fully reconstructed,
which is usually the case, the $D$ flight path and thus the $B$~decay
vertex are known.
The confidence level (C.L.) of the combined vertex fit  
is typically required to be greater than 1\%.
Furthermore, it is often required that the 
reconstructed $D$ decay vertex $V_D$ be positively displaced 
from the primary vertex, as projected along the direction of the
$D\ell$~momentum. 

The decay length $\lxy^B$ is then defined as the displacement 
of the secondary vertex $V_B$ from the primary event vertex $V_P$, measured in 
the plane transverse to the beam axis and  
projected onto the transverse momentum of the $D\ell$ system (see
Fig.~\ref{bdecay_sketch}): 
\begin {equation}
\lxy^B = \frac{(\vec V_B - \vec V_P) \cdot 
	\vec{\Pt}(D\ell)}{|\vec{\Pt}(D\ell)|}.
\end{equation}
$\lxy^B$ is a signed variable which can be negative for the
configuration, where the particle seems to decay before the point where
it is produced. 
The $B$~meson decay time is given by
\begin {equation}
c\,t\,(B) = \lxy^B \, \frac{m(B)}{\Pt(B)}, 
\end{equation}
where $m(B)$ is the mass of the $B$~hadron. In the case of a
fully reconstructed $B$~hadron, $\Pt(B)$ is known and $c\,t\,(B)$ is
used as input to fit directly for the $B$~lifetime.
In the case of a semileptonic $B$~decay, where 
the $B$~meson is not fully reconstructed, a
``pseudo-proper decay length'' is usually defined as
\begin {equation}
\lambda = \lxy^B \, \frac{m(B)}{\Pt(D\ell)}.
\end{equation}
Using single lepton trigger data, the reconstructed $B$~decay length
has a typical uncertainty of $\sim\!50$-$60~\mu$m, including the
contribution from the finite size of the primary event vertex. 
In addition, the ratio $K$ of the observed momentum to
the true momentum is introduced as
\begin {equation}
K = \frac{\Pt(D\ell)}{\Pt(B)},
\end{equation}
to correct between the reconstructed $\Pt(D\ell)$ and the unknown $\Pt(B)$ in
the data. The $B$~meson decay time is then given as
\begin {equation}
c\,t\,(B) = \lxy^B \, \frac{m(B)}{\Pt(D\ell)} \otimes K.
\end{equation}
The correction between $\Pt(D\ell)$ and $\Pt(B)$
is done statistically 
by smearing an exponential decay distribution with a Monte Carlo
distribution of the correction factor~$K$,
when extracting $c\tau(B)$
from the pseudo-proper decay length in the lifetime fit procedure as
described in the next Section~6.2.4.

\begin{figure}[tbp]
\centerline{
\put(14,50){\large\bf (a)}
\put(77,50){\large\bf (b)}
\put(21,0){\bf \boldmath{$\Pt(\Dsl)/\Pt(\Bs)$}}
\put(85,0){\bf \boldmath{$\Pt(\Dsl)/\Pt(\Bs)$}}
\epsfxsize=6.3cm
\epsffile[5 5 560 520]{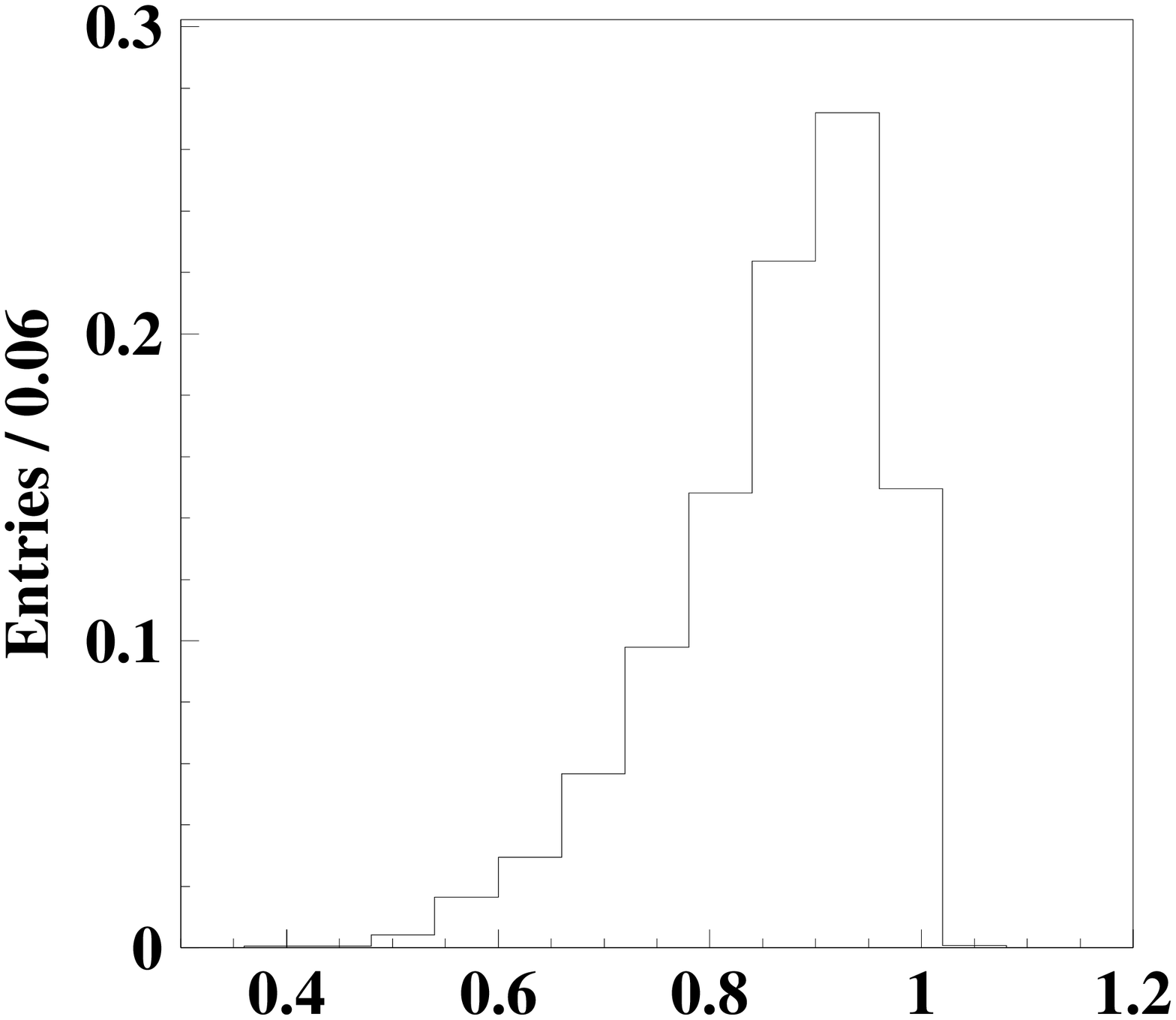}
\epsfxsize=6.3cm
\epsffile[5 5 560 520]{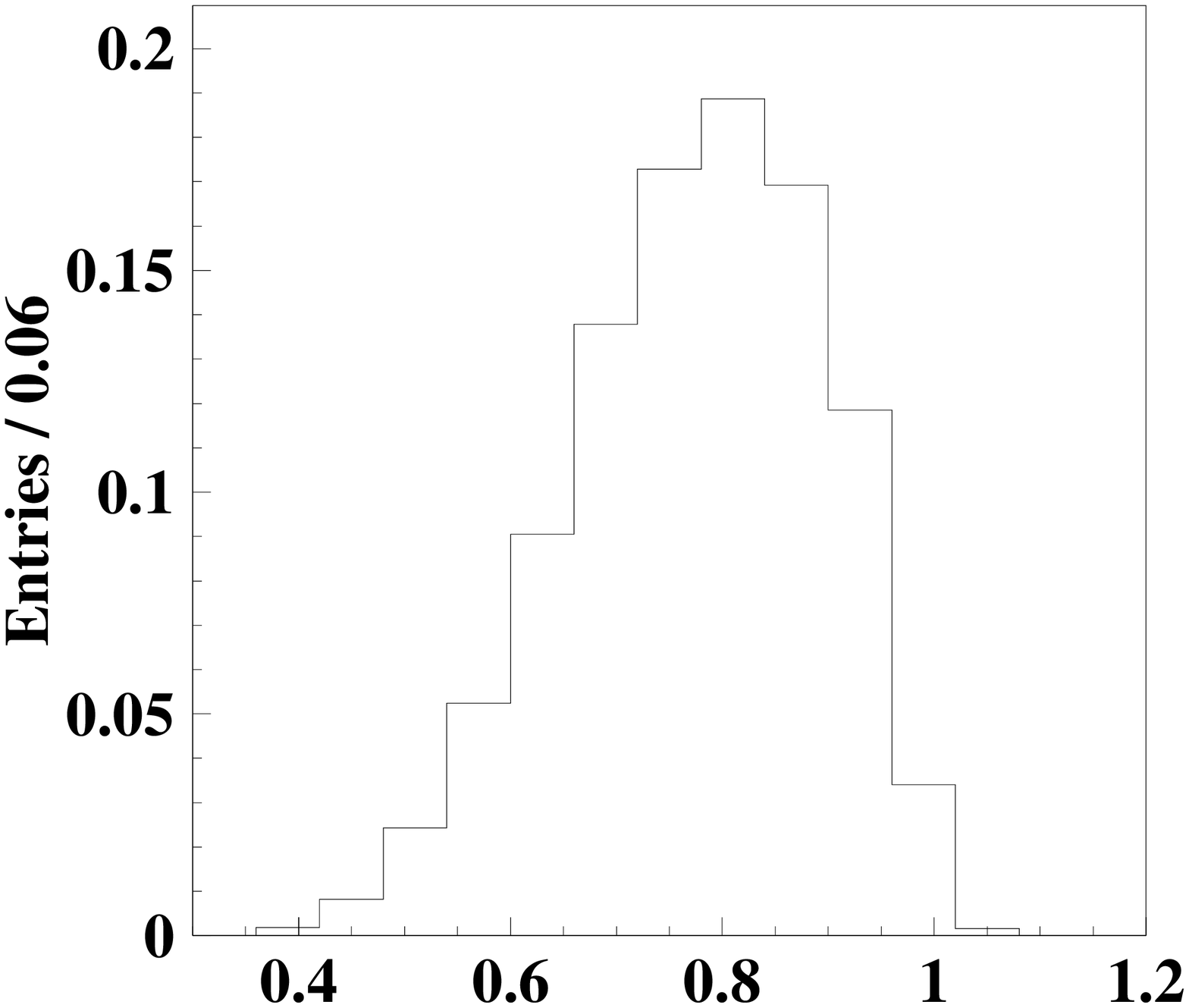}
}
\vspace*{0.2cm}
\fcaption{
Examples of normalized $K$-factor distributions $\Pt(\Dsl)/\Pt(\Bs)$, for 
$\Bs \ra \Dsl \nu X$ Monte Carlo decays with (a) $\Ds \ra \phipi$ and
(b) $\Ds \ra \phil$.}
\label{kdist}
\end{figure}

The $K$-distribution is obtained from $D\ell$~combinations which
originate from a Monte Carlo simulation of the semileptonic $B$~decay
of interest. As an example, the $K$-distribution 
is shown in Figure~\ref{kdist} for the $\Bs$~lifetime measurement
using the semileptonic 
decay $\Bs \ra \Ds \ell^+ \nu$ which is further described in Sec.~6.5. 
Figures~\ref{kdist}a) and \ref{kdist}b) show the $K$-distribution for
the cases where the \Ds~meson is reconstructed in the $\Ds \ra \phipi$ and   
$\Ds \ra \phil$ decay modes, respectively. 
These $K$-distributions have mean values of 0.86 and 0.77 with RMS
values of 0.10 and
0.12 for $\Ds \ra \phipi$ and  
$\Ds \ra \phil$, respectively.
The $K$-distribution is 
approximately constant as a function of $\Pt(D\ell)$ in the range of
interest, which is typically 15~\gevc\ to 25~\gevc\ for single lepton trigger
data.

To ensure a precise $B$ lifetime determination, usually only
$B$~candidates are considered for which the pseudo-proper decay length is
measured with an uncertainty of less than 0.1~cm.  
In addition, the $D$ candidates are often required to have a proper
decay length measured between $V_B$ and $V_D$ 
of less than 0.1~cm and an uncertainty on this proper decay length 
of less than 0.1~cm.  These requirements reject poorly measured decays
and reduce random track combinations.

\subsubsection{Description of lifetime fit procedure}
\noindent
We now describe the fit procedure used to extract the $B$~lifetime
from the pseudo-proper decay length distribution. We continue to use our
example of a semileptonic $B \ra D \ell \nu$ decay.
As input to the $B$~hadron lifetime fit, usually, a signal
sample is defined by a mass window around the respective $D$~mass
peak. To model the pseudo-proper decay length distribution of the
background events 
contained in the signal sample, a background sample is defined
consisting of events from the $D$~sidebands. See
Fig.~\ref{jpsi_mass_tau}a) for a better understanding of the
definition of signal events (light shaded area) 
and sideband samples (dark shaded area), shown here for a
$J/\psi \ra \mu^+\mu^-$ invariant mass distribution.
 
The pseudo-proper decay length distribution obtained from the signal sample
is then fit using an unbinned maximum
log-likelihood method. Both the $B$~lifetime, denoted as $c\tau$
below, and the background shape  
are determined in a simultaneous fit using the signal and background
samples. Thus the likelihood function ${\cal L}$ is a combination of
two parts  
\begin{equation}
{\cal L}  =  \prod^{N_S}_i\, [f_{\rm sig}{\cal F}^i_{\rm sig}
 + (1 - f_{\rm sig}){\cal F}^i_{\rm bg}\, ] 
\cdot \prod^{N_B}_j{\cal F}^j_{\rm bg},
\end{equation}
where $N_S$ and $N_B$ are the number of events in the signal and background 
samples.
Here, $f_{\rm sig}$ is the ratio of $D$~signal events 
to the total number of events
in the signal sample. To constrain $f_{\rm sig}$, which is usually a
free fit parameter, to the number of
$D$~signal events $\langle f_{\rm sig}\rangle$ with uncertainty
$\sigma_{\rm sig}$, 
obtained from the $D$~mass distribution, 
an additional
$\chi^2$~term 
$\chi^2=(f_{\rm sig}-\langle f_{\rm sig}\rangle)^2/
\sigma_{\rm sig}^2$
is factored into the  
likelihood function defined above.

The signal probability function ${\cal F}_{\rm sig}$ consists of a normalized 
decay exponential function convoluted with 
a Gaussian resolution function ${\cal G}$ and is smeared with a normalized
$K$-distribution ${\cal H}(K)$
\begin{equation}
{\cal F}^i_{\rm sig}(x) = 
\int dK\, {\cal H}(K)\, \left[
\frac{K}{c\tau} \exp\{-\frac{K x}{c\tau}\} 
\otimes {\cal G}(\lambda_i |\, x,s\sigma_i)
\right].
\end{equation}
Here, $\lambda_i$ is the pseudo-proper decay length measured for event
$i$ with uncertainty $\sigma_i$ and 
$x$ is the true pseudo-proper decay length.
The symbol ``$\otimes$" denotes a convolution and
${\cal G}(\lambda_i |\, x,s\sigma_i)$ is the Gaussian
distribution given by 
\begin{equation}
	{\cal G}(\lambda_i |\, x,s\sigma_i) = 
	\frac{ 1} { s \sigma_i \sqrt{ 2 \pi } }
	  \, \exp \left( - \frac {(x-\lambda_i)^2 } { 2 s^2 \sigma_i^2 } \right).
\end{equation}
Because of systematic uncertainties in the overall scale of the decay
length uncertainties, which are estimated on an event-by-event basis, 
a scale factor, $s$, is often introduced. It is usually a free
parameter in the $B$~lifetime fit. 
The integration over the momentum ratio $K$ is approximated by a
finite sum
\begin{equation}
\int dK\, {\cal H}(K) \ra
\sum_{i}\, \Delta K\,{\cal H}(K_i), 
\end{equation}
where the sum is taken over bin $i$ of a histogrammed distribution
${\cal H}(K_i)$ with bin width $\Delta K$ as shown e.g.~in Figure~\ref{kdist}.

The background probability function ${\cal F}_{\rm bg}$ is typically
parametrized by a Gaussian centered at zero, a negative exponential tail,
and a positive decay exponential to characterize the contribution of
heavy flavour decays in the background sample:
\begin{eqnarray}
{\cal F}^i_{\rm bg}(x) & = &
(1-f_{+}-f_{-}) \, {\cal G}(\lambda_i |\, x,s\sigma_i) 
 +  \frac{f_{+}}{\lambda_{+}} \exp\{-\frac{x}{\lambda_+}\} 
  \otimes {\cal G}(\lambda_i |\, x,s\sigma_i) + \nonumber \\
& + & \frac{f_{-}}{\lambda_{-}} \exp\{-\frac{x}{\lambda_-}\} 
  \otimes {\cal G}(\lambda_i |\, x,s\sigma_i).
\end{eqnarray}
Here, $f_{\pm}$ are the fractions of positive and negative lifetime
backgrounds and
$\lambda_{\pm}$ are the effective lifetimes of those backgrounds. 
The parameters usually allowed to float in the fit are the $B$~lifetime,
$f_{\rm sig}$, $\lambda_{\pm}$, $f_{\pm}$, and the overall scale
factor $s$. 

For a fully reconstructed $B$~decay, the smearing with a normalized
$K$-distribution ${\cal H}(K)$ does not apply but the convolution with 
the Gaussian resolution function ${\cal G}$ is essential to obtain the 
true pseudo-proper decay length $x$. In the case of a fully
reconstructed $B$~decay, the background is usually also
described by a Gaussian centered at zero, a negative exponential tail,
and a positive decay exponential.

\subsection{Measurement of average \boldmath{$B$}~hadron lifetime}
\noindent
The measurement of the average $B$~hadron lifetime is based on 
$B \ra J/\psi X$ decays where the $J/\psi$ is reconstructed through
its decay into $\mu^+\mu^-$. This measurement constitutes the
first publication of a $B$~lifetime at CDF\cite{inc_blife} using
only about half of the Run\,Ia data. This analysis demonstrated in 1993
the capabilities of the newly installed silicon vertex detector.
The final average $B$~hadron measurement\cite{lifetimeprd} uses a
subset of the $J/\psi$ data sample already shown in
Figure~\ref{jpsi_mass_tau}a). Only the Run\,Ib data and only
$J/\psi$~candidates, where both muons are matched to a CFT track at
Level~2, are used. Since the precision of the measurement is limited
by systematics, including more data would not improve the result. 

This analysis uses the two-dimensional decay length \lxy\ of the
$J/\psi$ vertex and corrects for the difference between the boost of
the $J/\psi$~meson and the $B$~hadron, using a Monte Carlo model which
compared the background subtracted \Pt-distribution of the data with
the $J/\psi$ momentum of the MC simulation (see Sec.~6.2.3.). Care is
taken in this analysis to avoid non-Gaussian components in the errors
on the decay length distribution. To ensure that the $J/\psi$ vertex
is well measured and to have a good understanding of the $\sigma$ of
the vertex fit, strict track and vertex quality cuts are
applied. These required, for example, both muons to have associated
clusters in all four layers of the SVX or
the calculated uncertainty on the decay length 
$\sigma_{\lxy}$ to be less than $150~\mu$m. This leaves 67,800
pairs of oppositely 
charged muons in the $J/\psi$~signal region and 7,900 pairs in the
combined sidebands, which are used to~determine the decay length
distribution of the background underneath~the~$J/\psi$~signal.

The pseudo-proper decay length distribution for the signal region, with
the result of the fit superimposed, is already shown in
Fig.~\ref{jpsi_mass_tau}b). The dark shaded area shows the
contribution from background, where the shape is determined from
the sidebands and the magnitude is derived by normalizing the
number of sideband events within Poisson fluctuations to the same range
in invariant mass as used for the signal sample. The light shaded
region shows the contribution due to adding the exponential
distribution from $B$~decay to the
background. The remaining unshaded region, which constitutes 
more than 80\% of the $J/\psi$~sample, shows the contribution from
prompt $J/\psi$~mesons produced at the primary interaction vertex.
 
The measured inclusive $B$~lifetime, which is the average over all
$B$~hadrons produced in $p\bar p$ collisions at $\sqrt{s} = 1.8$~TeV
weighted by their production cross section, branching ratios and
detection efficiencies, is
\begin{equation}
\langle\tau(B)\rangle = (1.533 \pm 0.015\ ^{+0.035}_{-0.031})\ {\rm ps}.
\end{equation}
The first uncertainty is statistical and the second is systematic.
This is the standard way to present results where
two uncertainties are quoted in this paper.

\subsection{Measurement of \boldmath{$B^+$} and \boldmath{$B^0$} lifetime} 
\noindent

\subsubsection{$B^+$ and $B^0$ lifetimes with fully reconstructed $B$~mesons}  
\noindent
The analysis principle for the 
$B$ lifetime measurement using fully reconstructed $B^+$ and $B^0$
mesons\cite{lifetimeprd} is as follows.  All Run\,I 
dimuons forming a $J/\psi$
candidate, as shown in Figure~\ref{jpsi_mass_tau}a), are used.
$B^+$~mesons are reconstructed in the decay modes 
$J/\psi K^+$, $J/\psi K^{*+}$, $\psi(2S) K^+$, and $\psi(2S) K^{*+}$, 
while $B^0$~mesons are reconstructed in the decay modes 
$J/\psi K^0_S$, $J/\psi K^{*0}$, $\psi(2S) K^0_S$, and $\psi(2S)
K^{*0}$ with $\psi(2S)\ra \mu^+\mu^-$.
The kaons are reconstructed using the decay channels 
$K^{*0} \ra K^+\pi^-$, $K^{*+} \ra K^0_S \pi^-$, and 
$K^0_S \ra \pi^+\pi^-$. $K^*$~candidates are accepted if their
invariant $K\pi$ mass is within $\pm 80$~\mevcc\ of the world average
$K^*$ mass\cite{PDG}. Since swapping the assignment of the kaon and
pion masses to the two tracks forming a $K^{*0}$~candidate can also
result in a $K^{*0}$ candidate within the $\pm80$~\mevcc\ mass window,
the $K^{*0}$~combination closest to the world average mass value is chosen.

\begin{figure}[tbp]
\centerline{
\put(53,55){\large\bf (a)}
\put(115,55){\large\bf (b)}
\epsfclipon
\epsfxsize=6.3cm
\epsffile[20 150 530 648]{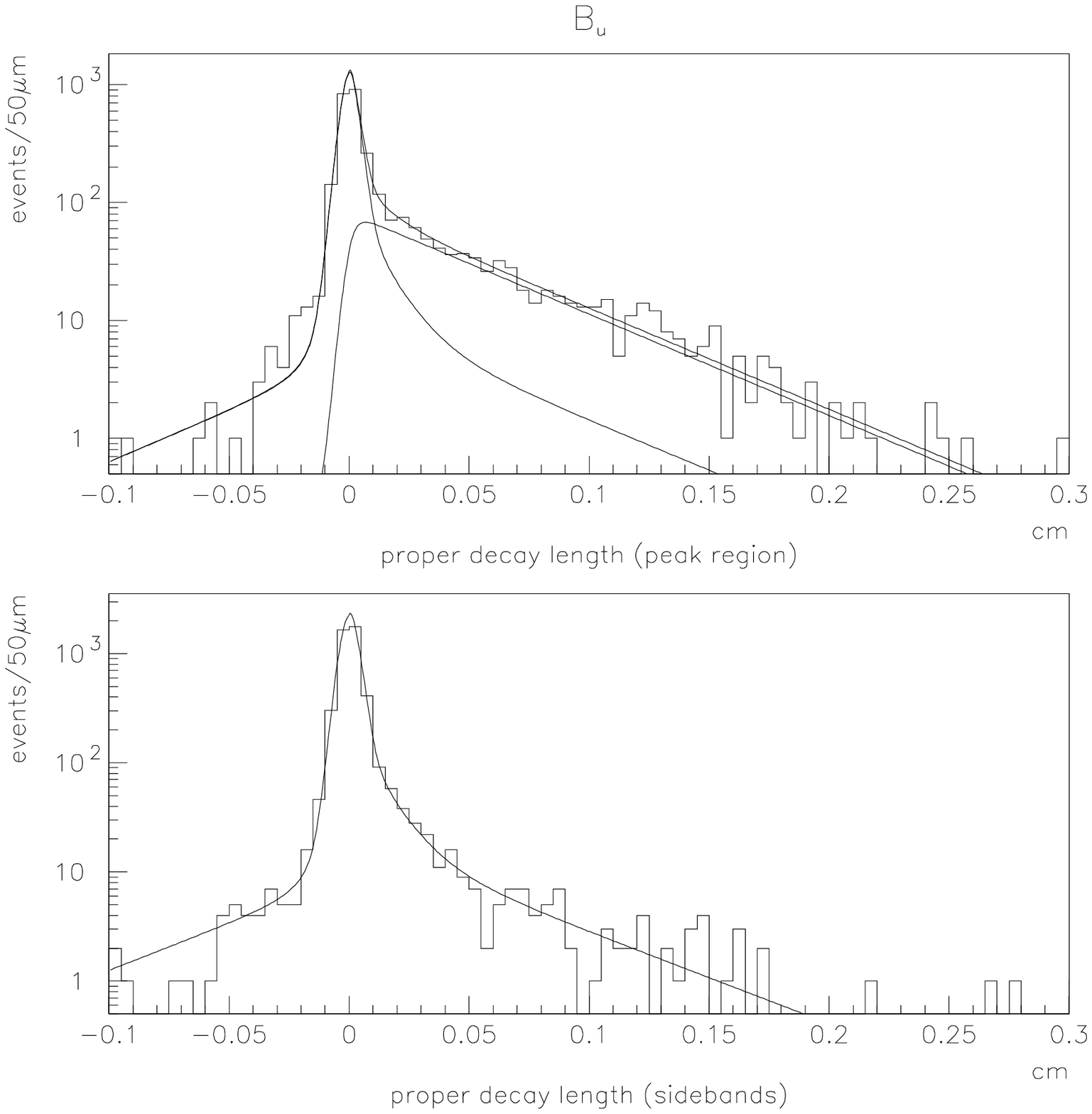}
\epsfxsize=6.3cm
\epsffile[20 150 530 648]{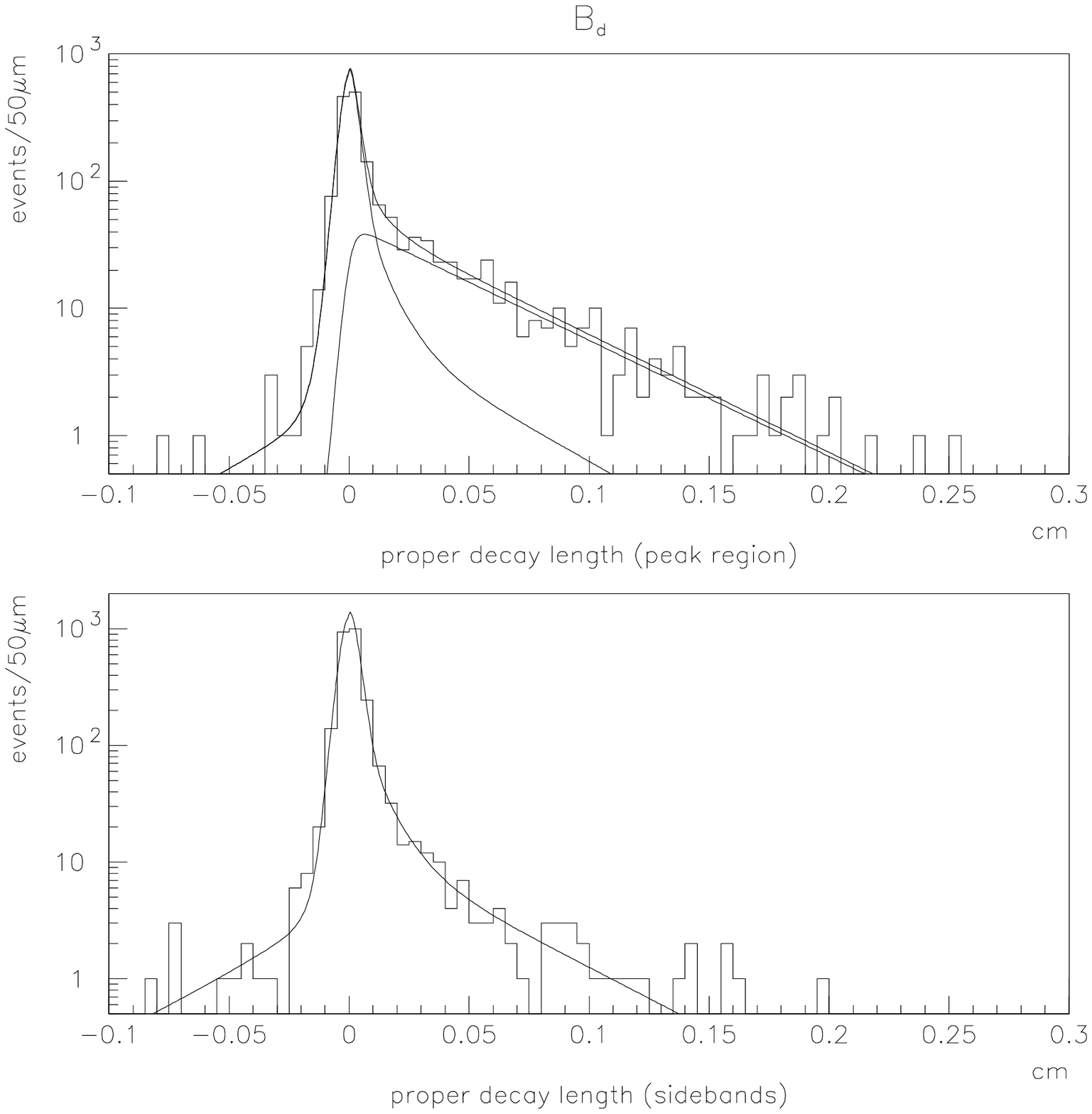}
\epsfclipoff
}
\vspace*{0.2cm}
\fcaption{
The proper decay length distribution for fully reconstructed (a)
$B^+$~mesons and (b) $B^0$~mesons. The upper (lower) histogram shows
the peak (sideband) region distribution. The superimposed curves are
the contributions from the signal, the background and their sum
determined by the log-likelihood fit.}
\label{lifetime_b0bp}
\end{figure}

The vertex and mass constraint $J/\psi$ and $\psi(2S)$~candidates are
fit with the kaon candidates to originate from a common vertex,
yielding the two-dimensional $B$~meson decay length $\lxy^B$. 
Together with the measured $B$~transverse momentum \Pt, 
$\lxy^B$ is used to obtain the proper time distributions 
shown in Fig.~\ref{lifetime_b0bp} for (a)~charged and (b)~neutral 
$B$~candidates.
The bottom plots represent the backgrounds as
obtained by fitting the $B$ sideband regions to a Gaussian with
exponential tails. Using this background shape, an unbinned
maximum log-likelihood fit of  
the signal, assumed to be an exponential convoluted with a Gaussian, is
performed. The following lifetimes are obtained:
\begin{eqnarray}
\tau(B^+) = (1.68 \pm 0.07 \pm 0.02)\ {\rm ps}, \nonumber \\
\tau(B^0) = (1.58 \pm 0.09 \pm 0.02)\ {\rm ps}, \nonumber \\
\tau(B^+)/\tau(B^0) = 1.06 \pm 0.07 \pm 0.02.
\label{eq:blife_jpsi}
\end{eqnarray}
When calculating the uncertainty on the lifetime ratio, the correlated
systematic errors are properly taken into account.

The exclusive $B$~lifetime measurement using fully reconstructed $B$~decays
is statistics dominated.
One way to increase the number of $B$ candidates is not to fully
reconstruct the $B$ meson. This is done in the semi-exclusive analysis 
described in the next section.
   
\subsubsection{$B^+$ and $B^0$ lifetimes with partially reconstructed
$B$~mesons}   
\noindent
The $B$ lifetime analysis using partially reconstructed $B$
mesons\cite{semiprd} 
exploits the semi\-leptonic decays $B \ra D^{(*)}\ell \nu X$ and follows the
description given in Sec.~6.2.3 and 6.2.4, as illustrated in
Fig.~\ref{bdecay_sketch}. The analysis starts with
the single lepton trigger data and searches for charm mesons in a cone
around the trigger 
electron or muon. $D^{(*)}$ meson candidates are reconstructed through their
decay modes: \\
\hspace*{0.5cm} (a)~$\bar D^0 \ra K^+ \pi^-$, 
	where the $\bar D^0$ is not from a
$D^{*-}$ decay, \\
\hspace*{0.5cm} (b)~$D^{*-} \ra \bar D^0 \pi^-,\ 
\bar D^0 \ra K^+ \pi^-$, \\
\hspace*{0.5cm} (c)~$D^{*-} \ra \bar D^0 \pi^-,\ 
\bar D^0 \ra K^+ \pi^-\pi^+\pi^-$, and\\
\hspace*{0.5cm} (d)~$D^{*-} \ra \bar D^0 \pi^-,\ 
\bar D^0 \ra K^+ \pi^- \pi^0$, 
where the $\pi^0$ is not reconstructed. 

\begin{figure}[tbp]
\centerline{
\epsfxsize=9.0cm
\epsffile[65 180 565 695]{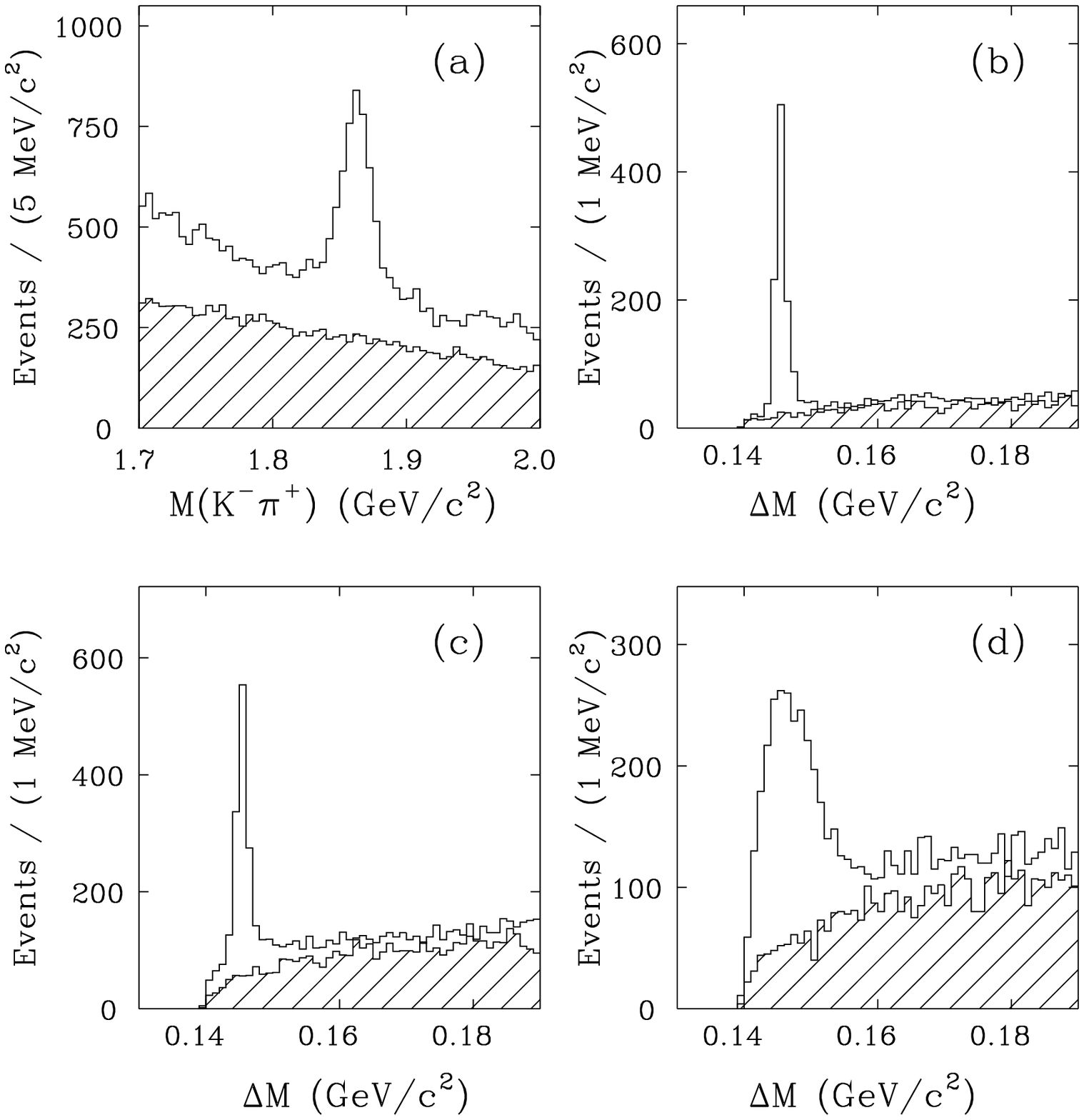}
}
\vspace*{0.2cm}
\fcaption{
Charm signals reconstructed in the 
$B$ lifetime measurement using partially reconstructed 
$B \ra D^{(*)}\ell\nu X$ decays with
(a) $\bar D^0 \ra K^+ \pi^-$, where the $\bar D^0$ is not from a
$D^{*-}$ decay, 
(b) $D^{*-} \ra \bar D^0 \pi^-,\ \bar D^0 \ra K^+ \pi^-$,
(c) $D^{*-} \ra \bar D^0 \pi^-,\ \bar D^0 \ra K^+ \pi^-\pi^+\pi^-$, and
(d) $D^{*-} \ra \bar D^0 \pi^-,\ \bar D^0 \ra K^+ \pi^- \pi^0$.
The shaded histograms show wrong-sign $D^{*+}\ell^+$ or $D^0\ell^+$
combinations.} 
\label{blife_semi_mass}
\end{figure}

The charm signals for the decay modes (a)-(d) above can be seen in
Fig.~\ref{blife_semi_mass}a)-d), respectively.  
Figure~\ref{blife_semi_mass}a) shows the $K^+ \pi^-$ mass spectrum
while in Fig.~\ref{blife_semi_mass}b)-d) the mass difference $\Delta m$ 
between the measured invariant masses of the $\bar D^0 \pi^-$ and
$\bar D^0$ candidates is displayed. 
The mass peak is broadened for mode (d) because of the missing $\pi^0$ meson.
Note, the charm signals in Fig.~\ref{blife_semi_mass} are
quite clean and rather competitive with $D^{(*)}$ signals found at $e^+e^-$
colliders. 
The number of events in the signal regions and the estimated
background fractions are summarized in Table~\ref{blife_semi_sig}.

\begin{table}[tbp]
\tcaption{
Summary of the definition of signal samples, numbers of charm
candidates and estimated background fractions.}
\centerline{\footnotesize\smalllineskip
\begin{tabular}{llccrc} 
\hline
 & & & & & \\
 \vspace*{-0.6cm} \\
 $B$ Mode & $\bar D^0$ mode & $\bar D^0$ mass range & $\Delta m$ range 
& Events & Background fraction \\
&			& [\gevcc]      & [\gevcc]
& & \\ 
\hline
 & & & & & \\
 \vspace*{-0.6cm} \\
$\ell^+\bar D^0$ 
& $K^+ \pi^- $ 
&  1.84 $-$ 1.88   &  Not $D^{*-}$
& 5198 & $0.526 \pm 0.018$ \\
$\ell^+ D^{*-}$ 
& $K^+ \pi^-$ 
&   1.83 $-$ 1.90 & 0.144 $-$ 0.147
& 935 & $0.086 \pm 0.011$ \\
$\ell^+ D^{*-} $
& $K^+ \pi^-\pi^+\pi^-$
&    1.84 $-$ 1.88 & 0.144 $-$ 0.147
& 1166 	& $0.183 \pm 0.015$ \\ 
$\ell^+ D^{*-} $
& $K^+ \pi^- \pi^0$
&    1.50 $-$ 1.70  & $< 0.155$
& 2858 & $0.366 \pm 0.016$ \\ 
\hline
\end{tabular}}
\label{blife_semi_sig}
\end{table}

\begin{figure}[tbp]
\centerline{
\put(16,58){\large\bf (a)}
\put(79,58){\large\bf (b)}
\epsfxsize=6.3cm
\epsffile[60 190 490 660]{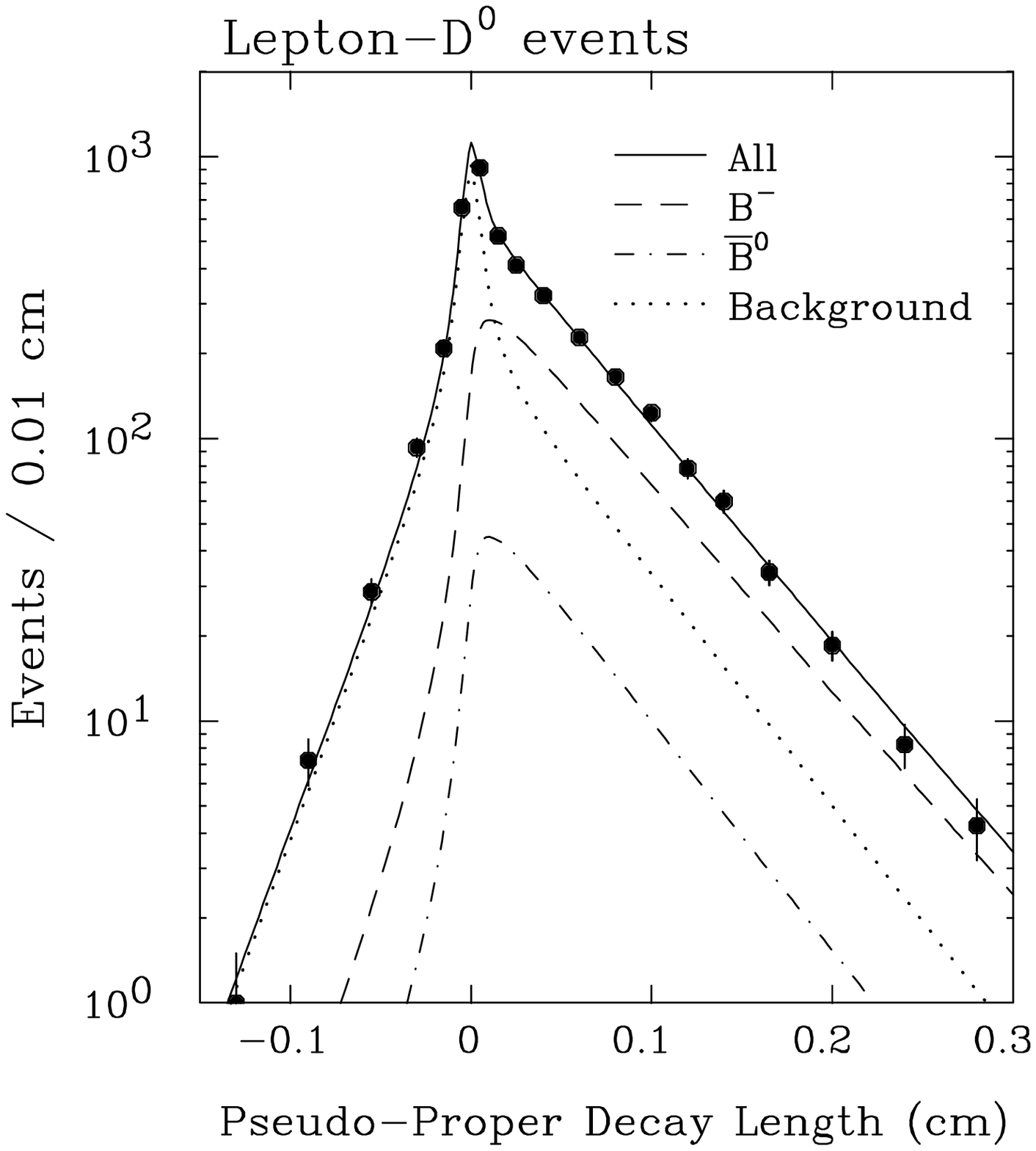}
\epsfxsize=6.3cm
\epsffile[60 190 490 660]{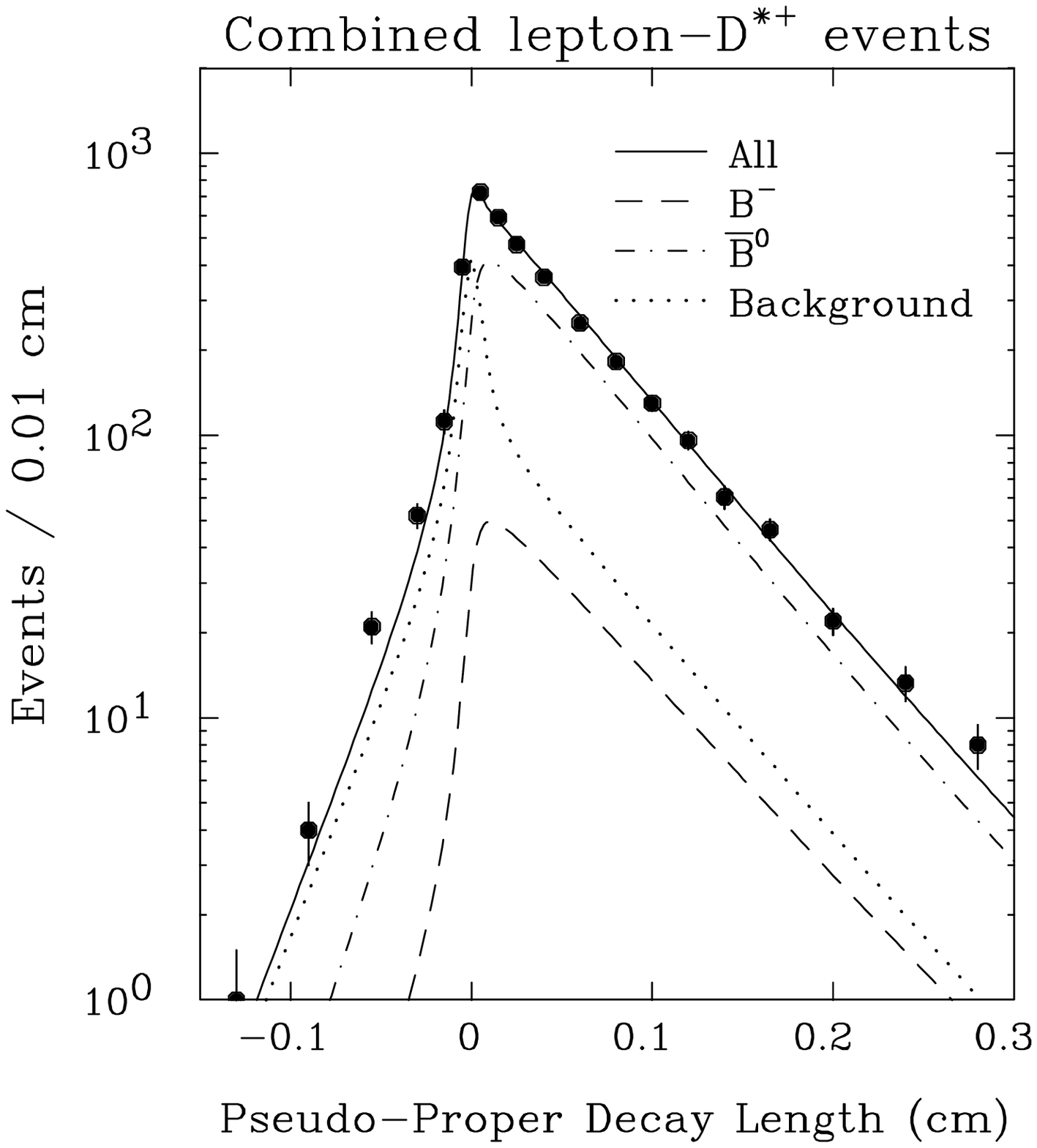}
}
\vspace*{0.2cm}
\fcaption{
Pseudo-proper decay length distribution 
of (a) $\bar D^0\ell^+$ and (b) $D^{*-}\ell^+$ candidates where the
three $\bar D^0$ decay modes are combined.
The curves show the result of the combined fit:
The $B^0$ component (dot-dashed curves),
the $B^+$ component (dashed curves),
and the background component (dotted curves).}
\label{blife_semi}
\end{figure}

The $\bar D^0$ and $D^{*-}$ candidates are then intersected with the
lepton to find the $B$ decay vertex. Since the $B$ meson is not fully
reconstructed, a $\beta\gamma$~correction
has to be applied to scale from the observed 
$D^{(*)} \ell$ momentum to $\Pt(B)$ as described in Sec.~6.2.3.    
The obtained lifetime distributions from $\bar D^0\ell^+$ and $D^{*-}\ell^+$ 
are used to determine the individual $B^+$ and $B^0$ lifetimes. 
A $\bar D^0\ell^+$ combination usually originates from a charged $B$ meson
while a $D^{*-}\ell^+$ candidate comes from a $B^0$. This simple
picture is complicated by the 
existence of $D^{**}$ states which are the source of $\bar D^0$ ($D^{*-}$)
mesons originating from a decay $B^0 \ra D^{**-} \ell^+ X,\ D^{**-}
\ra \bar D^0 X$
($B^+ \ra \bar D^{**0} \ell^+ X,\ \bar D^{**0} \ra D^{*-} X$). 
This cross talk from
$D^{**}$ resonances, decomposed using Monte Carlo, is an
important aspect of this analysis\cite{semiprd}. A combined
lifetime fit to the pseudo-proper decay length distributions, using an
unbinned maximum log-likelihood method, as described 
in Sec.~6.2.4, yields the results shown in Fig.~\ref{blife_semi}.
The contamination of the wrong $B$~species is only at the 10-15\%
level. It is indicated in Fig.~\ref{blife_semi}, in addition to the respective
signal and background contributions.

The final lifetimes of the
$B^+$ and $B^0$ mesons, using their
partially reconstructed semileptonic decays
$B^+ \ra \bar D^0 \ell^+ \nu X$  and 
$B^0 \ra D^{*-} \ell^+ \nu X$, are
\begin{eqnarray}
\tau(B^+) & = & (1.637 \pm 0.058 \, ^{+\, 0.045} _{ -\, 0.043}) \ {\rm ps}, 
\nonumber \\
\tau(B^0) & = & (1.474 \pm 0.039 \,^{+\, 0.052}_{ -\, 0.051}) \ {\rm ps}, 
\nonumber \\
\tau(B^+)  / \tau(B^0) & = & 1.110 \pm 0.056
\, ^{+\, 0.033} _{ -\, 0.030} , 
\label{eq:blife_semi}
\end{eqnarray}
where the main systematic error is from the estimate of the
$B$~meson~momentum.

\subsubsection{$B^0$ and $B^+$ lifetime summary} 
\noindent
We combine the $B$~lifetime measurement
using fully reconstructed decays, described in
Sec.~6.4.1 and given in Eq.~(\ref{eq:blife_jpsi}), with the CDF
measurement using partially reconstructed $B$~mesons, described in
Sec.~6.4.2 and given in Eq.~(\ref{eq:blife_semi}),
to derive the following CDF average $B$~lifetimes:
\begin{eqnarray}
 \tau(B^+) & = & (1.661 \pm 0.052) \;\; {\rm ps}, \nonumber \\
 \tau(B^0) & = & (1.513 \pm 0.053) \;\; {\rm ps}, \nonumber \\
 \tau(B^+) / \tau(B^0) & = & 1.091 \pm 0.050,
\end{eqnarray}
where the uncertainties include both statistical and systematic
effects.
There exists a small (about 0.01~ps) 
correlation in the systematic effects
between the two measurements. This
is taken into account in combining the results.
The ratio of the two $B$ meson lifetimes
differs from unity by about 
9\%, or almost two standard deviations.
This agrees with the small difference predicted by theory\cite{bigi}.

\begin{figure}[tbp]
\centerline{
\epsfysize=6.2cm
\epsffile[1 1 540 515]{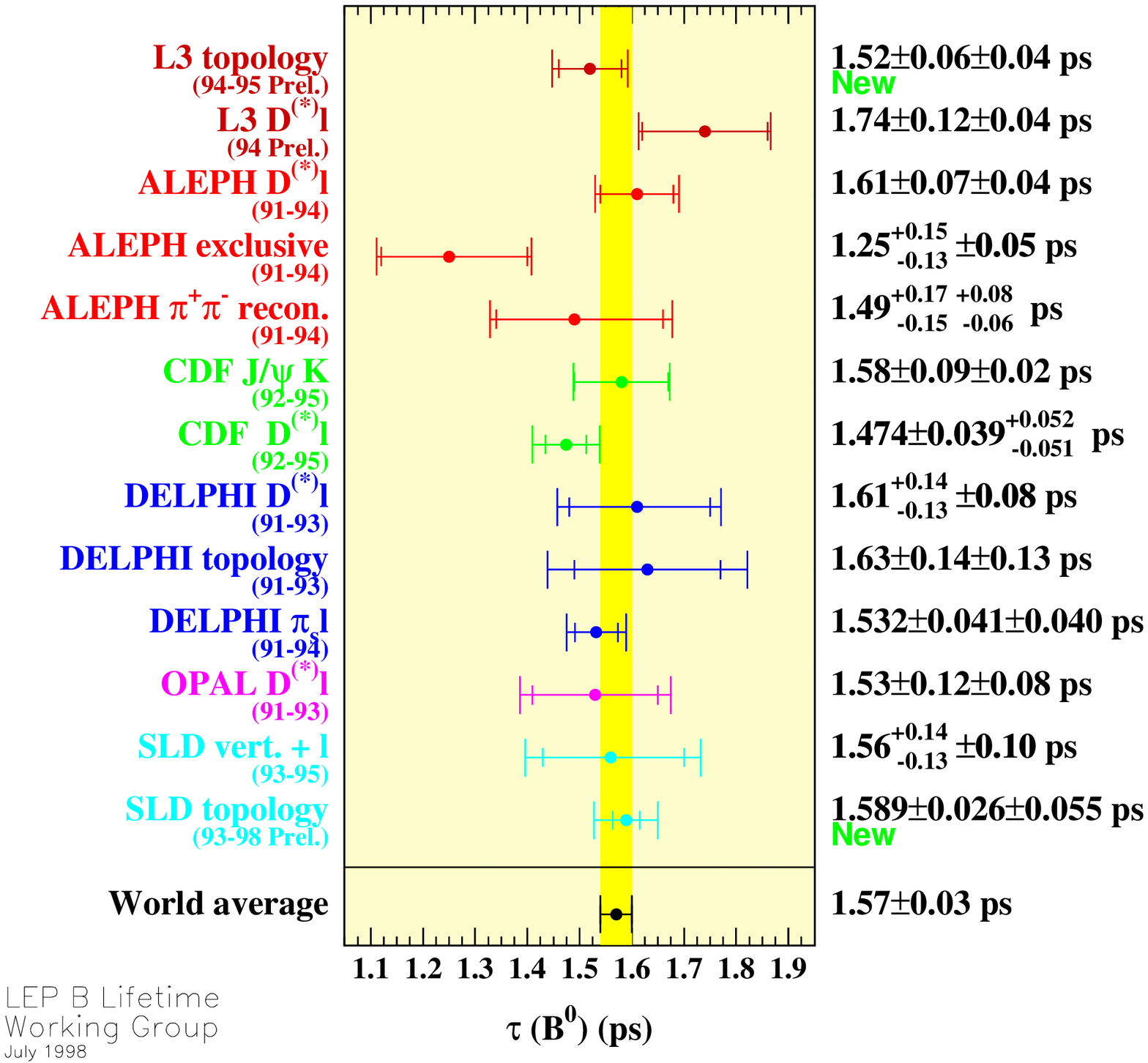}
\epsfysize=6.2cm
\epsffile[21 1 530 515]{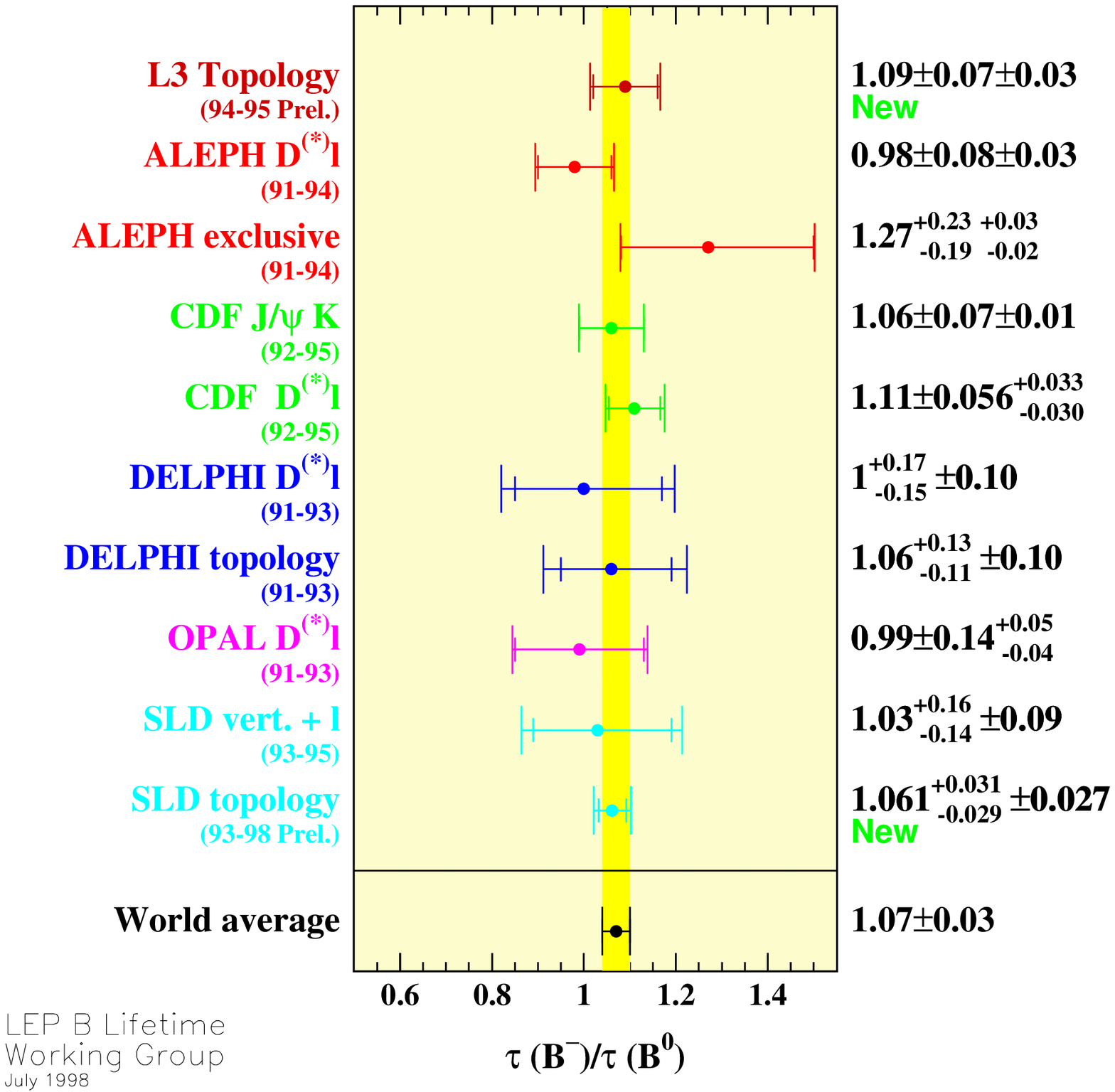}
}
\vspace*{0.2cm}
\fcaption{
Comparison of the CDF $B^0$ lifetime measurements (left) and the 
$\tau(B^+)/\tau(B^0)$ lifetime ratio
measurements (right) with other
experiments, compiled by the LEP $B$~lifetime working group as of July 1998.}
\label{blife_comp}
\end{figure}

As an example, a comparison of the CDF $B^0$~lifetime measurements with other
experiments, compiled by the LEP $B$~lifetime working group in July 1998
for the 29th International Conference on High Energy Physics,
Vancouver, Canada, 
can be found in Figure~\ref{blife_comp} on the left hand side.
A comparison of the CDF $B$~lifetime ratio measurements
compared to other experiments is
presented in Figure~\ref{blife_comp} on the right hand side. These
comparisons show that the CDF $B$~lifetime measurements are
very competitive with results from the $Z^0$ pole at LEP and SLC.
The current world average $B^0$ and $B^+$~lifetimes as of July 1998,
determined by the LEP $B$~lifetime working group, taking correlated
systematic uncertainties into account are:
\begin{eqnarray}
 \tau(B^+) & = & (1.67 \pm 0.03) \;\; {\rm ps}, \nonumber \\
 \tau(B^0) & = & (1.57 \pm 0.03) \;\; {\rm ps}, \nonumber \\
 \tau(B^+) / \tau(B^0) & = & 1.07 \pm 0.03.
\end{eqnarray}

\newpage
\subsection{Measurement of \boldmath{$\Bs$} lifetime} 
\noindent
CDF has measured the \Bs~lifetime using fully and partially
reconstructed \Bs~decays. A search for a lifetime difference \dgog\
in the \Bs~system is described in Section~6.5.3.

\subsubsection{\Bs\ lifetime with fully reconstructed \Bs~mesons}   
\begin{figure}[tbp]
\centerline{
\put(53,55){\large\bf (a)}
\put(115,55){\large\bf (b)}
\epsfxsize=6.3cm
\epsffile[15 145 525 675]{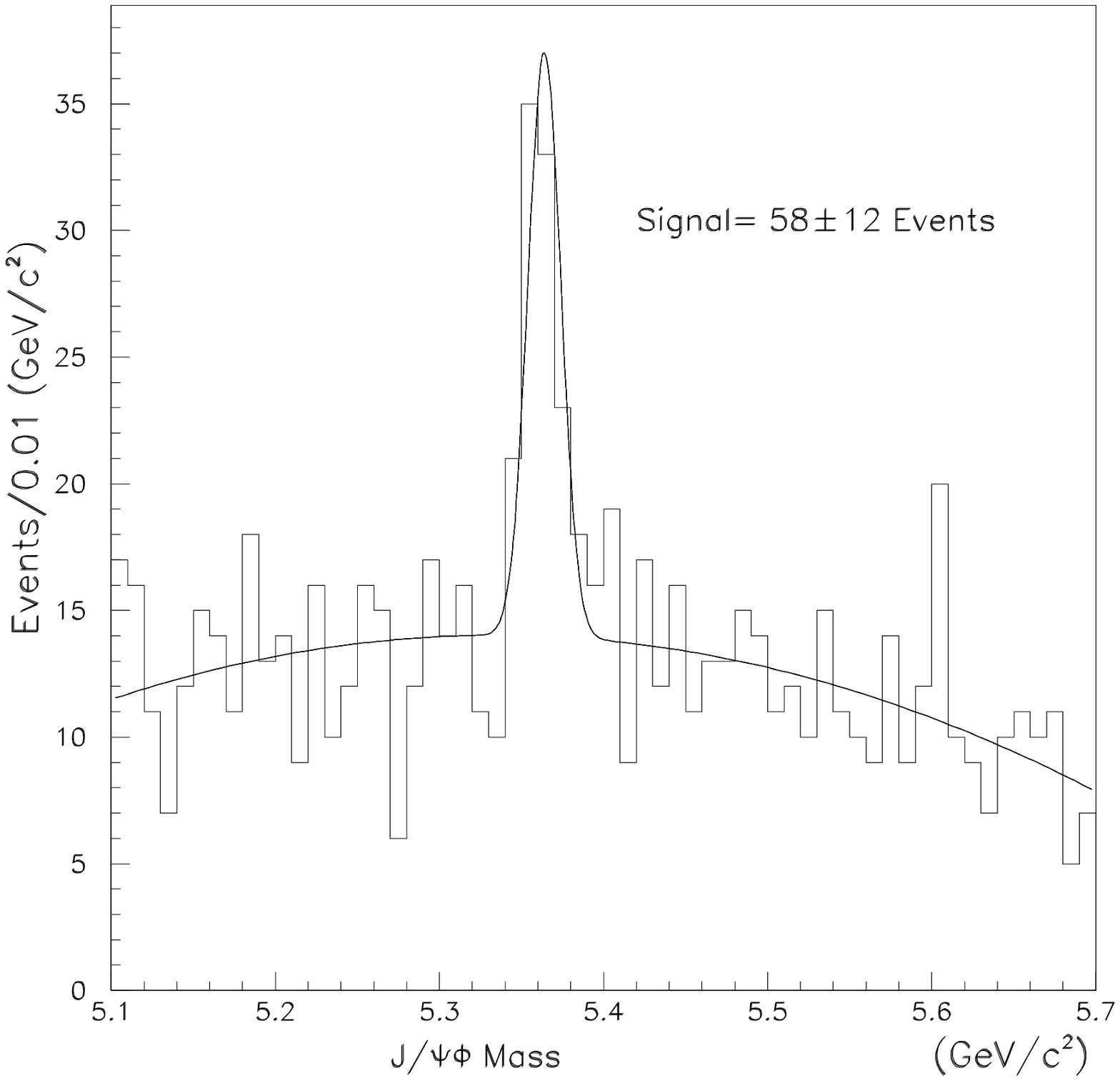}
\epsfxsize=6.3cm
\epsffile[15 145 525 675]{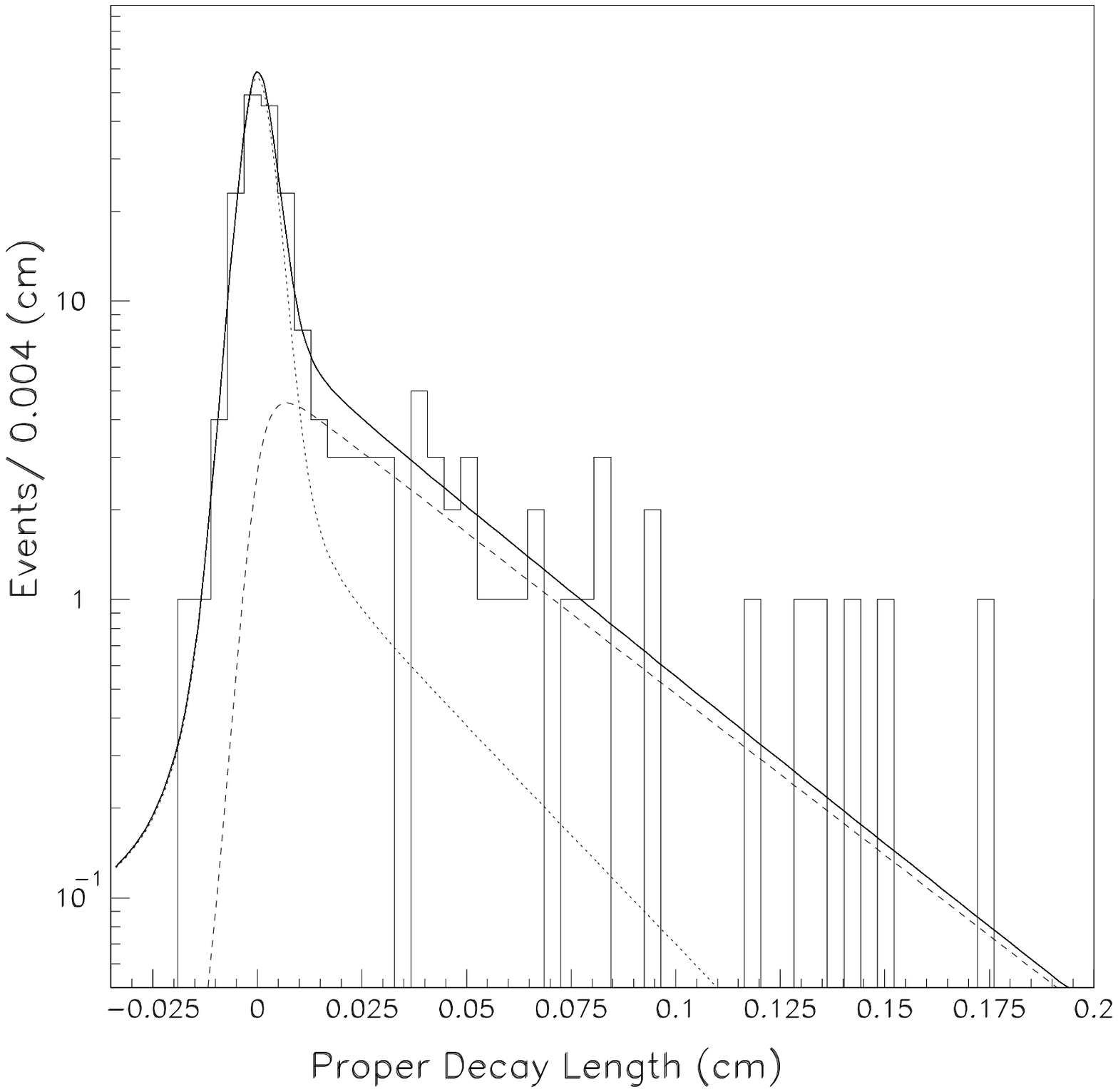}
}
\vspace*{0.2cm}
\fcaption{
(a) Invariant mass distribution of \Bs\ candidates from $\Bs \ra
J/\psi \phi$ fitted to a Gaussian plus a polynomial background.
(b) \Bs\ proper decay length distribution for events selected
within $\pm0.05$~\gevcc\ of the fitted \Bs~mass. The results of the
lifetime fit are shown: Signal (dashed line), background (dotted
line), and the sum of the two (solid line).}
\label{bs_psiphi}
\end{figure}

\noindent
In this exclusive \Bs~lifetime measurement\cite{lifetimeprd},
the \Bs\ candidates are reconstructed in the decay chain $\Bs \ra
J/\psi \phi$, with $J/\psi \ra \mu^+\mu^-$ and $\phi \ra K^+ K^-$. The
dimuon data sample used in the lifetime measurement
with fully reconstructed $B^0$ and $B^+$~mesons (see Sec.~6.4.1) is
also the starting 
point to reconstruct $J/\psi$ mesons in this analysis. Once a $J/\psi$
is found, $\phi \ra K^+K^-$ candidates are searched by selecting
oppositely charged track pairs assigning
each track the kaon mass. If the mass of the $\phi$~candidate lies
within $\pm 10$~\mevcc\ of the world average $\phi$ mass\cite{PDG},
all four tracks are constrained to come from a common vertex. The
$J/\psi\phi$ invariant mass spectrum is shown in
Fig.~\ref{bs_psiphi}a). A fit to a Gaussian plus a second order
polynomial background yields a signal of $(58\pm12)$ events.

Due to the more limited number of \Bs\ candidates in this analysis, a
maximum log-likelihood fit is simultaneously performed to the invariant mass 
and the proper decay length distributions, in order to better constrain the
number of \Bs\ signal events. A plot of the proper decay length
spectrum is given in Fig.~\ref{bs_psiphi}b) for candidates within $\pm
0.05$~\gevcc\ of the fitted \Bs~mass. The \Bs~lifetime is measured in
the exclusive mode $J/\psi\phi$ to be
\begin{equation}
\tau(\Bs) = (1.34\ ^{+0.23}_{-0.19}\ \pm 0.05)\ {\rm ps}.
\end{equation}
Since this $\Bs$~lifetime measurement, using fully reconstructed $\Bs$~decays,
is statistics dominated, using partially reconstructed \Bs\ decays
improves the statistical precision as described in the following section.

\subsubsection{\Bs\ lifetime with partially reconstructed \Bs~mesons}   
\noindent
The $\Bs$ lifetime measurement with partially reconstructed $\Bs$
mesons\cite{bsprd} uses the semileptonic decay 
$\Bs \ra \Dsl \nu$ and is very similar to the $B^+$ and $B^0$
lifetime analysis using partially reconstructed $B$~decays (see Sec.~6.4.2). 
The \Ds\ candidates are reconstructed in the decay modes \\ 
\hspace*{0.5cm} (a)~$\Ds \ra \phipi$, $\phi \ra K^+ K^-$, \\
\hspace*{0.5cm} (b)~$\Ds \ra \kstark$, $K^{*0}  \ra K^+ \pi^-$, \\
\hspace*{0.5cm} (c)~$\Ds \ra \ksk$, $K^0_S \ra \pi^+ \pi^-$, \\
\hspace*{0.5cm} (d)~$\Ds \ra \phil$, $\phi \ra K^+ K^-$. 

\begin{figure}[tb]
\centerline{
\put(11,82){\large\bf (a)}
\put(59,83){\large\bf (b)}
\put(14,37){\large\bf (c)}
\put(56,37){\large\bf (d)}
\epsfxsize=9.0cm
\epsffile[60 185 540 655]{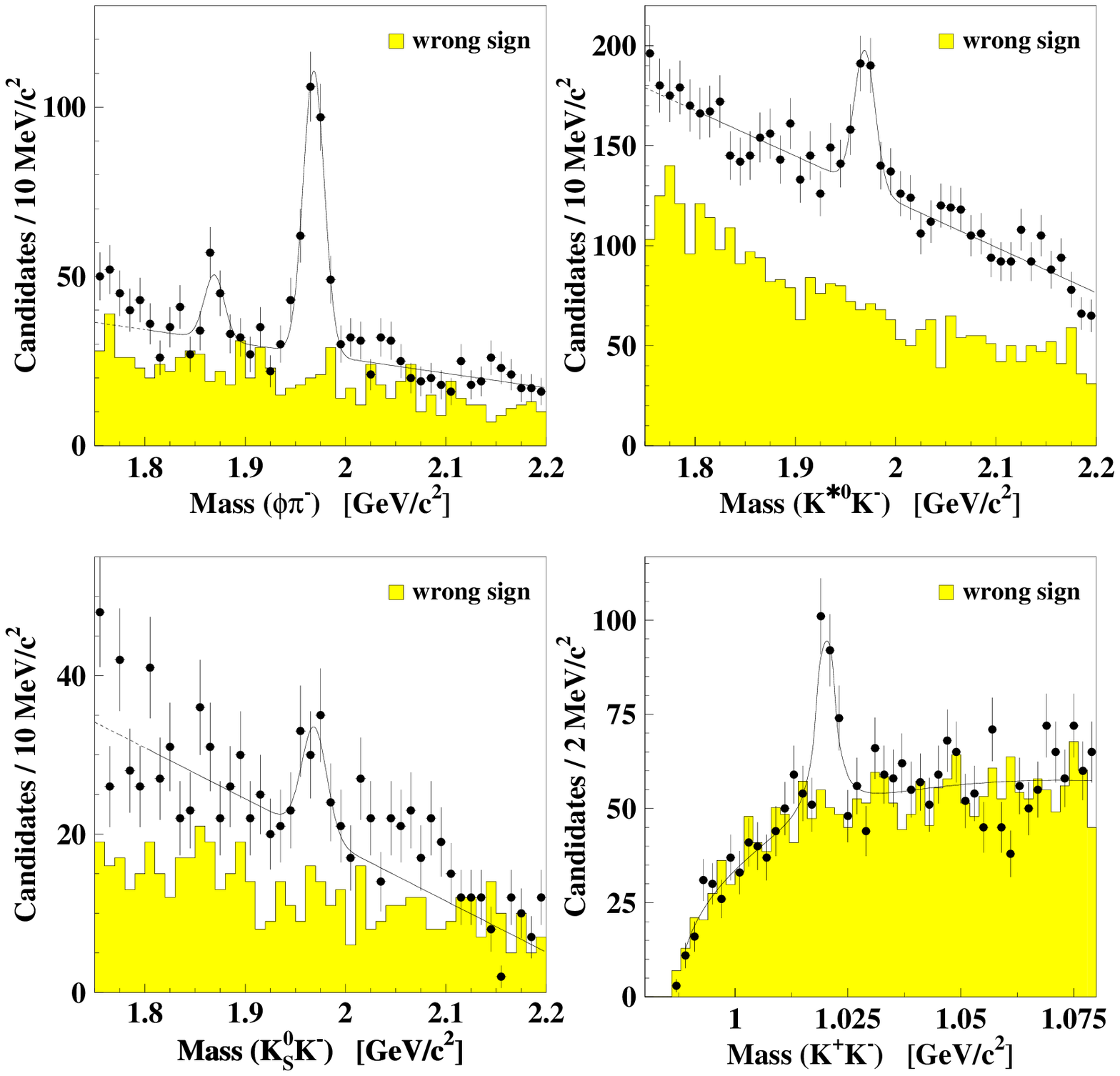}
}
\vspace*{0.2cm}
\fcaption{
Invariant mass distributions of candidates for 
(a) $\Ds \ra \phipi$, (b) $\Ds \ra \kstark$,
(c) $\Ds \ra \ksk$, and (d) $\phi \ra K^+K^-$ from $\Ds \ra \phil$. 
The dots with error bars are 
for right-sign \Dsl\ combinations while the
shaded histograms show
the corresponding wrong-sign distributions.
In (a) evidence of the decay $D^- \ra \phi \pi^-$ is also present.}
\label{ds_mass}
\end{figure}

For the first three decay modes the analysis
starts with a single lepton trigger data set, while the
semileptonic \Ds\ decay mode is based on a dimuon data sample obtained
with a trigger requirement of $m(\mu\mu) < 2.8$~\gevcc. 
\Ds\ candidates are searched for in a cone around the lepton and then   
intersected with the lepton to find the \Bs\ decay vertex 
(see also Fig.~\ref{bdecay_sketch}).
Since the \Bs\ meson is not fully
reconstructed, its $c\,t$ cannot be directly determined and a
$\beta\gamma$ correction is applied to scale from the \Dsl\ 
momentum to $\Pt(\Bs)$. Figure~\ref{ds_mass} shows the
\Ds\ invariant mass distributions for
(a) $\phipi$, (b) $\kstark$,
(c) $\ksk$, and (d) $K^+K^-$ from $\Ds \ra \phil$.
The dots with error bars are 
for right-sign \Dsl\ combinations while the
shaded histograms show
the corresponding wrong-sign $\Ds\ell^-$ distributions which show no signals.
The numbers of \Ds~signal events are compiled in Table~\ref{bs_life_sum}.

\begin{table}[tbp]
\tcaption{
Summary of the four \Ds\ decay modes: 
The estimated number $N(\Ds)$ of \Ds\ signal events,
the number $N_{\rm evt}$ of events in the signal samples, and
the fitted \Bs\ lifetimes $c\tau(\Bs)$, where the errors
shown are statistical only.} 
\centerline{\footnotesize\smalllineskip
\begin{tabular}{cccc}
\hline
 & & &  \\
 \vspace*{-0.6cm} \\
 \Ds\ Decay Mode & $N(\Ds)$ & 
 $N_{\rm evt}$ & $c\tau(\Bs)$ \\
 & & &  \\
 \vspace*{-0.6cm} \\
\hline
 & & &  \\
 \vspace*{-0.6cm} \\
 \phipi\  &  $220 \pm 21$  & 350  
	& $418\ _{-39}^{+43}\ \mu$m  \\
 & & &  \\
 \vspace*{-0.6cm} \\
 \kstark\  & $125 \pm 20$  & 820 
	& $411\ _{-66}^{+73}\ \mu$m \\
 & & &  \\
 \vspace*{-0.6cm} \\
 \ksk\  & $33 \pm 8$  & 146 
	&  $397\ _{-152}^{+161}\ \mu$m  \\
 & & &  \\
 \vspace*{-0.6cm} \\
 \phil\  &  $205 \pm 38$  & 635
	& $399\ _{-45}^{+50}\ \mu$m  \\
 & & &  \\
 \vspace*{-0.6cm} \\
\hline
\end{tabular}}
\label{bs_life_sum}
\end{table}

\begin{figure}[tbp]
\centerline{
\put(8,40){\bf (a)}
\put(40,40){\bf (b)}
\put(8,26){\bf (c)}
\put(40,26){\bf (d)}
\put(74,57){\bf (e)}
\put(74,26){\bf (f)}
\epsfxsize=6.3cm
\epsffile[5 15 515 510]{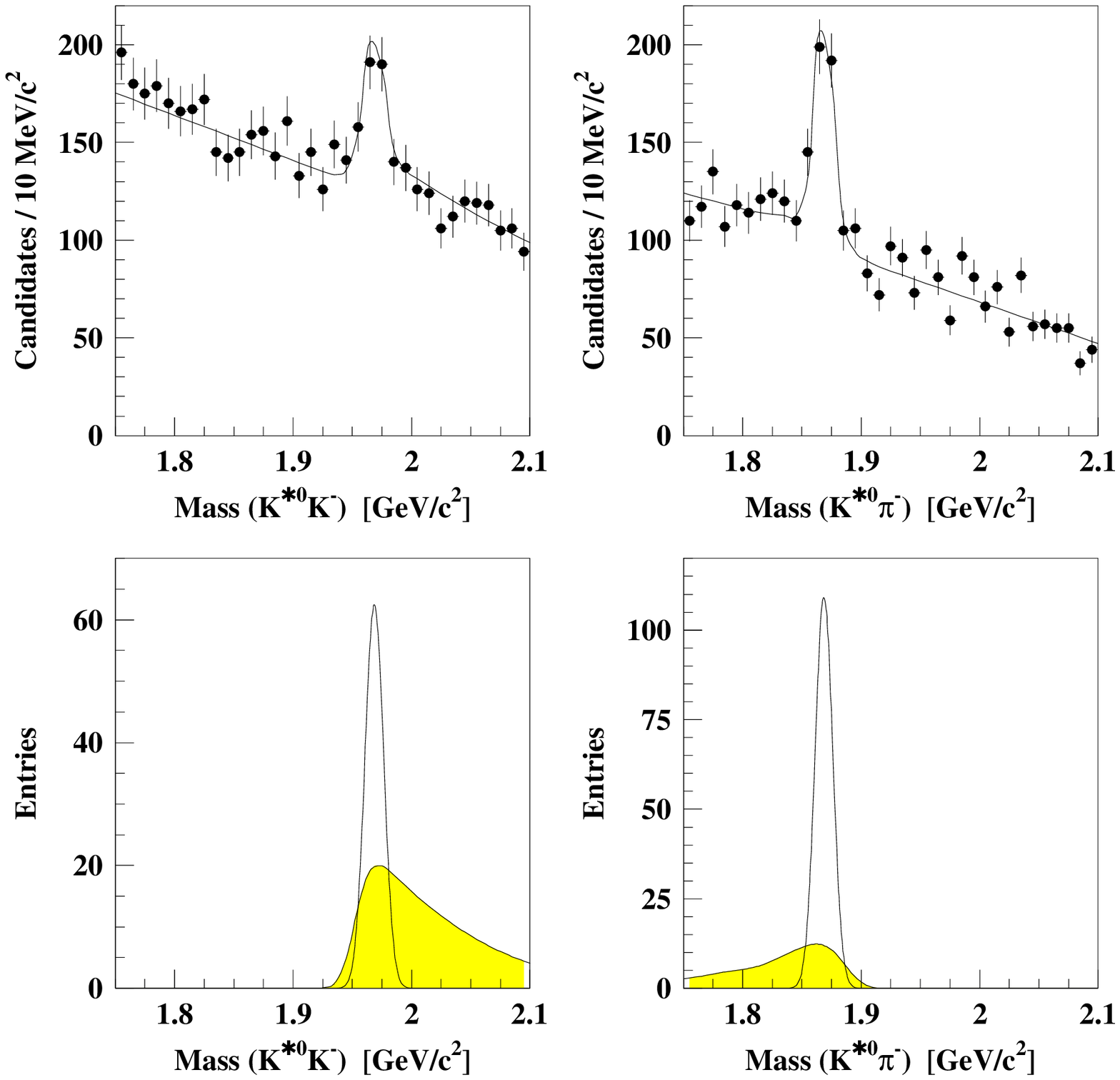}
\epsfxsize=6.3cm
\epsfysize=6.1cm
\epsffile[10 10 545 515]{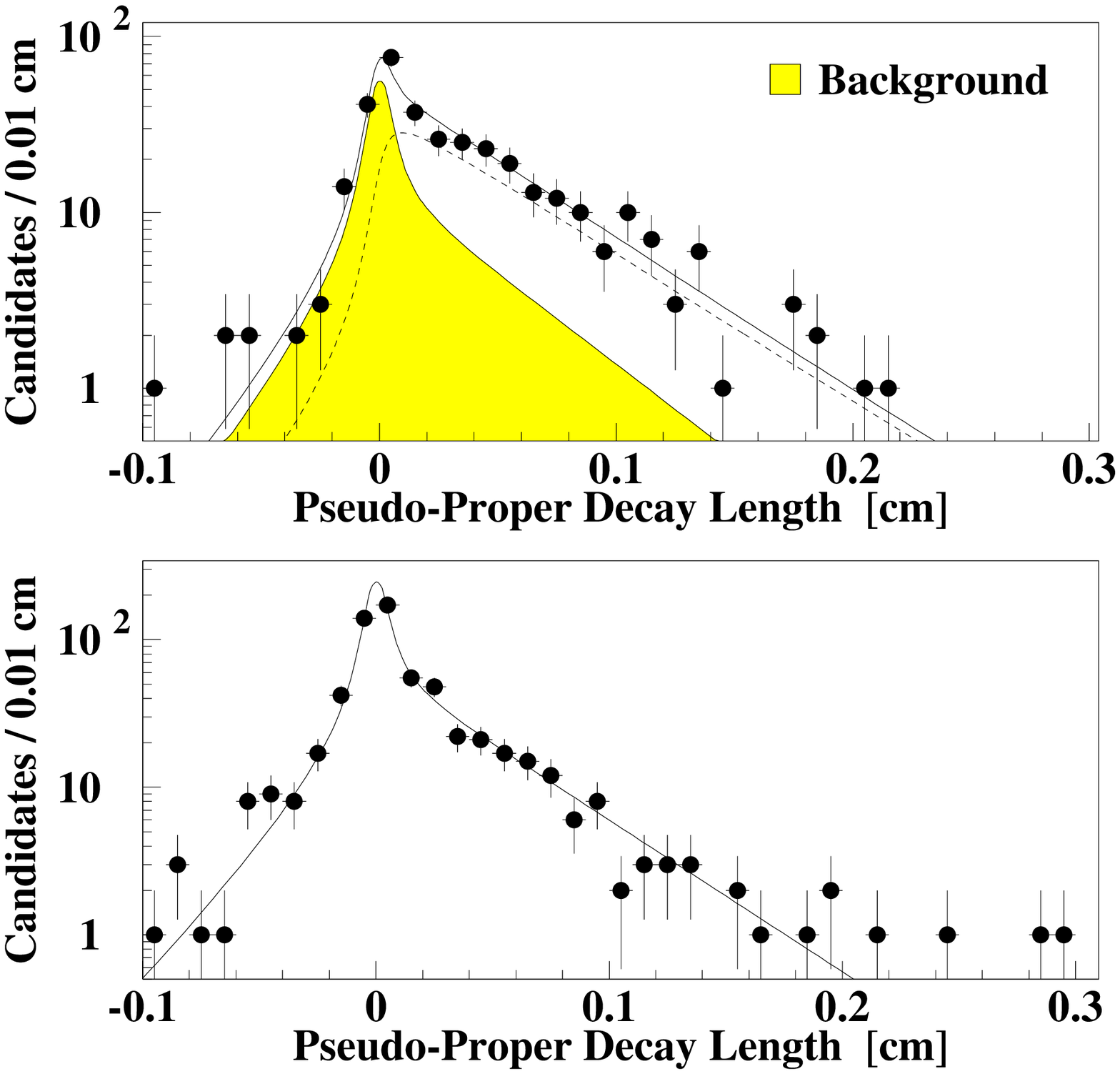}
}
\vspace*{0.2cm}
\fcaption{
(a) Mass distribution for candidates in the $\Ds \ra \kstark$ 
decay mode. (b) Mass distribution if these candidates are assumed to be
$D^- \ra \kstarpi$. 
(c) Distribution of the $\Ds \ra \kstark$ signal
and the reflection from $D^- \ra \kstarpi$ (shaded
area) as obtained from Monte Carlo simulations. Normalizations are determined
from the simultaneous fit described in the text. 
(d) Mass distribution of the corresponding $D^- \ra \kstarpi$ signal
and the reflection from $\Ds \ra \kstark$ (shaded area).
(e) \Bs\ pseudo-proper decay length distribution for $\Ds \ra \phipi$ 
with the results of the fit superimposed.
The dashed line is the \Bs\ signal contribution, while the
shaded area represents the contribution from the combinatorial background. 
(f) Pseudo-proper decay length distribution
for the background sample with the fit result superimposed.}
\label{dreflect}
\end{figure}

One of the crucial aspects of this analysis is that 
the reconstructions of the \Ds\ decay modes into \kstark\ and \ksk\
suffer from reflections 
of $D^- \ra \kstarpi$ and $D^- \ra K^0_S \pi^-$, respectively, where the
$\pi^-$ is incorrectly assigned the kaon mass. We will discuss this
reflection from $D^-$ and the determination of the true number of events from
the \Ds\ decay with the example of the $\Ds \ra \kstark$ mode.
The effect of this $K$-$\pi$ misassignment can be seen 
in Figure~\ref{dreflect}. Events from a $B \ra D^- \ell
\nu X$ Monte Carlo simulation with $D^- \ra \kstarpi$ yield an
invariant mass distribution 
indicated by the shape of the shaded area in Fig.~\ref{dreflect}c) if
they are reconstructed as $\Bs \ra \Ds \ell \nu X$ with $\Ds \ra
\kstark$, misinterpreting the $\pi^-$ as $K^-$. A significant portion
of this $D^-$ reflection lies at the \Ds\ mass peak.
Two methods are used to determine the $D^-$ reflection from the
data.

The first method performs a simultaneous fit to the \kstark\ and
\kstarpi\ invariant mass distributions, where the \kstarpi\ mass
distribution is created by switching the mass assignment on the
$K^-$ track to a pion. Figure~\ref{dreflect}a) shows the \kstark\
mass, while the corresponding \kstarpi\ mass 
is displayed in Fig.~\ref{dreflect}b). 
Each distribution is described by a Gaussian
for the corresponding $D^-$ and \Ds\ signal, as shown in
Figures~\ref{dreflect}c) and \ref{dreflect}d), plus a linear lineshape
to parametrize the combinatorial background. The shape of the
corresponding $D^-$ or \Ds\ reflection, as obtained from a Monte Carlo   
simulation, is also included in the fit as displayed in
Fig.~\ref{dreflect}c) and d) as the shaded areas. The two mass
distributions are fit
simultaneously, with the number of events in the Gaussian \Ds\ ($D^-$)
signal set equal to the number of events in the corresponding $D^-$
(\Ds) reflection. The fit result is shown in Figure~\ref{dreflect}a)--d). 

The second method for determining the amount of $D^-$ reflection in the
\Ds\ signal sample exploits the difference between the $D^-$~lifetime
[$\tau(D^-) = (1.057\pm0.0015)$~ps] and the \Ds~lifetime
[$\tau(\Ds) = (0.467\pm0.0017)$~ps]. Fitting for the 
\Ds~lifetime, the exponential describing the \Ds\ signal is replaced
by the sum of two exponentials, 
one with the \Ds~lifetime and one with the $D^-$~lifetime. 
The \Ds\ and $D^-$ lifetimes are fixed to their nominal values\cite{PDG} and 
the relative fractions of \Ds\ and $D^-$ are allowed to float in the
fit. The weighted average of \Ds\ events from both methods yields
a \Ds\ signal fraction of $(62\pm10)\%$ for the $\Ds \ra \kstark$
decay. 
Both methods are also used to calculate the number of \Ds\ events and
the contribution from the $D^-$ reflection in the $\Ds \ra
\ksk$ decay mode with the result displayed in
Table~\ref{bs_life_sum}.

The \Bs\ lifetime is determined for each of the four \Ds\ decay
channels individually, with the fit results and their statistical
uncertainties shown in Table~\ref{bs_life_sum}. As an example, 
the pseudo-proper decay length distribution 
of the $\Ds \ra \phipi$ signal sample with the result of the fit superimposed  
is shown in Figure~\ref{dreflect}e).
The dashed line represents the \Bs\ signal contribution, while
the shaded area shows the sum of the background probability function over
the events in the signal sample.
The same distribution of the background sample is displayed in 
Figure~\ref{dreflect}f).
The combined \Bs\
lifetime from all four \Ds\ decay modes is determined from a
simultaneous fit to~be 
\begin {equation}
  \tau(\Bs) = (1.36\ \pm 0.09 \ ^{+0.06}_{-0.05})\  {\rm ps}.
\end {equation}
This result is currently the world's best measurement of the \Bs\ lifetime
from a single experiment. 

Using the CDF average $B^0$ lifetime 
$\tau(B^0) = (1.513\pm0.053)$~ps (see Sec.~6.4.3),
we determine the $\Bs/B^0$ lifetime ratio to be $0.899\pm0.072$, taking
correlated systematic uncertainties into account. 
However, ignoring the correlated systematic uncertainties increases
the error on the $\Bs/B^0$ lifetime 
ratio only to $\pm 0.077$, since $\tau(\Bs)/\tau(B^0)$ is dominated by the
statistical error on the \Bs\ lifetime measurement.  

\subsubsection{Determination of \dgog}
\noindent
In the Standard Model, the \Bs~meson exists in two $CP$-conjugate states, 
$|\Bs\rangle = |\bar b s\rangle$ and $|\bar\Bs\rangle = |b \bar s\rangle$.
The two mass eigenstates of the \Bs\ meson, \Bsh\ and \Bsl\
($H =$ `heavy' and $L =$ `light'), are 
not $CP$ eigenstates but are mixtures of the two $CP$-conjugate quark
states:
\begin{equation}
|\Bsh\rangle = 1/\sqrt{2}\, (\,|\Bs\rangle - |\bar\Bs\rangle\,) 
\hspace*{0.5cm} {\rm and} \hspace*{0.5cm} 
|\Bsl\rangle = 1/\sqrt{2}\, (\,|\Bs\rangle + |\bar\Bs\rangle\,). 
\end{equation}
The mass and lifetime differences between the \Bsh\ and \Bsl\ 
can be defined as
\begin{equation}
\Delta m \equiv m_H - m_L,\ \ \dgam \equiv \Gamma_L - \Gamma_H,\ \ 
{\rm and} \ \ \Gamma = 1/2\,(\Gamma_H + \Gamma_L),
\end{equation}
where $m_{H,L}$ and $\Gamma_{H,L}$ denote the mass and decay width of \Bsh\
and \Bsl.
Please note, we have defined both $\Delta m$ and \dgam\ to
be positive quantities. In this case, the heavy state is the long-lived
state or the $CP$ odd state, while the light state is the short-lived
state or the $CP$ even state in analogy to the neutral kaon system.
Unlike the $B^0$~meson, 
the width difference in the \Bs\ system is expected to be 
large\cite{bigi_bs}.
Theoretical estimates\cite{isi_bs,bbd_bs} 
predict \dgog\ to be on the order of 10-20\%.
In the \Bs\ system the ratio $\Delta m / \dgam$ is related to the ratio of the
Cabibbo-Kobayashi-Maskawa\cite{ckm1,ckm2} matrix 
elements $|V_{cb} V_{cs}| / |V_{ts} V_{tb} |$, which is quite well
known, and depends only   
on QCD corrections 
within the Standard Model\cite{bbd_bs,datta}. 
A measurement of \dgam\ would therefore imply a determination of \dm\
and thus a 
way to infer the existence of \Bs\ meson oscillations, which will 
ultimately determine the ratio of the CKM matrix elements 
$|V_{td}| / |V_{ts}|$. 
The current limit on \dm\ in the
\Bs\ system\cite{PDG}, $\dms > 10.2$~ps$^{-1}$ at 95\%~C.L., gives rise
to expect a significant lifetime difference in the \Bs\ system. 

\begin{figure}[tb]
\centerline{
\put(0,49){\large\bf (a)}
\put(60,49){\large\bf (b)}
\epsfxsize=6.3cm
\epsffile[35 10 560 525]{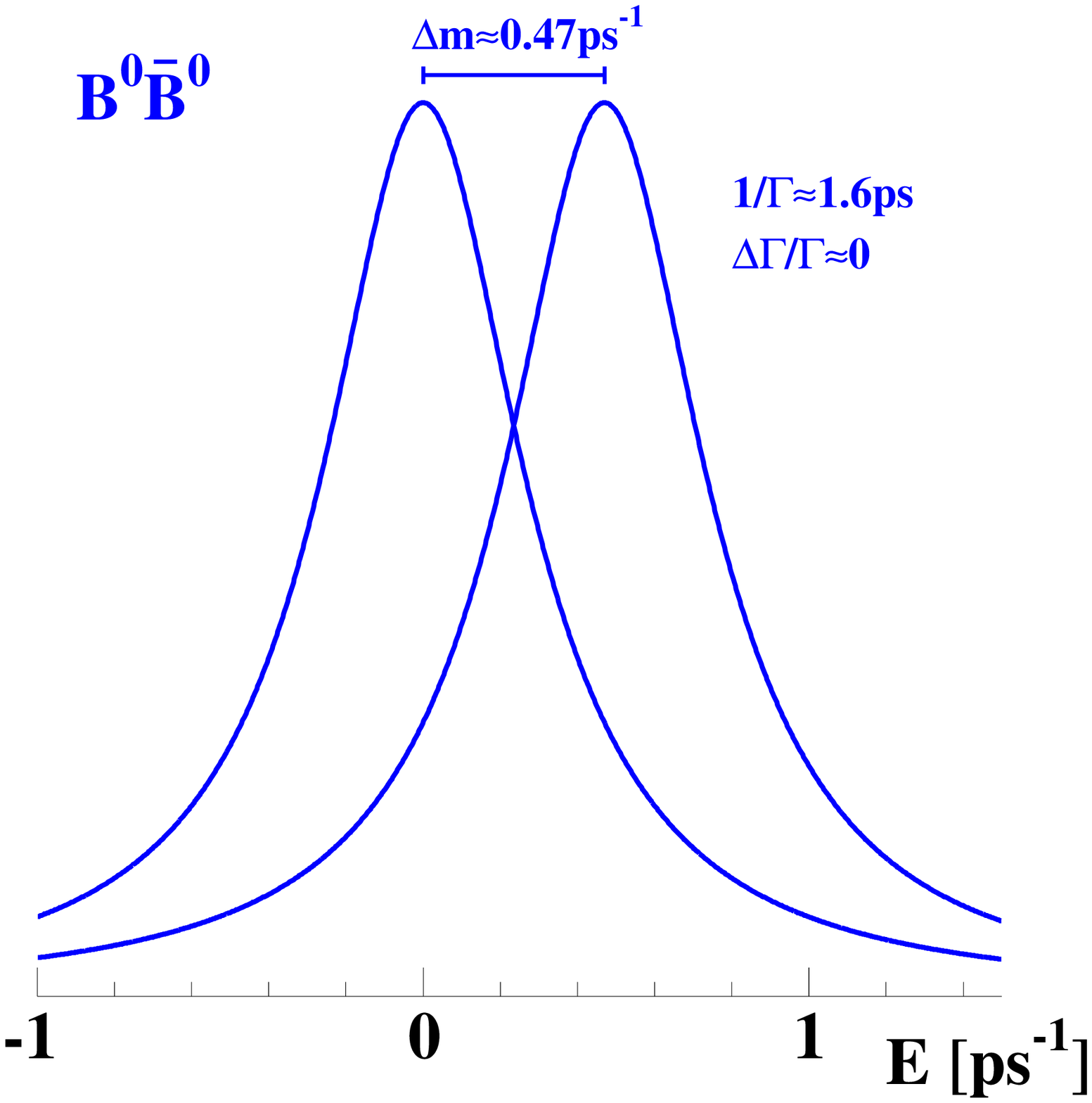}
\epsfxsize=6.3cm
\epsffile[35 10 560 525]{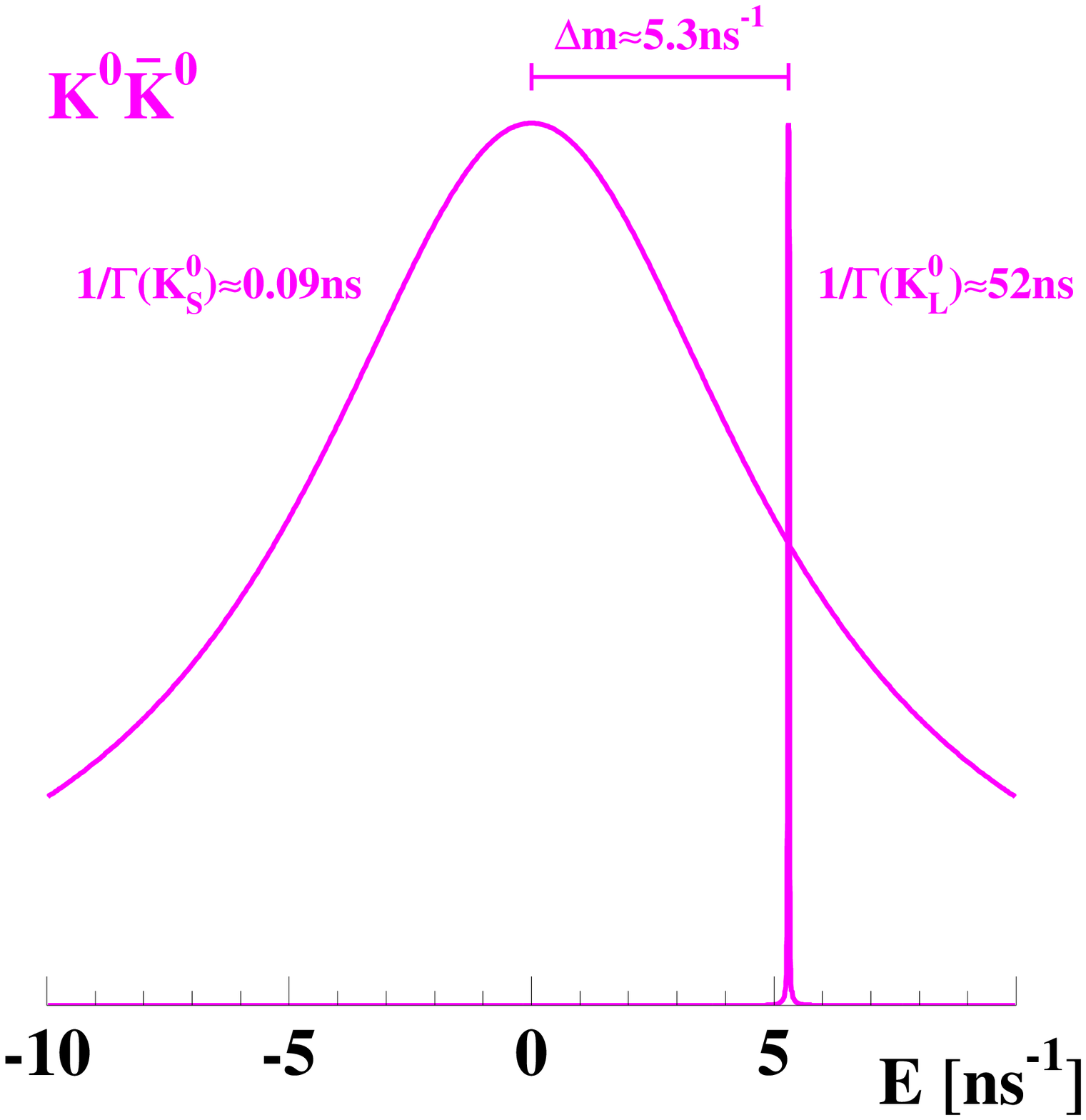}
}
\centerline{
\put(0,49){\large\bf (c)}
\put(65,49){\large\bf (d)}
\epsfxsize=6.3cm
\epsffile[35 10 560 525]{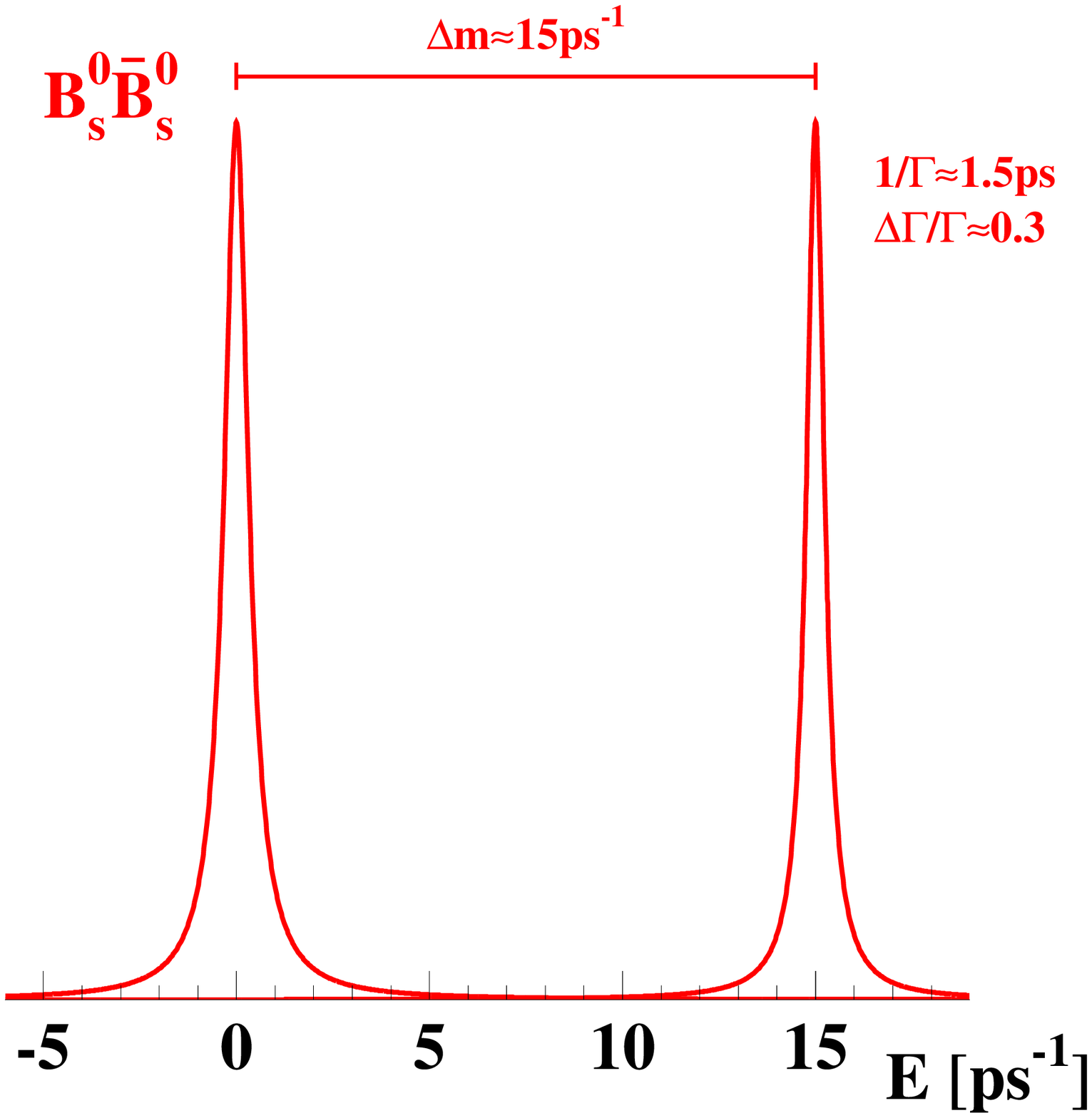}
\epsfxsize=6.3cm
\epsffile[35 10 560 525]{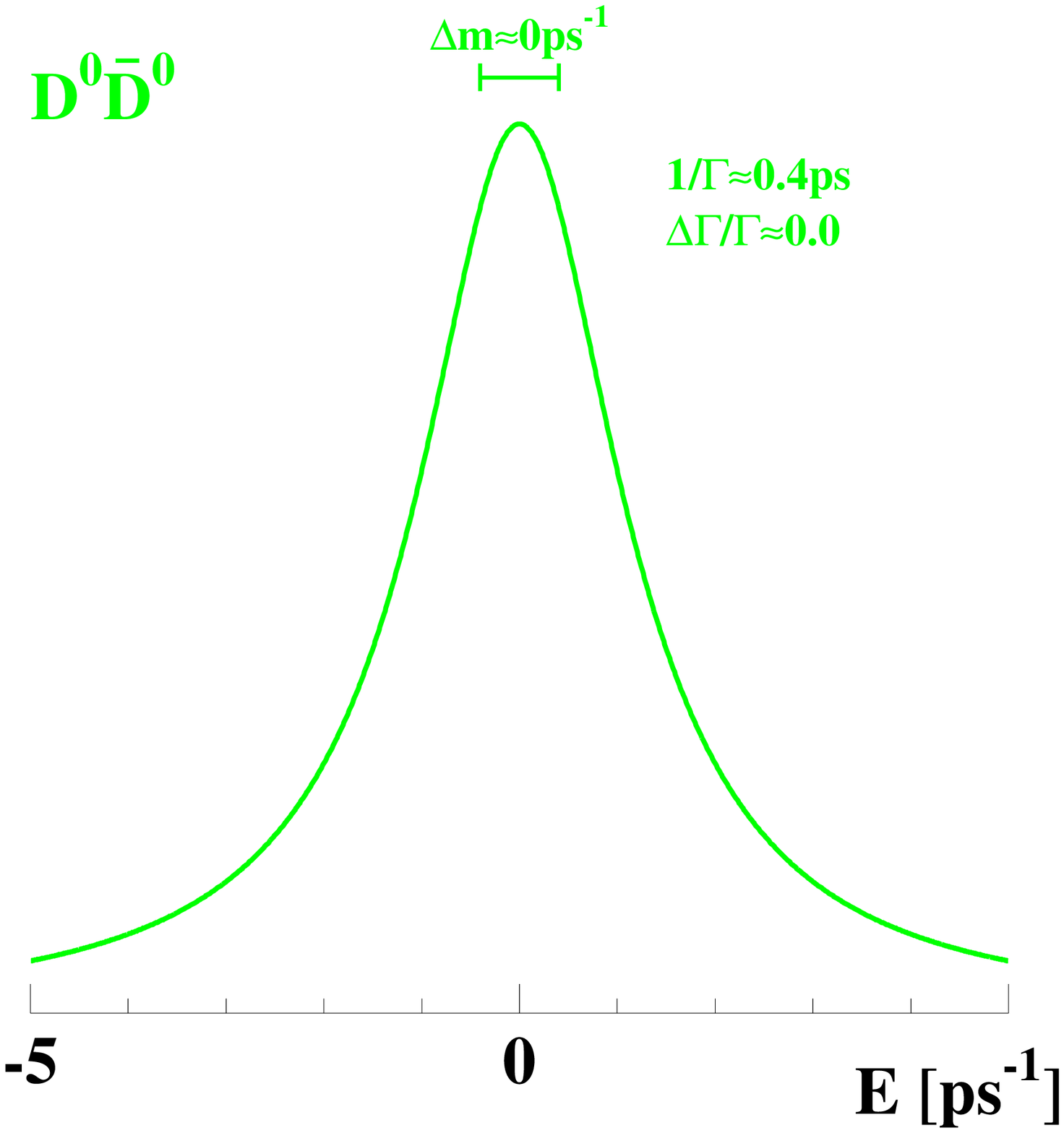}
}
\vspace*{0.2cm}
\fcaption{
Illustration of the relation between \dm\ and \dgam\ in 
(a) the $B^0 \bar B^0$ system,
(b) the $K^0 \bar K^0$ system,
(c) the $\Bs \bar \Bs$ system, and
(d) the $D^0 \bar D^0$ system.}
\label{dgam_dm}
\end{figure}

An illustrative way to better understand the relation between \dm\ and
\dgam\ in neutral meson systems\cite{gene_dm} is shown in
Figure~\ref{dgam_dm}. Let's 
at first concentrate on the well known $B^0\bar B^0$~system displayed
in Fig.~\ref{dgam_dm}a). Here, the $B^0$ state is illustrated as its
mass excitation curve, a Breit-Wigner lineshape, with the $x$-axis
representing an energy in ps$^{-1}$. The width of the Breit-Wigner
curve is given by the reciprocal of the $B^0$ lifetime $1/\Gamma(B^0) \sim
1.6$~ps. The second curve represents the heavy $B^0$ state which is
separated from the light $B^0$ state by a mass difference 
$\dm \sim 0.47$~ps$^{-1}$ 
which is well measured from $B^0\bar B^0$~oscillations. 
The lifetime difference \dgam\ is very small in the $B^0$ system
resulting in both curves appearing with the same width. 

The classic system to illustrate \dm\ and \dgam\ is the $K^0\bar
K^0$ system, displayed in Fig.~\ref{dgam_dm}b). Here, a short lived
state exists, the $K^0_S$, shown as the broad curve with a width of 
$1/\Gamma(K^0_S) \sim 0.09$~ns, and a long lived state exists, 
the $K^0_L$, shown as the narrow curve with a width of  
$1/\Gamma(K^0_S) \sim 52$~ns. The lifetime of the $K^0_L$ is about 500
times longer than the $K^0_S$ lifetime. The mass difference \dm\
between the states is about 5.3~ns$^{-1}$. Please note, the unit here
is in ns$^{-1}$ compared to ps$^{-1}$ for the $B^0 \bar B^0$ system. This
means particle antiparticle oscillations are much slower in the $K^0$
system compared to the $B^0$ system. 

The system of interest, in our case, is the $\Bs \bar \Bs$ system with a
large mass difference between the \Bsl\ and \Bsh\ state. We choose 
e.g.~$\dm = 15$~ps$^{-1}$ as drawn in Fig.~\ref{dgam_dm}c). In analogy
to the kaon 
system, the \Bsl\ is the shorter lived state, while the
\Bsh\ is the longer lived state. In Fig.~\ref{dgam_dm}c) a lifetime
difference of $\dgog \sim 0.3$ is assumed resulting in the \Bsl\ state
to appear slightly broader than the \Bsh. Finally, in the $D^0\bar
D^0$ system, shown in Fig.~\ref{dgam_dm}d), the mass difference \dm\
is very small in the Standard Model as is 
\dgam. This results in completely overlapping mass excitation curves for both
states which appear as one curve with a width of
$1/\Gamma \sim 0.4$~ps.

It is assumed that \Bs\ mesons are produced 
as an equal mixture\cite{bbd_bs} of \Bsh\ and \Bsl.
One way to search for \dgam\ in the \Bs\ system is to describe
the \Bs~meson decay length distribution by a function of the form  
\begin {equation}
{\cal F}(t) =
1/2\, (\Gamma_H\, {\rm e}^{-\Gamma_H t} +
           \Gamma_L\, {\rm e}^{-\Gamma_L t})\ \ \ \ \
{\rm with}\ \ \Gamma_{L,H} = \Gamma \pm \Delta\Gamma/2 ,
\end{equation}
rather than by just one exponential lifetime $\Gamma {\rm e}^{-\Gamma t}$.
Clearly, the integral over time  
$\int {\rm d}t\, {\cal F}(t)$ returns the same number of \Bsh\ and \Bsl\
mesons as produced. This would be the right functional form 
in a search for a lifetime difference \dgog, 
if a completely inclusive sample of \Bs\ mesons is used which is allowed to
decay into any final state. However, for a semileptonic final state,
the semileptonic decay widths are the same for \Bsh\ and \Bsl,
since the reason for a lifetime difference originates from 
hadronic \Bs\ decays. This can be easily seen in
the following way: In an excellent approximation, the semileptonic
\Bs\ decay is flavour specific, which means that a $\ell^+$ can only
originate from a $|\bar bs\rangle$ state. Since the fraction of $|\bar
bs\rangle$ states is the same for \Bsh\ and \Bsl, the
semileptonic widths are the same for \Bsh\ and \Bsl, thus 
the semileptonic branching ratios are different for \Bsh\ and \Bsl.
The correct functional form to describe the \Bs~meson decay length
distribution from semileptonic \Bs~decays, assuming a lifetime difference
\dgog, is therefore:
\begin {equation}
{\cal F}(t) =
\Gamma_H\Gamma_L/\Gamma \cdot
        \left( {\rm e}^{-\Gamma_H\,t} + {\rm e}^{-\Gamma_L\,t} \right)
\ \ \ \ \ 
{\rm with}\ \ \Gamma_{L,H} = \Gamma \pm \frac{\dgam}{2} 
= \Gamma \cdot (1 \pm \frac{1}{2} \frac{\dgam}{\Gamma}).
\label{eq:f}
\end{equation}
The parameter \dgog\ is the parameter fit for. 

In the case of a lifetime difference $\dgam \neq 0$, the total decay width
$\Gamma = 1/2 \cdot (\Gamma_H + \Gamma_L)$ and the mean \Bs\ lifetime 
$\tau_{\rm m}(\Bs)$ obtained from a fit assuming a single \Bs\
lifetime, are no longer reciprocal to each other but follow the relation
\begin{equation}
\tau_{\rm m}(\Bs) = \frac{1}{\Gamma} \cdot
         \frac{1+(\frac{\dgam}{2\,\Gamma})^2}
	      {1-(\frac{\dgam}{2\,\Gamma})^2}.
\label{eq:gamtau}
\end{equation}
The relation in Eq.~(\ref{eq:gamtau}) is incorporated into the
likelihood fitting function for \dgog, 
and the mean \Bs\ lifetime is fixed to the world average
$B^0$ lifetime, since both lifetimes are expected to
agree\cite{bigi,neubert} within 1\%.

The fit returns 
$\dgog = 0.34\ ^{+0.31}_{-0.34}$, where the given error is 
statistical only. This indicates that with the current statistics 
CDF is not sensitive to a \Bs\ lifetime difference. Based on this fit
result, the normalized likelihood is integrated as a function of \dgog\ and
the 95\% confidence level limit is found at
\begin {equation}
  \dgam/\Gamma < 0.83\ \ \ (95\%\ {\rm C.L.}).
\end {equation}

Using a theoretical value of
$\dgam/\dm = (5.6\pm2.6)\cdot 10^{-3}$
from Ref.\cite{bbd_bs} and setting $\tau_{\rm m}(\Bs)$ to the world
average $B^0$~lifetime\cite{PDG}, 
an upper limit on the \Bs\ mixing frequency
of $\dms < 96~{\rm ps}^{-1}$ (95\%~C.L.) can be determined within the
Standard Model. Including the
dependence on \dgam/\dm\ and $\tau_{\rm m}(\Bs)$ into the limit, we obtain
\begin {equation}
  \dms < 96~{\rm ps}^{-1} 
\times \left(\frac{5.6\cdot 10^{-3}}{\dgam/\dm}\right)
\times \left(\frac{1.55~{\rm ps}}{\tau_{\rm m}(\Bs)}\right)
 \ \ \ (95\%\ {\rm C.L.}).
\end {equation}

\subsection{Measurement of \boldmath{$\Lambda_b^0$} lifetime} 
\begin{figure}[tbp]
\centerline{
\put(50,50){\large\bf (a)}
\put(112,50){\large\bf (b)}
\epsfclipon
\epsfxsize=6.3cm
\epsffile[45 150 540 610]{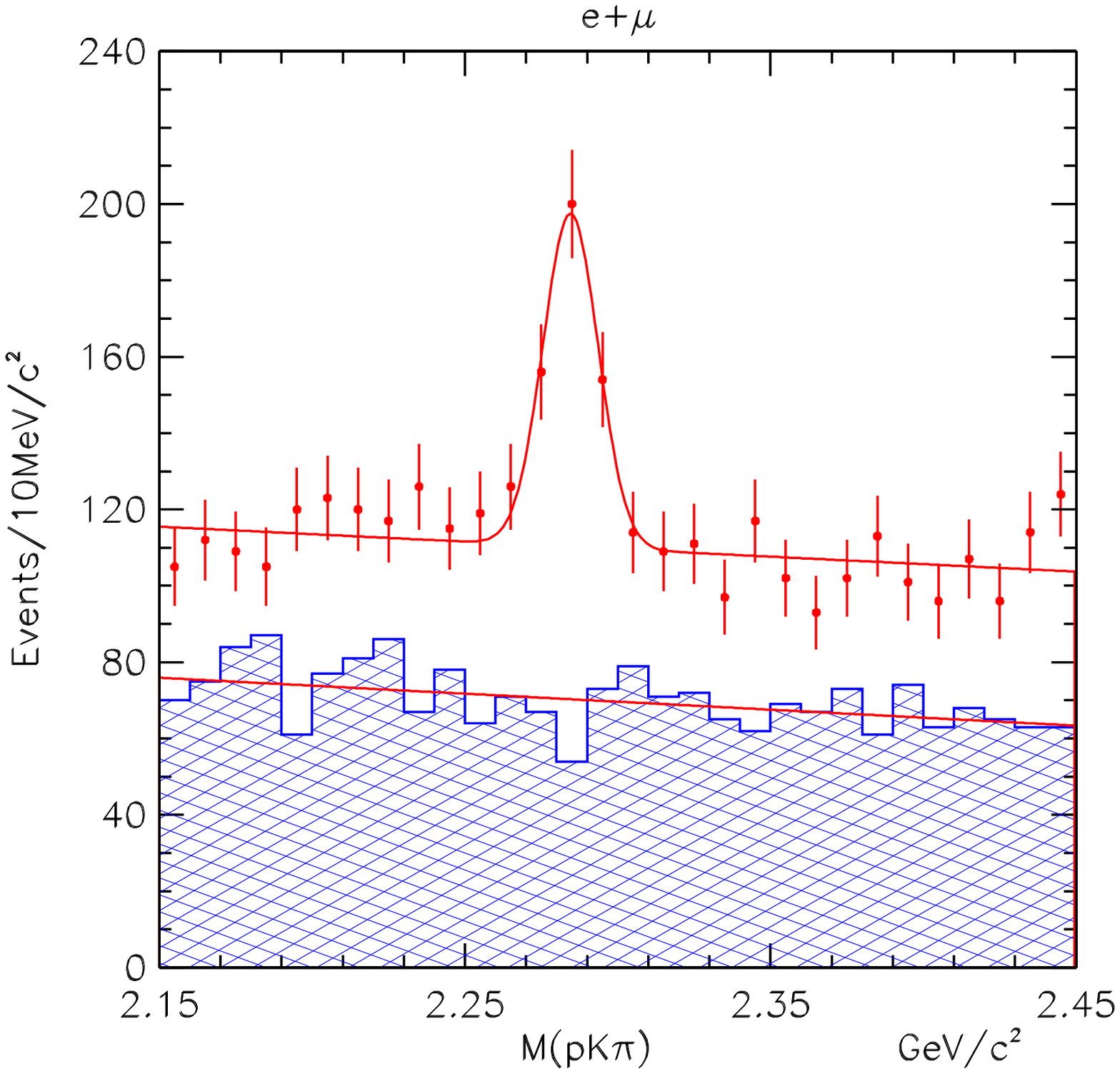}
\epsfclipoff
\epsfxsize=6.3cm
\epsffile[45 150 540 610]{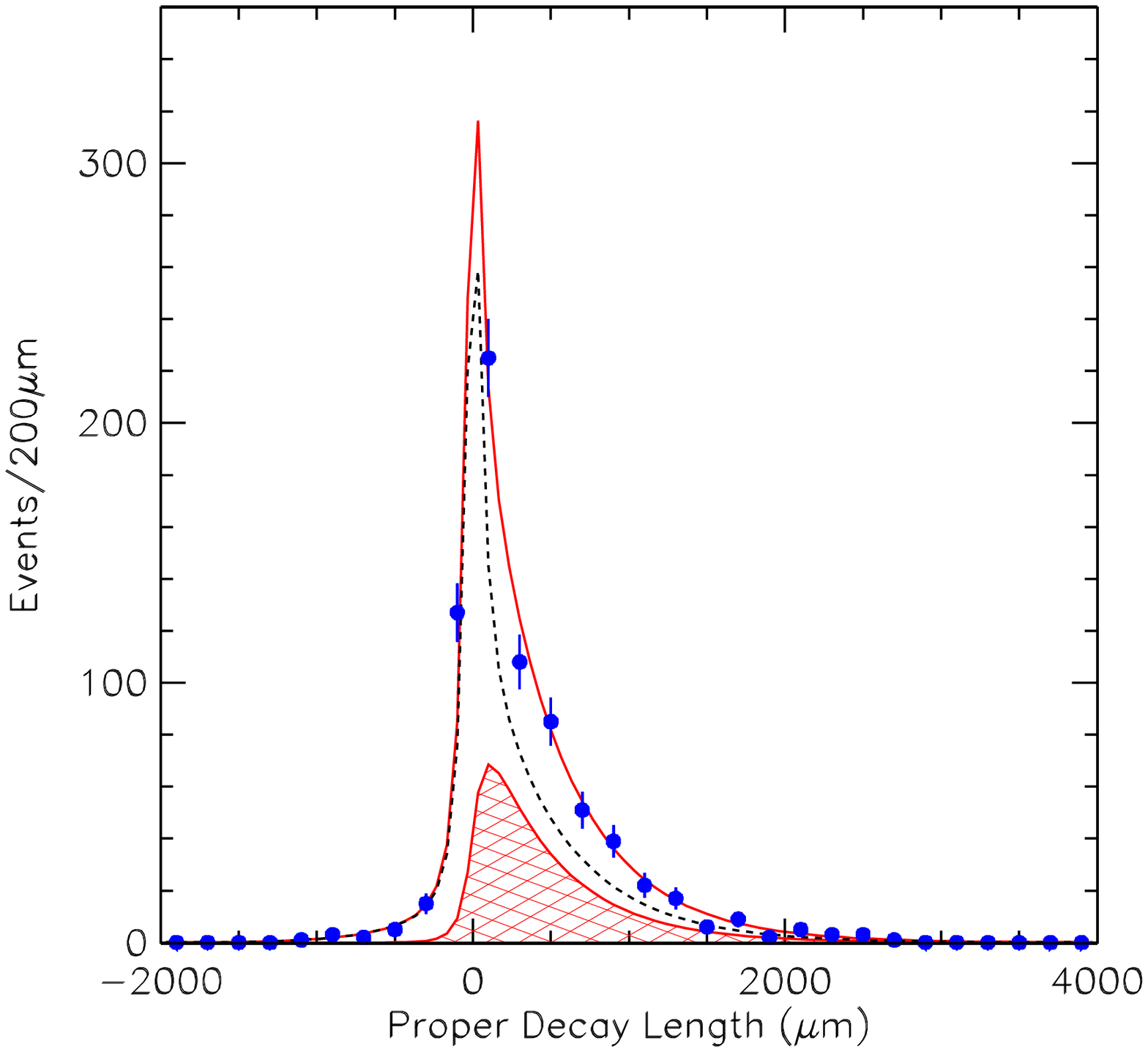}
}
\vspace*{0.2cm}
\fcaption{
(a) Invariant mass distribution of $pK^-\pi^+$ for right-sign (points
with error bars) and wrong-sign events (shaded area). 
(b) Pseudo-proper decay length distribution of events from the
$\Lambda_b^0$ signal region. The points with error bars are data, the
solid line is the fit result, the shaded area is the signal
distribution, and the dashed line is the background contribution.}
\label{lambdab}
\end{figure}

\noindent
The analysis principle for the $\Lambda_b^0$ lifetime
measurement\cite{lamb_prl} is 
very similar to the $B^0$, $B^+$, and \Bs\ lifetime analyses using
partially reconstructed decays. The $\Lambda_b^0$~baryon is
reconstructed through the semileptonic decay  
$\Lambda_b^0 \ra \Lambda_c^+ \ell^- \bar\nu X$, with the subsequent decay
$\Lambda_c^+ \ra p K^-\pi^+$.
The analysis again
uses the single lepton trigger data searching
for $\Lambda_c^+$
candidates in a cone around the lepton.   
The $\Lambda_c^+$ candidates are intersected with the lepton to find the
$\Lambda_b^0$ decay vertex.
A signal of $(197\pm25)$ $\Lambda_c^+$ events is obtained 
as shown in Fig.~\ref{lambdab}a), where the $p K^-\pi^+$ invariant mass
distribution for right-sign $\Lambda_c^+ \ell^-$ combinations is plotted. 
The shaded histogram shows the wrong-sign $\Lambda_c^+ \ell^+$ distribution.  

The pseudo-proper decay length distribution of events from the
$\Lambda_b^0$ signal region is shown in Fig.~\ref{lambdab}b). The points
with error bars are data, the 
solid line is the fit result, the shaded area is the signal
distribution, and the dashed line is the background contribution.
Using these events, the $\Lambda_b^0$ lifetime is determined to be
\begin{equation}
\tau(\Lambda_b^0) = (1.32\ \pm 0.15\ \pm 0.07)\ {\rm ps}. 
\end{equation}

The CDF
$\Lambda_b^0$ lifetime result is competitive with the LEP measurements in
precision\cite{PDG}.
Using the CDF average $B^0$ lifetime 
$\tau(B^0) = (1.513\pm0.053)$~ps (see Sec.~6.4.3),
we determine the lifetime ratio 
$\tau(\Lambda_b^0)/\tau(B^0) = 0.87\pm0.11$.
Theory favours the value for this ratio\cite{bigi} to be in the range
0.9-1.0 in good agreement with the CDF measurement.

\subsection{Measurement of \boldmath{\Bc} lifetime} 
\begin{figure}[tbp]
\centerline{
\epsfxsize=6.3cm
\epsffile[1 -15 380 350]{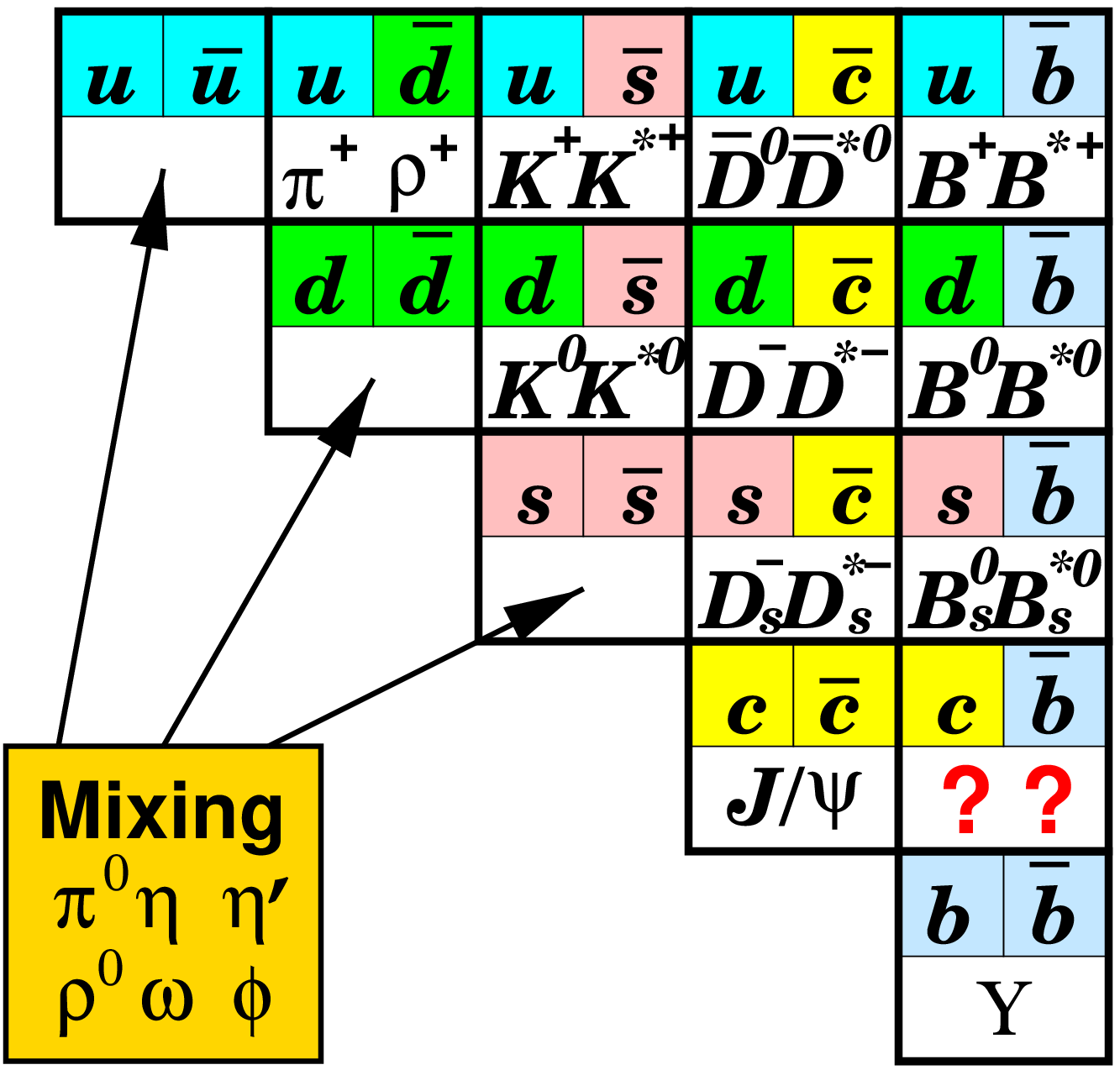}
\epsfxsize=6.3cm
\epsffile[15 40 515 545]{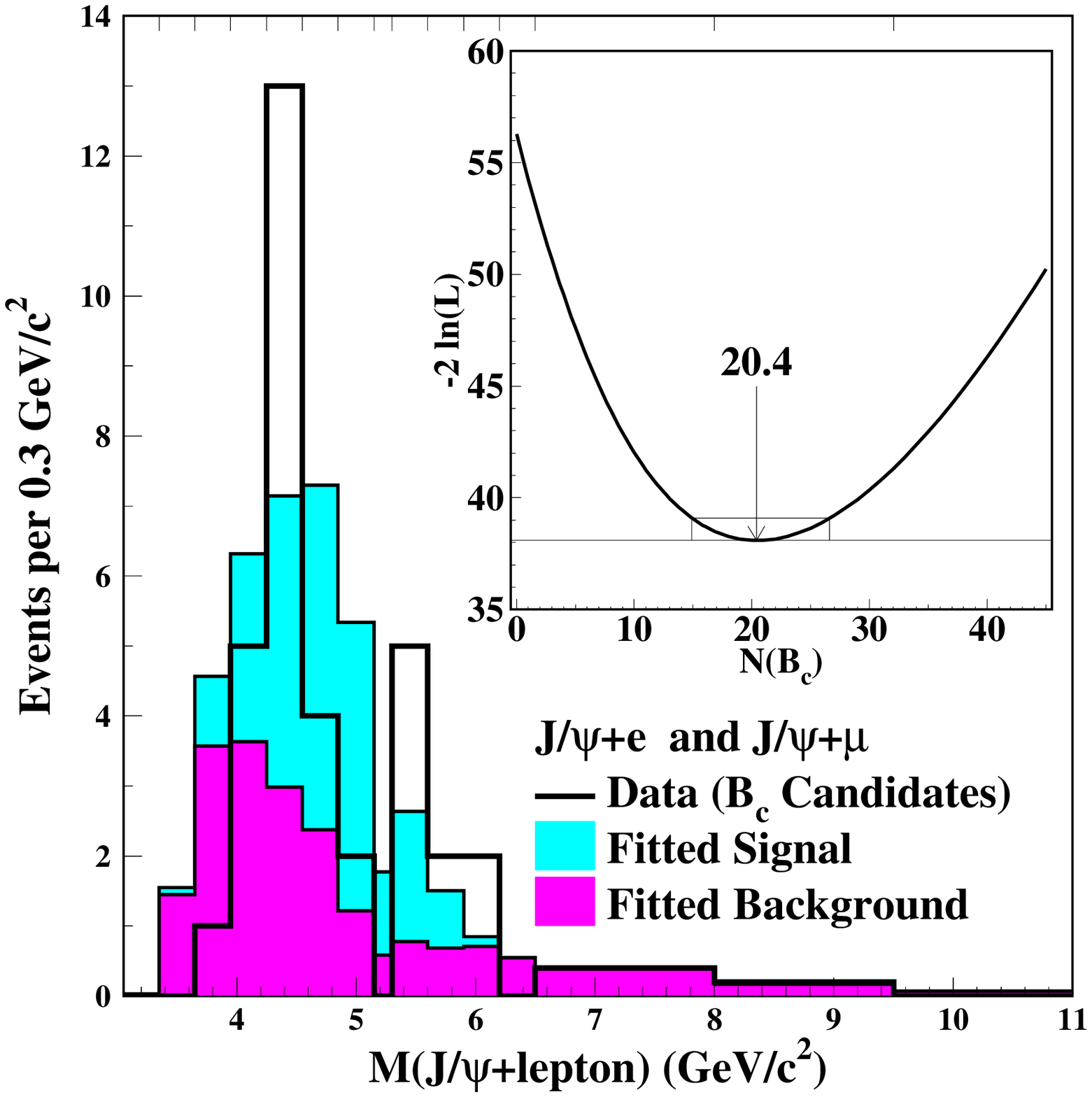}
\put(-126,46){\large\bf (a)}
\put(-55,58){\large\bf (b)}
}
\vspace*{0.2cm}
\fcaption{
(a) Sketch of quark periodic table. (b) Histogram of the
$J/\psi\ell$ mass distribution that compares the signal and background
determined in the likelihood fits to the combined data for $J/\psi e$
and $J/\psi\mu$. Note that the mass bins, indicated by tick marks at
the top, vary in width. The inset shows the log-likelihood function
versus the number of \Bc~mesons.}
\label{bc_mass}
\end{figure}

\noindent
Discussing the measurement of the \Bc~lifetime involves summarizing the
recent discovery of the \Bc~meson at CDF\cite{bcprd,bcprl}. 
The \Bc~meson is the
lowest-mass bound state of a bottom antiquark and a charm quark:
$|\Bc\rangle = |\bar b c\rangle$. This pseudoscalar ground state has
non-zero flavour and no strong or electromagnetic decays. 
It is the only state with two different heavy quarks where each can decay
weakly. 
As seen in Figure~\ref{bc_mass}a), the $|\bar b c\rangle$ state is
the last meson state predicted by the Standard
Model\cite{GSW}, which had not been discovered until early 1998. 
Figure~\ref{bc_mass}a) shows a quark periodic table
with all possible combinations of meson states which can be obtained
from the five quarks $u$, $d$, $s$, $c$, and $b$. 

Non-relativistic potential models predict a \Bc\ mass\cite{bcmass} around
6.2-6.3~\gevcc.
In these models, the $c$ and $\bar b$ quarks are tightly bound 
in a very compact system and have a rich spectroscopy of 
excited states. 
The predicted lifetime\cite{bclife} of the \Bc~meson is in the range 
0.4-1.4~ps.
Because of the wide range of predictions, 
a \Bc\ lifetime measurement is a test of the different
assumptions made in the various calculations. 

We attempt here a simple educated guess of the $\Bc$~mass and its
lifetime. We would estimate the \Bc~mass as 
\begin{equation}
m(\Bc) \sim m(b) + m(c) \sim 4.75~\gevcc + 1.55~\gevcc
= 6.3~\gevcc,
\end{equation}
where we have approximated the $b$ and $c$~quark masses with half of
the mass of the $\Upsilon(1S)$ and the $J/\psi$, respectively. To
estimate the \Bc~lifetime, we expect three major contributions to the
\Bc\ decay width: $\Gamma(\Bc) \sim \Gamma_b + \Gamma_c + \Gamma_{bc}$.
Decays of the $\bar b$ quark $\bar{b} \ra \bar{c}\, W^+$ with the $c$~quark
as a spectator, leading to final states such as $J/\psi \pi$\ or
$J/\psi \ell \nu$; $c$~quark decays $c \ra s W^+$,
with the $\bar b$ as spectator, leading to final states such as $\Bs \pi$\ 
or $\Bs \ell \nu$; and $c \bar{b} \rightarrow W^+$
annihilation, leading to final states like 
$D \, K$, $\tau \, \nu_{\tau}$\ or multiple pions.
In the simplest view, the $c$ and $\bar b$ decay like free quarks with
no annihilation contribution, and
the \Bc~lifetime would be:
\begin{equation}
\tau(\Bc) = 1/\Gamma(\Bc) \sim [\Gamma_b + \Gamma_c]^{-1} \sim 
[1/1.65~{\rm ps} + 1/0.47~{\rm ps}]^{-1} \sim 0.4~{\rm ps},
\end{equation}
where $\Gamma_b$ is approximated with the $B^+$~lifetime, while
we use $\tau(\Ds)$ to estimate~$\Gamma_c$.
  
\subsubsection{Discovery of \Bc~meson at CDF} 
\noindent
The \Bc~meson is reconstructed via its semileptonic decay $\Bc \ra
J/\psi \ell^+\nu X$ ($\ell = e,\ \mu$). In this analysis, advantage is
taken of a clean $J/\psi \ra \mu^+\mu^-$ signal (see
Fig.~\ref{jpsi_mass_tau} and Sec.~6.4.1) and
excellent lepton identification at CDF. To
reduce backgrounds from prompt $J/\psi$ production, 
tri-lepton vertices displaced from the primary vertex are searched for by
requiring the $J/\psi\ell$ pseudo-proper decay length to be greater
than 60~$\mu$m. 23 $\Bc \ra J/\psi e \nu$ candidates and 
14 $\Bc \ra J/\psi\mu\nu$ candidates are found.
Because of the missing neutrino the \Bc\ cannot be
fully reconstructed and no \Bc\ mass peak can be obtained. The
\Bc~signal is therefore found as an excess of events over expected backgrounds.
Great care is taken to correctly determine the different
backgrounds in this analysis\cite{bcprd}. 

\begin{table}[tbp]
\tcaption{
Summary of \Bc\ signal and background events.}
\centerline{\footnotesize\smalllineskip
\begin{tabular}{ lcc }
\hline
 & &  \\
 \vspace*{-0.6cm} \\
 & \multicolumn{2}{c}{ $3.35 < m(J/\psi \, \ell) <  11.0$\ GeV/$c^2$ } \\
  & $J/\psi\, e$ Events & $J/\psi\,\mu$ Events \\
\hline
False electrons	
	& $4.2 \pm 0.4$		& \\
Undetected conversions	& $2.1  \pm 1.7$ 		& \\
False muons
	& & $11.4 \pm 2.4$ \\
$B \bar{B}$ background 
	& $2.3 \pm 0.9$	
	& $1.44 \pm 0.25$ \\
Total background (predicted) 	& $8.6 \pm 2.0$		& $12.8 \pm 2.4$ \\
\hspace{2.3cm} (from fit)	& $9.2 \pm 2.0$	&	$10.6 \pm 2.3$ \\
Predicted $N(\Bc\ra J\psi e\nu)/N(\Bc\ra J\psi\ell\nu)$ 	
	& \multicolumn{2}{c}{$0.58 \pm 0.04$} \\
$e$\ and $\mu$\ signal (derived from fit)
	& $12.0^{+3.8}_{-3.2}$	
	& $8.4^{+2.7}_{-2.4}$ \\
Total signal (fitted parameter)
	& \multicolumn{2}{c}{\rule[-5pt]{0pt}{15pt}$20.4^{+6.2}_{-5.5}$} \\ 
\hline
  Signal + Background	
		& $21.2 \pm 4.3$	& $19.0 \pm 3.5$ \\
Candidates	& 23			& 14 \\
Probability for null hypothesis 
	& \multicolumn{2}{c}{ $0.63 \times 10^{-6}$ } \\
\hline 
\end{tabular}}
\label{bc_bg_sum}
\end{table}

Significant backgrounds in the \Bc\ candidates come from false leptons.
Hadrons reaching the muon detectors without being absorbed,
hadrons that decay in flight into a muon in advance of entering the muon
detectors, and hadrons falsely identified as electrons.
Background from photon conversions $\gamma \ra e^+e^-$ arises  
when one member of the pair remains undetected 
and the other accidentally intersects the $J/\psi$ decay point.
Background from $B\bar B$ decays arises when a $J/\psi$ from a $B$
decay and a lepton from a semileptonic decay of the $\bar B$
appear to originate from the same decay point.
A number of other backgrounds\cite{bcprd} are found
to be negligible.
Table~\ref{bc_bg_sum} summarizes the results of the
background calculation and of a simultaneous 
fit for the muon and electron channels 
to the mass spectrum over the region between 
3.35 and 11~\gevcc. Figure~\ref{bc_mass}b) shows the mass spectra
for the combined $J/\psi \ell$ candidates, the combined
backgrounds and the fitted  
contribution from $\Bc \ra J/\psi\ell\nu$ decays.  The fitted number of 
\Bc\ events is $20.4^{+6.2}_{-5.5}$. 

To test the significance of this result,
a number of Monte Carlo trials is generated with
the statistical properties of the 
backgrounds, but with no contribution from \Bc~mesons.
These are subjected to the same fitting procedure
to determine contributions consistent with the
signal distribution arising from background fluctuations.  
The probability of obtaining a yield of 20.4 
events or more is $0.63 \times 10^{-6}$, 
equivalent to a 4.8 standard deviation effect.  

To check the stability of the \Bc\ signal, 
the assumed \Bc\ mass is varied
from 5.52 to 7.52~\gevcc.
The signal template for each value of $m(\Bc)$\ and  
the background mass distributions are used 
to fit the mass spectrum for the data.  
The magnitude of the \Bc\ signal is stable over the
range of theoretical predictions for $m(\Bc)$.
The~minimum in the
log-likelihood function versus mass yields a \Bc~mass of
$m(\Bc) = (6.40 \pm 0.39 \pm 0.13)$~\gevcc.
 
\begin{figure}[tbp]
\centerline{
\put(21,57){\large\bf (a)}
\put(75,57){\large\bf (b)}
\epsfysize=6.3cm
\epsffile[25 5 535 485]{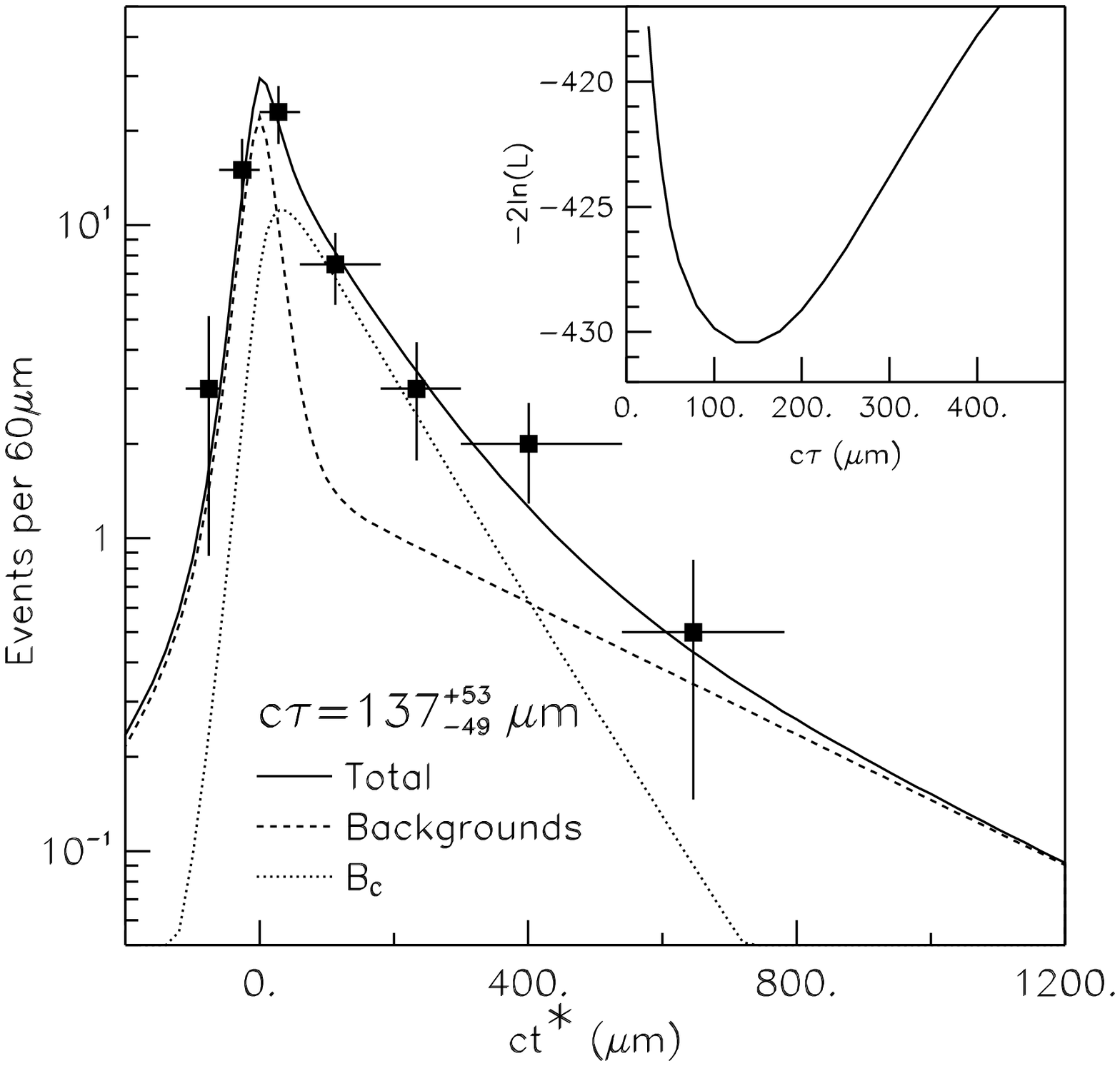}
\epsfysize=6.3cm
\epsffile[40 35 515 545]{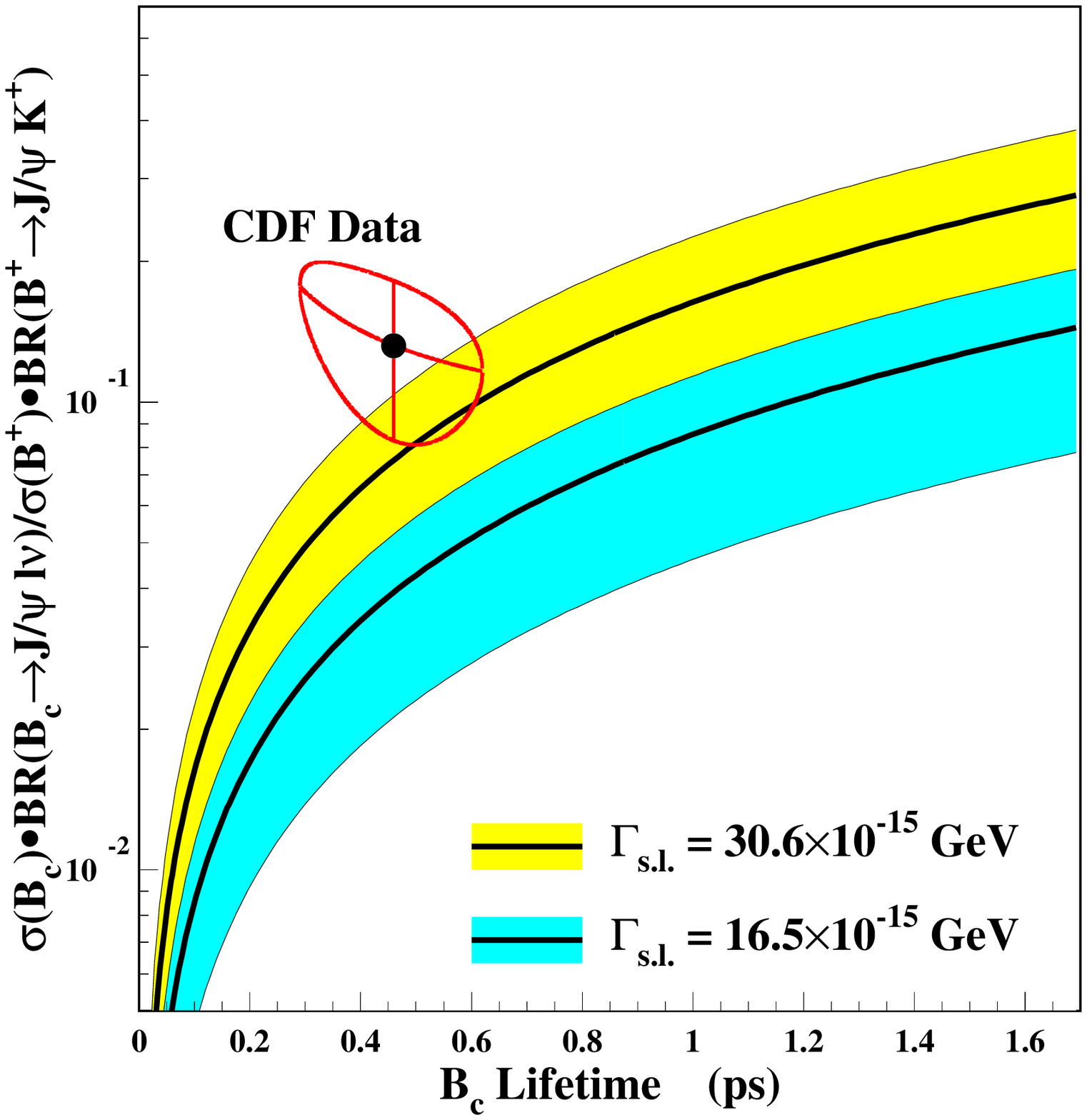}
}
\vspace*{0.2cm}
\fcaption{
(a) Pseudo-proper decay length distribution for the combined $J/\psi
e$ and $J/\psi\mu$ data with the fit result and the contributions from
signal and background overlaid. The inset shows the log-likelihood
function versus $c\tau$.
(b) Measured value of the $\sigma\times BR$ ratio shown as point
with one standard deviation contour and plotted at the measured value
for $\tau(\Bc)$. The shaded regions represent theoretical predictions
and their uncertainty bands for two different values of the 
\Bc~semileptonic decay width $\Gamma_{\rm sl}$.}
\label{bc_life}
\end{figure}

The lifetime of the \Bc~meson is obtained from the decay length
distribution using only events with 
$4.0 < m(J/\psi \ell) < 6.0$~\gevcc\ and relaxing the requirement
on the pseudo-proper decay length from $>60~\mu$m to
greater than $-100~\mu$m.
This yields a sample of 71 events, 42 $J/\psi e$ and 29 $J/\psi\mu$
candidates. An unbinned maximum log-likelihood fit yields
\begin{equation}
\tau(\Bc )  = (0.46\ ^{+0.18}_{-0.16}\ \pm 0.03)\ {\rm ps}. 
\end{equation}
The data, together with the signal and background 
distributions, are shown in Fig.~\ref{bc_life}a).

From the 20.4 \Bc\ events and a sample of 290 
$B^+ \ra J/\psi K^+$ events selected with the same
requirements as the \Bc~candidates,
the \Bc\
production cross section times the
$\Bc \ra J/\psi \ell^+ \nu$  branching fraction
$\sigma\times BR(\Bc\ra J/\psi \ell^+ \nu)$ is obtained, 
relative to that for the 
topologically similar decay $B^+ \ra J/\psi K^+$.
Many systematic uncertainties cancel in the ratio, while
Monte Carlo calculations yielded 
the values for the efficiencies that
do not cancel.
The detection efficiency for $\Bc \ra J/\psi \ell^+ \nu$ depends on $c\tau$ 
because of the cut on the pseudo-proper decay length at $>60\ \mu$m. We 
therefore quote a separate systematic uncertainty from the lifetime
uncertainty. We find
\begin{equation}
	\frac{\sigma(B_c) \times BR(B_c \rightarrow J/\psi \, \ell \nu)}
		   {\sigma(B) \times BR(B \rightarrow J/\psi \, K)}
	= 0.132 ^{+0.041}_{-0.037} \, ({\rm stat.}) \, 
		\pm 0.031 \, ({\rm syst.}) \,
	  ^{+0.032}_{-0.020} \, (c\tau)
\end{equation}
for \Bc\ and $B^+$\ with 
transverse momenta $\Pt > 6.0$~\gevc\ and rapidities $|y| < 1.0$.
Figure~\ref{bc_life}b) compares phenomenological predictions 
with the measurements of $\tau(\Bc)$ and the $\sigma\times BR$ ratio.
Within experimental and theoretical uncertainties, they are consistent.

\subsection{{\boldmath $B$} Lifetimes: Summary}   
\noindent
\begin{figure}[tb]
\centerline{
\epsfxsize=8.0cm
\epsffile[10 40 550 630]{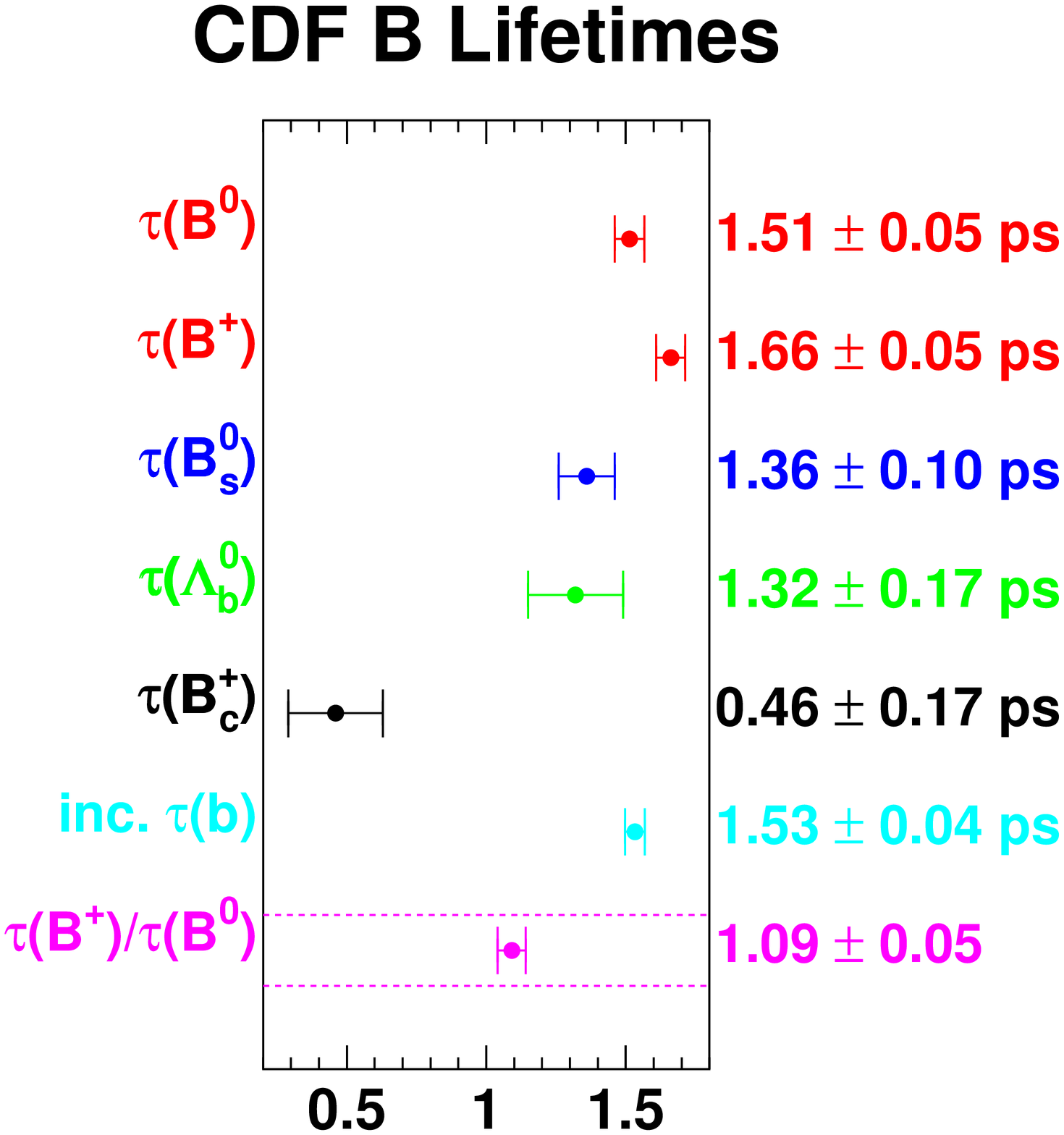}
}
\vspace*{0.2cm}
\fcaption{
Summary of CDF $B$ hadron lifetime results.}
\label{blife_comp_cdf}
\end{figure}
A summary of the $B$ lifetime measurements at CDF is given in
Figure~\ref{blife_comp_cdf}. As we have seen, CDF's $B$ lifetime results are
very competitive with the LEP and SLC measurements, where a precision
of a few percent is reached.  
Although CDF's measurement of the $B^+/B^0$ lifetime ratio appears to be
different from unity by almost two standard deviations, 
the precision is still not yet sufficient to distinguish between
theoretical approaches.
The $\Lambda_b^0$ lifetime lies closer to the $B^0$ lifetime 
at CDF ($\tau(\Lambda_b^0)/\tau(B^0) = 0.87\pm0.11$) compared to the 
measurements of the LEP experiments\cite{PDG}. Finally, the $B_c^+$ lifetime,
measured for the first time at CDF, is
clearly much shorter than the other $B$~hadron lifetimes and closer to
the $D^0$ or \Ds~lifetime.

\section{Measurements of \boldmath{$B^0\bar B^0$} Oscillations}
\runninghead{$B$ Lifetimes, Mixing and $CP$ Violation at CDF}
{Measurements of $B^0\bar B^0$ Oscillations}
\noindent
Oscillations between particles and antiparticles were
predicted\cite{pais} 
in 1955 and observed in the system of neutral
kaons\cite{kmix} in 1956.
Oscillations in the $B^0\bar B^0$ meson system were
observed for the first time by the ARGUS collaboration\cite{bmix_argus} in
1987.
This discovery signaled a large top quark mass and gave rise to new
prospects for the observation of $CP$ violation in $B$~meson decays
(see Sec.~8).
In this section, we describe measurements of the time dependence of
$B^0\bar B^0$ oscillations at CDF. After a short introduction in 
$B$~flavour oscillations, we review different approaches for measuring
the time dependence of $B^0\bar B^0$~oscillations at CDF. 

\subsection{Introduction to {\boldmath $B^0\bar B^0$} oscillations}   
\begin{figure}[tbp]
\centerline{
\epsfxsize=6.3cm
\epsffile[5 15 515 515]{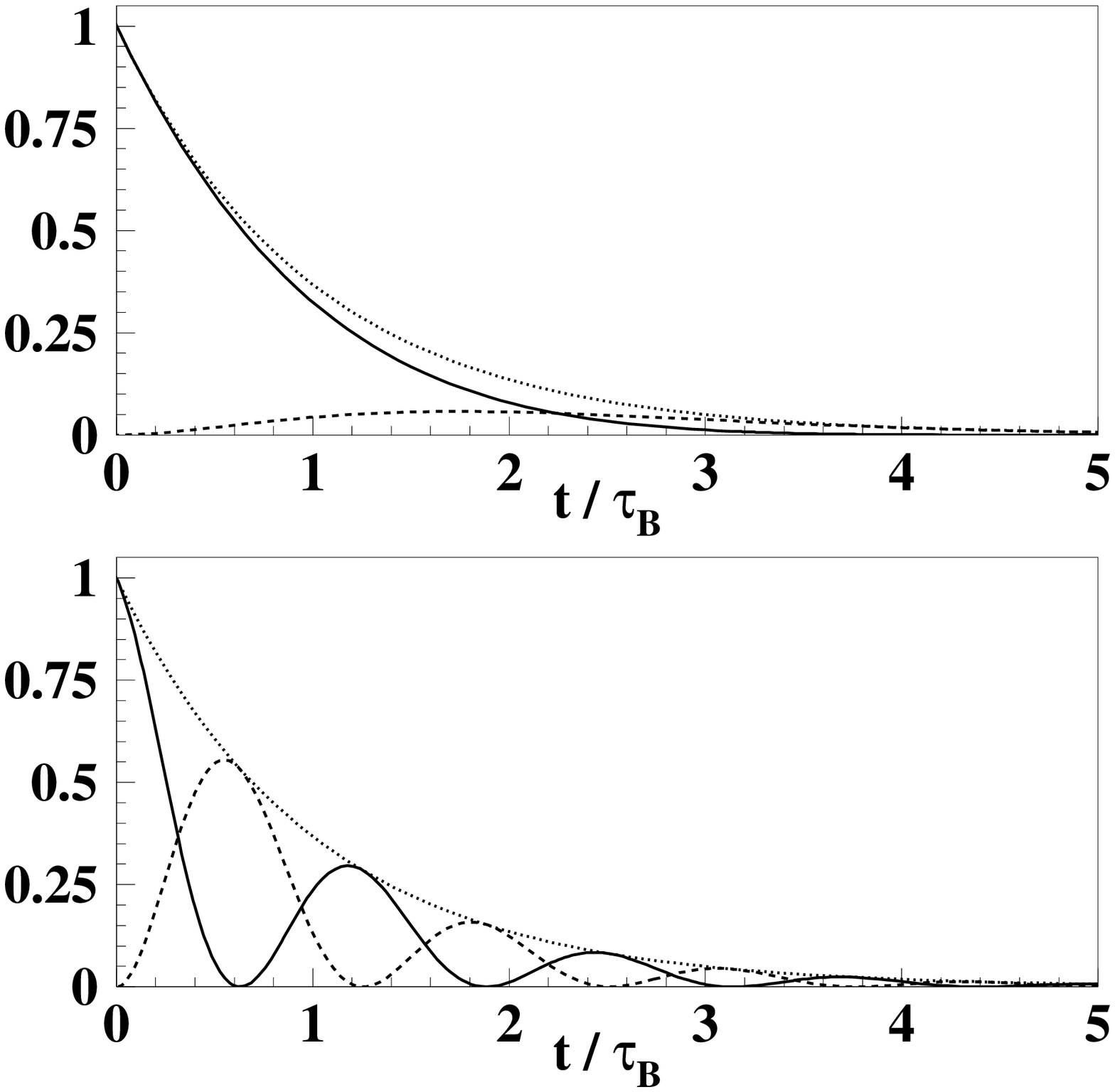}
\hspace*{0.3cm}
\epsfxsize=6.1cm
\epsffile[0 -25 340 290]{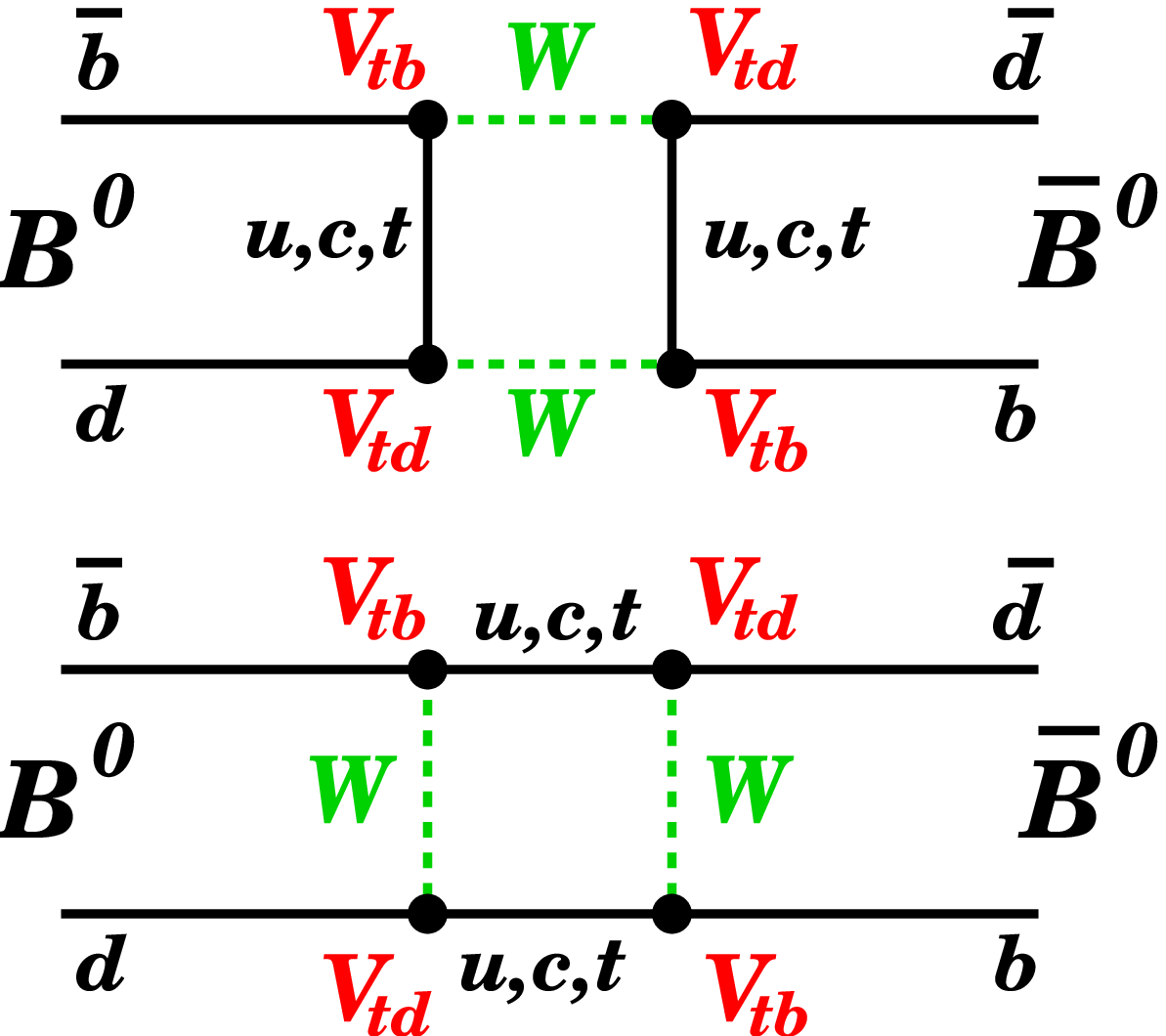}
}
\vspace*{0.2cm}
\fcaption{
Left: Time evolution of $B\bar B$ oscillations displaying the unmixed
(solid) and mixed (dashed) contribution as well as the sum of
both (dotted) for two different oscillation
frequencies. Right: Box diagram describing $B^0\bar B^0$ oscillations.}  
\label{bmix_box}
\end{figure}

\noindent
As already outlined in Sec.~6.5.3, the 
system of neutral $B$~mesons, $B^0$ and $\bar B^0$, can be
described in terms of states with well defined mass and lifetime,  
$B_H$ and $B_L$ ($H =$ `heavy' and $L =$ `light'),
\begin {equation}
  |B^0 \rangle = 1/\sqrt{2}\,\,(\,|B_L^0\rangle + |B_H^0\rangle\,)  
\hspace{2.0cm}
  |\bar B^0 \rangle = 1/\sqrt{2}\,\,(\,|B_L^0\rangle - |B_H^0\rangle\,).
\end{equation}
Here, we consider the effects of $CP$~violation to be small
compared to the expected mixing effects.
The difference in mass \dm\ between both mass
eigenstates leads to a time dependent phase difference between their
wave functions. The probability that an initially pure $B^0$ state  can
be observed as a $\bar B^0$ at proper time $t$ is given by
\begin{equation}
{\cal P}_{\rm mix} = Prob (B^0 \ra \bar B^0,t) = 1/2\,\, \Gamma
{\rm e}^{-\Gamma t} \, (1 - \cos \dm\, t),
\end{equation}
while the probability that it decays as $B^0$ is given by:
\begin{equation}
{\cal P}_{\rm unmix} =  Prob (B^0 \ra B^0,t) = 1/2\,\, \Gamma
{\rm e}^{-\Gamma t} \, (1 + \cos \dm\, t).
\end{equation}
Figure~\ref{bmix_box} shows, on the left side, the time evolution
of $B\bar B$ oscillations displaying the unmixed 
(solid) and mixed (dashed) contributions for two different oscillation
frequencies \dm. The sum of ${\cal P}_{\rm mix}$ and 
${\cal P}_{\rm unmix}$
is just the exponential particle decay $\Gamma{\rm e}^{-\Gamma t}$ and 
is displayed as a dotted line in Fig.~\ref{bmix_box}.

If the $B$~meson decay time $t$ cannot be measured, the time integrated
probability
which is usually referred to as $\chi$ can be obtained as
\begin{equation}
  \chi = \frac{{\cal P}_{\rm mix}} {{\cal P}_{\rm mix} + 
	{\cal P}_{\rm unmix} } = \frac{x^2}{2(1+x^2)}
\end{equation}
Here, the mixing parameter $x = \dm / \Gamma$ 
is introduced. It describes the oscillation period relative to the
$B$~meson lifetime $\tau_B = \hbar/\Gamma$.

If both neutral $B$ mesons, $B^0$ and \Bs, are produced, the time
integrated and flavour averaged mixing parameter $\bar{\chi}$ is
defined as
\begin{equation}
 \bar{\chi} = f_d\,\chi_d + f_{\mbox{\sl s}}\,\chi_{\mbox{\sl s}},
\end{equation}
where $f_d$ and $f_{\mbox{\sl s}}$ are the fractions
of $b$ hadrons that are produced as $B^0$ and \Bs~mesons,
respectively.

In the Standard Model\cite{GSW}, $B^0\bar B^0$ mixing occurs via
second order weak 
processes, as displayed in Fig.~\ref{bmix_box} on the right hand side. 
The mass difference \dmd\ can be
determined within the Standard Model by computing the electroweak box
diagram, where the dominant contribution is through top quark exchange:
\begin {equation}
  \dmd = \frac{{\rm G}_{\rm F}^2}{6\pi^2}\,
        m_B\, (f_B^2 B_B)\, \eta_{\rm QCD}\, 
        m_t^2\, F(\frac{m_t^2}{m_W^2})\, 
	|V_{tb}^*\,V_{td}|^2.
\end{equation}
Here, G$_{\rm F}$ is the Fermi coupling constant, $m_B$ the $B$~meson mass,
$f_B$ the weak $B$~decay constant, $B_B$ the
bag parameter of the $B$~meson, $\eta_{\rm QCD}$ are QCD corrections which
are in the order of one, 
$m_t$ is the top quark mass, 
$V_{tb}$ and $V_{td}$ are the two CKM matrix elements involved, 
and $F(z)$ is a slowly varying function of the top
quark mass and the $W$ boson mass $m_W$: 
\begin{eqnarray}
	F(z) & = & \frac{1}{4} + \frac{9}{4(1-z)} - \frac{3}{2}
 	\frac{1}{(1-z)^2} - \frac{3}{2} \frac{z^2 \ln z}{(1-z)^3}. 
\end{eqnarray}
In the case of the \Bs~meson the respective values of $B_{\Bs }$,
$f_{\Bs}$, $m_{\Bs}$, and $\eta_{\rm QCD} = \eta_{\Bs}$ have to be
used, and $V_{td}$ is replaced by $V_{ts}$. 

A measurement of \dmd\ or \dms\ would in principle determine the
Cabibbo-Kobayashi-Maskawa matrix elements $V_{td}$ or $V_{ts}$ but
theoretical uncertainties connected with the poor knowledge of 
the $B$~meson weak decay constant $f_B$ and the bag parameter $B_B$
limit the direct extraction of the CKM matrix elements from measurements of
\dmd\ and \dms. However, in the ratio \dmd/\dms\
several of the theoretical uncertainties cancel
\begin {equation}
  \frac{\dmd}{\dms} = \frac{m_{B^0}}{m_{\Bs}}\,
	\frac{\eta_{B^0}}{\eta_{\Bs}}\, \frac{f^2_{B^0} B_{B^0}} 
	{f^2_{\Bs} B_{\Bs}}\, \frac{|V_{td}|^2}{|V_{ts}|^2}. 
\end{equation}
and \dmd/\dms\ will ultimately determine one of the legs of the
unitarity triangle as further discussed in Sec.~8.1.

\subsubsection{Ingredients of $B^0\bar B^0$ mixing measurements}   
\noindent
In general, a measurement of the time dependence of $B^0\bar B^0$
oscillations requires the knowledge of
the proper decay time $t$ of the $B$~meson,
the flavour of the $B$~meson at production, and the flavour of the
$B$~meson at decay in order to determine whether the $B^0$ has
oscillated. The last two items
require identifying the flavour of a $B$~meson. They are the subject of the
next section 7.1.2, where we discuss $B$~flavour tagging. 
As outlined in Sec.~6.2.3, the $B$~meson decay time 
can be obtained from a measurement of the distance $\lxy^B$ between 
the primary interaction vertex, where the $B$~meson is produced, and
its secondary decay vertex. The decay time $t$ 
is related to the decay distance $\lxy^B$ by
\begin {equation}
c\,t\,(B) =  \frac{\lxy^B}{\beta \gamma} = \lxy^B \, \frac{m(B)}{\Pt(B)}.
\label{eq:tlxy1}
\end{equation}
If the $B$~meson is not fully reconstructed, a $\beta
\gamma$ correction is applied to scale from the only partially
measured $B$~momentum to the unknown $B$ momentum.

From Eq.~(\ref{eq:tlxy1}) the uncertainty on the decay time can be
calculated (in units of the $B$ lifetime $\tau_B$):
\begin {equation}
  \frac{\sigma_t}{\tau_B} = \sqrt{\left( \frac{\Delta\lxy^B}{\lxy^0}\right)^2
   + \left(\frac{t}{\tau_B}\frac{\Delta p_t}{p_t}\right)^2}
  \hspace{1.0cm} {\rm where} \hspace{0.5cm}
  \lxy^0 = \Pt/m_B \cdot c\tau_B. 
\label{eq:errors}
\end{equation}
The proper time resolution
$\sigma_t$ depends on the uncertainty to infer the decay length
from the primary to the $B$ decay vertex and on the 
$B$~momentum resolution which is dominated by the magnitude of the
$\beta\gamma$ 
correction. Note, the latter uncertainty scales with $t/\tau_B$, while the
vertexing resolution only adds an uncertainty constant with $c\,t$.

\subsubsection{$B$ flavour tagging}
\noindent
$B$~flavour tagging refers to the task of determining the
flavour of a $B$~meson either at production or at decay.  
Several methods of $B$~flavour tagging exist. Some methods
identify the flavour of the
other $B$~hadron produced in the initial collision along with the 
$B$~meson of interest. Since the dominant $b$~quark production mechanism
produces $b\bar b$ pairs, the flavours 
of both $B$~hadrons are assumed to be opposite at time of production. 
There are three common methods of opposite side flavour tagging.
One method, called ``lepton
tagging'', looks for a lepton from the semileptonic decay of the other
$B$~hadron in the event. The charge of this lepton is correlated with
the flavour of the $B$~hadron: 
An $\ell^-$ comes from a $b \ra c\, \ell^- \bar\nu X$ transition, 
while an $\ell^+$ originates from a $\bar b$ quark. Second, the charge of a
$K^{\pm}$ from the subsequent charm decay $c \ra s X$ is also
correlated to the $B$~flavour: A $K^-$ results from the decay chain 
$b \ra c \ra s$ while a $K^+$ signal a $\bar b$ flavour. Searching for a
charged kaon from the other $B$ hadron decay is usually referred to as
``kaon tagging''. Third, a method called ``jet-charge tagging''
exploits the fact that the sign of the momentum weighted sum of the
particle charges of the opposite side $b$~jet is the same as the
charge of the $b$~quark producing this jet. Finally, the flavour
of a $B$~meson can also be tagged on the same side as the $B$~meson of
interest, by exploiting correlations of the $B$~flavour with the charge
of particles produced in association with the $B$~meson. Such
correlations are expected to arise from $B$~quark hadronization and
from $B^{**}$~decays. We call this method ``same side tagging''. CDF
has results on opposite side lepton tagging, jet-charge tagging, and
same side tagging but has not studied kaon tagging due to the lack of
particle identification at CDF. These tagging methods are
described in more detail in the following sections where they are
applied in different $B^0\bar B^0$~mixing measurements.     

The figure of merit used to compare the ``tagging power'' of different
flavour tags is the so-called effective tagging efficiency \eD, where 
$\varepsilon$ is the efficiency of how often a flavour tag is
applicable. The dilution $\cal D$ is defined as the number of correctly
tagged events $N_R$ minus the number of incorrectly identified events
$N_W$ divided by the sum
\begin {equation}
  {\cal D} = \frac{N_R - N_W}{N_R + N_W}.
\end{equation}
To quantify the tagging power with the expression ``dilution'' can be
misleading. A flavour tag which always tags correctly has a dilution of
one, while a flavour tag giving the correct tag 50\% of the time
has a dilution of zero. This means, a tagging algorithm with a large
dilution is desirable, while a small dilution characterizes a less
powerful tagging method. 
The dilution is also related to the probability $p_R$ that the
flavour tag is correct and to the mistag probability $p_W = 1 - p_R$
that the flavour tag is incorrect 
\begin{equation}
  {\cal D} = 2 \, p_R - 1 = 1 - 2\, p_W.
\end{equation}

To illustrate the statistical significance of the product \eD,
we discuss an asymmetry measurement on a data sample with $N$ events,
where the flavour tagging method identifies whether the
event is of type $a$ or type $b$. 
Type $a$ and type $b$ could for example be ``mixed'' and ``unmixed'' decays
of a neutral $B$~meson.
The measured asymmetry~${\cal A}_{\rm meas}$ is
\begin{equation}
  {\cal A}_{\rm meas} = \frac{N_a - N_b}{N_a + N_b},
\end{equation}
where $N_a$ and $N_b$ are the number of events that are
tagged as type $a$ and type $b$, respectively.
The true asymmetry~${\cal A}_{\rm true}$ is
\begin{equation}
  {\cal A}_{\rm true} = \frac{N_a^{\rm t} - N_b^{\rm t}}
	{N_a^{\rm t} + N_b^{\rm t}},
\end{equation}
where $N_a^{\rm t}$ and $N_b^{\rm t}$ are the true number of events of type
$a$ and type $b$ in the sample.
The efficiency is simply
\begin{equation}
\varepsilon=\frac{N_a + N_b}{N_a^{\rm t} + N_b^{\rm t}}.
\end{equation}
It follows, from the definition of the dilution, that
the true asymmetry and the measured asymmetry are related by
${\cal A}_{\rm true} = {\cal A}_{\rm meas}/{\cal D}$,
and the statistical error on the true asymmetry is
\begin{equation}
  \sigma({\cal A}_{\rm true}) = \sqrt{ \frac{ 1 - 
	{\cal D}^2 {\cal A}^2_{\rm true} }
	{ \eD N } },
\end{equation}
where $N$ is the total number of events in the sample 
$N = N_a^{\rm t} + N_b^{\rm t}$.
The product $\eD N$ is the effective statistics of the data sample,
that is, it is the equivalent number of perfectly tagged events.
The statistical power of different flavour tagging methods varies
as \eD.
 
\subsubsection{Outline of $B^0\bar B^0$ mixing measurements at CDF} 
\noindent
All $B^0\bar B^0$ mixing measurements at CDF are based on lepton data
samples where the charge of the lepton determines the $B$~flavour at
decay assuming it originates from a semileptonic $B$~decay. The
$B$~flavour at production is determined by an opposite side lepton tag,
a jet-charge tag, or a same side tag in the various analyses
described in the next sections. The proper time at decay is determined
from the $B$~decay vertex, inferred from partially
reconstructed $B$~mesons or inclusive vertices. To illustrate the
effects discussed above, we consider a mixing measurement where an
opposite side lepton identifies the $B$~flavour at production.  

\begin{figure}[tb]
\centerline{
\put(0,41){\large\bf (a)}
\put(42,41){\large\bf (b)}
\put(84,41){\large\bf (c)}
\epsfxsize=4.2cm
\epsffile[5 5 545 515]{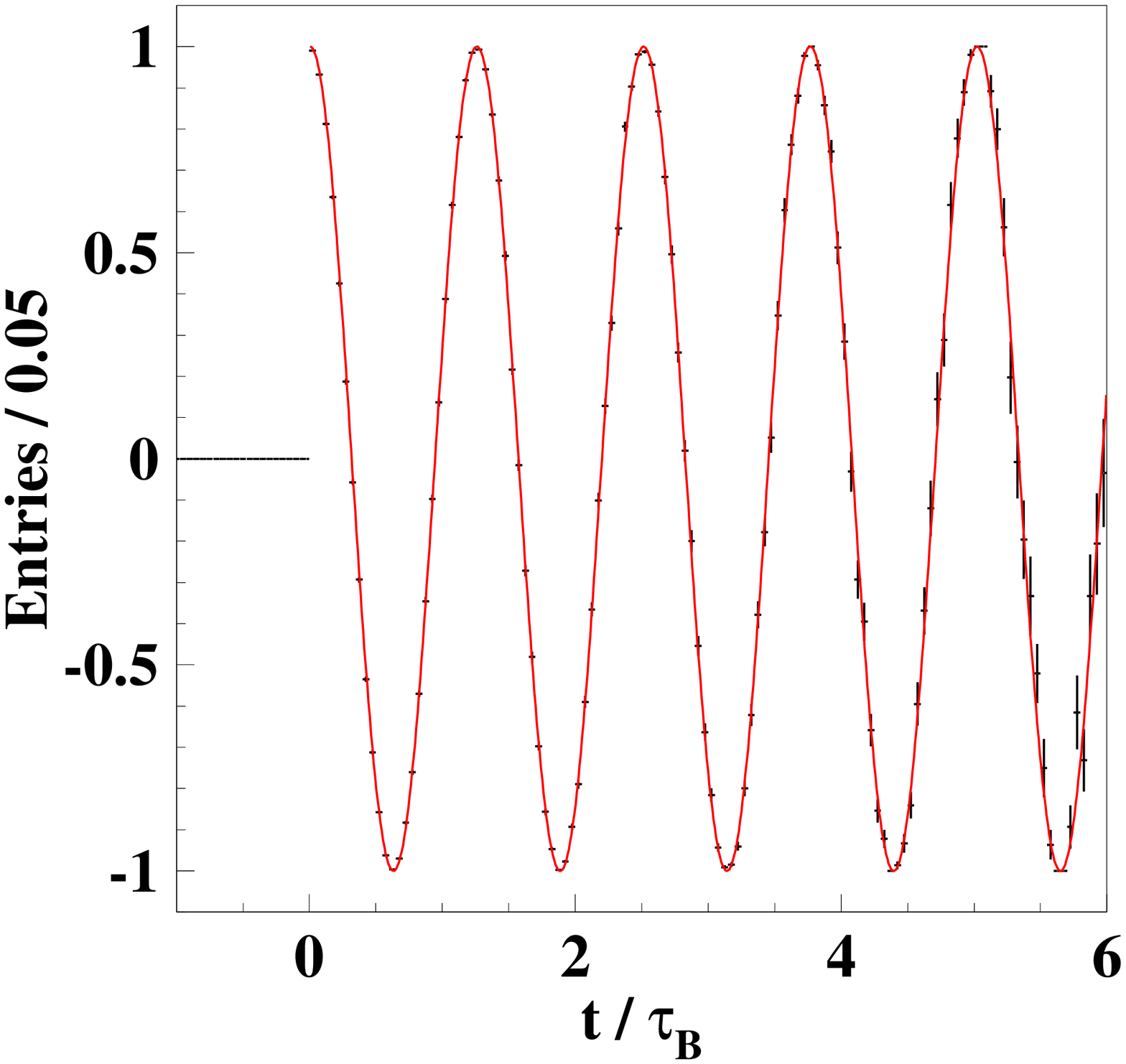}
\epsfxsize=4.2cm
\epsffile[5 5 545 515]{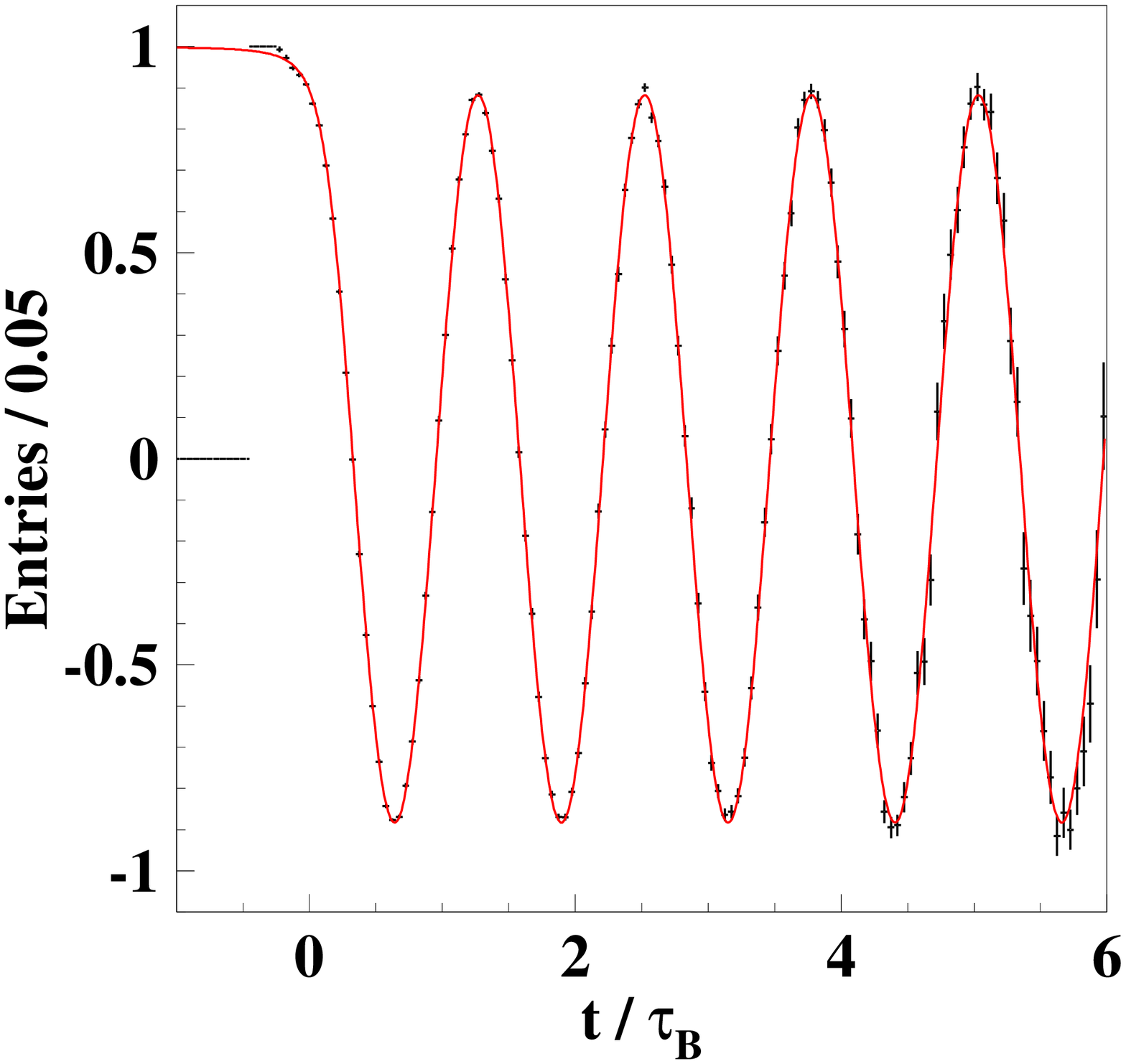}
\epsfxsize=4.2cm
\epsffile[5 5 545 515]{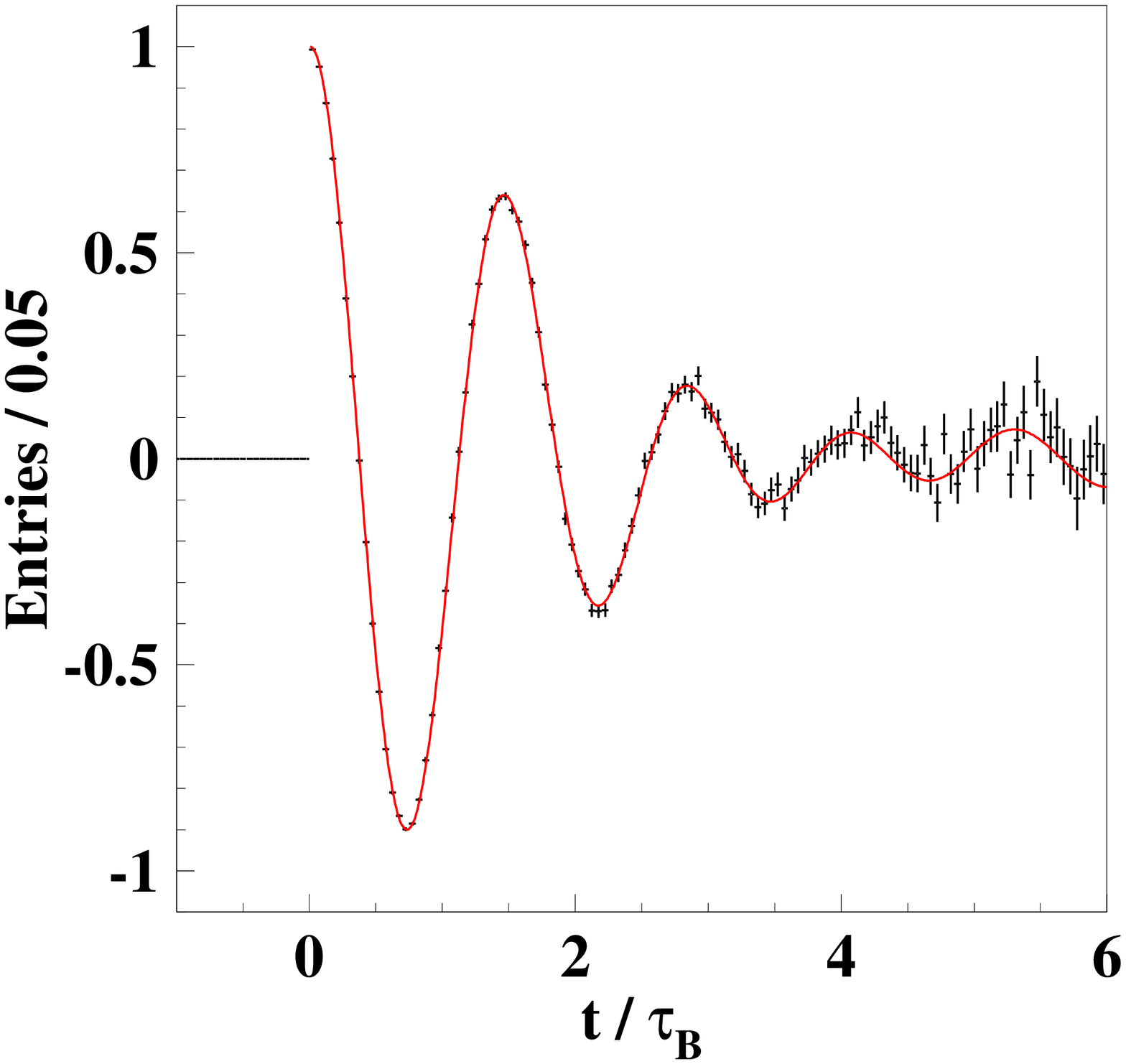}
}
\centerline{
\put(0,41){\large\bf (d)}
\put(42,41){\large\bf (e)}
\put(84,41){\large\bf (f)}
\epsfxsize=4.2cm
\epsffile[5 5 545 515]{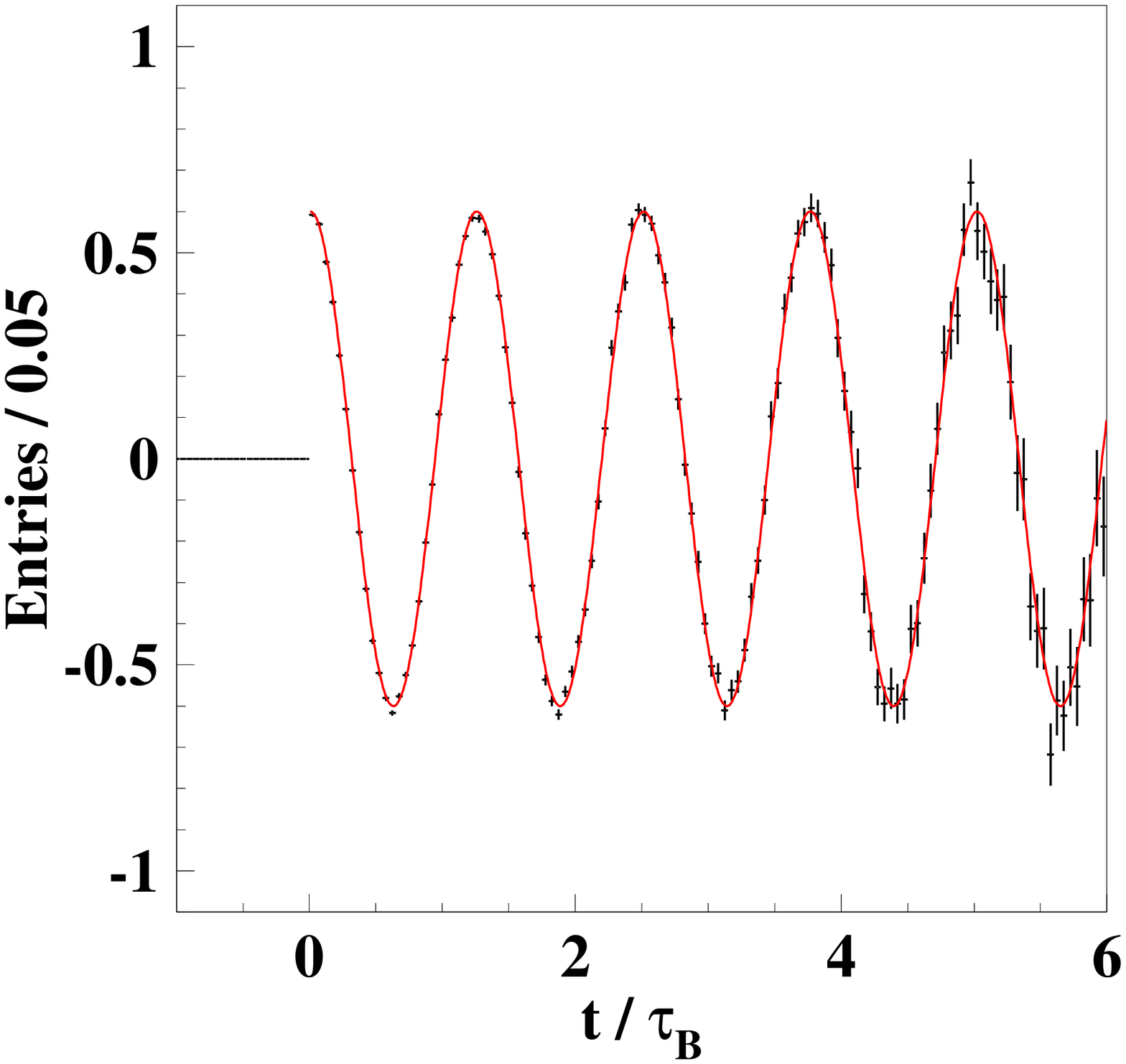}
\epsfxsize=4.2cm
\epsffile[5 5 545 515]{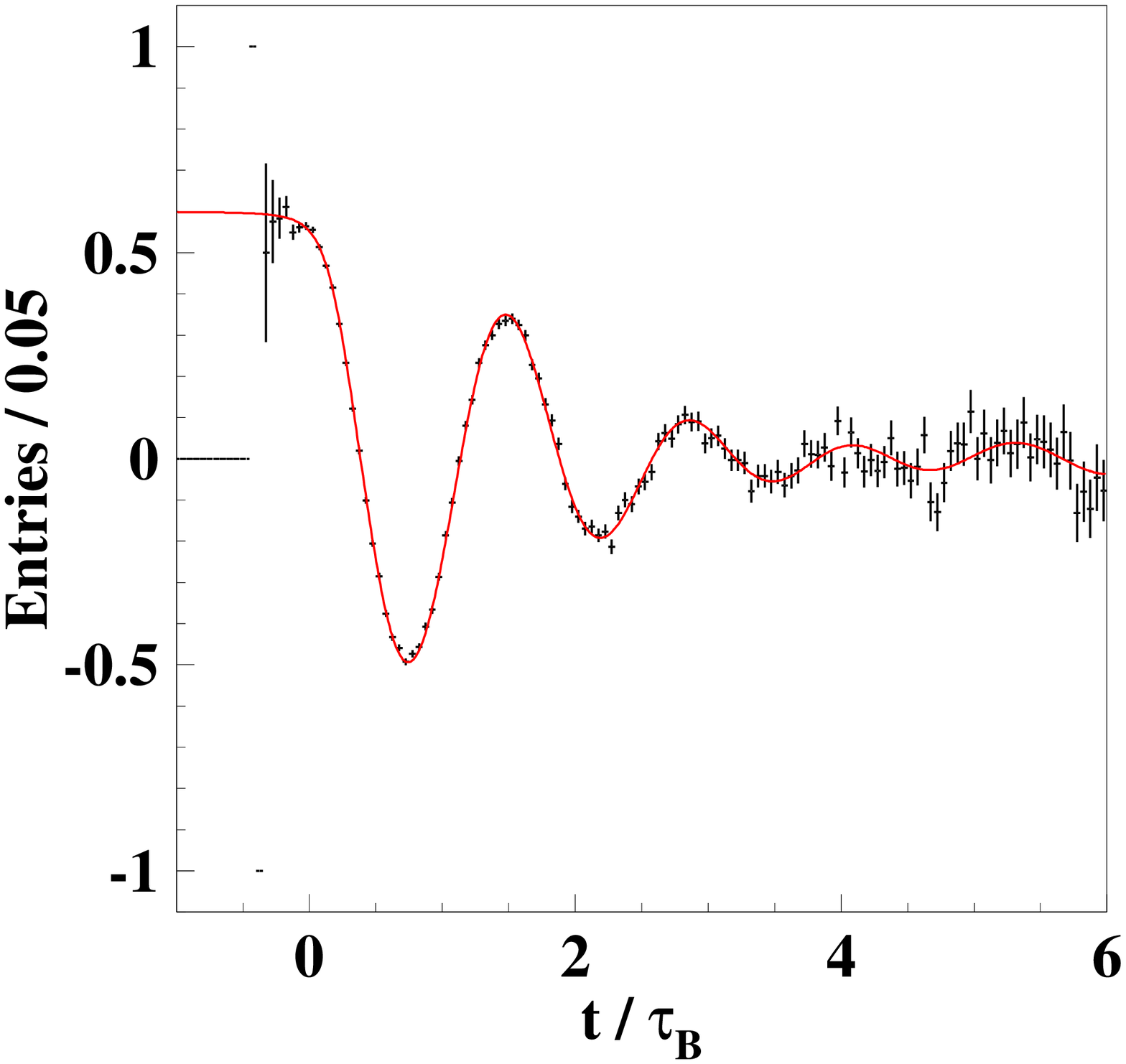}
\epsfxsize=4.2cm
\epsffile[5 5 545 515]{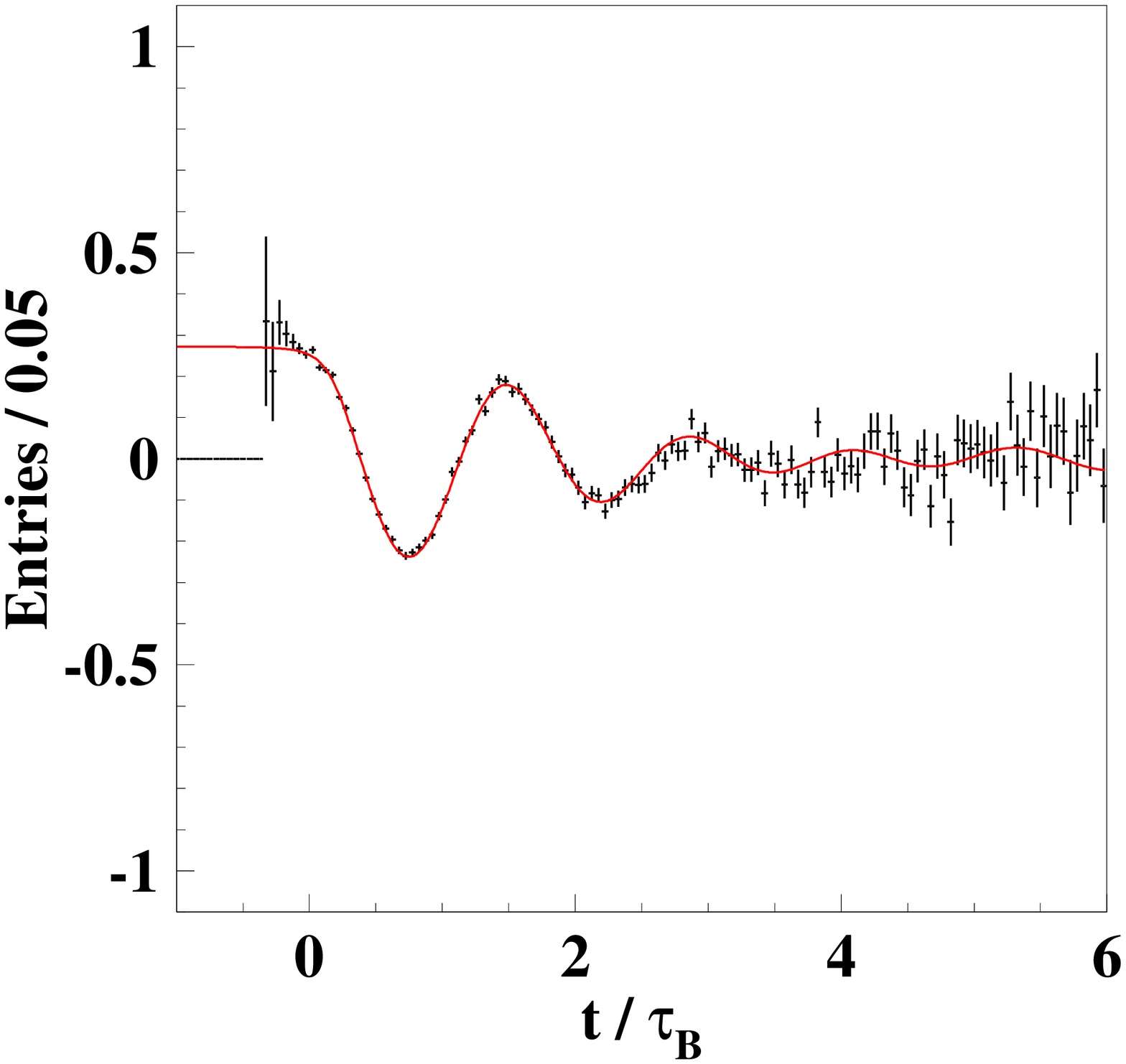}
}
\vspace*{0.2cm}
\fcaption{
Illustration of various effects on the asymmetry ${\cal A}_{\rm mix}$: 
(a) Perfect resolution, (b) vertex resolution, 
(c) momentum resolution, (d) mistag probability, 
(e) vertex and momentum resolution plus mistag
probability, and (f) vertex and momentum resolution plus mistag
including combinatorial background.}
\label{bmix_resol}
\end{figure}

We start with a pure sample of $B^0$~mesons and assume that the lepton
tag is always correct. In this case, an event with an opposite-sign
lepton pair signals an unmixed event, while a like-sign lepton pair
indicates a mixed event. The probabilities for an opposite-sign event 
${\cal P}_{OS}$ and a like-sign event ${\cal P}_{LS}$ are directly
related to the mixing probabilities
\begin{equation}
{\cal P}_{LS}(t) = {\cal P}_{\rm mix}(t) 
\hspace*{0.5cm} {\rm and} \hspace*{0.5cm} 
{\cal P}_{OS}(t) = {\cal P}_{\rm unmix}(t) 
\end{equation}
which are defined as
\begin{equation}
{\cal P}_{\rm unmix/mix}(t) =  
1/2\,\,\Gamma{\rm e}^{-\Gamma t} \,(1\pm\cos\dm\, t).
\end{equation}
To illustrate this behavior, we plot in Fig.~\ref{bmix_resol}a) the
asymmetry  
\begin{equation}
{\cal A}_{\rm mix} = \frac{{\cal P}_{OS} - {\cal P}_{LS}} 
{{\cal P}_{OS} + {\cal P}_{LS}} = \cos\dm\,t
\label{eq:asym}
\end{equation}
which in this case is a pure cosine like oscillation. We
have chosen $\dm = 5$. Next, we introduce a
vertexing resolution  
which smears the decay time measurement and effectively reduces the 
amplitude of the oscillation as shown in Fig.~\ref{bmix_resol}b). The
effect of introducing a momentum resolution, for example through a
$K$-factor distribution ${\cal H}(K)$ in case of a partially
reconstructed decay, is displayed in Fig.~\ref{bmix_resol}c). Since
the uncertainty on the $\beta\gamma$ resolution scales with proper
time $t$, as seen in Eq.~(\ref{eq:errors}), the oscillation damps with
proper time. In the case of a Gaussian vertexing resolution $\cal G$ and
a momentum resolution expressed through ${\cal H}(K)$, the mixing
probability would be modified as 
\begin{equation}
{\cal P}_{\rm mix}(t) =  
\int dK\, {\cal H}(K)\, \left[1/2\,\,K\Gamma\,
{\rm e}^{-K\Gamma t^{\prime}}\,(1-\cos\dm\, Kt^{\prime})
\right]\otimes {\cal G}(t^{\prime}|\, t,\sigma).
\end{equation}
${\cal P}_{\rm mix}$ would be modified accordingly.

If the lepton tag in our example does not always tag correctly but
with a mistag probability $p_W$, we would measure the following
like-sign and opposite-sign fractions
\begin{eqnarray}
{\cal P}_{LS}(t) &=& (1 - p_W)\,{\cal P}_{\rm mix}(t) +
p_W\,{\cal P}_{\rm unmix}(t)  \nonumber \\
{\cal P}_{OS}(t) &=& p_W\,{\cal P}_{\rm mix}(t) +
(1 - p_W)\,{\cal P}_{\rm unmix}(t).  
\label{eq:pmixsig}
\end{eqnarray}
The effect of a mistag probability $p_W$ is shown
in Fig.~\ref{bmix_resol}d), while the combined effects of vertexing
plus momentum resolution, together with a mistag probability are displayed in
Fig.~\ref{bmix_resol}e). In a real measurement there will be background
such as combinatorial background under a charm signal. We define 
${\cal P}_{LS}^{\rm sig}$ and ${\cal P}_{OS}^{\rm sig}$ as
${\cal P}_{LS}$ and ${\cal P}_{OS}$ as in Eq.~(\ref{eq:pmixsig})
and obtain
\begin{eqnarray}
{\cal P}_{LS}(t) &=& (1 - f_{\rm bg})\,{\cal P}^{\rm sig}_{LS}(t) +
f_{\rm bg}\,f_{LS}\,{\cal P}^{\rm bg}(t)  \nonumber \\
{\cal P}_{OS}(t) &=& (1 - f_{\rm bg})\,{\cal P}^{\rm sig}_{OS}(t) +
f_{\rm bg}\,(1 - f_{LS})\,{\cal P}_{\rm bg}(t).  
\end{eqnarray}
Here, $f_{\rm bg}$ is the fraction of background in a given sample,
$f_{LS}$ is the fraction of like-sign events in the background, while 
${\cal P}_{\rm bg}$ is the probability function to describe the
behavior of the background versus proper time $t$. The effect of a
background on the asymmetry, in addition to a vertexing and momentum
resolution, as well as a mistag rate are shown in
Fig.~\ref{bmix_resol}f) and exhibit a distribution as can be expected
in a real measurement. Some measurements demonstrate the time
dependence of $B\bar B$ oscillations by plotting the so-called mixed
fraction  
\begin{equation}
{\cal F}_{\rm mix}(t) = \frac{{\cal P}_{LS}} 
{{\cal P}_{OS} + {\cal P}_{LS}} = \frac{1}{2}\, (1 - \cos\dm\,t),
\end{equation}
which exhibits a sinusoidal behavior with ${\cal F}_{\rm mix}(0) = 0$. 

In summary, we would like to point out that a measurement of the time
dependence of $B\bar B$~oscillations is similar to a $B$~lifetime
measurement, the difference being that the entire sample is divided
into a mixed sample and an unmixed sample.
In the following section we review some of the $B^0\bar B^0$
mixing results at CDF. These measurements are an excellent proof that
$B$~flavour tagging works in a hadron collider environment.
Besides a measurement of \dm\ they allow, at the same time, the
determination of the 
effective tagging efficiency \eD\ of the applied flavour tag. 
The frequency of the oscillation will determine \dm, while the
amplitude characterizes~$\cal D$.

\subsection{$B^0\bar B^0$ mixing using same side tagging}
\begin{figure}[tbp]
\centerline{
\put(0,36){\large\bf (a)}
\put(69,36){\large\bf (b)}
\epsfysize=3.5cm
\epsffile[1 -40 600 280]{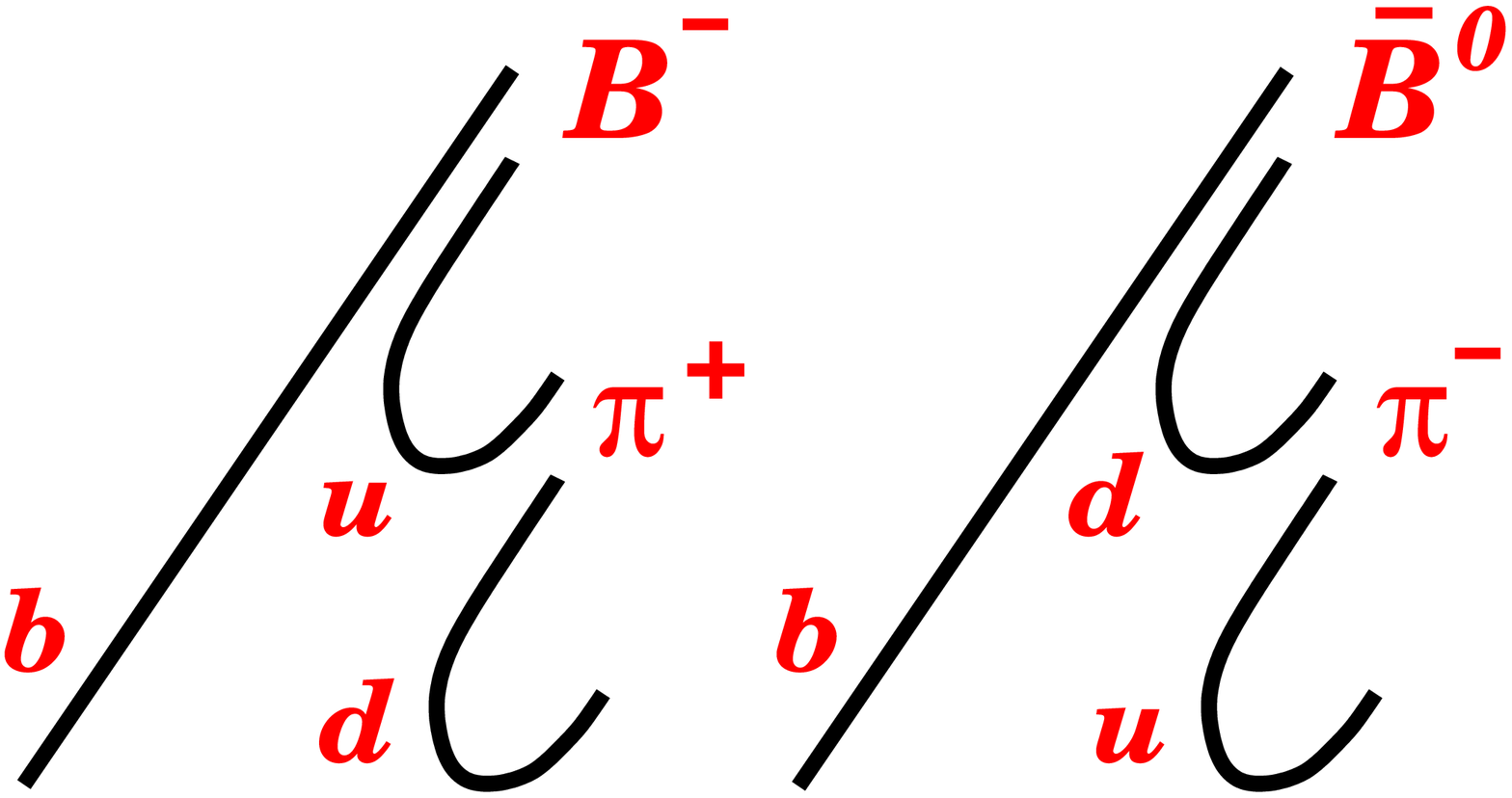}
\hspace*{0.5cm}
\epsfysize=4.0cm
\epsffile[1 1 357 277]{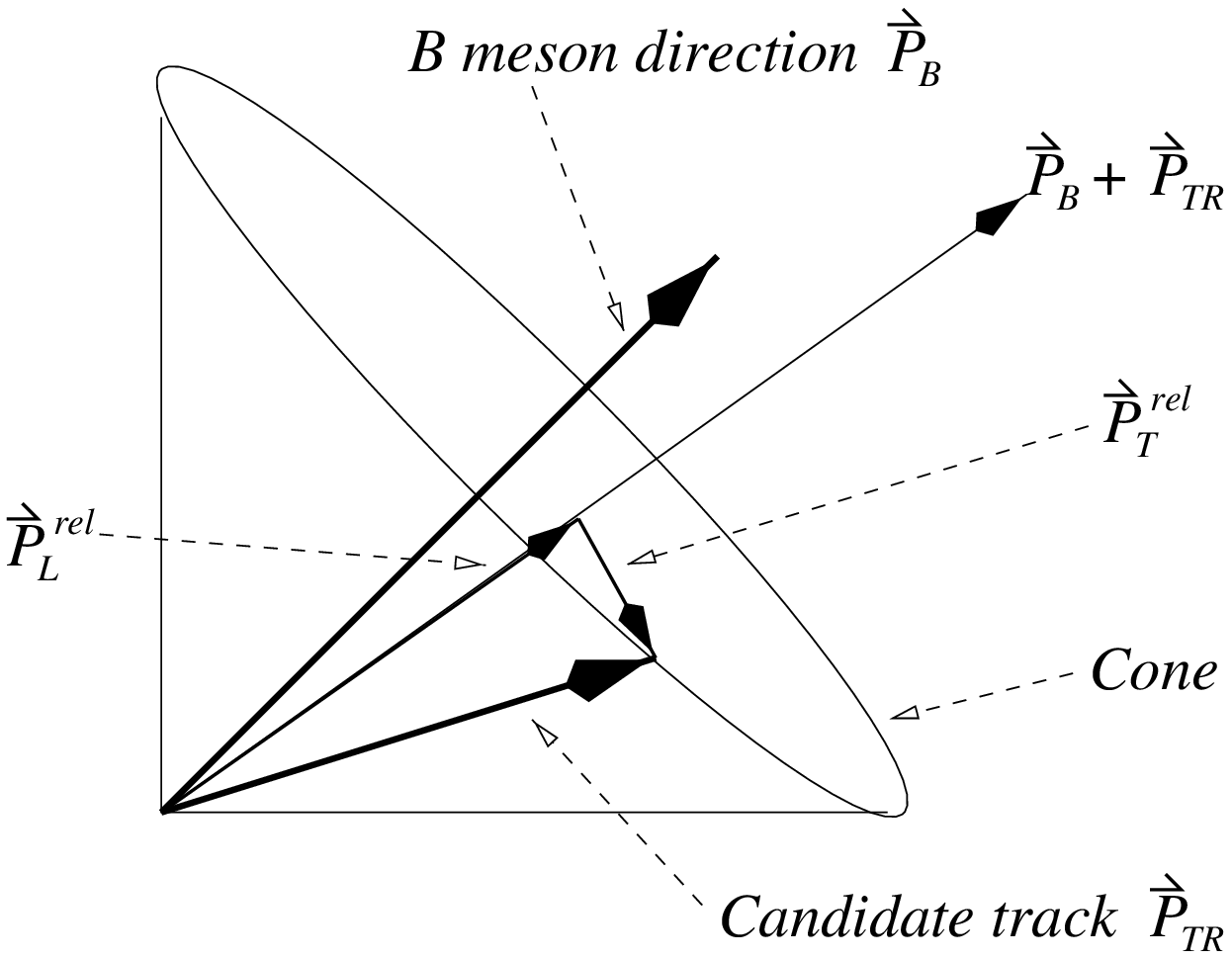}
}
\vspace*{0.2cm}
\fcaption{
(a) A simplified picture of $b$~quark fragmentation.
(b) Schematic drawing of the momentum vectors determining the \ptrel\ of
a same side tagging track candidate.} 
\label{bmix_sst_draw}
\end{figure}

\noindent
The measurement of the $B^0\bar B^0$
oscillation frequency \dmd, using a same side tagging
technique\cite{sst_prd,sst_prl} (SST)   
is based on partially reconstructed semileptonic $B$~decays to
$D^{(*)}\ell \nu X$. The lepton charge tags the $B$~flavour at decay time,
while a same side tag provides the $B$~flavour at production. 
It has been suggested\cite{Rosner_sst} that the electric charge of
particles produced near a $B$~meson can be used to
determine its initial flavour. This can be understood in a simplified
picture of fragmentation as shown in Figure~\ref{bmix_sst_draw}a).
For example, if a $b$~quark combines with a $\bar u$~quark to form a
$B^-$~meson, the remaining $u$ quark may combine with a
$\bar d$~quark to form a $\pi^+$.
Similarly, if a $b$~quark hadronizes to form a $\bar B^0$ meson, the
associated pion would be a $\pi^-$.
Another source of correlated pions are decays of the
orbitally excited ($L=1$) $B$~mesons ($B^{**}$)\cite{LEP_bdss},
$B^{**0} \ra B^{(*)+} \pi^-$ or
$B^{**+} \ra B^{(*)0} \pi^+$ (see also Sec.~7.2.1).
Here, no attempt is made to distinguish the 
hadronization pions from those originating from $B^{**}$ decays. 

\begin{figure}[tbp]
\centerline{
\put(77,54){\large\bf (e)}
\epsfxsize=6.3cm
\epsffile[30 5 530 490]{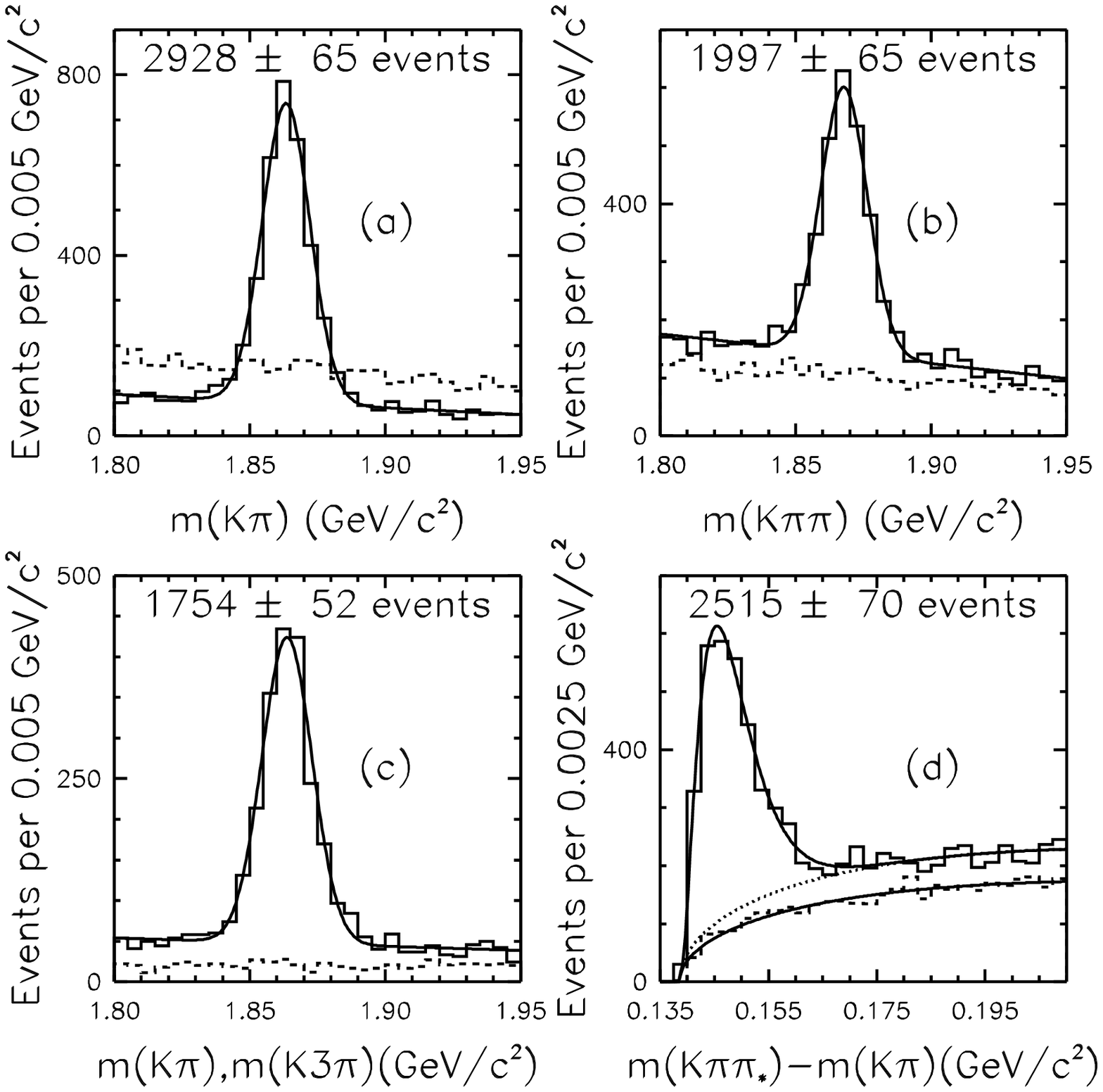}
\epsfysize=6.0cm
\epsffile[1 1 530 490]{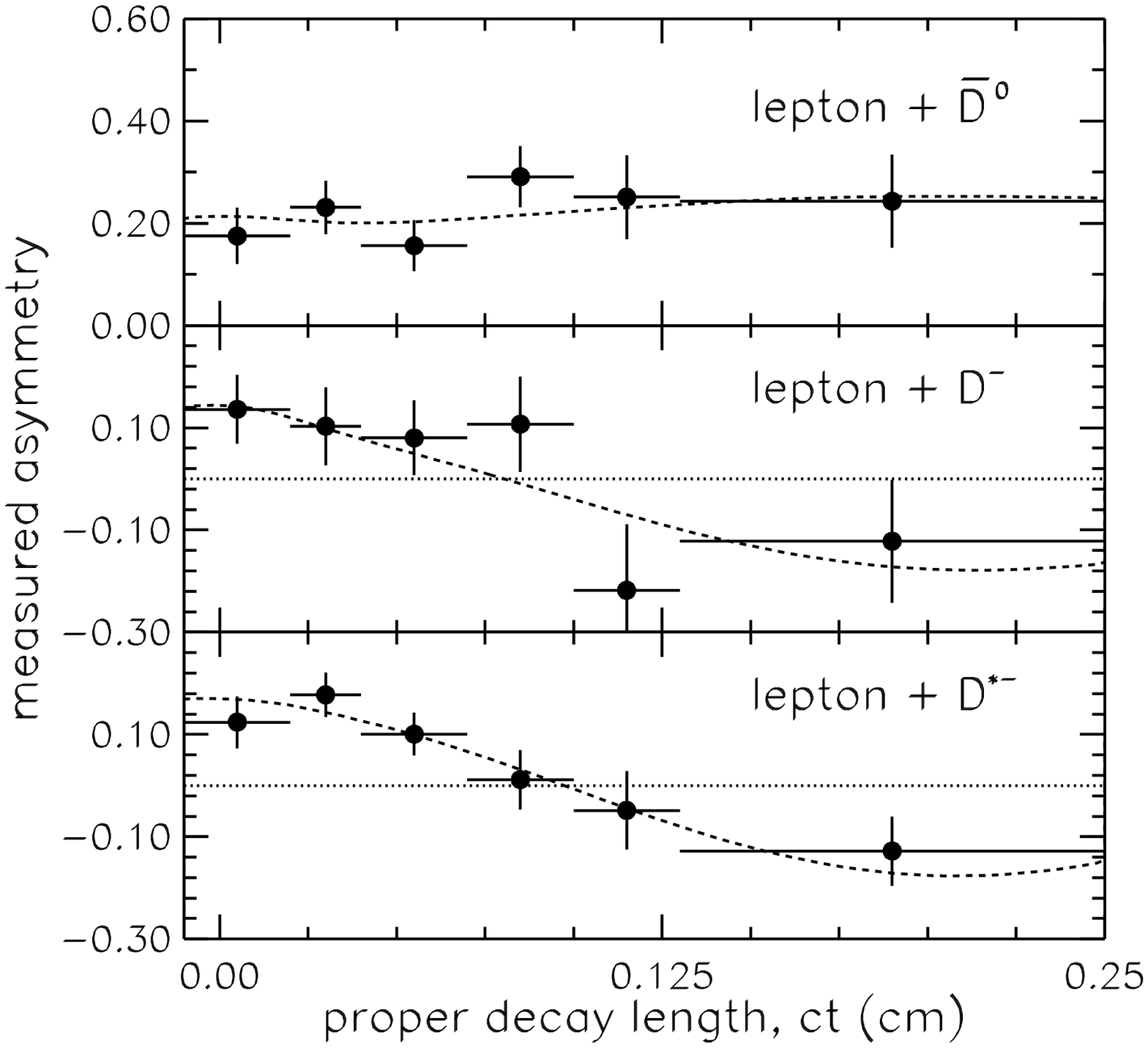}
}
\vspace*{0.2cm}
\fcaption{
Left: Invariant mass distributions of the five 
$B \ra D^{(*)} \ell^+ \nu X$ 
signatures used in the analysis:  
(a) $\bar D^0 \ra K^+\pi^-$ in $\bar D^0 \ell^+ X$,  
(b) $D^- \ra K^+\pi^-\pi^-$ in $D^- \ell^+ X$,  
(c) $D^{*-} \ra \bar D^0 \pi^-$ with 
$\bar D^0 \ra K^+\pi^-$ and $\bar D^0 \ra K^+\pi^-\pi^+\pi^-$ 
in $D^{*-} \ell^+ X$,  
(d) $K^+\pi^-\pi^-_*$ minus $K^+\pi^-$ mass difference for 
$D^{*-} \ra \bar D^0 \pi^-_*$ with $\bar D^0 \ra K^+\pi^-\pi^0$ 
in $D^{(*)-} \ell^+ X$.  
(e) Measured asymmetries as a function of proper decay length for the
decay signatures: $\bar D^0\ell^+$ (top), $D^- \ell^+$ (middle), and
the sum of all 
three $D^{*-} \ell^+$ (bottom). The dashed lines are the results of the fit.}
\label{bmix_sst_res}
\end{figure}

As in the $B$~lifetime analysis
with partially reconstructed $B$~mesons (see Section~6.4.2),
$B$~candidates are reconstructed using the decay chains
$B^0 \ra D^{(*)-}\ell^+ \nu$, with $ D^-\rightarrow K^+\pi^-\pi^-$,
or $D^{*-} \rightarrow \bar D^0 \pi_*^-$
followed by $\bar D^0$ decaying to $K^+\pi^-$, $K^+\pi^-\pi^+\pi^-$,
or $K^+\pi^-\pi^0$, where $\pi_*^-$ denotes
the low-momentum (soft) pion from the $D^{*-}$ decay.
Charged $B$~mesons are reconstructed through 
$B^+ \rightarrow \bar D^0 \ell^+ \nu$, with
$\bar D^0\rightarrow K^+\pi^-$, where the $\bar D^0$ is required not to form
a $D^{*-}$ candidate with any other $\pi$ candidate in the event.
The mass distributions of the four decay signatures with fully reconstructed
$D$~mesons are shown in Fig.~\ref{bmix_sst_res}a), b) and c), while 
the distribution of $m(K\pi\pi_*)-m(K\pi)$
for the signature with $D^{*-} \rightarrow \bar D^0 \pi_*^-$,
followed by $\bar D^0 \rightarrow K^+\pi^-\pi^0$ 
(the $\pi^0$ is not reconstructed)
is displayed in Fig.~\ref{bmix_sst_res}d).

To select the SST pion,
all tracks within a cone 
of radius $0.7$ in $\eta\,\varphi$-space, centered
around the direction of the $B$ meson, approximated by 
$\vec{p}(\ell)+\vec{p}(D)$ are considered (see Fig.~\ref{bmix_sst_draw}b).
SST candidate tracks should originate from
the $B$ production point (the primary event vertex),
and are therefore required 
to satisfy $d_0/\sigma_{d_0}<3$,
where $\sigma_{d_0}$ is the uncertainty on the track impact parameter $d_0$. 
String fragmentation models\cite{string_models} 
indicate that particles produced in the $b$~quark hadronization chain 
have low momenta transverse to the direction of the $b$~quark momentum.
We thus select as the tag the track that has the minimum
component of momentum, \ptrel, orthogonal to the momentum sum of
the track and the $B$~meson (see Fig.~\ref{bmix_sst_draw}b).  
The tagging efficiency, $\varepsilon$, is defined as 
the fraction of $B$ candidates with at least one track satisfying
the above requirements. It is measured as $\varepsilon \sim 70\%$
independent of the decay signature used.
On average, there are about $2.2$ SST candidate tracks per 
$B$ candidate.

For each of the five decay signatures, the $B$~candidates are
subdivided into six bins in proper decay length, $ct$, and 
the $B$-$\pi$ combinations are classified 
as right-sign (RS: $B^+\pi^-$ and $B^0\pi^+$) 
or wrong-sign (WS: $B^+\pi^+$ and $B^0\pi^-$).
In addition, the asymmetry in the RS and WS combinations is formed,
${\cal A}(ct)\equiv 
(N_{RS}(ct)-N_{WS}(ct))/( N_{RS}(ct)+N_{WS}(ct))$.
For $B^+$ mesons, we expect an asymmetry independent of $ct$:
${\cal A}_+(ct) = {\rm const.} \equiv {\cal D}_+$.
The dilution ${\cal D}_+$ is 
a direct measure of the SST purity, where $(1+{\cal D})/2$ is the
fraction of correctly tagged events.
Due to $B^0\bar{B}^0$ mixing, ${\cal A}_0(ct)$ for the neutral
$B$ mesons will vary as a function of $ct$.
It follows from Eq.~(\ref{eq:asym}) 
that the asymmetry is expected to oscillate as
${\cal A}_0(ct) = {\cal D}_0 \cdot\cos\dmd\, t$.
Mistags result in
a decrease of the oscillation amplitude by the dilution
factor ${\cal D}_0$.
The asymmetry is measured as a function of the
proper decay length $ct$
for both $B^+$ and $B^0$ mesons,
and fit with the expected time dependence, obtaining
$\dmd$, ${\cal D}_0$, and ${\cal D}_+$.

The measured asymmetries ${\cal A}^{\rm meas}(ct)$
are shown in Fig.~\ref{bmix_sst_res}e).
If the $\ell^+\bar D^0$ and $\ell^+D^{(*)-}$ signatures
are pure signals of $B^+$ and $B^0$ decays, \dmd\ could simply
be extracted using the time-dependence
of ${\cal A}_0(ct)$. However,
the signatures are mixtures of $B^+$ and $B^0$ decays,
and thus ${\cal A}^{\rm meas}(ct)$ is a linear combination of the
true asymmetries ${\cal A}_0(ct)$ and ${\cal A}_+(ct)$.
To extract $\Delta m_d$, ${\cal D}_0$ and ${\cal D}_+$,
it is necessary to determine 
the sample composition of each $D^{(*)} \ell^+$ signature,
which is
the fraction of $D^{(*)}\ell^+$ 
candidates originating from the decays of the $B^0$ and $B^+$ mesons.
Because a $B^+$ is associated with a $\pi^-$, whereas 
an unmixed $B^0$ is associated with a $\pi^+$, the observed
asymmetries are reduced by cross-contamination which
can arise
if the soft pion $\pi_*^-$ from the $D^{*-}$ decay is not
identified. The decay sequence $B^0 \rightarrow D^{*-} \ell^+ \nu$
will be reconstructed as $\ell^+ \bar{D}^0$, that is, as a $B^+$ candidate.
Another source of cross-contamination arises from semileptonic
$B$ decays involving $P$-wave $D^{**}$ resonances
as well as non-resonant $D^{(*)}\pi$ pairs,
which cannot be easily recognized and removed from the sample.
For example, the decay sequence $B^0 \rightarrow D^{**-}\ell^+ \nu$,
followed by $D^{**-} \rightarrow \bar{D}^0 \pi_{**}^-$ (by $\pi_{**}$ we 
denote the pion originating from a $D^{**}$ decay)
will be reconstructed as $\ell^+\bar{D}^0$, because of the
missed $\pi_{**}^-$. Again, a $B^0$ decay is
misclassified as a $B^+$ candidate.
The tagging is further complicated
when a $\pi_{**}^\pm$ from a $D^{**}$ decay is present. The $\pi_{**}^\pm$
may be incorrectly selected as the SST pion,
always resulting in a RS correlation.
The requirement $d_0/\sigma_{d_0}<3$, described above, reduces this
effect. The $\pi_{**}$ 
originates from the $B$~meson decay point, whereas the appropriate
tagging track comes from the $B$~meson production point.  

Taking these effects properly into account, 
the mass difference \dmd\ and 
the dilutions ${\cal D}_0$ and ${\cal D}_+$ are determined
from a $\chi^2$-fit to 
the measured asymmetries ${\cal A}^{\rm meas}(ct)$ with
the fit result overlaid, as shown
in Fig.~\ref{bmix_sst_res}e).  The oscillation in the
neutral $B$ signatures is clearly present.
The final result for the mixing frequency is 
\begin{equation}
\dmd = (0.471\ ^{+0.078}_{-0.068}\ \pm 0.034)\ {\rm ps}^{-1}. 
\end{equation}
In addition, the following values for the neutral and charged
SST flavour tagging dilutions are obtained: 
\begin{eqnarray}
B^0: & & \ \ \ 
{\cal D}_0 = (18\pm3\pm2)\%, \ \ \ 
\eD_0 = (2.4\pm0.7^{+0.6}_{-0.4})\%,
\\
B^+: & & \ \ \ 
{\cal D}_+ = (27\pm3\pm2)\%, \ \ \ 
\eD_+ = (5.2\pm1.2^{+0.9}_{-0.6})\%.
\end{eqnarray}
The fit indicates that
$\sim\!82\%$ of the $\bar{D}^0 \ell^+$ 
signature comes from $B^+$ decays, while $\sim\!80\%$ of the $D^-
\ell^+$  and
$\sim\!95\%$ of the $D^{*-}\ell^+$ originate from $B^0$.  
The $B^0$ component
of the $\bar{D}^0 \ell^+$ signature can be seen as a small 
anti-oscillation in Fig.~\ref{bmix_sst_res}e), top.

\subsubsection{Production of $B^{**}$ mesons}
\noindent
To shed some light on the question of how much of the observed tagging
power of same side tagging arises from fragmentation tracks compared to 
pions from $B^{**}$ decays, we briefly discuss the production of
$B^{**}$ mesons at CDF.
The term $B^{**}$ is a collective name for the four lowest lying $L=1$
states of $B$~mesons  which are usually labeled by the total angular
momentum $J$ of the light quark resulting in two doublets with 
$J= 1/2$ and $3/2$. The states in the $J=1/2$ doublet are expected to
be broad ($\sigma \sim 100~\mevcc$) since they can decay through a
$S$-wave transition while the 
$J=3/2$ states decay through a $D$-wave transition and are therefore
expected to be narrow ($\sigma \sim 20~\mevcc$).  
$B^{**}$ states have been observed at LEP\cite{LEP_bdss} with
properties in reasonable agreement with the expectations.

The procedure for selecting $B$~mesons follows the one described above
in Sec.~7.2 reconstructing semileptonic $B$~decays into 
$D^{(*)}\ell\nu X$. About 5500 $D^{*+}\ell$ and $D^+\ell$ candidates
associated with a $B^0$~signature and about 4200 $D^0\ell$ events
indicating a $B^+$ signature are selected. $B^{**}$ candidates are
constructed by combining the $B$~candidates with all tracks compatible
with those originating from the primary interaction vertex. Since the
resolution on the $B^{**}$ invariant mass is impaired by the unknown
momentum of the missing neutrino, a quantity 
$Q~=~m(B\pi)~-~m(B)~-~m(\pi)$ 
is constructed. A resolution of $\sim\!50~\mevcc$ on $Q$ is    
expected from $B^{**}$ decays. Since the charge of the pion from a
$B^{**}$~decay always matches the $B$~flavour, the sample is divided
into right-sign and wrong-sign combinations. We expect an excess of RS
combinations confined to $Q<500~\mevcc$ corresponding to the
$B^{**}$~states. 

The essential task of this analysis it to properly determine the 
backgrounds to the $B^{**}$~signal. These can be divided into
correlated and uncorrelated components. Three sources of uncorrelated
background are taken into account: Combinatorial background from fake
$B$~meson candidates, particles from the underlying event and tracks
from events with multiple hard $p\bar p$ collisions. All three
components are measured from data and subtracted from the
$Q$~distributions. The main correlated background arises from
fragmentation particles which contribute to the right-sign excess. The
shape of this background is predicted using Monte Carlo calculations
tuned to CDF data (see Sec.~4.5.2). To extract the $B^{**}$~production
fraction from the $Q$~distribution, cross contamination between $B^0$
and $B^+$ via $D^{**}$ and $B^*$ states is taken into account
as well as the known flavour mixing of $B^0$~mesons.  

\begin{figure}[tbp]
\centerline{
\epsfxsize=8.4cm
\epsffile[5 5 455 440]{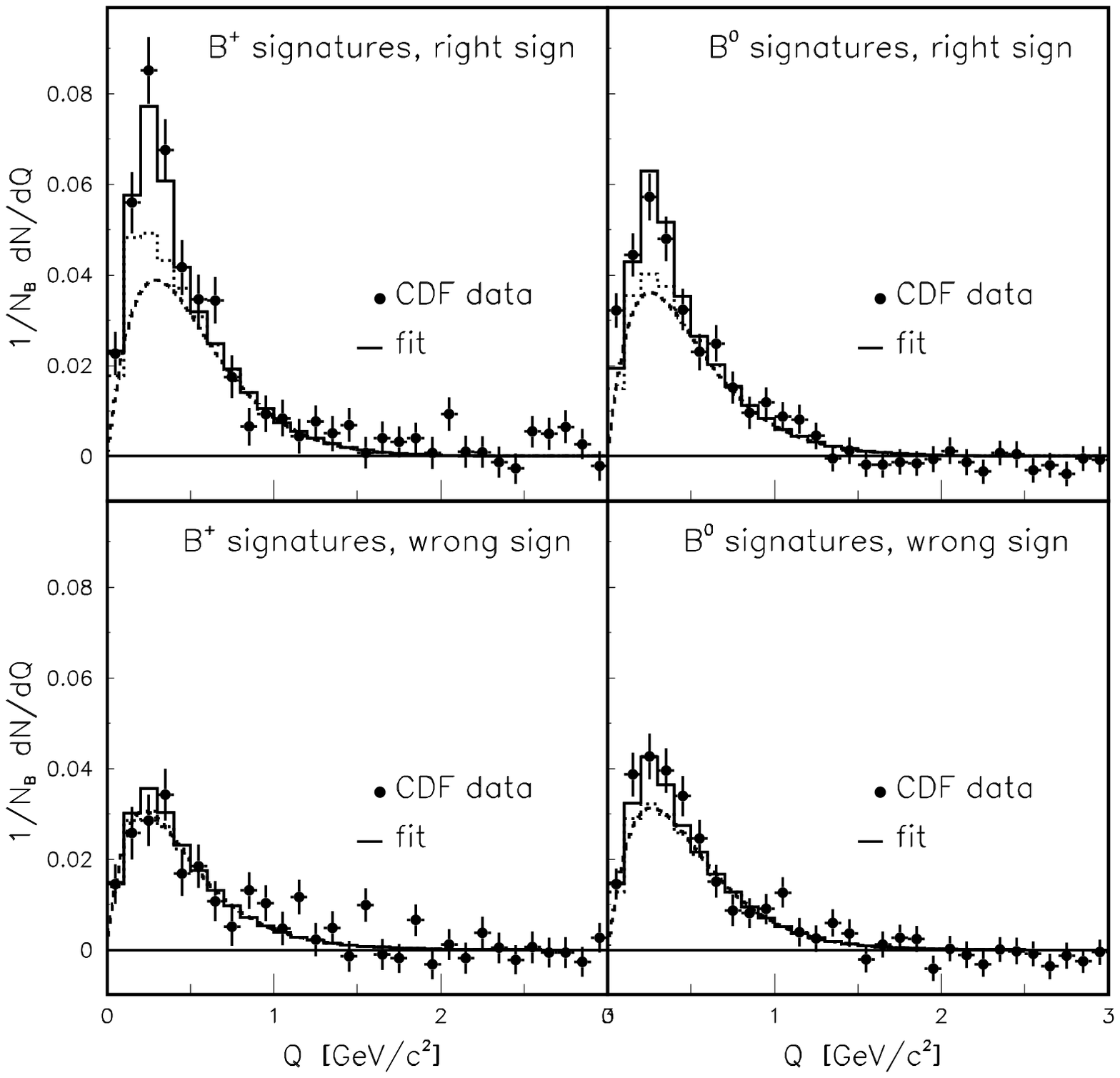}
}
\vspace*{0.2cm}
\fcaption{Result of the fit to the $Q$~distributions of the
$B\pi$ candidates. The dashed curves are the fitted hadronization
components, the dotted histograms include all backgrounds, and the
solid histograms are the sum including the fitted $B^{**}$ signal.} 
\label{b2star}
\end{figure}

The result of the fit to the $Q$~distributions is shown in
Fig.~\ref{b2star}. The points are the data, the
dashed curves are the fitted shapes of the hadronization
component, the dotted histograms include all backgrounds, and the
solid histograms are the sum including the fitted $B^{**}$ signal.
The $B^{**}$ production fraction, defined as the probability that a
$B$~quark hadronizes into a $B^{**}$~state, is measured as
$f^{**} = 0.28 \pm 0.06 \pm 0.03$. The experimental resolution does
not allow disentangling the four $B^{**}$~states and the average mass
of the ensemble is varied collectively to obtain 
$m(B_1) = (5.71 \pm 0.02)~\gevcc$ for $m(B_1)$ of the narrowest
state~($J = 3/2$).    

\subsection{$B^0\bar B^0$ mixing using jet-charge and lepton tagging}
\noindent
The analysis reported next\cite{qjet_mix} uses the same single lepton
data samples 
as the SST mixing result but increases the number of $B$~mesons by
over an order of magnitude 
by inclusively reconstructing $B$~hadrons decaying semileptonically
rather than using lepton-charm correlations. The inclusive
reconstruction is based on identifying the $B$~hadron decay point by
associating the trigger lepton with other $B$~decay products to form a
secondary vertex. 

First, charged particle jets are reconstructed in
the event using a track based jet clustering algorithm (see
Sec.~4.4). The trigger lepton is associated with a jet. The search
for the $B$~decay point in the trigger lepton jet is based on the
technique to identify $b$~quark jets coming from top quark
decays\cite{top_prd}. Some modifications are made to maintain good
efficiency for reconstructing displaced vertices, since $B$~hadrons in
this data sample have substantially lower \Pt\ than $B$~hadrons from
top decays. Tracks in the jet are selected for reconstructing the
secondary vertex based on the significance of their impact parameter
$d_0/\sigma_{d_0}$ with respect to the primary vertex, where 
$\sigma_{d_0}$ is the estimate of the error on $d_0$. First,
displaced vertices containing three or more tracks satisfying a loose
set of track quality requirements are searched for. If no such
vertices are found, two-track vertices which satisfy more stringent
quality requirements are accepted and the two-dimensional decay length
of the secondary vertex is calculated. 
To reduce background from false vertices formed from random combinations
of tracks coming from the primary interaction vertex,
$|\lxy/\sigma_{\lxy}|>2.0$ is required,
where $\sigma_{\lxy}$ is the estimated error on \lxy.
This leaves 
243,800 events: 114,665 from the electron data sample
and 129,135 from the muon data sample.
Secondary vertices are also searched for in the other jet in the
event. If an additional displaced vertex is found, the event is
classified as a ``double-vertex'' event.

\begin{figure}[tbp]
\centerline{
\epsfysize=4.6cm
\epsffile[85 58 509 510]{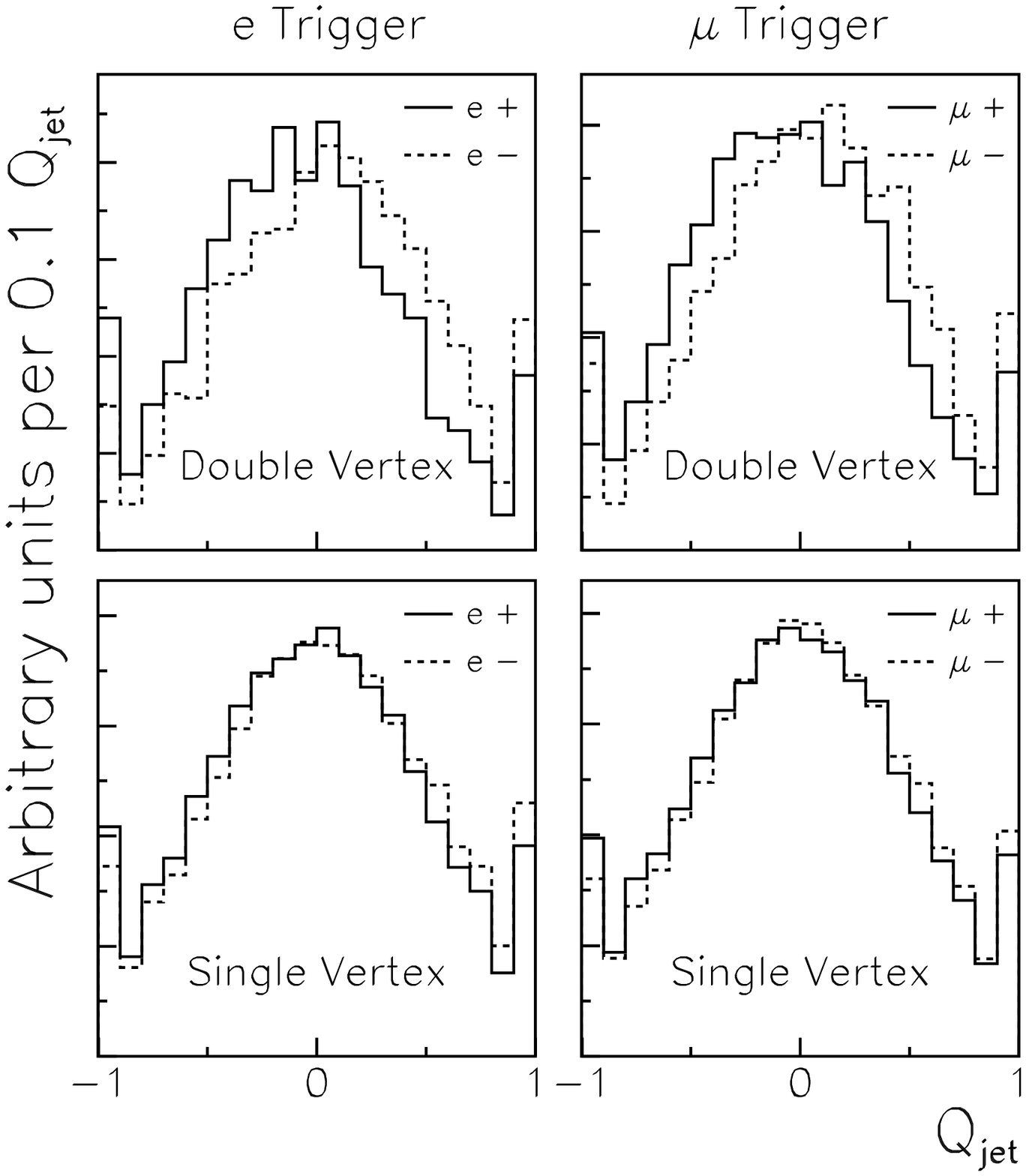}
\hspace*{1.4cm}
\epsfysize=4.6cm
\epsffile[113 227 496 510]{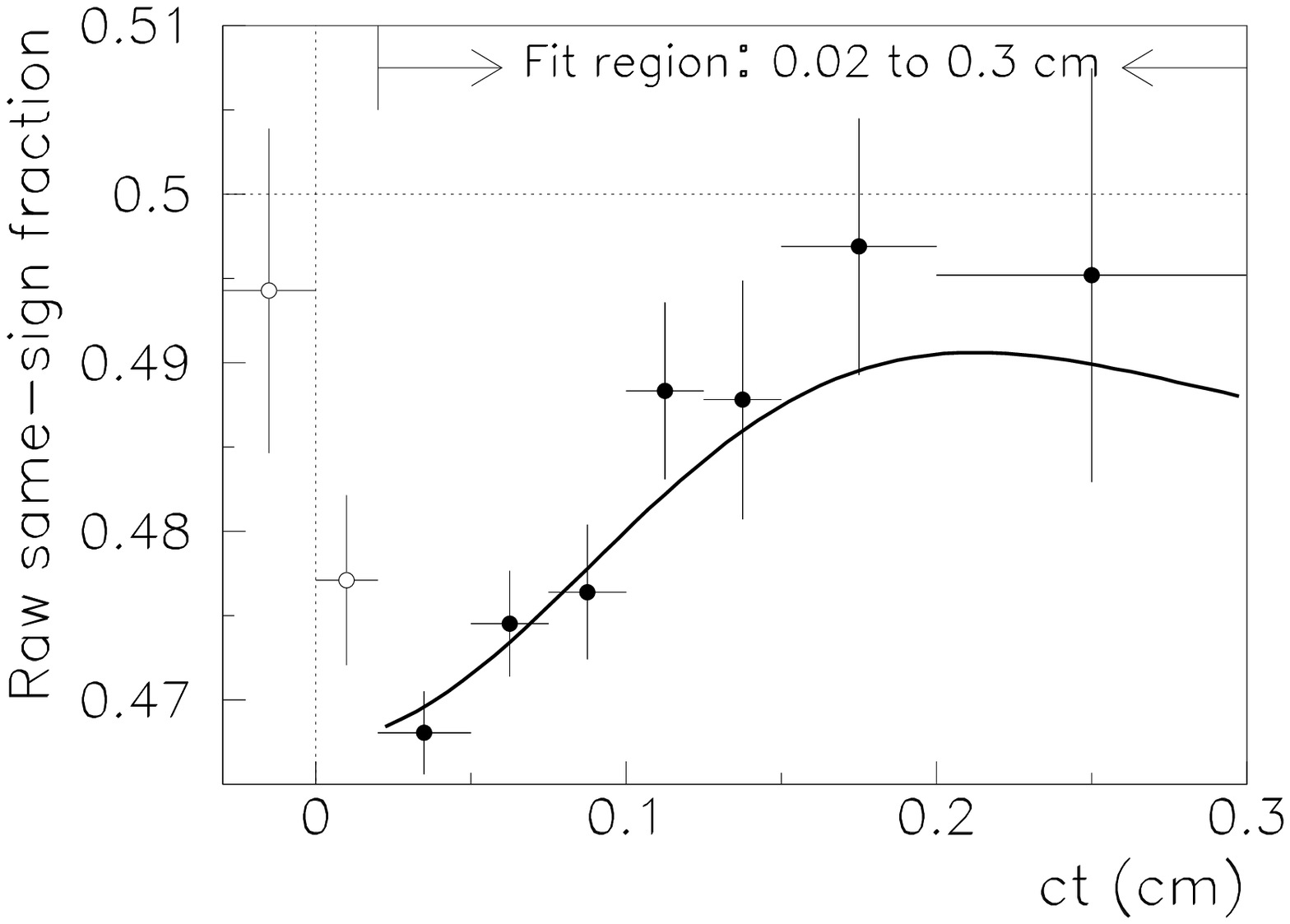}
}
\vspace*{1.0cm}
\fcaption{
Left: Jet-charge distributions for double-vertex (top) and
single-vertex (bottom) 
events for the $e$- and $\mu$-trigger data. Right: Same-sign
fraction as a function of proper decay length. A representation of the
fit result is superimposed on the data.}
\label{bmix_qjet}
\end{figure}

The flavour of the $B$~hadron producing the trigger lepton is
identified using a jet-charge tag and an opposite side lepton tag. 
First, an additional lepton ($e$ or $\mu$) from the semileptonic decay
of the opposite side $B$~hadron is searched for in the event. 
The invariant mass of this lepton and the trigger lepton must be
greater than 5~\gevcc\ to reject leptons that come from the same
$B$~hadron producing the trigger lepton. 
Approximately 5.2\% of the 243,800 events contain an opposite side lepton.
If such a lepton is not found, 
the jet produced by the opposite $b$~quark is identified by
calculating a quantity called the jet-charge $Q_{\rm jet}$,
\begin{equation}
  Q_{\rm jet} = \frac{ \sum_i q_i \cdot (\vec{p}_i \cdot\hat{a}) }
                     { \sum_i \vec{p}_i \cdot\hat{a}  },
\end{equation}
where $q_i$ and $\vec{p}_i$ are the charge and momentum
of track $i$ in the jet and $\hat{a}$ is
a unit-vector defining the jet axis.
On average, the sign of the jet-charge is the same as the sign
of the $b$~quark that produced the jet. In the case of a double-vertex
event the jet containing the other
secondary vertex is used to calculate $Q_{\rm jet}$.
About 7.5\% of the sample consists of
jet-charge double-vertex events and 42\%
are jet-charge single-vertex events.
Figure~\ref{bmix_qjet} shows, on the left side, the jet-charge
distributions for single-vertex and double-vertex events. 
The degree of separation between the solid and dashed distributions is
related to the tagging power of the jet-charge flavour tag. For
double-vertex events, the presence of the second displaced vertex
increases the probability that the selected jet in fact originates
from the other $B$~hadron in the event. This translates into a
significantly larger separation in jet-charge for double-vertex~events.

Unlike the case of the SST mixing analysis,
where a charm peak in the invariant mass spectrum indicates the amount
of signal and background, the number of trigger
leptons from $b\bar b$ production must be determined differently
in this analysis. Other sources of the selected events include
$c\bar c$ production and light quark or gluon production that can
result in fake 
leptons or vertices. The fraction of events from false vertices and
fake leptons is found to be small. 
The fraction of events due to $b\bar{b}$ and $c\bar{c}$ production
is determined using two kinematic quantities:
the trigger lepton $\ptrel$ and the invariant mass $m^{\rm cl}$ of the
cluster of secondary vertex tracks.  
The quantity $\ptrel$ is defined as the magnitude of the 
component of the trigger lepton momentum 
perpendicular to the jet axis.
The trigger lepton is removed from the jet, and the jet-axis
is recalculated to determine $\ptrel$.
To calculate $m^{\rm cl}$, the pion mass is assigned to all of the tracks
used to form the secondary vertex (except the lepton). 
The trigger lepton is included in this calculation even
if it is not attached to the secondary vertex.
Both kinematic quantities are effective in discriminating between $b\bar{b}$
and $c\bar{c}$ events because of the significant
mass difference between hadrons containing $b$ and $c$~quarks.
Template $\ptrel$ and $m^{\rm cl}$ distributions are obtained from
$b\bar{b}$ and $c\bar{c}$ Monte Carlo samples (see Sec.~4.5.2) and fit
to the data.
The determination of the sample composition yields the
data being primarily ($> 90\%$) from $b\bar b$~production.

According to the flavour tag and the charge of the trigger lepton, the
data are divided in a like-sign and opposite-sign sample. The proper
decay time of the $B$~hadron is determined from the \lxy\ of the
secondary vertex, combined with an estimate of the $B$~momentum
obtained from Monte Carlo calculations. From an unbinned maximum
log-likelihood fit to 
the time dependence of the OS and LS sample, 
the $B^0$ oscillation frequency is determined to be
\begin{equation}
\dmd = (0.500\ \pm0.052\ \pm 0.043)\ {\rm ps}^{-1}. 
\end{equation}
The distribution of the like-sign
fraction as a function of proper decay length is shown in
Fig.~\ref{bmix_qjet} on the right hand side. A representation of the
fit result is superimposed on the data.
In addition, the effective tagging efficiencies of the jet-charge and
opposite side lepton tags are obtained: 
\begin{eqnarray}
\mbox{jet-charge tag}: & & \ \ \ 
\eD = (0.78 \pm 0.12 \pm 0.08)\%,
\\
\mbox{lepton tag}: & & \ \ \ 
\eD = (0.91 \pm 0.10 \pm 0.11)\%.
\end{eqnarray}
These values for \eD\ are much lower than those achieved
in experiments on the $Z^0$~resonance\cite{qjet_lep}. On the other
hand, the much higher $b\bar b$~cross section at the Tevatron
can be used to compensate for the
disadvantage in~\eD.

\subsection{$B^0\bar B^0$ mixing using dilepton data}
\noindent
The previous analysis searches single lepton events for additional
leptons used as opposite side flavour tags. However, there exist
large dilepton data sets at CDF where the lepton tag has already
been triggered on. These data are also used for proper time
dependent measurements of $B\bar B$~oscillations. 
In these data sets both leptons
are assumed to come from the semileptonic decay of both $B$~hadrons in
the event: $b \ra \ell_1X$ and $\bar b \ra \ell_2 X$. 
This means, the flavour of
the $B$ meson at decay is tagged by its semileptonic decay, while the
lepton from the semileptonic decay of the other $B$~hadron in the
event tags the $B$ flavour at production.
The requirement $m({\ell_1\ell_2}) > 5$ \gevcc\ ensures
that both leptons originate from two $B$~hadrons and not from a
sequential decay of one $B$~hadron: $b \ra c\, \ell_1  X$, with $c \ra
\ell_2 X$.  
Here, we briefly describe
two mixing results using the $e\mu$~trigger data as well as dimuon
trigger events.

The analysis using $e\mu$~data\cite{meros} also searches for an
inclusive secondary 
vertex associated with one of the leptons. As in the previous analysis 
in Sec.~7.3, secondary vertices are reconstructed using a modified
version of the algorithm used to find displaced vertices in the
search for the top quark\cite{top_prd}.
The decay length \lxy\ of this
vertex and the momenta of the tracks associated with the lepton
provide an estimate of the $ct$ of the $B$ meson. 
Again, the important task of this analysis is to determine the sample
composition, the fraction of events coming from $b\bar b$
decays with respect to events from $c\bar c$ or background events.
The sample composition is estimated from several kinematic
quantities, like \ptrel\ or the invariant mass of
the tagged secondary vertex. Here, \ptrel\ is defined as the
transverse momentum of the lepton with respect to the highest \Pt\
track in a cone around the lepton. The determination of the sample
composition finds
that more than 80\% of the events originate from $b\bar b$ decays. 

\begin{figure}[tbp]
\centerline{
\put(10,44){\large\bf (a)}
\put(75,44){\large\bf (b)}
\epsfxsize=6.3cm
\epsffile[0 10 525 525]{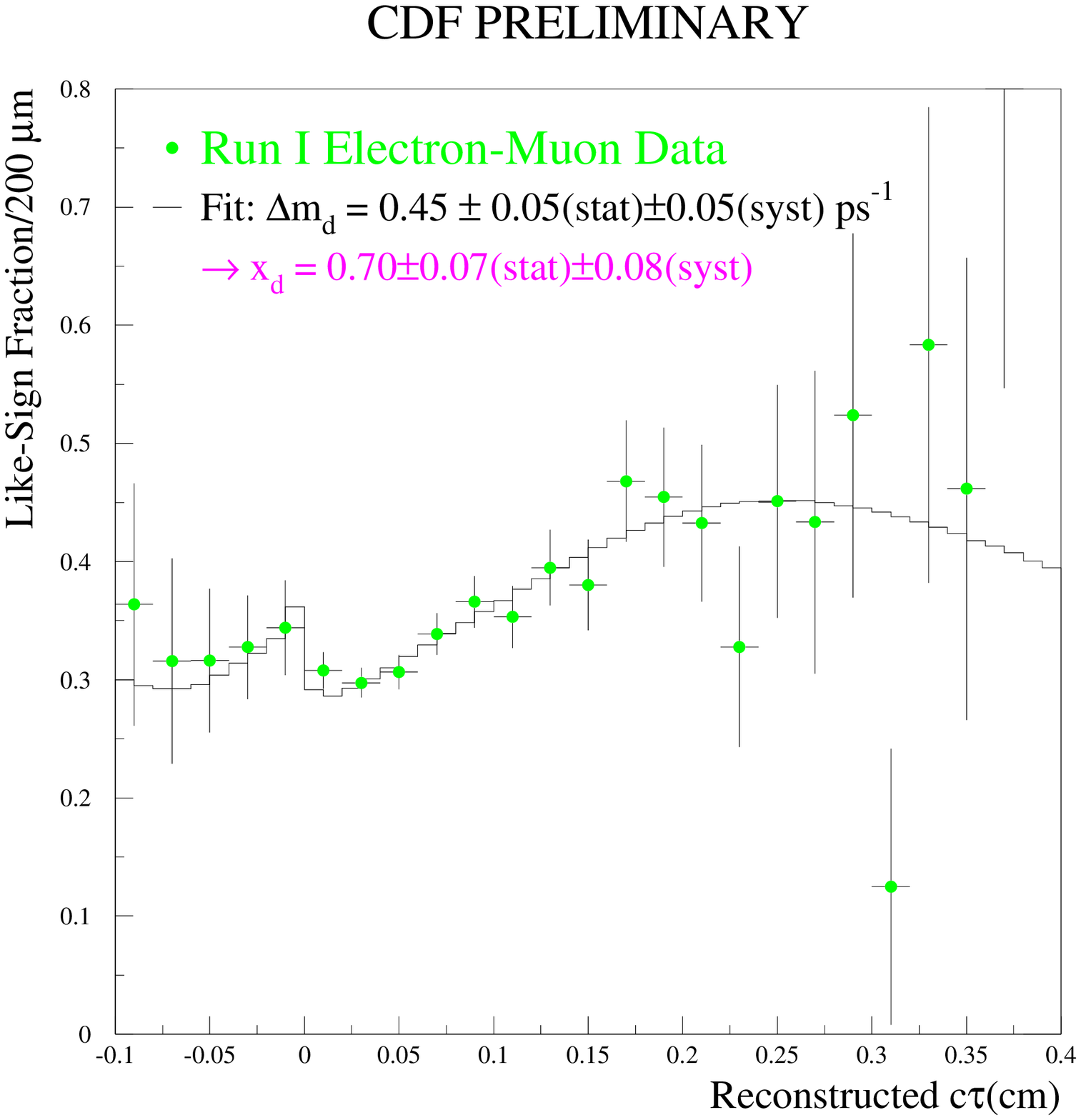}
\epsfxsize=6.3cm
\epsffile[1 1 515 515]{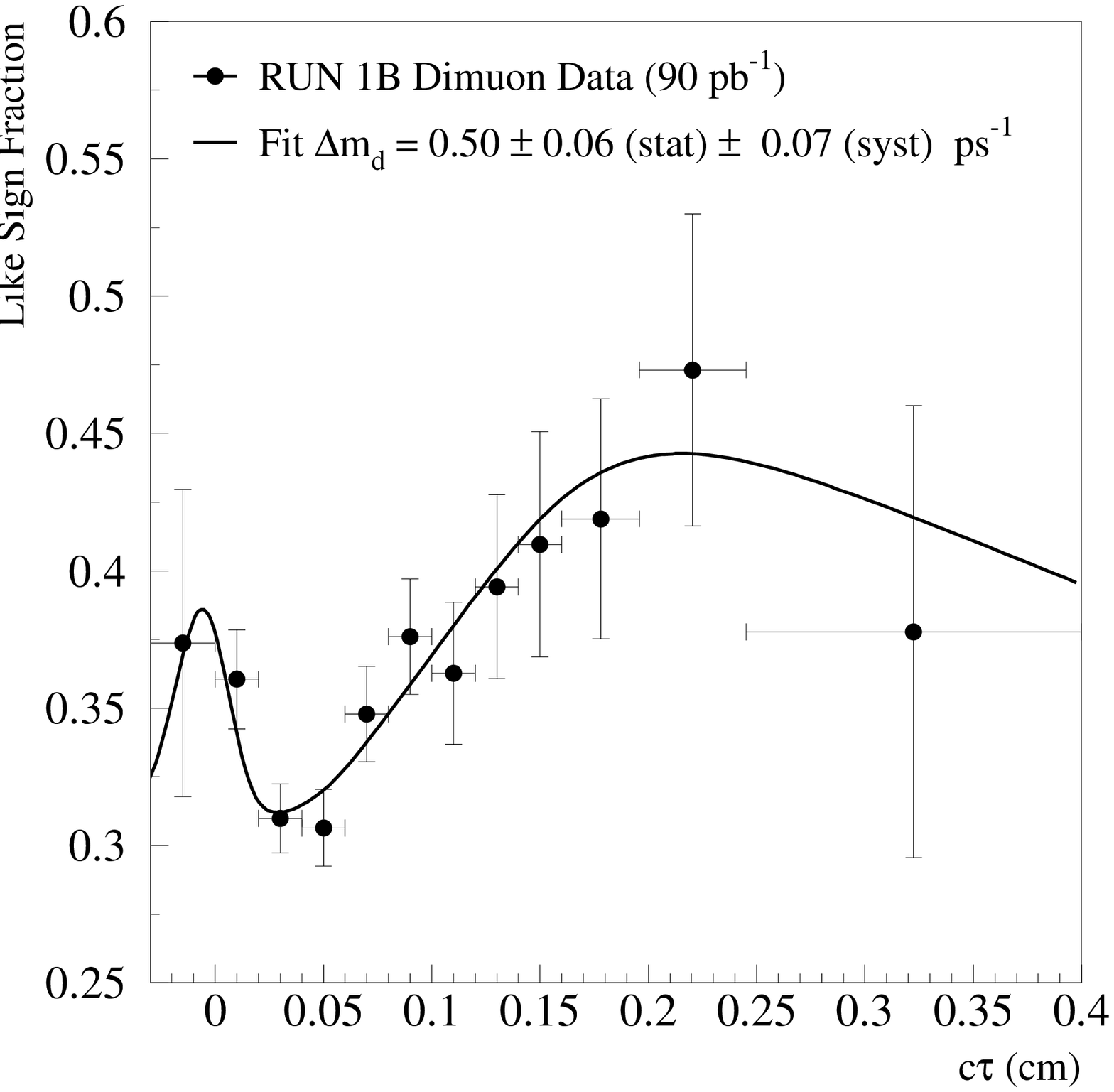}
}
\vspace*{0.2cm}
\fcaption{
Fraction of (a) like-sign $e\mu$ events and (b) like-sign dimuon
events as a function of pseudo-proper decay length. The results of the
\dmd\ fits are superimposed.} 
\label{bmix_dilep}
\end{figure}

From a fit to the like-sign lepton fraction, as a function of $c\,t$, the
mixing frequency \dmd\ is extracted as shown in
Fig.~\ref{bmix_dilep}a). The fit includes components for direct 
and sequential $b$ decays, $c\bar c$, and fake events.  
In about 16\% of the events with a secondary vertex around one lepton,
a secondary vertex is also found around the other lepton. These events
enter the like-sign fraction distribution twice, where we allow for a
statistical correlation between the two entries. The final fit result is
\begin{equation}
\dmd = (0.450\ \pm0.045\ \pm 0.051)\ {\rm ps}^{-1},
\end{equation}
where the dominant systematic error arises from the uncertainty in 
the sample composition.

The measurement of the $B^0\bar B^0$ oscillation frequency using
dimuon data\cite{mumu_mix} is very similar to the
$e\mu$~analysis. Here, the algorithm used 
to find displaced vertices around one of the muons is based on the
correlation between the impact parameter $d_0$ and the azimuthal angle
$\varphi$ of tracks coming from these vertices. A vertex is formed by
three or more correlated tracks with a significance $d_0/\sigma_{d_0}>2$.
Tracks from a secondary vertex form a line in the 
$\varphi\,d_0$-plane with non-zero slope, while tracks from the primary
interaction vertex have small $d_0$ values and show no 
correlation~with~$\varphi$. 

The determination of the sample composition again uses different
kinematic variables and finds the fraction of events from $b\bar b$
production to be larger than 80\%. The final sample amounts to 2044
like-sign and 3924 opposite-sign dimuon events. The fraction of events
with like-sign muons is shown in Fig.~\ref{bmix_dilep}b) as a function
of proper time of the $B$~hadron associated with the identified
secondary vertex. From a $\chi^2$-fit to the like-sign fraction, 
the $B^0$ oscillation frequency is determined to be
\begin{equation}
\dmd = (0.503\ \pm0.064\ \pm 0.071)\ {\rm ps}^{-1},
\end{equation}
where the main systematic error arises from the determination of the
background and the fraction of muons from sequential charm decays.

\subsection{$B^0\bar B^0$ mixing using charm decays with a lepton tag}
\noindent
We briefly describe two more $B^0\bar B^0$ mixing results\cite{meros}
also using a 
lepton tag to identify the $B$~flavour at production. Both preliminary
analyses
allow the determination of the dilution or mistag probability of the
lepton tag. One analysis uses higher momentum leptons than the other
which can shed some light on the momentum dependence of the tagging
dilution. 

The first analysis uses the dimuon as well as $e\mu$ trigger data and
searches for a charm decay around one of the leptons identifying
semileptonic $B$~decays. In particular, leptons associated with a
$D^{*+}$ meson are used as signature for $B^0$ decays. The
$D^{*+}$~mesons are reconstructed in the decay mode $D^{*+} \ra D^0
\pi^+$, followed by $D^0 \ra K^- \pi^+$, $K^-\pi^+\pi^+\pi^-$ or
$K^-\pi^+\pi^0$. A signal of about 500 such decays is found 
on low backgrounds. The decay vertices of the $D^{*+}\ell^-$
combinations are reconstructed
and the proper decay length of the $B^0$~meson is estimated using the
$D^{*+}\ell^-$ momentum. The charge of the final state identifies the
$B^0$~flavour at decay time, while the other lepton in the event
provides the $B$~flavour at production. As in the $B$~lifetime
analysis using partially reconstructed $B$~decays (Sec.~6.4.2),  
$D^{*+}\ell^-$ combinations are a good signature for neutral
$B$~mesons, but $D^{**}$ decays allow a contamination of charged
$B$~decays into the $D^{*+}\ell^-$ sample, diluting the charge
correlation between the final states and the parent $B$~meson. Using
Monte Carlo techniques as in Sec.~6.4.2, the cross contamination
from $B^+$ decays is 
found to be $0.19^{+0.08}_{-0.10}$ for the $D^{*+}\mu^-$ sample and   
$0.14^{+0.06}_{-0.08}$ for the $D^{*+}e^-$ sample.

\begin{figure}[tbp]
\centerline{
\put(52,49){\large\bf (a)}
\put(74,56){\large\bf (b)}
\put(74,26){\large\bf (c)}
\epsfysize=6.3cm
\epsffile[30 150 530 660]{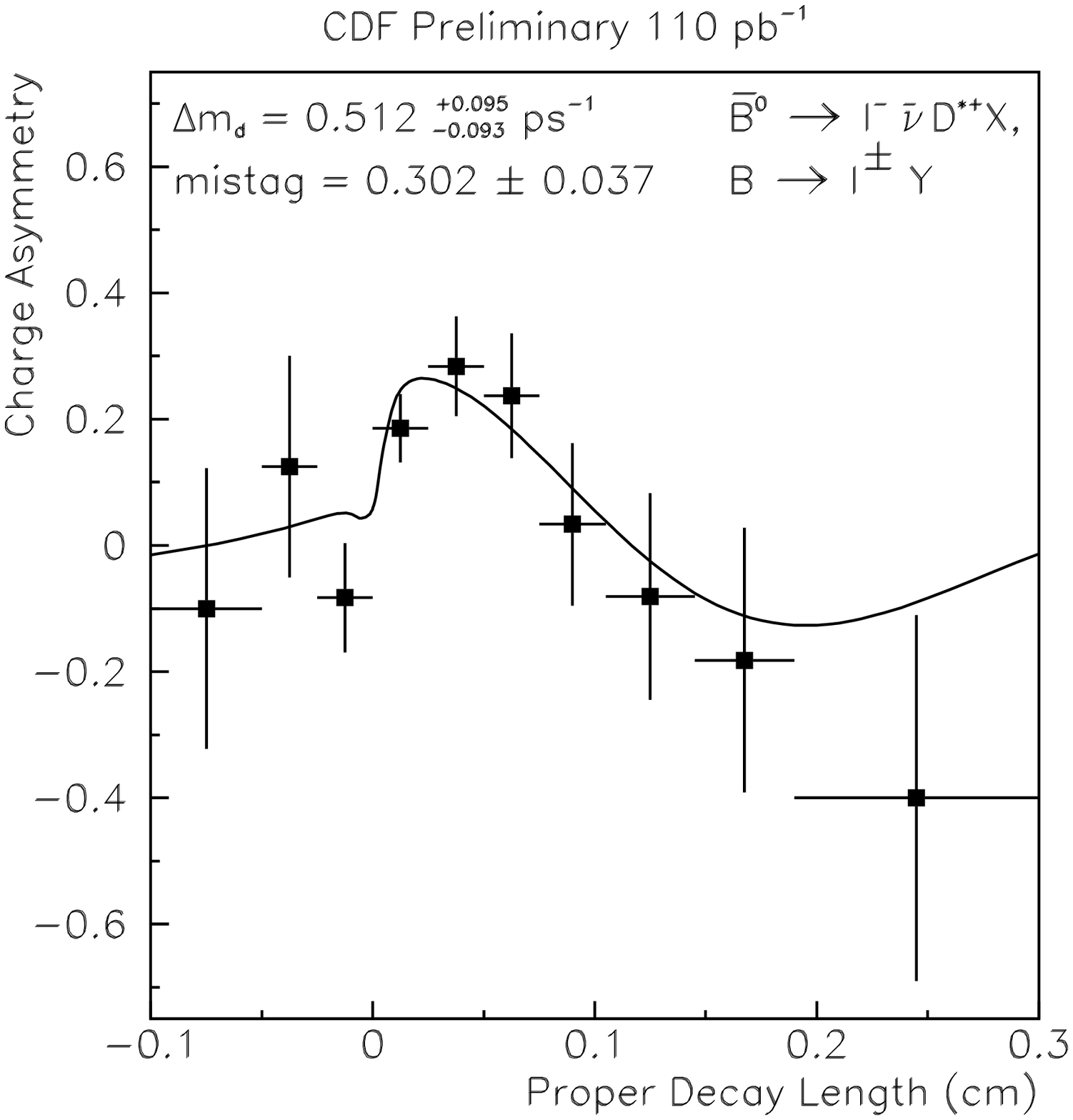}
\epsfysize=6.3cm
\epsffile[45 145 530 616]{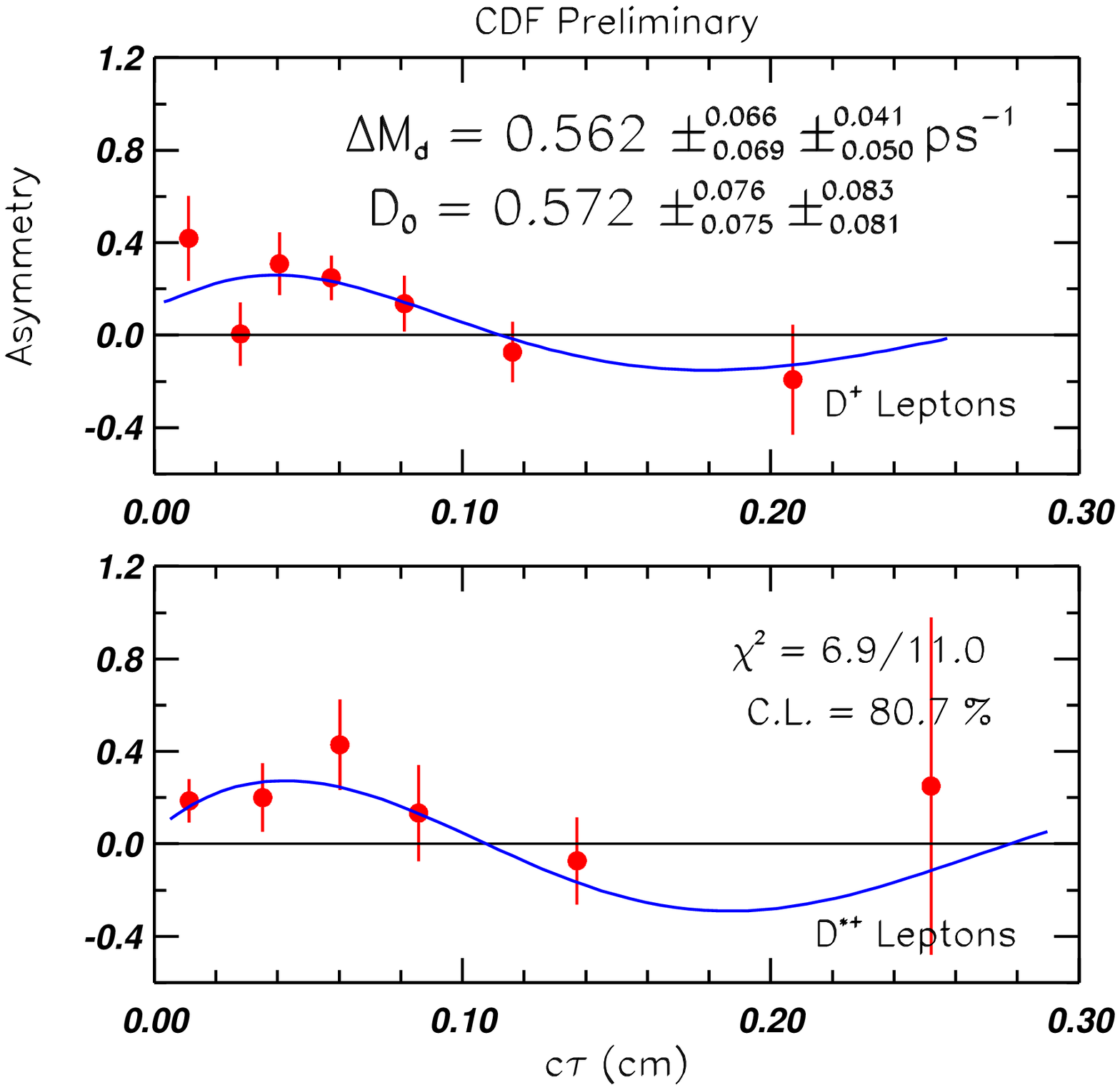}
}
\vspace*{0.2cm}
\fcaption{
Asymmetry distribution of (a) $D^{*+}\ell^-$ candidates, 
(b) $D^{-}$ candidates, and 
(c) $D^{*-}$ candidates 
as a function of proper decay length. The points
are the data and the curves show the result of the \dmd\ fits.}
\label{bmix_lepcharm}
\end{figure}

The signal sample is divided in OS and LS events and the corresponding
asymmetry distribution is shown in
Fig.~\ref{bmix_lepcharm}a). From the time dependence of the OS and LS
sample, the $B^0$ oscillation frequency \dmd\ as well as the mistag
probability $p_W$ of the lepton tag are simultaneously extracted. The
unbinned log-likelihood fit yields
\begin{equation}
\dmd = (0.512\ ^{+0.095}_{-0.093}\ ^{+0.031}_{-0.038})\ {\rm ps}^{-1},
\end{equation}
while the flavour misidentification probability is 
$p_W = 0.302 \pm 0.037 ^{+0.005}_{-0.012}$, 
yielding a dilution of the lepton tag of
${\cal D} = 0.396 \pm 0.074 ^{+0.024}_{-0.010}$.

The second analysis uses the trigger lepton from the single lepton
trigger data as $B$~flavour tag. The transverse momentum of these
leptons is greater than 8~\gevc, in comparison to the dilepton data
where $\Pt(\ell)$
is greater than about 2~\gevc. The $B^0$~meson is inferred from a
$D^{(*)-}$~decay opposite the trigger lepton, where
the charge of the $D^{(*)-}$~meson tags the
$B^0$~flavour at decay. This means, this analysis   
searches for events with $b \ra \ell^- X$ and $\bar b \ra B^0 \ra  
D^{(*)-} X$. A $\ell^- D^{(*)-}$ combination is therefore a signature
for an unmixed event, while a $\ell^+ D^{(*)-}$ pair signals a mixed event.   
$D^{(*)-}$ candidates are reconstructed as 
$D^- \ra K^+\pi^-\pi^-$ and $D^{*-} \ra \bar D^0\pi^-$, with $\bar D^0
\ra K^+ \pi^-$. The decay length of the charm
meson provides an estimate of the $B^0$ decay time together with the
measurement of $\Pt(D^{(*)-})$ used to infer $\Pt(B^0)$. 

Signals of $(460\pm31)$ $D^-$ events and $(358\pm32)$ $D^{*-}$ events are
selected. The $D^{(*)-}$~mesons may originate from $b\bar b$ and $c\bar c$
production. In addition, cross contamination of $D^{(*)-}$ mesons from
charged $B$~mesons via $D^{**}$~decays must be considered, as
discussed above and in Sec.~6.4.2. Using MC techniques and fits to
kinematic variables, it is found that approximately 65\% of the signal
sample comes 
from $B^0$~decays, $\sim\!20\%$ from $B^+$, and $\sim\!15\%$ from 
$c\bar c$~production. The time dependent asymmetry distribution of  
$\ell^+ D^{(*)-}$ and $\ell^- D^{(*)-}$ pairs is plotted in
Fig.~\ref{bmix_lepcharm}b) for $D^-$ (top) and $D^{*-}$ (bottom). The
fit for the $B^0$ mixing frequency results in  
\begin{equation}
\dmd = (0.562\ \pm 0.068\ ^{+0.041}_{-0.050})\ {\rm ps}^{-1},
\end{equation}
while the dilution of the lepton tag is determined to be 
${\cal D} = 0.572 \pm 0.080 ^{+0.083}_{-0.081}$.

\subsection{Summary of \boldmath{$B^0\bar B^0$} mixing measurements}
\noindent
\begin{figure}[tbp]
\centerline{
\put(0,58){\large\bf (a)}
\put(68,58){\large\bf (b)}
\epsfysize=6.4cm
\epsffile[25 55 625 625]{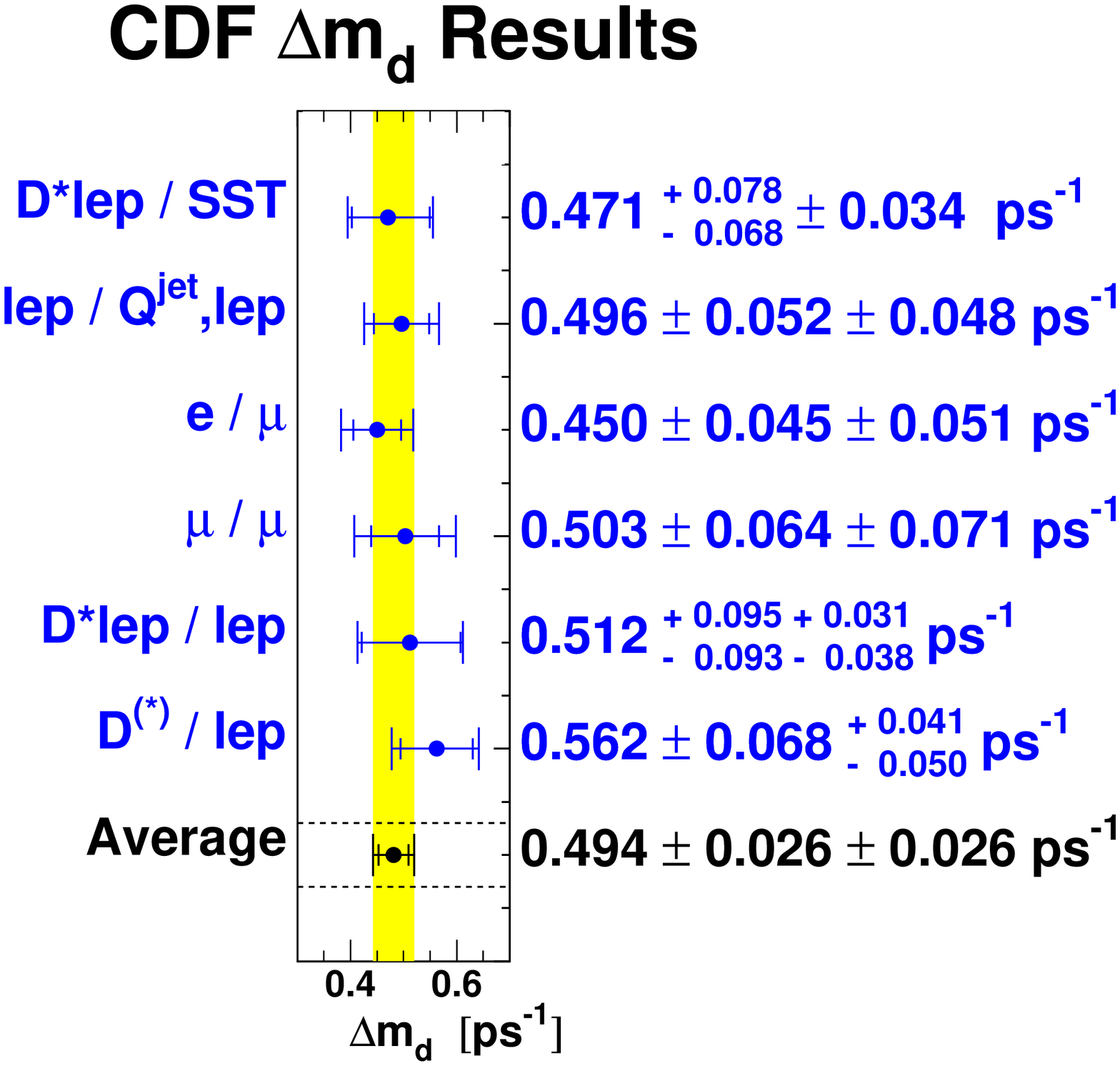}
\hspace*{0.2cm}
\epsfysize=6.4cm
\epsffile[20 45 600 750]{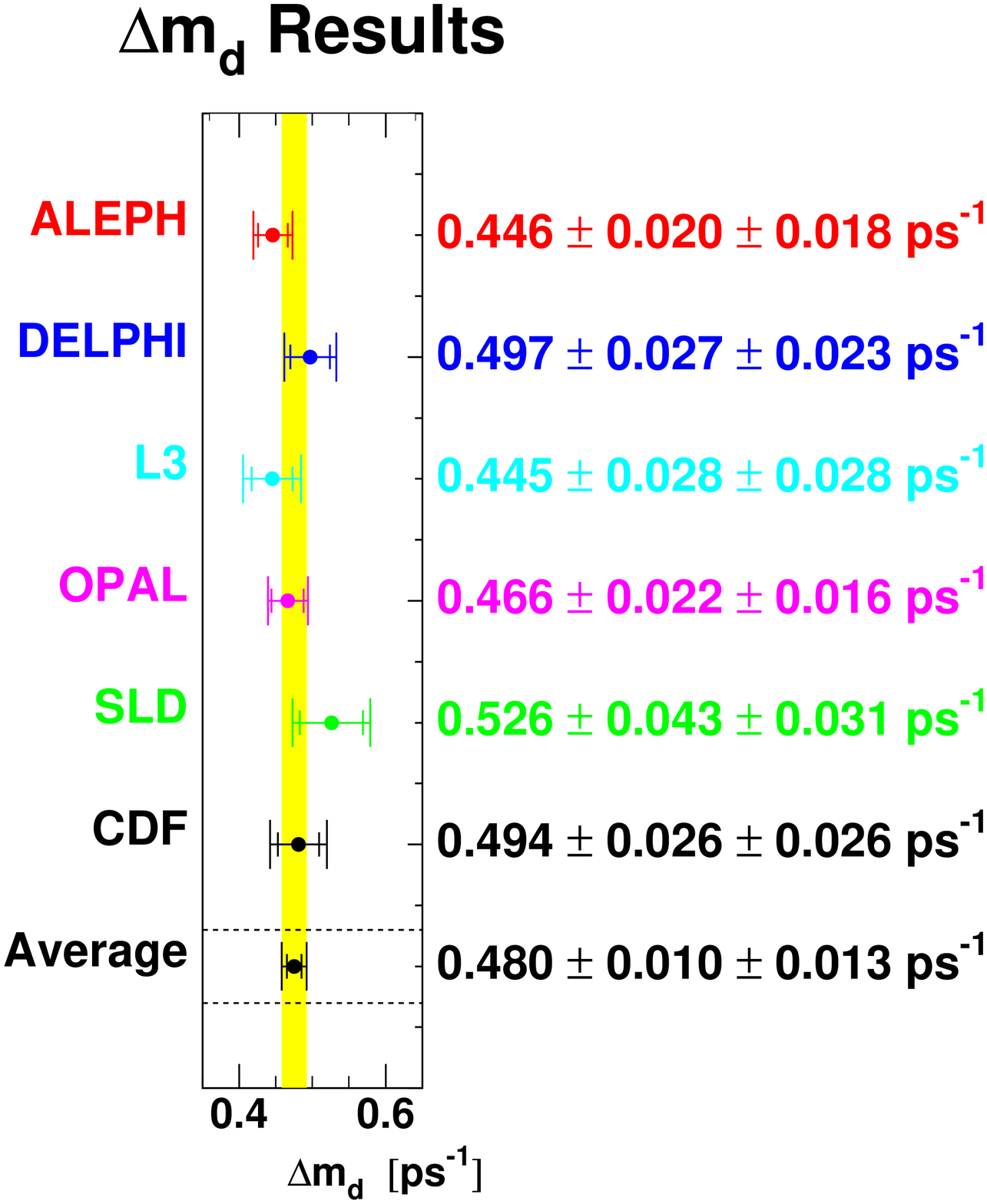}
}
\vspace*{0.2cm}
\fcaption{Compilation of (a) the CDF time dependent $B^0\bar B^0$ mixing
measurements and (b) the \dmd\ average results for the LEP, SLD, and
CDF experiments as of July 1998.}
\label{bmix_sum}
\end{figure}
A compilation of all \dmd~measurements at CDF is found in
Figure~\ref{bmix_sum}a). We use these measurements to determine a CDF
average result for \dmd.
Since the measurements of \dmd\ are quite precise, it is important
to correctly take the correlated systematic uncertainties into account
in forming the average. In addition, most of the presented results
depend on common physics 
parameters, where different values are used in the original
analyses. In order to correctly handle these issues, we use the
procedure developed by the LEP $B$ Oscillation Working Group
and adjust the measurements of \dmd\ on the
basis of a common consistent set of input values. First,
the procedure linearly rescales each measurement of \dmd\ for each 
common physics parameter, in accordance
with the difference between the originally used parameter value and
its new common value, and with the corresponding systematic uncertainty
determined in the analysis.
This systematic uncertainty is also adjusted, if the uncertainty
on the parameter is different from that originally used.
The combination procedure then performs a
common fit of $\dmd$ and of the physics parameters related to the sources of
the common systematic uncertainties. We also take into account a statistical
correlation of about 20\% between the \dmd\ result using jet-charge
and lepton tagging and the \dmd\ analyses using dilepton data. We
determine the average of all \dmd\ results at CDF to be
\begin{equation}
\dmd = (0.494\ \pm 0.026\ \pm 0.026)\ {\rm ps}^{-1}.
\end{equation}
A comparison of this CDF average with the \dmd\ averages of the   
LEP and SLD experiments as of July 1998 is displayed in
Figure~\ref{bmix_sum}b). The CDF \dmd\ results are
quite competitive with other experiments. 

\section{Search for \boldmath{$CP$} Violation}
\runninghead{$B$ Lifetimes, Mixing and $CP$ Violation at CDF}
{Search for $CP$ Violation}
\noindent
In 1964, Christenson, Cronin, Fitch, and Turlay discovered\cite{cp_kl}
that the laws of nature are not invariant under the combined action of charge
conjugation~$C$ and parity~$P$. They found that the $K^0_L$~meson,
associated with the
$CP$-odd state in the neutral kaon system, decays with a branching
fraction of about $2\cdot 10^{-3}$ into $\pi^+\pi^-$, which 
is a $CP$-even state. This result can briefly be illustrated in the
following way: 
As in the $B$~meson system, discussed in the Sec.~6.5.3
and Sec.~7.1, neutral kaons exist in two $CP$-conjugate flavour states
$|K^0\rangle = |\bar s d\rangle$ and $|\bar K^0\rangle = |s \bar d\rangle$.
Applying the $CP$ operation to these states, we define the following
phase convention:
\begin{equation}
CP\, |K^0\rangle = - |\bar K^0\rangle \ \ \ \ {\rm and} \ \ \ \
CP\, |\bar K^0\rangle = - |K^0\rangle.
\end{equation}
The physical kaon states with well defined mass and lifetime, which
decay through the weak interaction, are linear combinations of the two
$CP$-conjugate quark states:
\begin{equation}
|K^0_1\rangle = 1/\sqrt{2}\,\, (|K^0\rangle - |\bar K^0\rangle)
\ \ \ \ {\rm and} \ \ \ \
|K^0_2\rangle = 1/\sqrt{2}\,\, (|K^0\rangle + |\bar K^0\rangle).
\end{equation}
If $CP$ is conserved, the physical kaon states are $CP$ eigenstates:
\begin{equation}
CP\,|K^0_{1}\rangle 
= CP\, [1/\sqrt{2}\, (|K^0\rangle - |\bar K^0\rangle)\,] = 
1/\sqrt{2}\, (-|\bar K^0\rangle + |K^0\rangle) =
+|K^0_{1}\rangle.
\end{equation}
We call $K^0_{1}$ the ``$CP$-even'' state. For $K_2^0$, we obtain
\begin{equation}
CP\,|K^0_{2}\rangle 
= CP\, [1/\sqrt{2}\, (|K^0\rangle + |\bar K^0\rangle)\,] = 
1/\sqrt{2}\, (-|\bar K^0\rangle - |K^0\rangle) =
-|K^0_{2}\rangle
\end{equation}
and call $K^0_{2}$ the ``$CP$-odd'' state. From experiment, we know
that neutral kaons decay either into two or three
pions. Since $CP\,|\pi^+\pi^-\rangle = +1$ and 
$CP\,|\pi^+\pi^-\pi^0\rangle = -1$,
we associate $|K^0_{1}\rangle$ with $K^0_S \ra \pi^+\pi^-$
and $|K^0_{2}\rangle$ with $K^0_L \ra \pi^+\pi^-\pi^0$. 
If $CP$ is conserved, a $K^0_L$ is not allowed to decay into two
pions. The discovery of $K^0_L$ decays into $\pi^+\pi^-$ thus manifests the
existence of $CP$~violation and states that the laws of nature are not
exactly the same for matter and antimatter. 

More than 30 years have passed since the surprising discovery of
$CP$~violation, and the kaon system is still the only place where
$CP$~violation has been observed in nature. However, the system of neutral
$B$~mesons is expected to yield large $CP$~violating effects. This is
the subject of this section, where we report on a search for
$CP$~violation in $B^0 \ra \jpks$ decays at CDF. 
First, we briefly introduce the 
Cabibbo-Kobayashi-Maskawa mixing matrix and the CKM unitarity
triangle.

\subsection{CKM matrix and unitarity triangle}
\noindent
$CP$~violation in the kaon system can be explained in an elegant way in
the Standard Model\cite{GSW} with three generations, as originally
suggested by Kobayashi and Maskawa\cite{ckm2} in 1973. Note, this was
still before the discovery of the $J/\psi$ resonance\cite{jpsidisc} in 1974. 
In the Standard Model with SU(2)$\times$U(1) as the gauge group of
electroweak interactions, the quark mass eigenstates are not the same
as the weak flavour eigenstates. The matrix which relates the mass
eigenstates to the weak eigenstates was defined for six quarks and
given an explicit parametrization by Kobayashi and Maskawa\cite{ckm2}.
It generalizes the four-quark case, where the corresponding matrix is
parametrized by a single parameter, the Cabibbo angle\cite{ckm1}.
The mixing is often expressed in terms of a $3\times 3$
unitarity matrix $V_{\rm CKM}$, called the  
Cabibbo-Kobayashi-Maskawa mixing matrix, which operates by convention
on the charge $-1/3$ quarks $d$, $s$, and $b$:
\begin{eqnarray}
 \left(\begin{array}{c} d'  \\ s' \\ b' \end{array} \right) =
 \left( \begin{array}{lcr}
  V_{ud}   & V_{us}   & V_{ub}   \\
  V_{cd}   & V_{cs}   & V_{cb}   \\
  V_{td}   & V_{ts}   & V_{tb}
 \end{array} \right)
 \left(\begin{array}{c} d  \\ s \\ b \end{array} \right).
\end{eqnarray}
The individual matrix elements can, in principle, all be determined from
weak decays of the relevant quarks or from deep inelastic neutrino
scattering. 

There are several parametrizations of the CKM matrix. In the standard
para\-metrization, three rotation angles and an imaginary phase must be
used. This phase being non-zero gives rise to $CP$~violation in the
weak interaction. Another popular approximation of the CKM matrix
which emphasizes the hierarchy in the size of the matrix elements is,
due to Wolfenstein\cite{wolfenstein}
\begin{eqnarray}
 V_{\rm CKM} = 
 \left( \begin{array}{ccc}
  1 - \lambda^2/2   & \lambda   & A \lambda^3(\rho - i \eta)   \\
  -\lambda   & 1 - \lambda^2/2   & A \lambda^2   \\
  A \lambda^3(1 - \rho - i \eta)   & -A \lambda^2   & 1
 \end{array} \right) + {\cal O}(\lambda^4).
\end{eqnarray}
Here, the CKM matrix is parametrized in powers of $\lambda$, the sine
of the Cabibbo angle $\lambda = \sin \theta_{\rm c} = |V_{us}|$. 
The Cabibbo angle is well measured from leptonic kaon decays 
$K \ra \pi \ell \nu$ yielding $\lambda = 0.2196 \pm 0.0023$. The other
parameters used in the Wolfenstein parametrization are $A$, $\rho$,
and $\eta$. The parameter $A$ is related to the CKM matrix element
$V_{cb} = A \lambda^2$ and can be determined from semileptonic 
$b \ra c$ transitions. As we will see below, the parameters $\rho$ and
$\eta$ are related to $CP$~violation in the $B$~meson system and are
connected to the matrix elements $V_{ub}$ and $V_{td}$.

\begin{figure}[tbp]
\centerline{
\put(-10,20){\large\bf (a)}
\put(53,20){\large\bf (b)}
\hspace*{0.2cm}
\epsfysize=1.8cm
\epsffile[145 63 486 244]{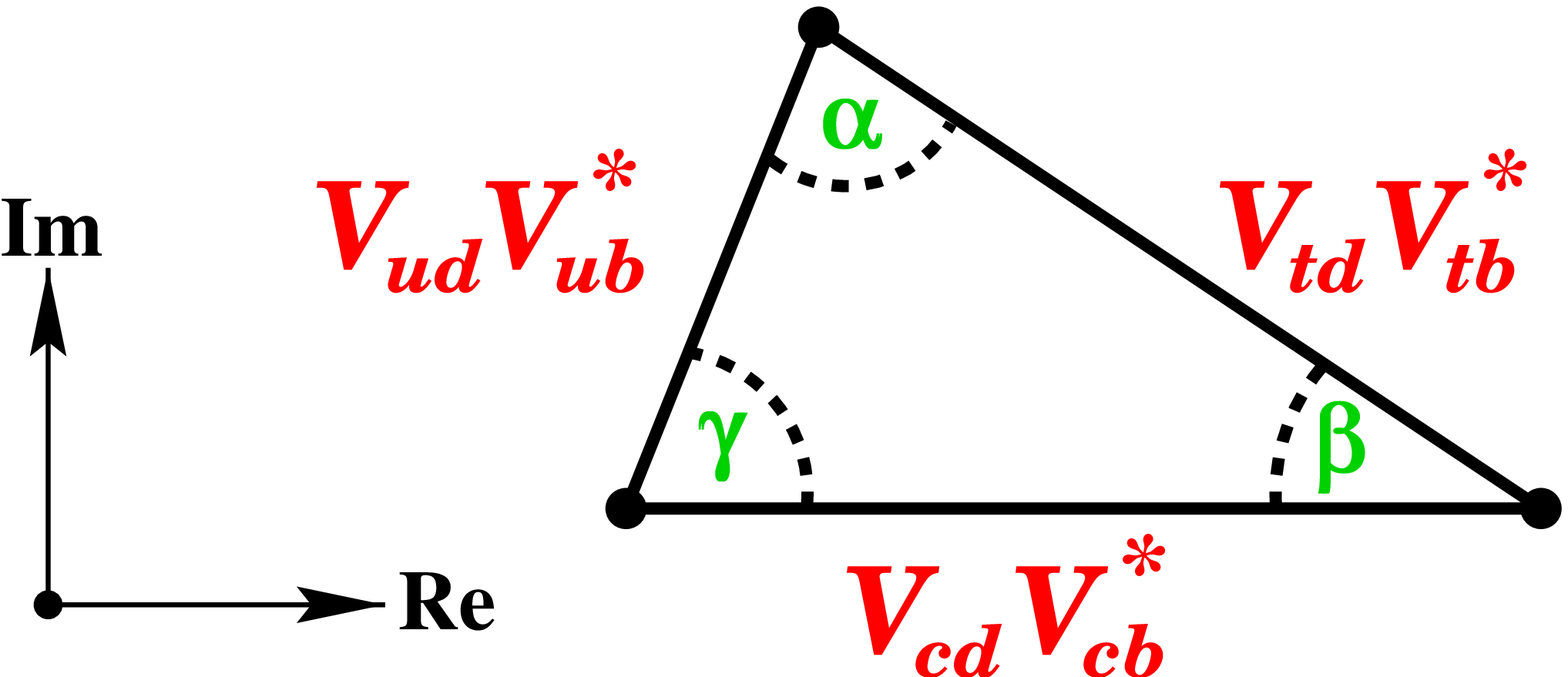}
\hspace*{3.2cm}
\epsfysize=1.8cm
\epsffile[179 36 522 217]{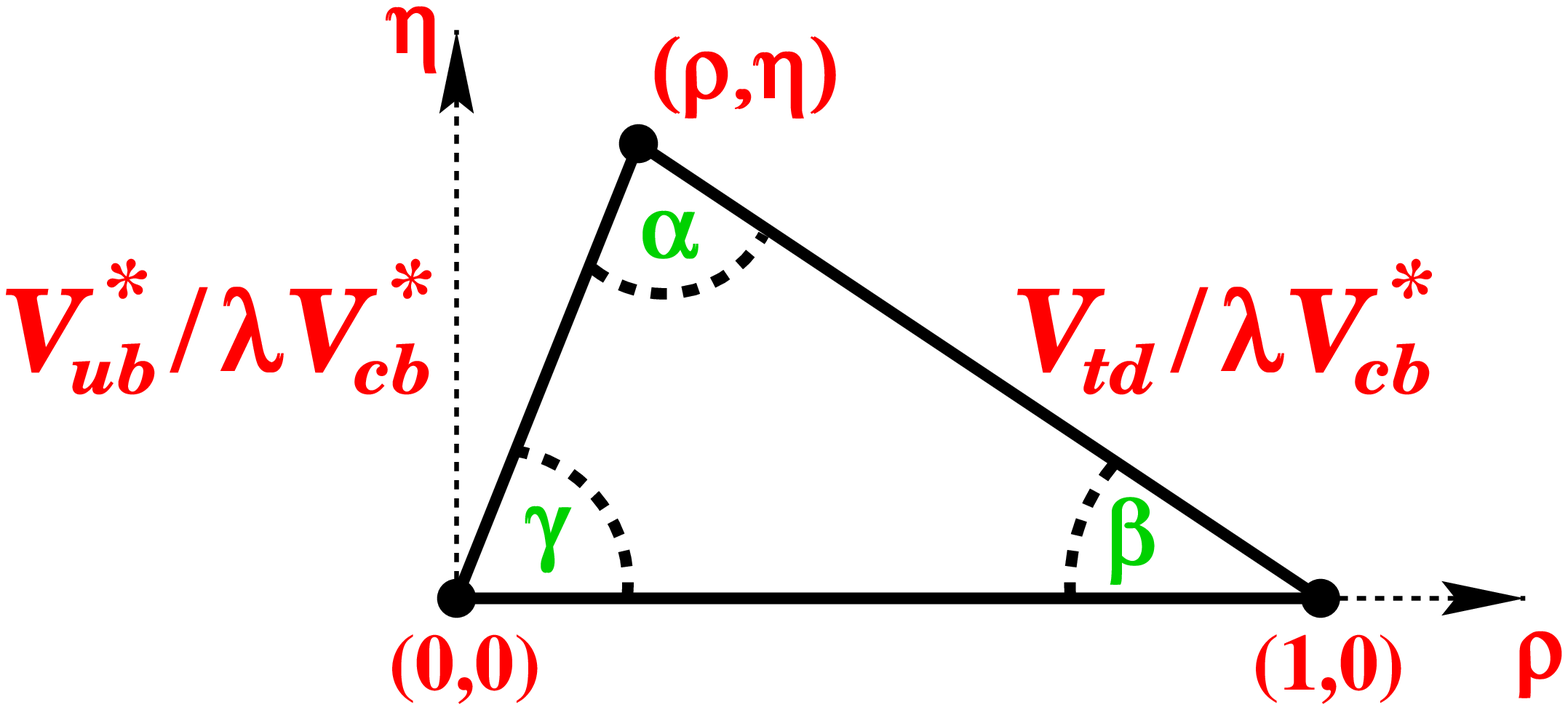}
}
\vspace*{0.9cm}
\fcaption{The CKM unitarity triangle (a) in standard representation
and (b) in the $(\rho,\eta)$-plane using the Wolfenstein parametrization.}
\label{ckm_triangle}
\end{figure}

Because of the unitarity of the CKM matrix derived from conserving
probabilities, we 
obtain a relation involving the two smallest elements $V_{ub}$ 
and $V_{td}$ of~$V_{\rm CKM}$
\begin{equation}
V_{ud} V_{ub}^* + V_{cd} V_{cb}^* + V_{td} V_{tb}^* = 0.
\end{equation}
This is a triangle relation in the complex plane as shown in
Fig.~\ref{ckm_triangle}a). The Cabibbo-Kobayashi-Maskawa matrix
elements determine the size 
of the legs of this triangle, often called the CKM unitarity
triangle. The angles $\alpha$, $\beta$, and $\gamma$ are related to
$CP$~violating asymmetries in $B$~decays. For 
example, an asymmetry in $B^0 \ra \jpks$ decays measures \stb, while
$B^0 \ra \pi\pi$ decays are related to \sta. For
$CP$~violation to be permitted in the Standard Model, the area of the
unitarity triangle must be non-zero and the
angles of the triangle different from zero or $\pi$.     

\begin{figure}[tbp]
\centerline{
\epsfysize=5.5cm
\epsffile[15 35 530 320]{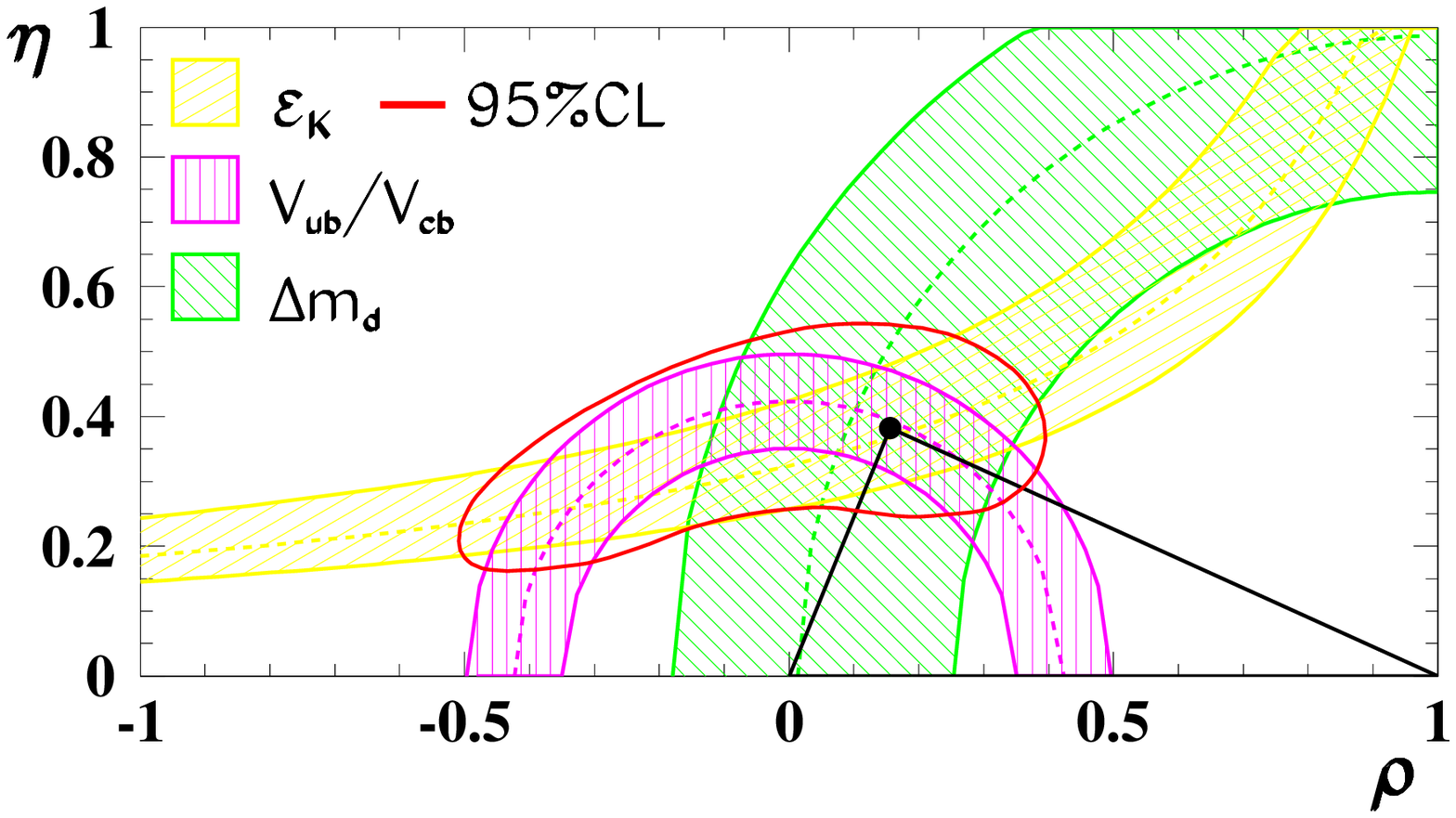}
}
\vspace*{0.2cm}
\fcaption{Constraints on the $(\rho,\eta)$-plane arising from
measurements of $|V_{ub}/V_{cb}|$, $B^0\bar B^0$ mixing (\dmd), and
$CP$~violation in kaon decays $(\epsilon_K)$.}
\label{ckm_constraint}
\end{figure}

We briefly illustrate the current knowledge about the CKM unitarity
triangle. For this purpose, we use the Wolfenstein parametrization,
where the unitarity triangle is described in the
$(\rho,\eta)$-plane, as shown in Fig.~\ref{ckm_triangle}b). We make the
following approximations $|V_{tb}^*| = 1$, $V_{cd} = -\lambda$, 
and set $V_{ud} = 1 - \lambda^2/2 \approx 1$. We also normalize the 
legs of the triangle by $|V_{cb}|$ keeping the base of the
triangle of unit length. The other two legs are then described by
$|V_{ub}^*\,/\lambda V_{cb}^*|$ and $|V_{td}\,/\lambda V_{cb}^*|$.  
A measurement of $|V_{ub}/V_{cb}|$ constrains one leg of the triangle to
lie between the two half circles centered at the origin, as shown in
Figure~\ref{ckm_constraint}. The other leg of the triangle is
constrained by our current knowledge of $V_{td}$ obtained from 
$B^0\bar B^0$~mixing measurements. The large uncertainty with which
the $B$~meson decay constant $f_B$ and the bag
parameter (see Sec.~7.1) are known, allows this leg of the unitarity
triangle to 
lie in the broad area between the two circles centered around the
point $(1,0)$. Finally, the knowledge of $CP$ violation in the kaon
system adds another constraint on the CKM unitarity triangle indicated
by the third band in Fig.~\ref{ckm_constraint}. We fit the three
bands to originate from a common area and obtain the 95\% confidence
level contour, 
also shown in Fig.~\ref{ckm_constraint}. We see that the apex of the
unitarity triangle, described by the coordinates $(\rho,\eta)$, 
can be placed in a fairly wide range. Direct measurements of the
angles $\alpha$, $\beta$, or $\gamma$ would better constrain the CKM
unitarity triangle.

\subsection{Measurement of \boldmath{$CP$} violation parameter 
\boldmath{$\stb$} in \boldmath{$B^0 \ra \jpks$}}
\begin{figure}[tbp]
\centerline{
\put(9,41){\large\bf (a)}
\put(74,46){\large\bf (b)}
\put(-3,57){\large\bf ${\cal A}_{\rm mix}$}
\put(-3,26){\large\bf ${\cal A}_{CP}$}
\epsfxsize=6.3cm
\epsffile[5 5 525 515]{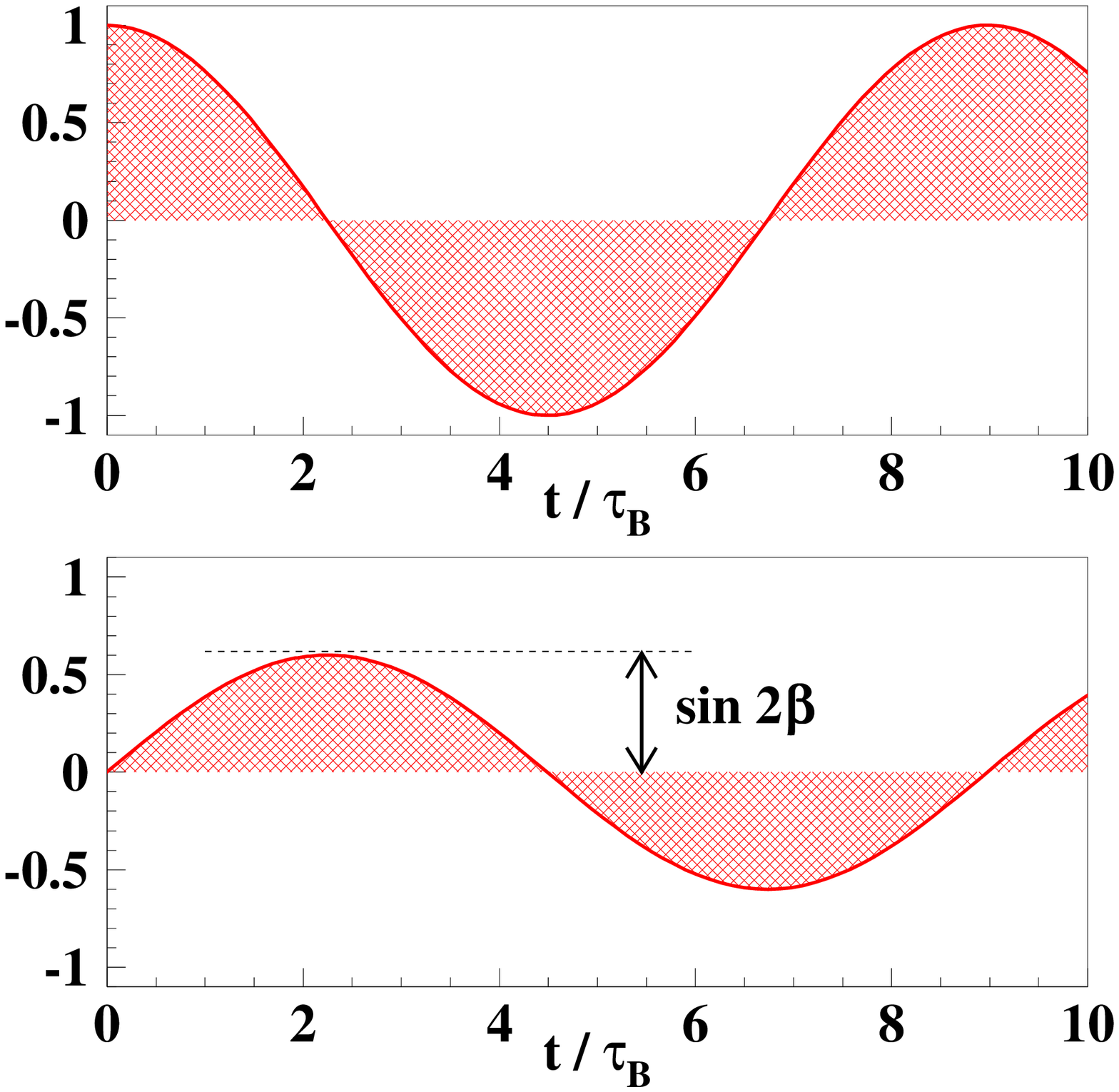}
\epsfxsize=6.3cm
\epsffile[20 5 550 530]{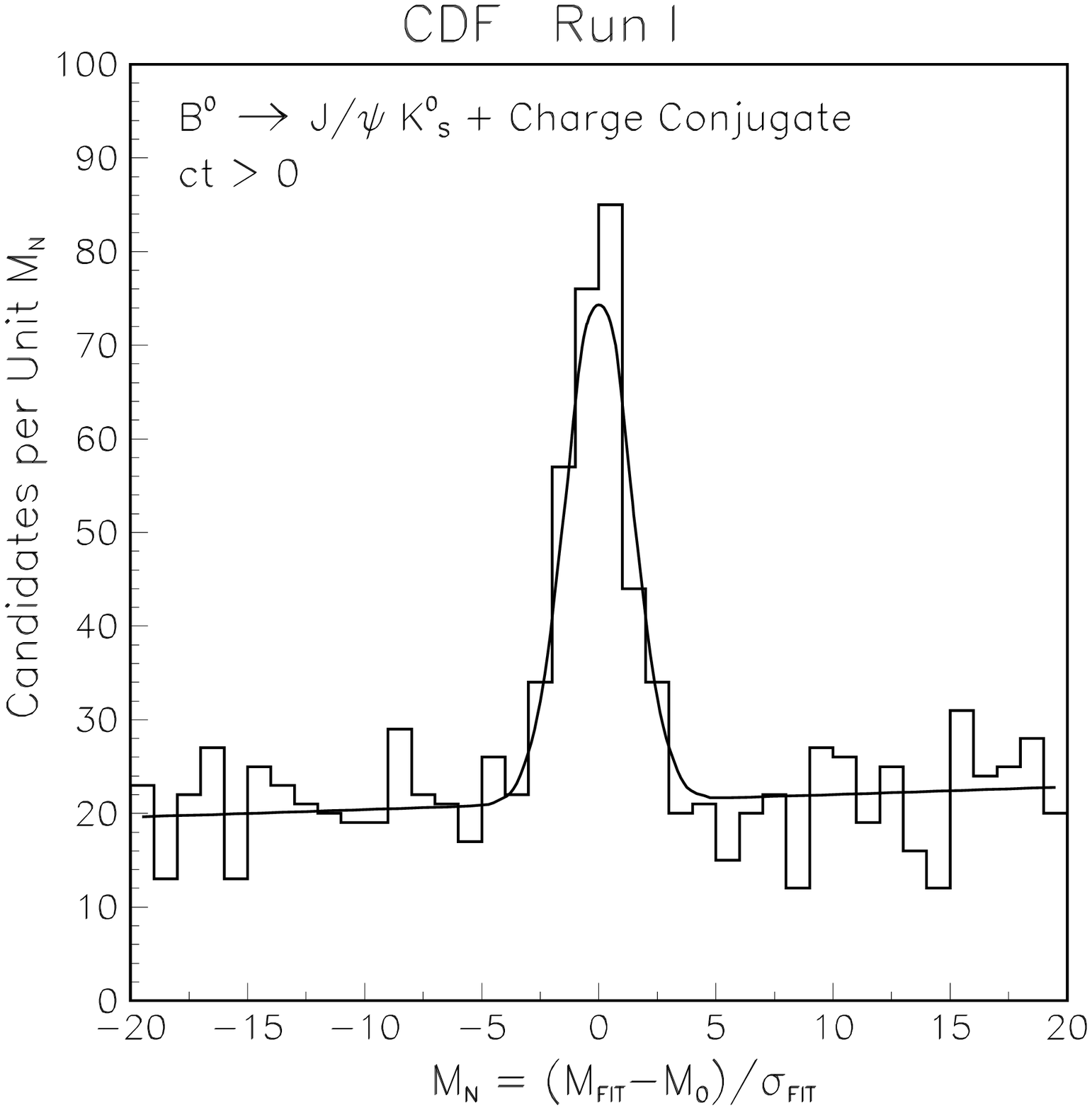}
}
\vspace*{0.2cm}
\fcaption{
(a) Illustration of the connection between the time dependence of
the $B^0\bar B^0$ mixing asymmetry (top) and the $CP$~asymmetry (bottom) in 
$B^0 \ra \jpks$ decays. (b) Normalized mass distribution of \jpks\
candidates. The curve is the Gaussian signal plus linear background.}
\label{sin2b_ks}
\end{figure}

\noindent
One way to observe $CP$~violation in $B$~meson decays is to use the
interference between the direct decay of a $B^0$ into a $CP$~final
state $f_{CP}$ and the process $B^0 \ra \bar B^0 \ra f_{CP}$ which can
provide a second amplitude to interfere with the direct decay 
$B^0 \ra f_{CP}$.
A large $CP$ violating effect is expected\cite{kshort} 
in $B^0/\bar{B}{^0}$ decays to the $CP$~eigenstate
$J/\psi K^0_S$.
The interference of direct 
decays, $B^0 \ra \jpks$, versus those that undergo 
mixing, $B^0 \ra \bar{B}{^0} \rightarrow \jpks$,
gives rise to a decay asymmetry 
\begin{equation}
{\cal A}_{CP}(t) \equiv \frac{\bar{B}{^0}(t)-B^0(t) }
                        {\bar{B}{^0}(t)+B^0(t) } = 
                       \stb \cdot \sin \dmd t,
\label{eq:cp_asym}
\end{equation}
where $B^0(t)$ ($\bar{B}{^0}(t)$) is the number of decays
to $\jpks$ at proper time $t$ given that the 
produced meson was a $B^0$ ($\bar{B}^0$) at $t=0$.
The $CP$ phase difference between the two decay paths
appears via the factor \stb, and the 
$B\bar B$~flavour oscillation through the mass difference \dmd\ between
the two $B^0$ mass eigenstates.
Figure~\ref{sin2b_ks}a) illustrates the connection between the
time dependence of the 
$B^0\bar B^0$ mixing asymmetry ${\cal A}_{\rm mix}$ (top) and the
$CP$~asymmetry ${\cal A}_{CP}$ (bottom) in 
e.g.~$B^0 \ra \jpks$ decays.
At $t=0$ we start with a pure $\bar B^0$ state. Since no $B^0$ mesons are
present to interfere with, the $CP$~asymmetry is zero. After about two
$B$~lifetimes, we have the same amount of $B^0$ and $\bar B^0$ due to
mixing (${\cal A}_{\rm mix} = 0$) and the $CP$~asymmetry is
maximal. From this we see that the $CP$ asymmetry follows a sinusoidal
time behavior with \stb\ characterizing the amplitude of this sine
curve as shown in Eq.~(\ref{eq:cp_asym})
while the mixing asymmetry follows a cosine like time
curve as expressed~in~Eq.~(\ref{eq:asym}).

The measurement of the $CP$~violation parameter \stb\ using
$B^0/\bar B^0 \ra \jpks$ decays\cite{sin2b_prl} 
reconstructs $J/\psi$
mesons through $\mu^+\mu^-$ and searches for $K^0_S \ra \pi^+\pi^-$
candidates. 
The $J/\psi$ and $K^0_S$ daughter tracks are combined in a four particle fit 
to originate from a common $B^0$ vertex.
The decay length of the $B^0$ is used to calculate its proper 
decay length $ct$.
We define $M_N \equiv (m_{\rm fit} - m_0)/\sigma_{\rm fit}$, where 
$m_{\rm fit}$ is the
mass of the $B$ candidate from the fit described above, 
$\sigma_{\rm fit}$ is its uncertainty (typically $\sim\!9\;{\rm MeV}/c^2$),
and $m_0$ is the central value of the $B^0$ mass peak.
The normalized masses  $M_N$ 
for the selected candidates with $ct >0$
are shown in Fig.~\ref{sin2b_ks}b) along with
the result of the maximum log-likelihood fit described later.
The fit yields $(198 \pm 17)$ $B^0/\bar{B}{^0}$ mesons
for all $ct$.

\begin{figure}[tbp]
\centerline{
\put(10,55){\large\bf (a)}
\put(75,51){\large\bf (b)}
\epsfxsize=6.3cm
\epsffile[5 5 530 515]{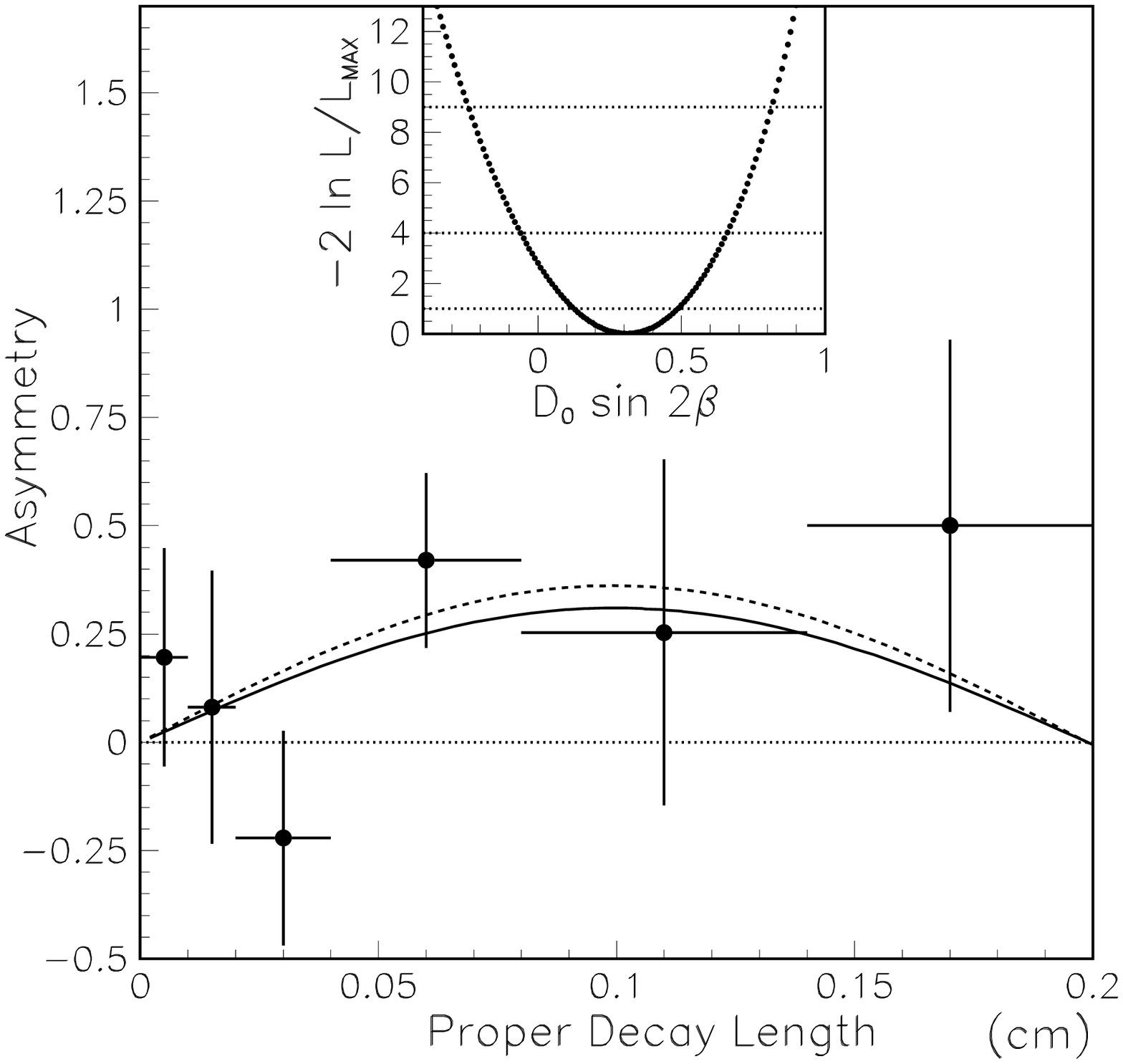}
\epsfxsize=6.3cm
\epsffile[5 5 545 515]{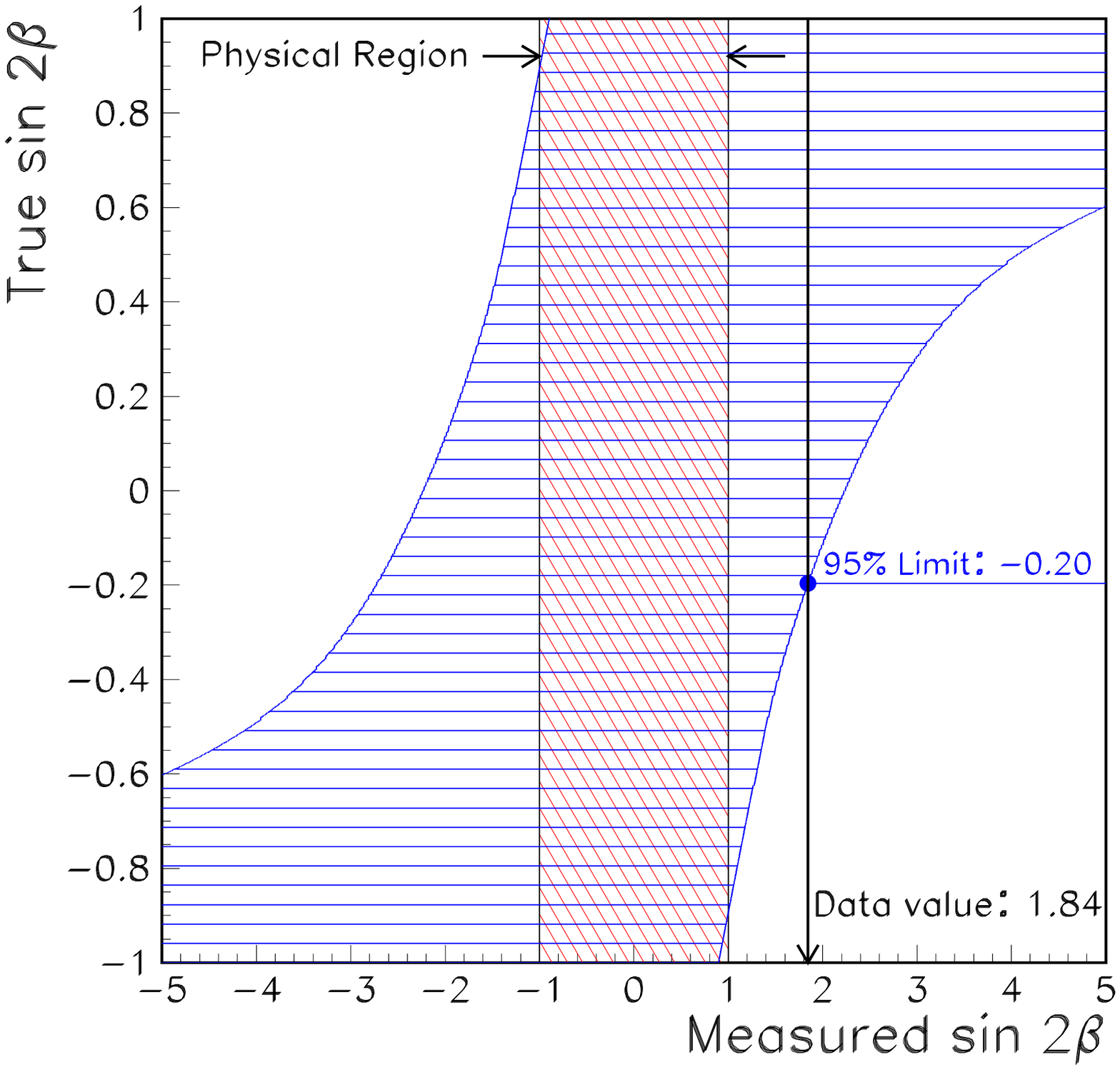}
}
\vspace*{0.2cm}
\fcaption{
(a) The sideband-subtracted tagging asymmetry as a function 
of the reconstructed \jpks\ proper decay length  (points).
The dashed curve is the result of a simple $\chi^2$-fit to 
${\cal A}_0\sin\dmd t$.
The solid curve is the likelihood fit result, and the inset shows a
scan through the log-likelihood function as ${\cal D}_0\stb$ is
varied about the best fit value.
(b) ``Confidence belt'' of the true value of \stb\ versus the measured
value of \stb\ used to extract the limit on \stb.}
\label{sin2b_asym}
\end{figure}

To measure ${\cal A}_{CP}(t)$, we need to know
whether the production flavour of the $B$~meson
is $B^0$ or $\bar{B}{^0}$. 
We determine this with a same side tagging method 
described in Sec.~7.2. 
The effectiveness of this method is demonstrated by tagging 
$B \ra D^{(*)}\ell\nu$ decays and observing
the time dependence of $B^0\bar{B}{^0}$ oscillations
measuring \dmd.
Applying the SST method to the $J/\psi K^0_S$ sample yields a 
tagging efficiency of $\sim\!65\%$.
Since negative (positive) tags are associated 
with $\bar{B}{^0}$'s ($B^0$'s), the asymmetry 
\begin{equation}
{\cal A}(ct) \equiv \frac{N^-(ct)-N^+(ct) }
                         {N^-(ct)+N^+(ct) }
\label{eq:meas_asym}
\end{equation}
is formed analogous to Eq.~(\ref{eq:cp_asym}),
where $N^\pm(ct)$ are the numbers of positive and negative tags in 
a given $ct$ bin. 
The sideband-subtracted asymmetry is displayed
in Fig.~\ref{sin2b_asym}a) along with a $\chi^2$-fit (dashed curve) to
${\cal A}_0\sin\dmd t$, where \dmd\ is fixed\cite{PDG} to 
$0.474 \;{\rm ps}^{-1}$.
The amplitude, ${\cal A}_0 = 0.36\pm 0.19$, 
measures \stb\ attenuated by the dilution factor
${\cal D}_0 = 2p_R-1$, where $p_R$ is the probability
that the tag correctly identifies the $B$ flavour.

The fit is refined
using an unbinned maximum log-likelihood method.
This fit makes optimal use of the low 
statistics by fitting signal and background distributions in $M_N$ and $ct$,
including sideband and $ct<0$ events which help constrain the background.
The likelihood fit also incorporates resolution effects and
corrections for systematic biases, such as the
inherent charge asymmetry favouring positive tracks
resulting from the
wire plane orientation in the main drift chamber.  
The solid curve in Fig.~\ref{sin2b_asym}a) is the result of the likelihood
fit, which yields 
${\cal D}_0\stb=0.31 \pm 0.18\pm 0.03$ including systematic effects.  
As expected, the
two fits give similar results, indicating that the result is dominated
by the sample size and that the corrections and improvements of the
likelihood fit introduce no dramatic effects.
Also shown in the Fig.~\ref{sin2b_asym}a) inset
is the relative log-likelihood as a function of ${\cal D}_0\stb$.
The shape is parabolic, indicating Gaussian errors.  

To obtain \stb, dilution measurements 
from other $B$ samples are used.
CDF's best single ${\cal D}_0$  measurement
from a large $B \rightarrow D^{(*)}\ell X$ sample\cite{sst_prd,sst_prl}
is $0.181 ^{+0.036}_{-0.032}$ (see Sec.~7.2). 
Because of differing lepton
\Pt\ trigger thresholds, the average \Pt\ of the 
semileptonic $B$ sample is $\sim\!21$~\gevc, but it is only $\sim\!12$~\gevc\
in the \jpks\ data. 
This difference is corrected for using Monte Carlo studies.
The ${\cal D}_0$ appropriate for the
\jpks\ sample is found to be $0.166\pm 0.018\pm 0.013$,
indicating a small shift from 0.181.
The first error is due to the uncertainty in the dilution measurement,
and the second  is due to the Monte Carlo extrapolation.

Using this value of ${\cal D}_0$, \stb\ is determined to be
\begin{equation}
\stb=1.8\pm 1.1\pm 0.3.
\label{eq:s2b_sst}
\end{equation}
The central value is  unphysical
since the amplitude of the measured 
asymmetry is larger than  ${\cal D}_0$.
To express this result in terms of a confidence
interval in \stb, 
the frequentist construction by Feldman and Cousins\cite{Cousins} is followed.
This approach gives proper confidence intervals even for
measurements in the unphysical region. Figure~\ref{sin2b_asym}b) shows
the ``confidence belt'' of the true value of \stb\ versus the measured
value of \stb\ using the frequentist method. 
This measurement thereby corresponds to excluding $\stb<-0.20$
at 95\% confidence level.
This result favours current Standard Model expectations 
of a positive value of \stb\ and establishes
the feasibility of measuring $CP$ asymmetries in $B$ meson decays at a hadron
collider. 

\subsubsection{Updated measurement of $CP$ violation parameter 
$\stb$ in $B^0 \ra \jpks$}
\noindent
Just prior to completion of this article, the CDF collaboration
released an updated measurement of the $CP$~violation parameter 
$\stb$ in $B^0 \ra \jpks$ decays. This preliminary result 
combines the same side tagging method with a lepton and jet charge
flavour tag (see Sec.~7.3). In addition, it includes $\jpks$ events
which are not fully  
contained within the acceptance of the SVX detector.

\begin{figure}[tbp]
\centerline{
\put(10,55){\large\bf (a)}
\put(73,55){\large\bf (b)}
\epsfxsize=6.3cm
\epsffile[5 10 515 515]{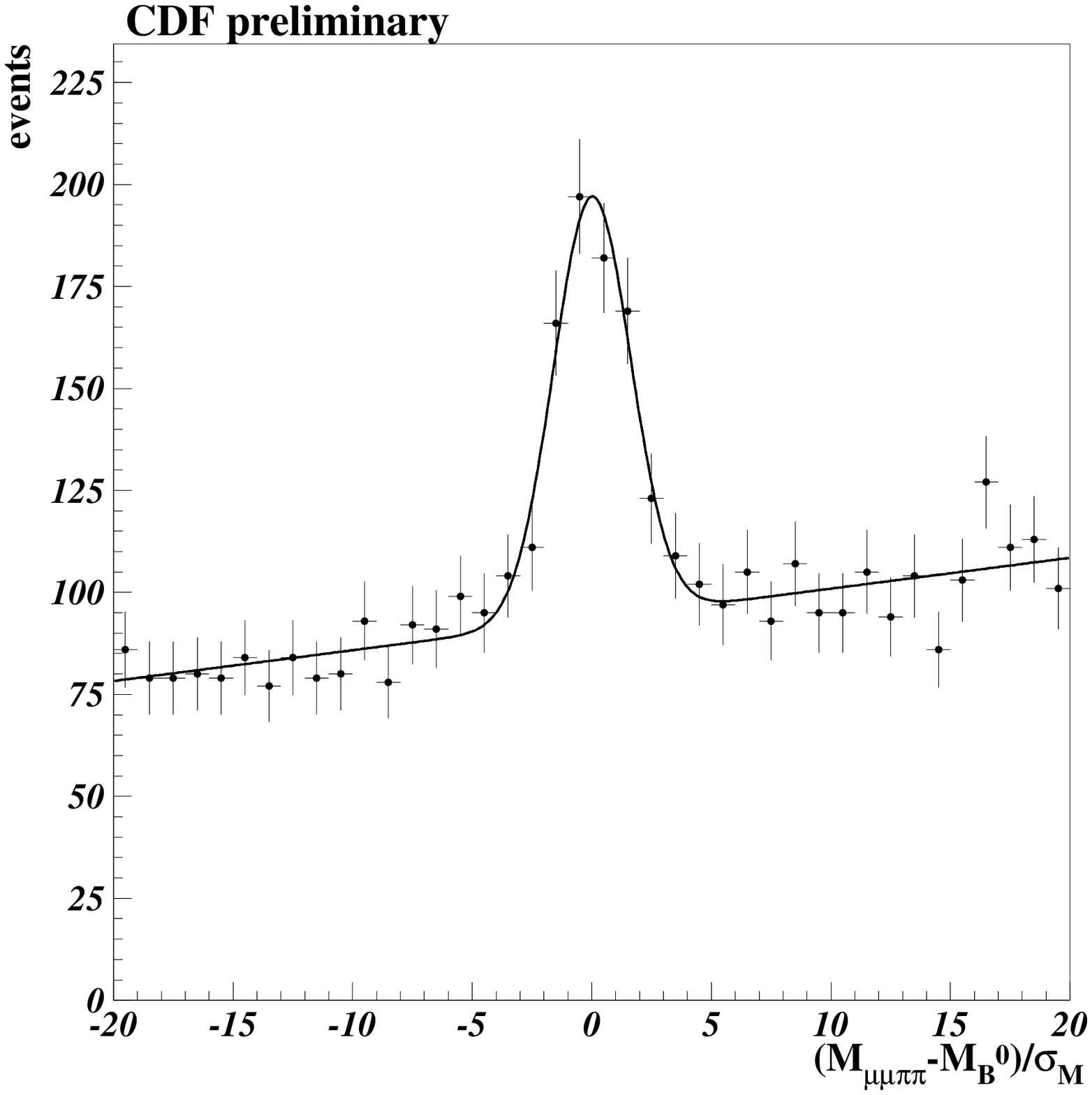}
\hspace*{0.1cm}
\epsfxsize=6.3cm
\epsffile[20 5 535 500]{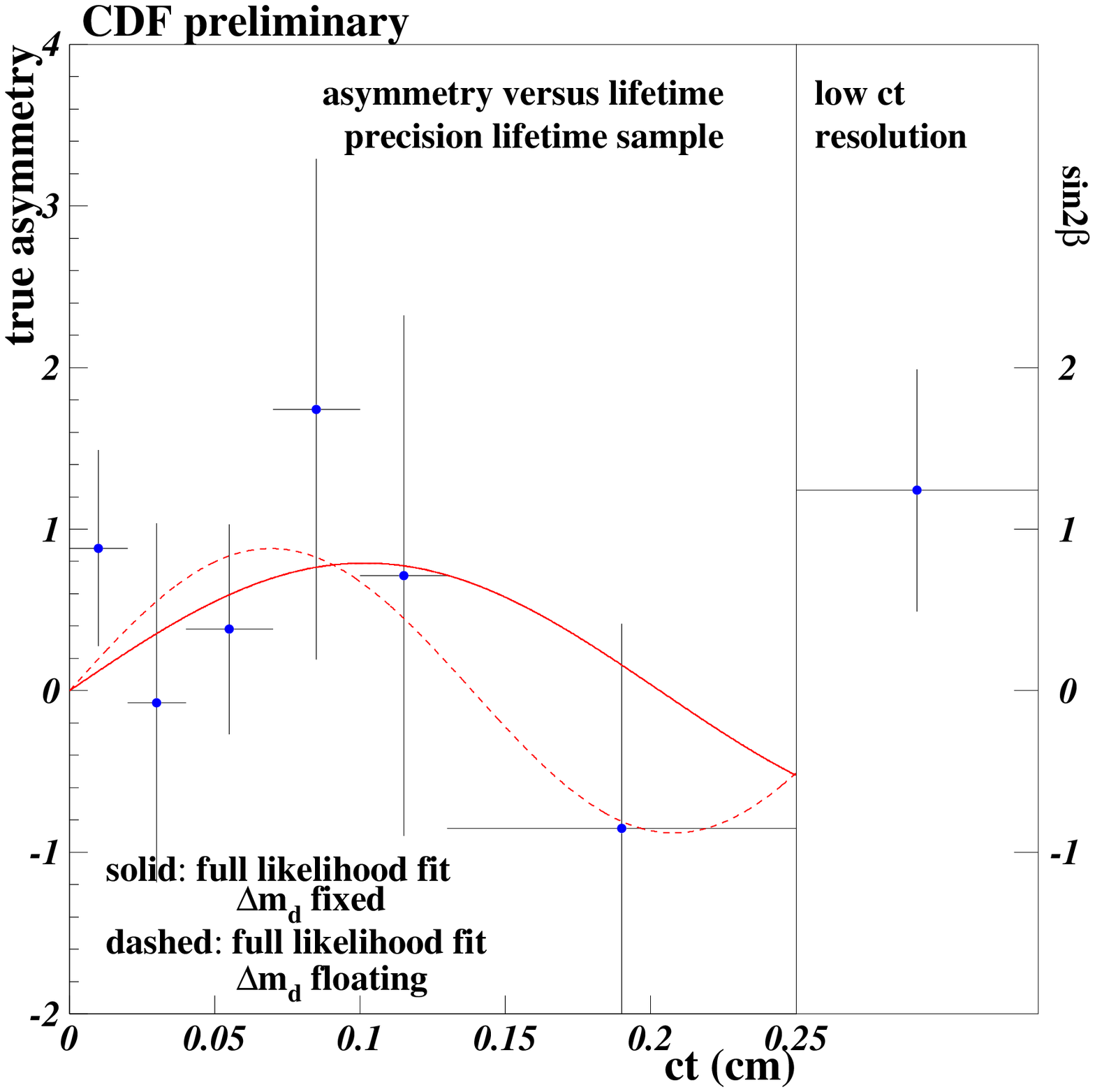}
}
\vspace*{0.2cm}
\fcaption{
(a) Normalized mass distribution of \jpks\
candidates. The curve is the Gaussian signal plus linear
background from the log-likelihood fit. 
(b) The true asymmetry \stb\ as a function 
of the reconstructed \jpks\ proper decay length.
The data points are sideband subtracted and are combined
according to the effective dilution for single and double tags. The
non-SVX events are shown on the right.}
\label{sin2b_update}
\end{figure}

The selection criteria for the $B^0 \ra \jpks$ sample with 
$J/\psi \ra \mu^+\mu^-$  and $K^0_S \ra \pi^+\pi^-$ are very similar to
the requirements used in the analysis in Sec.~8.2. To increase the
sample size, both
muon candidate tracks are no longer required to be measured in the silicon
vertex detector. The data are divided into two samples. 
The non-SVX sample contains events where both muons are not required
to have SVX information, accepting events that do not have a precise
decay length measurement. However, about 30\% of the events in this
sample have one muon track with SVX information. The other sample of
\jpks\ decays, called
SVX sample, requires both muons to be well measured in the SVX
providing precise decay length information. This sample is very
similar to the data used in the \stb\ analysis described in Sec.~8.2. 
The normalized mass distribution (see Sec.~8.2) for all \jpks\
candidates is shown in Fig.~\ref{sin2b_update}a)
comprising a sample of $(395\pm31)$ signal events. The SVX sample
contains $(202\pm18)$ signal events while the non-SVX
sample consists of $(193\pm26)$ events. 

This analysis combines all three flavour tagging methods studied at
CDF. In addition to the same side tagging algorithm, the two opposite
side tagging methods, jet charge tagging and lepton tagging, as
described in Sec.~7.3, are added. The tagging dilutions and
efficiencies for both opposite side tags are determined from a sample
of $(985\pm46)$ $B^+ \ra J/\psi K^+$ events and are presented in
Table~\ref{s2b_new_tag}. The SST tagging dilution for the SVX sample 
(${\cal D} = (16.6\pm2.2)\%$) is taken from 
the \stb\ analysis summarized in Sec.~8.2. 
For the non-SVX sample, the SST algorithm is slightly modified to
include non-SVX tracks as candidate tagging tracks. A dilution scale
factor, which relates the SST tagging performance from the SVX sample
to the non-SVX sample, is derived from the $J/\psi K^+$ data
resulting in ${\cal D} = (17.4\pm3.6)\%$ for the non-SVX sample (see
also Tab.~\ref{s2b_new_tag}). 

\begin{table}[tb]
\tcaption{Summary of efficiency $\varepsilon$ and dilution $\cal D$ of
the tagging algorithms. The quoted efficiencies are
relative to the entire \jpks\ sample. To compare the SST efficiency with
the number given in Sec.~8.2, it is necessary to double the SST
efficiency for the SVX sample.}
\centerline{\footnotesize\smalllineskip
\begin{tabular}{lcccc}
\hline
 & & & & \\
 \vspace*{-0.6cm} \\
 Flavour tag & Data & Efficiency $\varepsilon$ & Dilution $\cal D$ & \eD  \\
\hline
 & & & & \\
 \vspace*{-0.6cm} \\
 Same side tag & SVX sample & $(35.5\pm3.7)\%$ & $(16.6\pm2.2)\%$ &
 $(1.0\pm0.3)\%$ \\
   & non-SVX sample & $(38.1\pm3.9)\%$ & $(17.4\pm3.6)\%$ &
 $(1.2\pm0.5)\%$ \\
 Lepton tag & all events & $(5.6\pm1.8)\%$ & $(62.5\pm14.6)\%$ &
 $(2.2\pm1.2)\%$ \\
 Jet charge tag & all events & $(40.2\pm3.9)\%$ & $(23.5\pm6.9)\%$ &
 $(2.2\pm1.3)\%$ \\
\hline
\end{tabular}}
\label{s2b_new_tag}
\end{table}

Each event can be tagged by as many as two tags: One same side tag and one
opposite side 
tag. If both the lepton and jet charge tags are available, only the 
lepton tag is used, following the $B^0\bar B^0$ mixing analysis
described in Sec.~7.3. The procedure used to combine double tagged
events calculates a combined dilution weighted by the individual
dilutions and combines the efficiencies in a similar way. Defining the
two tags $q_1$ and $q_2$ as $+1$ if they identify a $B^0$ meson, as
$-1$ if they tag a $\bar B^0$ and as 0 if the tag is not applicable,
the individual tags are weighted by the dilutions 
${\cal D}_{1}$ and ${\cal D}_{2}$ 
according to ${\cal D}_{1}^{\prime} = q_{1}\,{\cal D}_{1}$
and ${\cal D}_{2}^{\prime} = q_{2}\,{\cal D}_{2}$, respectively. 
With the individual efficiencies $\varepsilon_1$ and $\varepsilon_2$,
the combined dilution ${\cal D}_{\rm com}$ and efficiency 
$\varepsilon_{\rm com}$ are defined as 
\begin{equation}
{\cal D}_{\rm com} = \frac{{\cal D}_1^{\prime} + {\cal D}_2^{\prime}}
{1 + {\cal D}_1^{\prime} {\cal D}_2^{\prime}}
\hspace*{0.5cm} {\rm and} \hspace*{0.5cm} 
\varepsilon_{\rm com} = \varepsilon_1\varepsilon_2\,
(1 + {\cal D}_1^{\prime} {\cal D}_2^{\prime}),
\end{equation}
where the sign of ${\cal D}_{\rm com}$ is the combined tag with
dilution $|{\cal D}_{\rm com}|$.

In a similar way as summarized in Sec.~8.2, a maximum log-likelihood method is
used to make optimal use of the 
statistics by fitting signal and background distributions in
normalized mass and $ct$,
including sideband events. 
The true asymmetry as a function of the reconstructed \jpks\ proper
decay length is shown in Fig.~\ref{sin2b_update}b) separately for the
SVX and non-SVX sample with the result of the fit overlaid. 
The non-SVX sample contribution is included as a single point
because of the low decay length resolution. The full log-likelihood fit
uses both the SVX and non-SVX sample and properly treats the decay length and
error for each event. The final fit yields
\begin{equation}
\stb=0.79\pm 0.39\pm 0.16,
\label{s2b_new_res}
\end{equation}
where the systematic error reflects the uncertainty in the dilution
parameters. Although the individual dilutions are not precisely
determined due to the limited statistics of the $B^+ \ra J/\psi K^+$
sample, this error does not dominate the uncertainty on \stb. 
Removing the constraint that fixes \dmd\ to its
world average value\cite{PDG}, the fit determines \stb\ and \dmd\
simultaneously to be $\stb = 0.88^{+0.44}_{-0.41}$ and $\dmd =
(0.68\pm0.17)$~ps$^{-1}$ as indicated by the dashed line in
Fig.~\ref{sin2b_update}b). 
The result given in Eq.~(\ref{s2b_new_res}) leads to the confidence
interval $\stb<-0.08$ (95\%~C.L.) using the
frequentist method by Feldman and Cousins\cite{Cousins}.
This result excludes $\stb < 0$ at the 93\% confidence level and
is the best direct hint for $CP$~violation in the neutral $B$~meson
system to date.

\section{Future \boldmath{$B$} Physics at CDF}
\runninghead{$B$ Lifetimes, Mixing and $CP$ Violation at CDF}
{Future $B$ Physics at CDF}
\noindent
The Fermilab accelerator complex is currently undergoing an upgrade to
produce an order of magnitude higher luminosities in the Tevatron
collider. The
largest change is the replacement of the Main Ring with the new Main
Injector, which will 
provide higher proton intensity onto the antiproton production target,
and larger aperture for the antiproton transfer into the Tevatron 
increasing the antiproton current and thus the luminosity of the Tevatron.
After the completion of the Main Injector, 
the Tevatron is scheduled to run again in April 2000 at a
centre-of-mass energy of 2.0 TeV. In this so-called Run\,II,
luminosities of $2.0\cdot 10^{32}$ cm$^{-2}$s$^{-1}$ will be reached
yielding an integrated 
luminosity of 2~fb$^{-1}$ delivered to the collider experiments
within two years.
To handle the higher luminosities and shorter bunch spacing of
initially 396~ns and later 132~ns, the CDF experiment is currently
undergoing a major detector upgrade.
In this section, we first briefly describe the CDF detector upgrade
and then summarize the prospects for measuring \stb\ and \sta\ at CDF
in Run\,II.  

\subsection{CDF\,II detector upgrade}
\begin{figure}[tbp]
\centerline{
\epsfxsize=6.6cm
\epsffile[10 35 630 480]{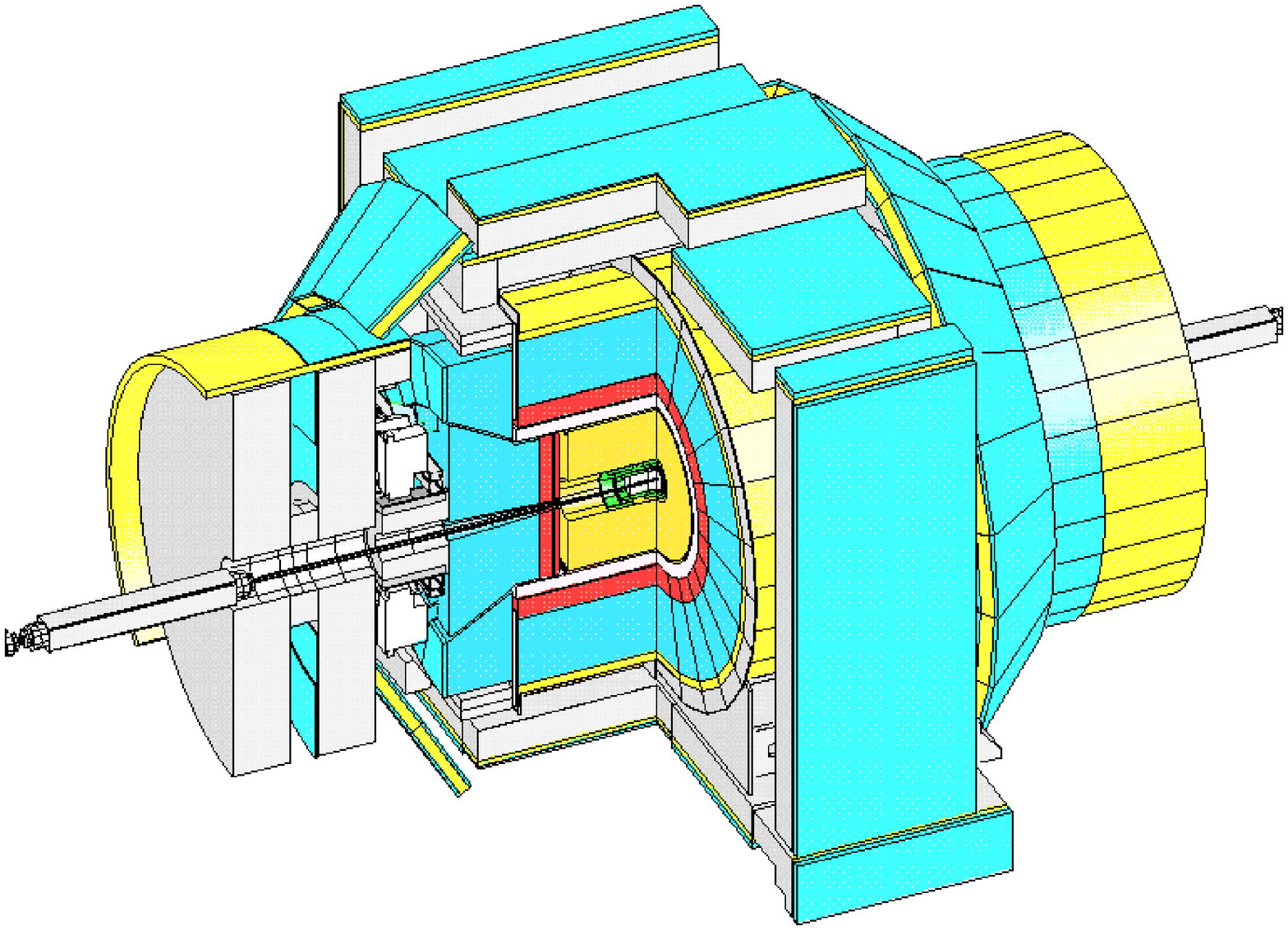}
\epsfxsize=6.0cm
\epsffile[35 310 570 700]{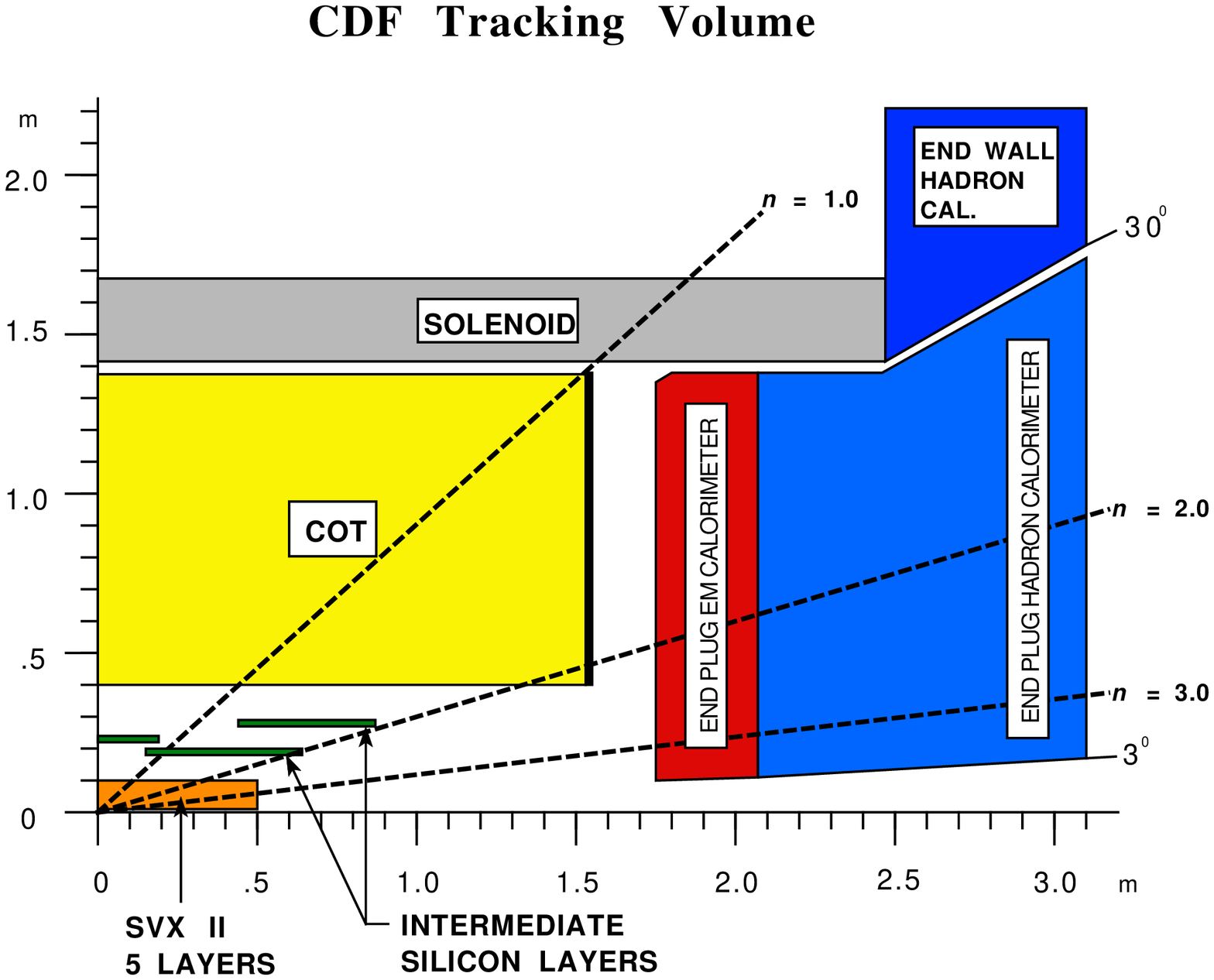}
\put(-124,41){\large\bf (a)}
\put(-51,41){\large\bf (b)}
}
\vspace*{0.2cm}
\fcaption{
(a) Schematic cut-away view of the CDF\,II detector. (b) Longitudinal view
of the CDF\,II tracking system.}
\label{cdf2_det}
\end{figure}

\noindent
The CDF detector improvements for Run\,II are motivated by the shorter
accelerator bunch spacing and the increase in luminosity by an order
of magnitude.
The primary upgrade goal is to maintain detector
occupancies at Run\,I levels, although many of the detector changes
also provide qualitatively improved detector capabilities.
The CDF\,II upgrade is described in detail
elsewhere\cite{cdfup}. 
A schematic view of the CDF\,II detector is shown in
Figure~\ref{cdf2_det}a). 
We briefly summarize here the changes most
relevant for $B$~physics in Run\,II.

One major upgrade is to the charged particle tracking system (see
Fig.~\ref{cdf2_det}b),
vital for the CDF $B$~physics program. 
A new silicon vertex detector (SVX\,II) will consist of five layers of
double sided silicon from radii of 2.9~cm to 10~cm arranged in five axial
layers, two small angle ($1.2^{\circ}$) and three $90^{\circ}$ stereo
layers. SVX\,II will consist of three modules covering the entire
$p\bar p$~luminous 
region and provide stand alone 3-dimensional tracking. The silicon sensors
are read out by a radiation hard chip in deadtimeless mode. In
addition, an intermediate silicon layer (ISL) consisting of two
double sided silicon layers at larger radii permits stand alone
silicon tracking out to $|\eta|<2$. The inner layer provides complete $\eta$
coverage while the outer layer has partial coverage ($1<|\eta|<2$).
A new open cell drift chamber (COT) will operate at a beam crossing time
of 132~ns without having overlapping events in a single cell. This is
accomplished with a fast gas and shorter drift cells of 0.9~cm. The COT
consists of 96 layers arranged in four axial and four stereo
superlayers. It increases the number of stereo layers from 24 in the
CTC to 48 and
provides d$E$/d$x$ information for particle identification.

The upgrades to the muon system almost double the central
muon coverage. The CMP coverage increases by $\sim\!17\%$ while the CMX
system is completed, increasing its coverage by $\sim\!45\%$.
A new intermediate muon system (IMU) extends the muon coverage up to 
$|\eta|<1.5$ with fine granularity and provides coverage
sufficient to identify isolated high \Pt\ tracks as muons or hadrons
between $|\eta|$ of 1.5 and 2.0. In addition, a new scintillating tile plug
calorimeter will allow good electron identification up to $|\eta| < 2$.

All front-end electronics is designed to handle the 132~ns beam
crossing period. The trigger and data acquisition upgrade allows for
higher data rates such as a 50~kHz Level~1 accept rate and increases the
sophistication of the trigger decision. Data are stored in a 42 cell
pipeline while awaiting the Level~1 trigger decision and can be
transfered to Level~2 with no deadtime to the Level~1 trigger. Finally,  
the CDF collaboration has recently proposed two ``beyond the
baseline'' projects 
which will significantly enhance the $B$~physics capabilities of the
CDF\,II detector. These include the installation of a low-mass radiation
hard single sided silicon detector, with axial strips at a very small radius of
$\sim\!1.6$~cm, just outside the beam pipe as well as the installation of a
time-of-flight (TOF) system, employing 216 three meter long
scintillator bars with fine mesh photomultiplier tubes on each end,
to be located between the outer radius of the COT and the
superconducting solenoid magnet.
 
\subsection{Measurement of \boldmath{$\stb$} in Run\,II}
\noindent
CDF has the advantage of being an existing experiment with plenty of
data and experience from Run\,I. We will make use of the Run\,I knowledge to
estimate the Run\,II expectations for measuring $CP$~violation. For
the measurement of \stb\ in the $B^0 \ra \jpks$ channel, CDF expects
10,000 \jpks\ events with $J/\psi \ra \mu^+\mu^-$ and $K^0_S \ra
\pi^+\pi^-$. This number is estimated in the following way: Starting
with $\sim\!200$ \jpks\ events reconstructed in the SVX in Run\,I and
multiplying it by 2~fb$^{-1}$/110~pb$^{-1}$ for the total Run\,II integrated
luminosity, by a factor of 1.5 for the extended coverage of SVX\,II,
and by two for  
a lower \Pt\ threshold of 1.5~\gevc\ in the muon trigger, as well as an
improved muon coverage with the completed CMX, one obtains
10,800 reconstructed \jpks\ events. CDF also plans to trigger on
$J/\psi \ra e^+e^-$ which would increase the number of \jpks\ events
by $\sim\!50$\%. However, these data are not included in this estimate.

In Run\,II, CDF expects to improve the effective tagging efficiencies \eD\ of
the $B$~flavour tagging methods, as summarized in Table~\ref{ed_run2}. 
The extended lepton coverage with the completed CMX, IMU, and the plug
calorimeter results in a total \eD\ of 1.7\% for lepton tagging.   
A significant improvement in $\eD \sim 3\%$ is possible for jet charge
tagging. The extended coverage of SVX\,II and ISL as well as their
added pattern recognition capabilities will substantially enhance the
purity of the algorithm. Finally, for our Run\,II extrapolation a
value of $\eD\sim2\%$ is 
assumed for same side tagging.

\begin{table}[btp]
\tcaption{Summary of flavour tagging methods used in the measurement
of \stb\ in Run\,II and the data samples used to calibrate the tagging
algorithms.} 
\centerline{\footnotesize\smalllineskip
\begin{tabular}{lccc}
\hline
 & & &  \\
 \vspace*{-0.6cm} \\
 Flavour tag & \eD\ & Calibration sample & Sample size \\
\hline
 & & &  \\
 \vspace*{-0.6cm} \\
 Same side tag & 2.0\% & $J/\psi K^{*0}$ & $\sim\!20,000$ \\
 Jet charge tag & 3.0\% & $J/\psi K^+$ & $\sim\!40,000$ \\
 Lepton tag & 1.7\% & $J/\psi K^+$ & $\sim\!40,000$ \\
\hline
\end{tabular}}
\label{ed_run2}
\end{table}

To estimate the uncertainty on a \stb\ measurement in Run\,II, we 
extrapolate from the measured error of the Run\,I result given in
Eq.~(\ref{eq:s2b_sst}) which uses SVX events only. Defining the
$CP$~asymmetry as ${\cal A} = {\cal D}\,\stb$, we obtain 
$\stb = {\cal A} / {\cal D}$ and write the error on \stb\ as
\begin{equation}
\sigma^2(\stb) = \left(\frac{\sigma({\cal A})}{\cal D}\right)^2 +
\left(\stb\,\frac{\sigma({\cal D})}{\cal D}\right)^2,
\label{eq:errstb}
\end{equation}
where we substitute ${\cal A}/{\cal D}$ with \stb\ in the second
term. This way, the error on \stb\ can be broken up into a statistical part
from the measurement of the $CP$~asymmetry $\cal A$ and a systematic part due
to the dilution uncertainty. In Run\,II, the flavour tag dilutions will
be calibrated with large samples of about 40,000 $J/\psi K^+$ and
$\sim\!20,000\ J/\psi K^{*0}$ events. Assuming $\stb=1$ in the second
term of Eq.~(\ref{eq:errstb}) as worst case and the same signal to
noise ratios for the $J/\psi K$ samples as in Run\,I, the error on a
measurement of \stb\ can be estimated to
$\sigma^2(\stb)=0.078^2 + 0.031^2$. This will allow the observation of
$CP$ violation in Run\,II and a measurement of \stb\ from $B^0\ra\jpks$ 
with a precision of $\pm0.08$ comparable to $e^+e^-$ machines.

\subsection{Measurement of \boldmath{$\sta$} in Run\,II}
\noindent
Another goal of $B$~physics in Run\,II is the observation of
$CP$~violation in $B^0 \ra \pi^+\pi^-$  measuring \sta. 
The key to measuring the $CP$ asymmetry in $B^0 \ra \pi^+\pi^-$ is to
trigger on this hadronic decay 
mode in $p\bar p$ collisions. CDF plans to do this with its three level
trigger system. 
On Level~1, two oppositely charged tracks with $\Pt > 2\ \gevc$ found with a
fast track processor yield an accept rate of
about 22~kHz at luminosities of $\sim\!1.0\cdot 10^{32}$ cm$^{-2}$s$^{-1}$. This rate will be reduced to less than $\sim\!25$~Hz on
Level~2 using track impact 
parameter information ($d_0 > 100\ \mu$m). On Level~3, the full event
information 
is available further reducing the trigger rate to about 1~Hz. With this
trigger, CDF expects about 10,000 $B^0 \ra \pi^+\pi^-$ events in 
2~fb$^{-1}$, assuming $BR(B^0 \ra \pi^+\pi^-) = 1.0\cdot10^{-5}$. With the 
same effective tagging efficiency as in the \stb\ measurement above (see
Tab.~\ref{ed_run2}), CDF estimates an uncertainty of 0.10
on \sta.
Backgrounds from $B \ra K \pi$ and $B \ra K K$ decays can be extracted from
the untagged signal by making use of the invariant mass distribution,
as well as CDF's d$E$/d$x$ capability with the COT.

\section{Conclusions}
\runninghead{$B$ Lifetimes, Mixing and $CP$ Violation at CDF}
{Conclusions}
\noindent
The CDF collaboration has shown that it is possible to take advantage
of the high production rate for $B$~hadrons at the Tevatron
Collider. The key elements for a broad $B$~physics program at a hadron
collider are the successful operation of a silicon vertex detector to
identify $B$~decay vertices displaced from the primary $p\bar p$
interaction vertex, the 
excellent tracking capabilities of CDF's central tracking chamber,
together with the silicon vertex detector and the use
of specialized triggers. Due to these features of the CDF detector
together with 
the high yield of $B$~hadrons, the hadron collider $B$~physics
program complements that at $e^+e^-$ machines. 

In this article, we gave a brief overview of
heavy quark production in $p\bar p$ collisions and 
described some of the features of $B$~physics at a hadron collider.
We reviewed the $B$~hadron lifetime measurements at
CDF, all very competitive with results from LEP and SLC.   
We discussed several   
proper time dependent measurements of $B^0\bar B^0$ oscillations 
as well as 
$B$~tagging methods to identify the $B$~flavour in hadronic
collisions. 
The application of all $B$~flavour tags to the current data set of
$B^0 \ra \jpks$ decays at CDF established the feasibility of measuring
$CP$ asymmetries in $B$ meson decays at a hadron collider. 
With many years of experience in $B$~physics at the Tevatron, the CDF
collaboration plans to observe $CP$ violation in $B^0 \ra \jpks$ in
the next run of the Tevatron collider scheduled to start in the year
2000 and will
measure \stb\ with a precision comparable to the dedicated $e^+e^-$
$B$ factories. 

\nonumsection{Acknowledgments}
\noindent
The results assembled in this article are the work of the
CDF~collaboration. 
It is a pleasure to thank all friends and colleagues at CDF and
especially the CDF $B$~physics group 
for their excellent work.
I wish to thank especially 
F.~Bedeschi, J.~Boudreau, K.~Burkett, B.~Carithers,
F.~DeJongh, L.~Galtieri, 
M.~Garcia-Sciveres, C.~Gay, A.~Goshaw, C.~Haber, J.~Kroll, 
C.~Lacunza, N.~Lockyer,
O.~Long, J.~Lys, P.~Maksimovi\'c, M.~Mangano, V.~Papadimitriou, M.~Peters, 
K.~Pitts, M.~Schmidt, M.~Shapiro, P.~Sinervo, P.~Sphicas, 
D.~Stuart, S.~Tkaczyk, F.~Ukegawa, as well as B.~Wicklund. 
This article is dedicated to my wife Ann, a constant source of
inspiration and support, and to our daughter Emma, born during the 
preparation of this manuscript. I would like to thank them both for
their continuous understanding about the life of a physicist. 
The standard acknowledgments of the CDF collaboration follow:
We thank the Fermilab staff and the technical staffs of the
participating institutions for their vital contributions. This work was
supported by the U.S. Department of Energy and National Science Foundation;
the Italian Istituto Nazionale di Fisica Nucleare; the Ministry of Education,
Science and Culture of Japan; the Natural Sciences and Engineering Research
Council of Canada; the National Science Council of the Republic of China;
the Swiss National Science Foundation; and the A.P.~Sloan Foundation.

\nonumsection{References}
\runninghead{$B$ Lifetimes, Mixing and $CP$ Violation at CDF}
{References}

\end{document}